\documentclass[12pt]{report}
\usepackage[online]{suthesis-2e}

\newcommand{\D}{\mathcal{D}}
\newcommand{\be}{\begin{equation}}
\newcommand{\ee}{\end{equation}}
\usepackage{graphicx,epsf,amssymb,amsbsy,amsfonts,amsmath,appendix}
\usepackage{xcolor}
\RequirePackage[numbers,sort&compress]{natbib}

\pdfoutput=1
\usepackage[colorlinks=true,urlcolor=blue,anchorcolor=blue,citecolor=blue,filecolor=blue,linkcolor=blue,menucolor=blue,linktocpage=true,pdfproducer=medialab]{hyperref}
\usepackage{feynmp}
\newcommand{\bea}{\begin{align}}
\newcommand{\eea}{\end{align}}

\newcommand{\IR}{\relax{\rm I\kern-.18em R}}
\def\blfootnote{\xdef\@thefnmark{}\@footnotetext}

\newcommand{\IC}{{\relax\hbox{$\inbar\kern-.3em{\rm C}$}}}
\newcommand{\IP}{\relax{\rm I\kern-.18em P}}
\newcommand{\IZ}{\relax\ifmmode\mathchoice
{\hbox{\cmss Z\kern-.4em Z}}{\hbox{\cmss Z\kern-.4em Z}}
{\lower.9pt\hbox{\cmsss Z\kern-.4em Z}} {\lower1.2pt\hbox{\cmsss
Z\kern-.4em Z}}\else{\cmss Z\kern-.4em Z}\fi} \font\cmss=cmss10
\font\cmsss=cmss10 at 7pt
\newcommand{\inbar}{\,\vrule height1.5ex width.4pt depth0pt}

\begin{document}
 \setlength{\unitlength}{1mm}
\def\Re{\mathrm{Re}}
\def\Im{\mathrm{Im}} 
 \def\M{{\mathcal M}}
\def\hat{\widehat}
 \def\tilde{\widetilde}
\def\V{{\mathcal V}}
\def\C{{\mathbb C}}
\def\F{{\mathcal F}}
\def\O{{\mathcal O}} 
\def\Z{{\mathbb Z}}
\def\Tr{{\mathrm{Tr}}}
\def\R{{\mathbb R}}
\def\RR{{\mathcal R}}
\def\S{{\mathbb S}}
\def\I{{\mathcal I}}
\def\D{{\mathcal D}}
\def\A{{\mathcal A}}
\def\bar{\overline}
\def\T{{\mathcal T}}
\def\J{{\mathcal J}}
\def\b{{\hat{b}}}
\def\tilde{\widetilde}
\def\P{{\text{LegendreP}}}
\def\Q{{\text{LegendreQ}}}
\def\p{{\delta\varphi}}
\def\t{{\delta\theta}}

\def\Re{\mathrm{Re}}
\def\Im{\mathrm{Im}} 
 \def\M{{\mathcal M}}
\def\hat{\widehat}
 \def\tilde{\widetilde}
\def\V{{\mathcal V}}
\def\C{{\mathbb C}}
\def\CC{{\mathcal C}}
\def\F{{\mathcal F}}
\def\O{{\mathcal O}} 
\def\Z{{\mathbb Z}}
\def\Tr{{\mathrm{Tr}}}
\def\R{{\mathbb R}}
\def\RR{{\mathcal R}}
\def\S{{\mathcal S}}
\def\I{{\mathcal I}}
\def\D{{\mathcal D}}
\def\A{{\mathcal A}}
\def\bar{\overline}
\def\T{{\mathcal T}}
\def\J{{\mathcal J}}
\def\tilde{\widetilde}


\def\th{\theta}
\def\expp#1{{\rm exp}\big ( {#1} \big )}
\def\uu{^}
\def\llo{_}
\def\sqd{^2}
\def\hh{{\frac{1}{2}}}
\def\fakeindent{\hskip .2in}
\def\llsk{\fakeindent}


\definecolor{Sorbus}{rgb}{0.996094, 0.429688, 0.0273438}
\def\SHCommentColor{Sorbus}
\definecolor{FrenchRose}{rgb}{0.96875, 0.292969, 0.5625}
\def\SBCommentColor{FrenchRose}
\definecolor{Violet}{rgb}{0.5,0,1}
\def\JMCommentColor{Violet}
\definecolor{red}{rgb}{1,0,0}

\def\redlowdash{{\color{red}{\rule[-0.5ex]{2pt}{0.4pt}}}}
\def\reduline{\bgroup \markoverwith \redlowdash \ULon}
\def\shc{\bgroup \markoverwith \redlowdash \ULon}

\definecolor{Blue}{rgb}{0, 0.429688, 0.8}

\newcommand{\shg}[1]{{\color{\SHCommentColor} \it NOTE: {#1} -- SH ~~}}

\newcommand{\jmc}[1]{{\color{\JMCommentColor} \it JONATHAN SUGGESTION: {#1}  ~~}}

\title{Towards a String Theory Model of de Sitter Space and Early Universe Cosmology}
\author{Jonathan David Maltz}
\principaladvisor{Leonard Susskind}
\secondreader{Stephen Shenker}
\thirdreader{Shamit Kachru}
\dept{Physics}
\onlinefalse
\beforepreface

\prefacesection{Preface}
\hspace{0.25in}This thesis and the work contained within is primarily based on the following articles \cite{Harlow:2011ny,Maltz:2012zs,Banerjee:2012gh}. Many of the ideas and insights are due to my collaborators and coauthors in these papers and this work should not be cited without reference to these papers.
\newpage
\prefacesection{Acknowledgments}
\hspace{0.25in}There are to many people to thank.
First off I would like to thank my family, my father Steven Ira Maltz, my mother Guylaine Dorvil Maltz, and my brother Jason Elliot Maltz, who without their love and support I would never have gotten here. I want to thank my research advisor Leonard Susskind who has been a second father to me during my six years at Stanford. I don't know any other way to say it. I also would like to thank the rest of my reading committee Stephen Hart Shenker and Shamit Kachru who might actually read this thing, and who's knowledge and guidance have helped me through my time at the SITP I would also like to thank the rest of my defense committee Sean Hartnoll and Perci Diaconis who listened to me defend my this thesis. I want to thank my co-authors Daniel Lord Harlow, Edward Witten, Stephen Hart Shenker, Shamik Banerjee, and Simeon Hellermann who I will be joining at the IPMU in Japan this fall, from whom I have learned so much and without which, this work would not be here. I would like to thank my colleagues, collaborators, and friends at the SITP and SLAC theory. You have all influenced me and provided richness to my life. Some people of note are  Daniele Spier Moreira Alves, Dionysis Anninos, Richard Anantua, Ning Bao, Masha Baryakhtar, Siavosh Rezvan Behbahani, Kassahun Betre, Adam Brown, John Joseph Carrasco, George Coss, Savas Dimopolous,  Xi Dong, Daniel Freedman, Sarah Harrison, Anson Hook, Bartholomew Horn, Kiel Howe, Sonia El Hedri, Ahmed Ismail, Eder Izaguirre, Lampros Lamprou, Andrew Larkoski, Chao-Ming Jian, Renata Kallosh, Andrei Linde, Matt Lewandowski, Victoria Lynn Martin, Natalie Paquette, Michael Peskin, David Ramirez, Tomas Rube, Micheal Salem, Prashant Saraswat, Dusan Simic, Edgar Shaghoulian, Eva Silverstein, Douglas Stanford, Vitaly Vanchurin, Jay Wacker, Alexander Westphall, Daniel Whalen, Sho Yaida, Ying Zhao, Jiecheng Zhang
and the rest of the SITP. 

I also want to thank some friends in the Physics Department and at Stanford at large who influenced my time here and were not mentioned above like Marcos Belfiore, Katie McGrath Bluett, Maria Elizabeth Frank, Conrad Hutchenson, George Karakonstantakis, Yvonne Kung, Joshua Lande, George Ndabashimiye, Andrew Rodrigo Nigrinis, Benjamin Shank, Micheal Shaw, James from the 750 pub, and Karen Slinger. 

I must thank all my friends at the Stanford Kenpo Karate Club especially
Brett Allard, Jeff Arnold, Vanessa Baker, Richard Ehrman Wade Gupta, Laurie Hubert, Sarah Kongpachith, Erin McNear, Barbara Minnetti, Robert Neivert, David Trowbridge, and Valicia Trowbridge. I want to thank you all for being the great people who were the cast for my Twenties.
Lastly I want to thank PBS, Star Trek, and Alice in Wonderland for getting me interested in Physics and Mathematics in the first place.
If I didn't add you it wasn't intentional and I do thank you but this thesis is getting around 300 pages long.

\afterpreface

\part{Motivations}
\chapter{Introduction}

\hspace{0.25in}This thesis describes various technical aspects of constructing a mathematical description of de Sitter space within the String Theory framework. Specifically focusing on components of the proposed FRW - CFT correspondence, with some applications that affect dS - CFT.
\subsection{String Theory/M-Theory And General Relativity's Quantum Banes}

\hspace{0.25in}String theory, is a framework in theoretical physics which currently seeks to provide a quantum description of gravity \cite{Polyakov:1981rd,Green:1987sp,Green:1987mn,Polchinski:1998rq,Polchinski:1998rr,Schwarz1999107,klebanov:28}. It is an extension of the formalism of Quantum Field Theory in which basic objects are one-dimensional extended objects; the  ``\emph{Strings}"\footnote{In the more modern viewpoint, higher dimensional non-perturbative extended objects, \emph{D-branes} share this role with the strings and there is no current reason to believe that the strings are any more fundamental then the D - branes. Both can be used to describe the theory's dynamics. This is termed \emph{Brane Democracy}\cite{Townsend:1995gp}.}. The theory contains Quantum Field Theory (QFT) and General Relativity (GR) in appropriate low energy limits. It can therefore potentially provide a unified quantum framework of gravity, (which is currently described classically and with high experimentally accuracy by General Relativity), with a description of all known non-gravitational interactions; which are currently described, with exceptional experimental and theoretical accuracy in the Standard Model of Particle Physics\footnote{The literature on these two spheres of knowledge is far too vast to cite here. For classic texts of QFT and the Standard Model, the reader is advised to refer to \cite{0521670535,Weinberg:1996kr,Weinberg:2000cr,0201503972,0521318270,citeulike:4667395}, and for GR \cite{weinberg:1972,Misner1973,0226870332,carroll2003spacetime}.}.

Gravitational interactions, which are classically described by Einstein's Theory of General Relativity\cite{Einstein:1915by,Einstein:1915ca}, have been known since the 1960's to resist quantization through the standard methods of Quantum Field Theory \cite{Feynman:1963ax,1967PhRv..160.1113D,PhysRev.171.1834.3,PhysRev.162.1239}. The first issue is, General Relativity when treated perturbatively as a spin-2 gauge field quantized on a curved background, develops large numbers of Ultra-Violet (UV) divergences at higher loop order which prevent the theory from being renormalizable in the standard sense \cite{'tHooft:1974bx}. In other words the gravitational interaction is \emph{irrelevant} in the Wilsonian sense \cite{Wilson:1973jj} implying that at very high energies (or short distances), near the Planck scale, some new ``UV" physics becomes important and must exist to generate the interaction.

Secondly, the gauge symmetry of General Relativity is diffeomorphism invariance (coordinate invariance). Since the field to be quantized in gravity, the metric, determines whether space-time points are spacelike, timelike, or null separated. The standard method of quantizing local fields in a QFT, which is to have the fields commute at space-like separation, is not a gauge invariant statement in a quantum GR.  The separation between points can change depending on the gauge choice used \cite{PhysRev.171.1834.3}. This ambiguity shows up in different guises in preventing quantization of GR, i.e. the quantum time problem when forming the Wheeler-DeWitt equation in temporal gauge in GR\cite{1967PhRv..160.1113D,Wheeler:1988zr,Banks:1984np}.

Lastly, it seems QFT as a formalism is incompatible with the phenomenon that occur in GR, primarily black holes. In a theory of gravity, there is a limit to how much matter and energy can be put into a region of space before it collapses into a black hole. Adding more energy makes the black hole bigger, since the Schwarzschild radius is proportional to the black hole's mass. Hence the Schwarzschild radius is the minimum size that can contain a mass-energy $m$. This puts a restrictions on what types configurations the degrees of freedom describing gravity can have. QFT, because of its manifest locality (space-like separated operators commute) has no such restriction; as one can have modes of arbitrarily high energy separated at arbitrarily small spacelike separations\footnote{A Planck mass black hole would have a  Schwarzschild radius of twice the Planck length. There is nothing to stop a well defined renormalizable QFT from having modes separated in energy by more than a Planck mass within a Planck length. }. In a sense gravitational theories must be holographic \cite{'tHooft:1993gx,Susskind:1994vu} and standard QFT or QFT on Curved Space-times are not.
These issues are not necessarily independent.

 The contradictions in quantizing General Relativity came to a head in the 1970's when the semi-classical
behavior of black holes was meshed with the ideas of Statistical Mechanics and it was shown that the black holes could be treated as thermal systems obeying Thermodynamics \cite{MR0334798}. It was shown by Bekenstein \cite{PhysRevD.7.2333} that the area of the black hole horizon in Planck Units\footnote{Meaning $c$, $\hbar$, $G$,$\frac{1}{4\pi\epsilon_{0}}$,and $k_b$ are all set to 1. These units will be employed throughout this thesis unless otherwise mentioned.} could be interpreted as an entropy,
\be\label{bharea}
S_{BH} = \frac{\Omega_{D-2}r^{D-2}_{s}}{4}  .
\ee

Here $r_s$ is the the black hole's Schwarzschild radius, $D$ is the space-time dimension, and $\Omega_{D-2}$ is the area of the unit $D-2$ sphere.
When Bekenstein came up with this formula for a black hole's entropy, he conjectured that entropy was counting the micro-states of the black hole. This lead to a small and very large paradox. The small paradox was, when computing the entropy through direct field theory methods in order to describe the entropy in terms of micro-states, that near black hole horizon the field theory calculation of the entropy was found to be UV divergent \cite{1985NuPhB.256..727T}\footnote{The first controlled finite calculation of black hole entropy was carried out by Strominger and Vafa in 1995 \cite{Strominger199699}. Here they computed a controlled counting of micro-states of extremal black holes using their BPS degeneracies. This was followed my many other calculations of near and non-extremal black holes agreeing with \ref{bharea}}.

The main paradox occurred when Hawking showed, using methods of QFT on Curved Space-time, that the black hole would radiate as a Black Body at a temperature in Planck units \cite{MR0381625}
\be\label{hawkingtemp}
T_{H} = \frac{D - 3}{4\pi r_s}.
\ee
It was further conjectured that since the black hole was obeying the laws of thermodynamics and radiating energy in the form of radiation, that the black hole would evaporate \cite{hawking1975}. 
This leads to a serious paradox, if the Hawking Radiation is thermal or close to thermal as Hawking's reasoning would suggest, then it does not have enough information (non-trivial correlations) to describe the formation of the black hole in the first place. Due to the unitarity of Quantum Mechanics, different initial states must evolve into different final states, so this lack of information as to which initial state lead to the final state radiation is inconsistent in a quantum formalism \cite{1984NuPhB.244..125B}. 

The question of whether information is lost when a black hole evaporates lead Hawking at the time to argue that General Relativity and Quantum Mechanics were inconsistent with each other, leading to a huge debate throughout the theoretical physics community during a good portion of the history of String Theory \cite{1984NuPhB.244..125B,Preskill:1992tc,Giddings:1993vj,Giddings:1995gd,susskind09}.

Much of the study of Quantum Gravity has been centered about this Information Paradox of Hawking's.

Beginning its life in 1960's as an effort to describe the Strong Interaction in nuclear and hadronic matter \cite{Veneziano:390478,Susskind:1970xm,Frye:1970bu,1974NuPhB..81..118S}, String Theory has since found new life as a quantum theory of gravity after it was supplanted from its original purpose by the far more successful QCD \cite{1974NuPhB..81..118S}.
One of the features that made it fail as a theory of the Strong Interaction made it perfectly suited as a theory of Quantum Gravity, this is that the formalism necessarily contained a massless spin-two mode implying it must contain GR in a low energy limit \cite{Feynman:1963ax,Misner1973,Weinberg:1996kr,FeynmanGravity}. 

In the 1980's it was found that  in 10 flat space-time dimensions, the interacting theory of super-symmetric strings is perturbatively finite to all orders and at low energies reproduces 10 dimensional Supergravity (The super-symmetric version of General Relativity)\footnote{The String Theory S- Matrix does this without having to rely on other fields, as the extended nature of the string acts as internal ``UV" cutoff, removing the non-renormalizabity issues of other attempted quantizations of gravity \cite{Polchinski:1998rq}.}. This can then  be compactified to lower dimensions, i.e. our 4 dimensional world. The perturbative formulation of String Theory is almost unique in 10 dimensions up to a few discrete choices\footnote{Hence the specific types, IIA, IIB, Heterotic E8$\times$ E8 and so on.}. Compactification to 4 dimensions leads to a massive number of meta-stable vacua, the low energy dynamics, fields, interactions, couplings, and sign of the cosmological constant depending on how the space-time was compactified \cite{Bousso:2000xa,Kachru:2003aw,Susskind:2003kw,Douglas:2003um}. This space of meta-stable vacua is commonly referred to as the Landscape and its extent is still not well understood, except that it is large as some estimates place the number of vacua at $\sim10^{500}$ \cite{Douglas:2003um}. Although there is no direct proof or line of argument for this, String Theory being our only theory of quantum gravity, might suggest that any theory of Quantum Gravity would posses a landscape of some extent and hence this notion of many vacua should be taken seriously. 

The 1990's brought other developments, including Polchinski's discovery of  D - Branes \cite{Polchinski:1995mt} (extended non-perturbative objects on which open strings can end). The first controlled finite calculation of black hole entropy was carried out by Strominger and Vafa in 1995 \cite{Strominger199699}, when they computed a controlled counting of micro-states of extremal black holes using their BPS degeneracies. This provided an explicit realization of the Bekenstein-Hawking entropy formula, \ref{bharea}.

 Beginning in the mid-nineties a web of dualities between the various perturbative string theories was discovered by Witten and others. One example being S - Duality, where a perturtabive string theory in the strong coupling regime could be described by another in a weak coupling regime. This suggested that all the string theories were all different aspects of one \cite{Witten:1995zh,Witten199585}, an 11 dimensional ``UV" completion of Maximal Supergravity termed M - theory. Different non-perturbative formulations of M - theory have been proposed, the most establish being the BFSS proposal of Matrix Theory \cite{Banks:1996vh}.  

Championed by Susskind and 't Hooft, the 
Holographic Principal and Black Hole Complementarity \cite{'tHooft:1993gx,Susskind:1993if,Susskind:1994vu} seemed to answer the question as to whether information was lost when a black hole evaporated. Here the information of what fell into the black hole is encoded in the Hawking Radiation of the evaporating hole\footnote{In the standard description of Black Hole Complementarity which prescribes a dual description of in falling observers' physics within the correlations of the out going Hawking Radiation, information is not cloned as no observer can see both frames, due to the horizon.}\cite{1742-6596-171-1-012009}. The radiation must be in a pure state and not thermal as Hawking suggested, so information is not lost \footnote{There have recently been issues regarding whether the standard assumptions of Black Hole Complementarity are consistent \cite{Almheiri:2012rt,Braunstein:2009my}. A lot has been written in the literature about the Firewall issue \cite{Susskind:2012rm,Susskind:2012uw,Larjo:2012jt,Harlow:2013tf,VanRaamsdonk:2013sza,Almheiri:2013hfa,Maldacena:2013xja,Marolf:2013dba}, and the matter is not yet settled, some possible resolutions include \cite{Harlow:2013tf,Maldacena:2013xja}. While the question of what the in falling observer experiences while crossing the horizon might still be up for debate, the many examples of the AdS - CFT correspondence, implies that Complimentary and Holography as a concept is still essential, even if our notions of it particular formulations must be modified.  The question is largely irrelevant as far as this thesis is concerned as the horizons discussed here are cosmological much like Rindler Space, which do not possess firewalls \cite{Gary:2013oja}. This can be intuitively shown due to the fact that we are passing through these types of horizons every moment and we are not burning up.}. 

Complimentary seems to be essential ingredient in the Holographic Principal, a nice review of which can be found in \cite{Bousso:2002ju}. In late 90's, the first explicit example of the Holographic Principle was shown in the AdS -CFT correspondence.

\subsection{AdS - CFT}
\hspace{0.25in}  In the late 90's Maldcena \cite{Maldacena:1997re} provided the first mathematically consistent realization of the Holographic Principle when he discovered the AdS - CFT correspondence. This correspondence shows that low energy Type IIB String Theory (really type IIB Supergravity, which is the low energy effective theory of type IIB String Theory) propagating in the bulk of an $AdS_{5}\times\mathbb{S}^{5}$ space-time is dual to a non-gravitational Conformal Field Theory (CFT) on the space-time's boundary. This theory being $\mathcal{N} = 4$ $SU(N)$ Super Yang-Mills theory.

Besides being an explicit realization of the Holographic Principal in String Theory, this correspondence provided a non-perturbative definition of String Theory on the AdS background
\cite{Gubser:1998bc,Witten:1998qj}. 
There have been many realizations of AdS - CFT beyond Maldecena's original work on multiple backgrounds including an extension of the correspondence to the 11 dimensional M-theory regime in the form of the ABJM theory  \cite{Aharony:2008ug}. The literature on the subject is too vast to cover here.  Common reviews on the subject include \cite{Aharony:1999ti,Aharony:2008ug}.

The above discussion  was to show how String Theory addresses for the most part all the issues of gravity mentioned in the beginning of this section, which is why it is taken seriously as a quantum theory of gravity. To summarize, in the formalism of String Theory it is possible to directly compute the entropy \cite{Strominger199699} of sufficiently super-symmetric black holes reproducing the Bekenstein- Hawking entropy \ref{bharea}. Secondly and perhaps more importantly in 10 flat space-time dimensions the interacting theory of super-symmetric strings is perturbatively finite to all orders and at low energies reproduces 10 dimensional Supergravity (The super-symmetric version of General Relativity) which can be compactified to lower dimensions, i.e. our 4d world.
Finally Black Hole Complimentary and more concretely the AdS-CFT correspondence provided explicit realizations of the Holographic Principle in String Theory.

There has been a deluge of results and developments since String Theory's inception. The subject  has a vast literature and I will not attempt a complete or even semi-complete reference list,  some common introductions and reviews on the subject are \cite{Green:1987sp,Green:1987mn,Polchinski:1998rq,Polchinski:1998rr,Tong:2009np,RevModPhys.60.917,Aharony:1999ti,citeulike:2222766}.

\subsection{de Sitter Space, Theoretical, And Observational Motivations}

\hspace{0.25in}During the early nineties a revolution occurred in observational cosmology when supernova Ia data revealed that the universe possesses a small positive Cosmological Constant (CC) \cite{Spergel:2003cb}. This contradicted the long standing theoretical belief that the Cosmological Constant was zero \cite{RevModPhys.61.1,Carroll:2000fy,Polchinski:2006gy}\footnote{In fact, the famous statement of ``Einstein's Greatest Blunder" references just this point \cite{1970george}. Specifically, that when he was formulating GR, Einstein found that the theory predicted that the universe was not naturally static, and added a cosmological constant to make it so. by doing this he missed the opportunity to predict the expansion of the universe before it proposed by Hubble \cite{1929PNAS...15..168H} and possibly Lema\^{i}ter \cite{2011JRASC.105..151V}, followed by its discovery by Hubble nearly a decade later \cite{1936ApJ....84..517H,1937MNRAS..97..506H}. Once the universe's expansion was discovered it was the motivated theoretical preference to set the constant to zero. As a CC was no longer needed to keep the universe static. The blunder refers to Einstein adding the constant in the first place, and has nothing to do with the fact that we currently measure a small positive CC.}. Around the same time observational experiments and surveys revealed that the Cosmic Microwave Background (CMB) possessed a scale invariant power spectrum \cite{Smoot:1992td,Perlmutter:1998np,Fixsen:1996nj} which matched up nicely with the theoretical idea that the universe experienced an era of exponential expansion referred to as the Inflationary Epoch. Here the effective CC was large compared to the scales of the Standard Model Physics. The work in the previous paragraph suggests, that on the largest scales the universe is asymptotically a de Sitter space.  

de Sitter (dS) space is a maximally symmetric solution of the vacuum Einstein's Equations with positive cosmological constant, which is characterized by exponentially expanding spatial slices. A metric that describes the global geometry of dS is \cite{Anninos:2012qw} 

\begin{equation}
ds^{2} = -dt^{2} + \emph{l}^{2}_{dS}\cosh^{2}\Bigg[\frac{t}{\emph{l}_{dS}}\Bigg]\big(d\psi^{2} + \sin^{2}[\psi]d\Omega^{2}_{2}\big).
\end{equation}

The Penrose Diagram associated with dS is shown in Figure \ref{dS}\footnote{ When the dS is written in terms of flat slicing coordinates, which cover the half of the space,\begin{equation}
ds^{2} = -dt^{2} + e^{2t/l_{dS}}d\vec{x}^{2}
\end{equation} the exponetial expansion is apparent.}
\begin{figure}[ht]
\begin{center}
\hspace{0.5in}\includegraphics[width=8cm]{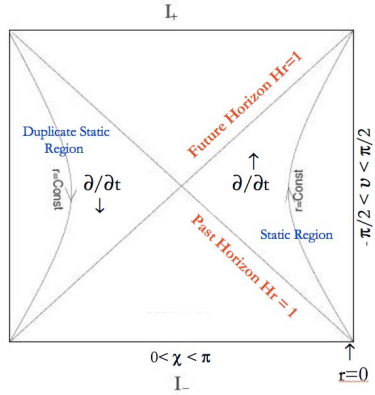}
\caption{ Conformal (Penrose) Diagram for de Sitter Space \cite{Mottola:2010gp}.}
 \label{dS}
\end{center}
\end{figure}

\subsection{de Sitter Space Issues}
\hspace{0.25in}The AdS-CFT correspondence provides a non-perturbative formulation of String Theory on AdS space times, one would expect that the same trick could be pulled on dS.  
String Theory, although a very powerful theory of Quantum Gravity is currently only formulated to work only on very specific backgrounds. Much like QFT, String Theory correlation functions are determined by computing S - Matrix components in flat space or boundary correlators in AdS. This depends on having exact in and out states at asymptotic infinity in flat space or on the boundary in AdS. This property is colloquially referred to as \emph{asymptotic coldness} \cite{Susskind:2007pv}; the property being that the energy density and therefore fluctuations in geometry go to zero at the spatial infinity.

de Sitter Space however is an entirely different beast. 
\begin{itemize}
\item de Sitter space, because it is has accelerating expansion, possesses horizons, unlike AdS space and Minkowski space. An observer can only access a small region of the space-time. Since objects are accelerated away from each other causing them to fall out of causal contact. 
\item Due to the presence of these horizons, dS has a temperature throughout its space-time, which makes S-matrix or boundary methods ill defined. It also has an entropy and therefore, it is not clear whether it even makes sense to talk about exact in and out states on an asymptotically dS background \cite{Dyson:2002nt}.
\item Even though it has been shown that positive CC vacua exist with in the String Landscape \cite{Kachru:2003aw}, de Sitter space is unstable to vacuum decay \cite{Goheer:2002vf}. Because of the positivity of the vacuum energy (the CC), dS spaces are unstable to vacuum decay to vacua of different CC; unlike zero CC spaces which are stable or negative CC spaces like AdS. Therefore the structure of timelike future infinity is dependent on the evolution of the cosmology, and it is not even clear on what space to even construct out states if they can be defined.
\item The fact the dS is unstable to vacuum decay implies that the evolution of a dS depends on its initial conditions. Since most initial states are singular \cite{Hawking27011970}, String Theory or some form of Quantum Gravity is needed to determine the initial state, this is not the case of AdS or Minkowski space.

\end{itemize}
\hspace{0.25in}As was argued for, in context of field theories in \cite{Coleman:1977py,Callan:1977pt} and in the gravitational context in \cite{Coleman:1980aw}; a theory with meta-stable (false) vacua and a stable (true) vacuum of necessarily lower energy, will have the field locally tunnel from a meta-stable vacuum to a meta-stable vacuum of lower energy and eventually the true vacuum. In the gravitational context with the Landscape of String Theory Vacua \cite{Bousso:2000xa,Kachru:2003aw,Susskind:2003kw,Douglas:2003um}, the inflaton field or fields can locally tunnel from a meta-stable vacuum to a vacuum of lower vacuum energy. This is viewed as a bubble nucleation of a child vacuum within a region of the parent vacuum. The child vacua is characterized by a different vacuum energy than the parent. Since the vacuum energy determines the Cosmological Constant (CC) within the frame work of String Theory and Supergravity, the bubble of child vacuum, which has a different CC then the parent vacuum will expand at a different rate. In a de Sitter space with a positive vacuum energy and CC, the Coleman de-Luccia tunneling process  will cause the vacuum to locally decay to a region of space-time of lower positive, zero, or negative CC \cite{Goheer:2002vf}. This coupled with the notion of the Landscape has been lead to a picture of Eternal Inflation \cite{Linde1986395,doi:10.1142/S0217732386000129,Guth:2007ng}. 

Here with a sufficiently low nucleation rate, a de Sitter vacuum even though it is decaying to regions of lower CC will expand faster than it is being eaten up by the slower inflating or crunching child vacuums. The parent de Sitter inflates \emph{eternally}, and is populated by vacua of all different values and hence explores the String Theory Landscape
\cite{1982Natur.295..304G,1983veu..conf..251S,Linde1986395,doi:10.1142/S0217732386000129,Susskind:2003kw}. The collection of the eternally inflating dS and all its child vacua is usually referred to as the Multiverse, with our own universe being one of the child vacua \footnote{In fact if this line of reasoning is correct, our own de Sitter vacuum will decay become such an eternally inflating vacuum.}. 
This leads to one more difficulty.  The decay of the Eternally Inflating dS vacuum to vacua of lower CC \footnote{Even though the process is highly suppressed, there can be transitions to bubbles of higher CC. These \emph{up transitions} must be addressed as well in a de Sitter description.}, is constantly nucleating bubbles, anything that can happen will happen and infinitely often \cite{Guth:2007ng}. This makes interpreting probabilities on the eternally inflating space-time extremely difficult and subtle, requiring some new formalism to deal with these infinities \cite{Linde:1993xx,Guth:2011ie}.

A description of dS should address all of this phenomenon. 

\subsection{FRW - CFT}

\hspace{0.25in}In the papers \cite{Freivogel:2004rd,Freivogel:2006xu,Susskind:2007pv,Sekino:2009kv} it was argued that in a parent de Sitter which contains a bubble of an open FRW vacuum with zero CC, the dual CFT of the FRW bubble not only contains information from the FRW bulk but also of the parent de Sitter\footnote{One reason for this is that the central charge of the CFT fields is proportional to the parent de Sitter's CC.} and Multiverse at large. The conjecture of FRW - CFT is that the correlation functions seen on the late time sky of such a stable zero CC bubble are computed from a non-unitary euclidean CFT on the $\mathbb{S}^{2}$, with metric $g_{ij}$ at spacelike infinity. See Figure \ref{CDL}.

The FRW-CFT is a holographic Wheeler-deWitt theory \cite{Freivogel:2004rd,Freivogel:2006xu,Susskind:2007pv,Sekino:2009kv,Harlow:2010my}. In \cite{Harlow:2010my} it was argued that the FRW-CFT can be viewed as a dimensionally reduced dS-CFT, which is UV complete in the hat. FRW-CFT, if the conjecture is valid, would then be a microscopic description of the de Sitter space much in the same sense as Matrix theory or AdS-CFT.

An open FRW bubble will have hyperbolic space like slices $\mathcal{H}_{3}$ which is identical to EADS$_{3}$ \cite{Susskind:2007pv}. In ADS - CFT, the boundary field theory  defines the ultraviolet (UV) degrees of the bulk system. The analogous boundary in the FRW - CFT is the spacelike $\mathbb{S}^{2}$, which is labeled $\Sigma$ at spacelike infinity of the FRW bubble, representing the sky of a late time observer in the FRW bubble. The $O(3,1)$ of these EADS$_{3}$ slices acts as 2D conformal transformations on $\Sigma$ \cite{Susskind:2007pv}.

 In ADS - CFT, the boundary is one dimension lower than the bulk space-time and there is a natural time ordering. In FRW  - CFT different time slices end on the same asymptotic $\Sigma$ and the metric on $\Sigma$ is dynamical\footnote{One way to way to see this is that the free energy of the system does not go to zero at space-like infinity in the FRW bubble, gravitational degrees of freedom propagate all the way to the boundary which is the asymptotic 2-sphere of spacial infinity.}.  The fact that the metric on $\Sigma$  is dynamical accounts for this mismatch in dimensions as it implies that dual CFT would be a collection of matter CFT fields with central charge proportional to the parent dS and a  Timelike Liouville field possessing a canceling central charge, the Liouville field accounting for the missing dimension.

That the dual CFT necessarily has Timelike Liouville Theory in it can be seen in the argument given in \cite{Harlow:2010my}. If the fields of the CFT are collectively called $A$  and the metric on the $\mathbb{S}^{2}$ called $g$ the WdW wave function would have the form 
\begin{align}\label{wavefunction}
\Psi(g) &= \int DA\, e^{-\int L(A;g)}\\
\Psi(g)^{*} &= \int DB^{*}\, e^{ -\int L^{*}(B^{*};g)}
\end{align}
 and expectation values of functionals of the metric will have the form
 \begin{align}
 \langle F(g)\rangle &=\int D\,g \Psi^{*}(g)F(g)\Psi(g)\\
 &=\int Dg\, \,DA\,\,DB^{*}\,\, F(g)e^{-\int[L(A;g) + L^{*}(B^{*};g)]}.
 \end{align}.
 
 From this one can see that the expectation values of functions of the metric depend on the integration over the CFT fields. When the ``UV" modes of $A$ and $B^{*}$ are integrated out, a kinetic term for the metric is generated. If $g$ is written in conformal gauge, the effective action for $g$ becomes that of a Liouville theory with a central charge that cancels that of $A$ and $B^{*}$ which have central charge proportional to that of the ancestor de Sitter \cite{Freivogel:2006xu,Susskind:2007pv,Harlow:2010my}. This means the central charge of the Liouville theory is large and negative. This type of theory is call a Timelike Liouville theory\footnote{It should be noted that dS-CFT methods were used in this argument. In this view FRW-CFT is a dimensionally reduced dS-CFT.}.
  The Timelike Liouville Theory is an analytic continuation of the Standard Liouville Theory where the kinetic term possesses the wrong sign. In this context, gravitational degrees of freedom on the boundary are represented by the Timelike Liouville field and it is the presence of this field that takes the place of the missing dimension of the boundary. 

 \begin{figure}[ht]
\begin{center}
\includegraphics{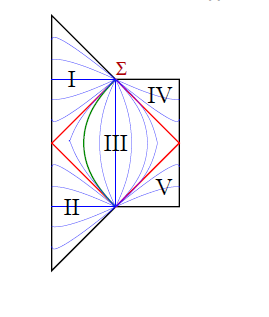}
\caption{The Penrose diagram of of the Lorentzian continuation of CDL instanton solution \cite{PhysRevD.21.3305,Freivogel:2004rd,Freivogel:2006xu}. Region I is an open (k = -1) FRW universe wihich is asymptotically flat. Region IV is asymptotically de Sitter. $\Sigma$ is the conformal 2-sphere defined by the intersection of light-like infinity of region I and the space-like infinity of region IV. The curves indicate orbits of the $SO(3,1)$ symmetry, which as the conformal group on $\Sigma$ \cite{Freivogel:2006xu}.}\label{CDL}
\end{center}
\end{figure}This section was  put here to show how Timelike Liouville theory fit in the FRW - CFT formalism, and to give a justification for studying Liouville Theory which is the focus of most of this thesis. 
Far better and more complete reviews on the FRW -CFT can be found here \cite{Freivogel:2006xu,Susskind:2007pv,Sekino:2009kv,Harlow:2010my,Harlow:2011az} and the reader is encouraged to read them in order to better understand the state of the field.

\subsection{Comments On Other de Sitter Duals And dS - CFT}
\hspace{0.25in}Other than FRW - CFT, there are other proposed dual descriptions of dS, which have met with some level of success \cite{Strominger:2001pn,Witten:2001kn,Maldacena:2002vr,Alishahiha:2004md,Dong:2010pm}. One of the more prominent ones being dS - CFT \cite{Strominger:2001pn,Witten:2001kn,Maldacena:2002vr}, which tries to directly extend and use the AdS - CFT formalism to describe de Sitter space  in the context of String Theory. Here the Wheeler-de Witt wave-function of the cosmology is computed via the partition function of a non-unitary euclidean CFT which is defined by sources simply related to the parameters of the wave function \cite{Harlow:2011az}. Typically via the analytic continuation of various parameters in AdS Correlators \cite{Anninos:2011jp,Anninos:2011ui}\footnote{Heuristically, spacelike infinity of the AdS space becomes the timelike future infinity of dS}. The dual CFT is thought to be non-unitary and describes de Sitter Space  by placing the dual theory on future infinity $\mathcal{I}^{+}$ \cite{Strominger:2001pn}. dS - CFT has the added advantage that there are some nice examples in the literature that can be directly calculated in its formalism, and recently an explicit example of this was shown in the context of Vasilev Gravity \cite{Anninos:2011jp} which is one of the motivations for better understanding possible Vasilev duals in this work.

There have been arguments against the possibility of an direct exact de Sitter dual like pure dS-CFT, based on entropy arguments \cite{Dyson:2002nt}. The statement being that since dS has a temperature and entropy, it cannot posses and exact dual description in the same way as AdS, due to dS possessing a finite entropy. Another issue is that the boundary theory in dS - CFT is on timelike future infinity $\mathcal{I}^{+}$. No one observer in the dS, can actually measure correlators of operators at different points since they are spacelike separated. Only a meta-observer that could view the entire dS would be able to view the correlators of dS - CFT\footnote{In our own bubble's cosmology, this may be a feature and not a bug of a dS - CFT. As was mentioned earlier, the universe underwent a period of inflation very early in its history, where the effective CC was much higher. We are in a sense meta-observers to that de Sitter epoch since the ``late time"  $\mathcal{I}^{+}$ of that period would be when the universe exited inflation. The surface of last scattering in effect represents such a spacelike slice and it is possible that dS-CFT dual of the inflationary portion of our bubble would be described in terms of correlators within the CMB. This is however far beyond the scope of this thesis or the state of knowledge of the field.}.  Also as was mentioned previously, dS at best is meta-stable so $\mathcal{I}^{+}$ is not purely de Sitter and it is unclear how the dS -CFT would deal with this situation \cite{Harlow:2011az}. 
While it is the personal prejudice of the author that the FRW-CFT or some evolution of it will turn out to be the actual microscopic  description of de Sitter space; much like Matrix Theory, it is currently unwieldy and it is very difficult to untangle intuitive notions or relevant observables from its formalism. dS -CFT on the other hand is far more wieldy and is also a serious component in the study of de Sitter space as well as FRW -CFT. This is one the reasons behind suggested studies of vector like holography \cite{Anninos:2011ui,Anninos:2012ft}. 
Due to this, a particular aspect of it is studied in part \ref{chern-simons paper} of this thesis. That being what the effects of non-trivial topology are to proposed duals of these Higher Spin Gravity theories.

This thesis will remain agnostic as to whether dS-CFT or FRW-CFT or both provides a description of dS. Since the two are intimately related, (FRW -CFT is in a sense a dimensionally reduced dS-CFT \cite{Harlow:2011az}), the work in this thesis is relevant to both of them.

\subsection{Summary And Outline}

\hspace{0.25in}In this thesis I will describe some of the work and results that have been taken by my collaborators and myself to better understand some of the technical aspects of proposed String Theory descriptions  of de Sitter Space via the FRW - CFT and other models of de Sitter space at large.

A study of Liouville theory, its analytic properties as well as what gauge invariant quantities can be perturbatively computed in the Timelike regime is necessary for a more complete formulation of the FRW - CFT and other aspects of String Theory. It is therefore the focus of this work to shed some light on these two aspects of of Liouville theory. Also since the FRW-CFT can be described as a dimensionally reduced dS - CFT and is built on the formalism of ADS-CFT correspondence, as is the case with most de Sitter duals; a better understanding of specific limits of ADS-CFT that may be pertinent to cosmological applications is useful.  Pursuing interest in vector like models of holography it has been conjectured that Vasiliev Higher Spin Gravity theories \cite{Vasiliev:1992av,Vasiliev:1995dn,Vasiliev:1999ba,Vasiliev:2003ev,Sezgin:2002ru,Sezgin:2003pt,PROP:PROP2190351103} describe ADS -CFT in various strongly coupled limits \cite{Giombi:2010vg,Chang:2012kt,YinSlides}. It has been shown that dS-CFT correlators may be obtained by analytically continuing ADS-CFT correlators \cite{Anninos:2011ui} and hence that Vasiliev might shed some light on de Sitter dual theories. Hence finding a dual to this higher spin theory might help in the understanding of a de Sitter dual. Chern-Simons Theory coupled to Matter in fundamental representations (CSM) have been proposed to be a dual to Vasiliev Theory on certain backgrounds \cite{Klebanov:2002ja,Giombi:2011kc}. In this work, it is shown that the connection may not be as easy as originally intended. When the background is topologically non-trivial there are light states in CSM that are not present in the Vasiliev dual. These light states must be explained in the Vasiliev context for the duality to be true.
\begin{itemize}
\item In Part \ref{ed liouville paper} it will be discussed how correlation functions in Liouville theory are meromorphic functions of the Liouville momenta, as is shown explicitly by the DOZZ
formula for the three-point function on $\sf S^2$.  In a certain physical region, where a real classical solution exists, the semiclassical limit of the DOZZ formula is known to agree with what one would expect from the action of the classical solution.  Here,
we ask what happens outside of this physical region.  
Perhaps surprisingly we find that, while in some range of the Liouville momenta the semiclassical limit is associated to complex saddle points, in general Liouville's equations do not have enough complex-valued solutions to account for the semiclassical behavior.  For a full picture, we either must
include  ``solutions'' of Liouville's equations in which the Liouville field is multivalued (as well as being complex-valued), or else we can reformulate
Liouville theory as a Chern-Simons theory in three dimensions, in which the requisite solutions exist in a more conventional sense.  
We also study the case of ``timelike'' Liouville theory, where we show that a proposal of Al. B. Zamolodchikov for the exact three-point function on
$\sf S^2$  can be computed by the original Liouville path integral evaluated on a new integration cycle. This will show how Liouville can be analytically continued and how to define the Timelike Liouville field theory for the FRW-CFT.
\item  In Part \ref{Gauge Fixing Paper} it is discussed how,
Timelike Liouville theory admits the sphere $\mathbb{S}^{2}$ as a real saddle point, about which quantum fluctuations can occur. An issue occurs when computing the expectation values of specific types of quantities, like the distance between points. The problem being that the gauge redundancy of the path integral over metrics is not completely fixed even after fixing to conformal gauge by imposing $g_{\mu\nu} = e^{2\b\phi}\tilde{g}_{\mu\nu}$, where $\phi$ is the Liouville field and $\tilde{g}_{\mu\nu}$  is a reference metric. The physical metric $g_{\mu\nu}$, and therefore the path integral over metrics still possesses a gauge redundancy due to invariance under $SL_{2}(\mathbb{C})$ coordinate transformations of the reference coordinates. This zero mode of the action must be dealt with before a perturbative analysis can be made.  

This part shows that after fixing to conformal gauge, the remaining zero mode of the linearized Liouville action due to $SL_{2}(\mathbb{C})$ coordinate transformations can be dealt with by using standard Fadeev-Popov methods. Employing the gauge condition that the ``dipole" of the reference coordinate system is a fixed vector, and then integrating over all values of this dipole vector. The ``dipole" vector referring to how coordinate area is concentrated about the sphere; assuming the sphere is embedded in $\mathbb{R}^{3}$ and centered at the origin, and the coordinate area is thought of as a charge density on the sphere. The vector points along the ray from the origin of $\mathbb{R}^{3}$ to the direction of greatest coordinate area.

 A Green's function is obtained and used to compute the expectation value of the geodesic length between two points on the $\mathbb{S}^{2}$ to second order in the Timelike Liouville coupling $\b$. This quantity doesn't suffer from any power law or logarithmic divergences as a na\"{i}ve power counting argument might suggest.
Which will demonstrate that once the all the gauge redundancies are fixed, a perturbative calculation can be carried out in the Timelike Liouville regime for FRW - CFT.
\item In Part \ref{chern-simons paper} it is discussed how other string theory models of de Sitter space namely dS - CFT and aspects of AdS - CFT in various limits are affected by some results in describing proposed duals of Vasiliev Higher Spin Gravity models.
Motivated by developments in vectorlike holography, we study SU(N) Chern-Simons theory coupled to matter fields in the fundamental representation on various spatial manifolds.  
On the spatial torus $T\uu 2$, we find  light states at small `t Hooft coupling $\lambda=N/k$, where $k$ is the Chern-Simons level, taken to be large.
In the free scalar theory the gaps are of order $\sqrt{\lambda}/N$ and in the critical scalar theory and the free fermion theory they are of order $\lambda/N$.  The entropy of these states grows like $ N \log(k)$.   We briefly consider spatial surfaces of higher genus.  Based on results from pure Chern-Simons theory,
it appears that there are light states with entropy that grows even faster,  like $N^2 \log(k)$.  This is consistent with the log of the partition function on the three sphere $S\uu 3$, which also behaves like $N^2 \log(k)$.   These light states require bulk dynamics beyond standard Vasiliev higher spin gravity to explain them. 
This will section brings to light that duality between Vasiliev Higher Spin Gravity and Chern-Simons Matter theories needs to be augmented or improved if the duality is valid, which has an impact on current understanding of dS - CFT and AdS - CFT.
\end{itemize}
\subsection{An Appologia For The Motivations}

\hspace{0.25in}While FRW - CFT and de Sitter duals where the motivations for this work, it should be noted that the grand majority of what is here is actually applicable to a much broader range of ideas and constructs in String Theory and Theoretical Physics at large. In fact most of what is here is still true independent of a FRW - CFT or other de Sitter dual description. 
\\

As was previously mentioned the work presented in this thesis is primarily presented in
 \cite{Harlow:2011ny,Maltz:2012zs,Banerjee:2012gh}.
\part{}\label{ed liouville paper}
\chapter{Introduction}
\hspace{0.25in}Quantum Liouville theory has been studied extensively since it was first introduced by 
Polyakov several decades ago in the context of non-critical string theory 
\cite{Polyakov:1981rd}.  Since then it has been invoked as a model for 
higher-dimensional Euclidean gravity, as a noncompact conformal field theory, 
and as a dilaton background in string theory.  Among more
recent developments, Liouville theory   has been found \cite{AGT} to have a 
connection to four-dimensional gauge theories with extended supersymmetry and 
has emerged as an important component of speculative holographic duals of de Sitter 
space and the multiverse \cite{Freivogel:2006xu,Sekino:2009kv,Harlow:2010my}.  In many of these applications the Liouville objects of interest are evaluated at complex values of their parameters.  The goal of this part is to understand to what extent these analytically continued objects are computed by appropriately continued versions of the Liouville path integral.
  
Liouville theory has been studied from many points of view, but the essential point for studying the question of analytic continuation is that several nontrivial quantities are, remarkably, exactly computable.  A basic case is the expectation value on a two-sphere $\sf S^2$
of a product of three primary fields of Liouville momentum $\alpha_i,\,i=1,2,3$:
  \begin{equation}\label{toffo}\left\langle\prod_{i=1}^3 
e^{2\alpha_i\phi}(z_i)\right\rangle.\end{equation}
For this correlation function,   there is an
 exact formula -- known as the DOZZ formula  \cite{Dorn:1994xn,Zamolodchikov:1995aa}. 
  The existence of such exact
formulas makes it possible to probe questions that might otherwise be out of reach.  
We will exploit this in studying the semiclassical limit of Liouville theory in the present part.

Liouville theory is conveniently parametrized in terms of a coupling constant $b$, 
with the central charge being $c=1+6Q^2$, where $Q=b+b^{-1}$.  For a semiclassical limit, 
we take $b\to 0$, giving two interesting choices for the Liouville momenta.
We can consider ``heavy'' operators, $\alpha_{i} = \eta_{i}/b$, with $\eta_i$ fixed as $b\to 0$.
The insertion of a heavy operator changes the saddle points which 
dominate the functional integral.  Alternatively, we can consider
``light'' operators,
$\alpha_i=b\sigma_i$, where 
now we keep $\sigma_i$ fixed as $b\to 0$.  Light operators do not affect
a saddle point; they just give us functions 
that have to be evaluated at a particular saddle point.  We will consider both cases
in the present part.  

A real  saddle point in the Liouville path integral is simply a real solution 
of the classical equations of motion
\begin{equation}\label{murko}
-\partial_a\partial^a\phi+Q\tilde\RR+8\pi b\mu e^{2b\phi}=0.\end{equation}
Such a solution is a critical point\footnote{When heavy
operators are present, they add delta function terms in (\ref{murko}) and also
make contributions to the action $S$.  In this introduction, we will omit such details.} 
of the classical action $S$.
Path integrals are most simple if they are dominated by a real saddle point. 
For the Liouville correlation function of three heavy fields on $\sf S^2$, there is a real  
saddle point  that dominates the semiclassical
limit of the path integral if and only if the $\eta_i$ are real, less than $1/2$, and 
obey $\sum_i\eta_i>1$.  These inequalities, which define
what we will call the physical region, 
were described in \cite{Seiberg:1990eb} and the explicit solution was described 
and its action computed in \cite{Zamolodchikov:1995aa}.    
The action evaluated at the classical solution 
 is of the form $S=G(\eta_1,\eta_2,\eta_3)/b^2$
where the function $G$ can be found in 
(\ref{region1act}).    In \cite{Zamolodchikov:1995aa}, it was shown that, in 
the physical region, the weak coupling limit of the three-point function of heavy fields is
\begin{equation}\label{lango}
\left\langle\prod_{i=1}^3
\exp(2\eta_i\phi /b)(z_i)\right\rangle 
\sim \exp(-S)=\exp\left(-G(\eta_1,\eta_2,\eta_3)/b^2+\O(1)\right),\end{equation}
as one would expect.

\subsection{Analytic Continuation}\label{ancon}

\hspace{0.25in}What happens when we leave the physical region?  
There is no problem in 
continuing the left hand side of (\ref{lango})  beyond the physical region.  
Indeed, the DOZZ formula shows
that the left hand side of the three-point function (\ref{toffo}) is, for fixed $b$, a 
meromorphic function of the variables $\alpha_i$, and
in particular one can analytically continue the $\eta_i$ to arbitrary complex values.  
Similarly, the DOZZ formula is analytic in $b^2$ for $b^2>0$.

 What happens on the right hand side of
eqn. (\ref{lango})?  If continued outside the physical region, the 
function $G$ extends to a {\it multivalued} function of complex
variables $\eta_i$.  
This multivaluedness takes a very simple form.  
There are branch points at integer values of the $\eta_i$ or of simple
sums and differences such as $\eta_1+\eta_2-\eta_3$.  The monodromy 
around these branch points takes the form
\begin{equation}\label{dofo}G\to G+2\pi i
\left(n+\sum_{i=1}^3m_i\eta_i\right),\end{equation}
where $n$ and the $m_i$ are integers and the $m_i$ are either all even or all odd.

There actually
is one specific region of complex $\eta_i$ -- the case that $\Re\,\eta_i=1/2$ and $\Im\,\eta_i>0$, so that
the external states are normalizable states in the sense of \cite{Seiberg:1990eb} --
in which a semiclassical interpretation of the DOZZ formula is available
\cite{HJ1,HJ2} in terms of real singular solutions of Liouville's equations that have a natural
interpretation in terms of hyperbolic geometry.  The action of these singular solutions
is given by a particular branch of the multivalued function $G$, for the values of the $\eta_i$ in question.  
(The solutions themselves are given by an analytic continuation of those constructed in \cite{Zamolodchikov:1995aa}
and thus are a special case of the solutions we discuss later.)
In the present part, we aim to interpret the DOZZ formula semiclassically for arbitrary complex $\eta_i$.

Our investigation started with the question of how to interpret the 
multivaluedness (\ref{dofo}).  The most obvious interpretation
is that the branches of $G$ might correspond 
to the actions of {\it complex} solutions of the Liouville equation.  Outside the physical
region, the correlation function of a product 
of heavy fields does not have a real saddle point, but one might hope that it would
have one or more complex saddle points. 

The most obvious notion of a complex saddle point is simply a complex-valued
solution of the classical Liouville equation (\ref{murko}).  Such a solution is a critical
point of the Liouville action $S$, interpreted now as a holomorphic function of a complex 
Liouville field $\phi$.  Optimistically, one would
think that Liouville theory for the case of three heavy operators on $\sf S^2$
has complex-valued solutions parametrized by the integers
$n$ and $m_i$ that appeared in eqn. (\ref{dofo}).  Then one would hope that
for any given values of the $\eta_i$, the path integral could be expressed as a sum of
contributions from the complex saddle points.  Which saddle points must be included
(and with what coefficients) would in general depend on the $\eta_i$, as Stokes phenomena
may intrude.

To appreciate the analytic continuation of path integrals,
one needs to know that to a given saddle point one can
associate, in principle, much more than a perturbative expansion.
The basic machinery of complex saddle points and Stokes phenomena says the 
following.\footnote{See for example section 2 of \cite{Analytic} for an elementary 
explanation, much more detailed and
precise than we can offer here.  Some familiarity with this material is necessary for a full
appreciation of the present part.}
Let $\S$ be the set of complex saddle points; these are also known as critical
points of the complexified action.  To each $\rho\in \S$, one associates an integration
cycle\footnote{An integration  ``cycle'' is simply the 
multi-dimensional analog of an integration ``contour.''  For simplicity, we assume that the critical
points are isolated and nondegenerate.}
 $\CC_\rho$ in the  complexified path integral.  Roughly speaking, $\CC_\rho$
is defined by steepest descent, starting at the critical point $\rho$ and descending by gradient
flow with respect to the ``Morse function'' $h=-\mathrm{Re}\,\,S$.  The $\CC_\rho$
are well-defined for generic values of the parameters; in our case, the parameters
are $b$ and the $\eta_i$.  The definition of $\CC_\rho$ fails if two critical points $\rho$
and ${\tilde\rho}$ have the same value of $\mathrm{Im}\,S$ and unequal values of $\mathrm{Re}\,S$. In this case, the difference 
$S_\rho-S_{\tilde\rho}$ is either positive or negative (we write $S_\rho$ for the value of 
$S$ at $\rho$ and similarly $h_\rho$ for the value of $h$); 
if for example it is positive, then the jumping of integration cycles
takes the form
\begin{equation}\label{jumping}\CC_\rho\to \CC_\rho+ n\CC_{\tilde\rho},\end{equation}
for some integer $n$.  

Any integration cycle $\CC$ on which the path integral converges can always be expressed
in terms of the $\CC_\rho$:
\begin{equation}\label{expres}\CC=\sum_{\rho\in\S}a^\rho \CC_\rho,\end{equation}
with some coefficients $a^\rho$.  In particular -- assuming that the machinery
of critical points and Stokes walls applies to Liouville theory, which is the hypothesis
that we set out to test in the present part -- the integration cycle for the Liouville path
integral must have such an expansion.    The subtlety is that the coefficients in this expansion
are not easy to understand, since one expects them to jump in crossing Stokes walls.
However, there is one place where the expansion (\ref{expres}) takes a simple form.
In the physical region, one expects Liouville theory to be defined by an integral
over the ordinary space of real $\phi$ fields. On the other hand, in the physical
region, there is a unique critical point $\rho_0$ corresponding to a real solution.
Starting at a real value of $\phi$ and performing gradient flow with respect to $h$, $\phi$ remains real.  (When $\phi$ is real, $h$ is just the ordinary real Liouville action.)
So $\CC_{\rho_0}$ is just the space of real $\phi$ fields as long as the $\eta_i $ are
in the physical region.  In the physical region, the expansion (\ref{expres}) collapses therefore to
\begin{equation}\label{nexpres}\CC=\CC_{\rho_0}.\end{equation}
In principle -- if the machinery we are describing does apply to Liouville theory -- the expansion (\ref{expres}) can be understood for any values of the $\eta_i$ and $b$
by starting in the physical region and then varying the parameters at will, taking Stokes
phenomena into account.

If one knows the coefficients in the expansion (\ref{expres}) for some given values of the
parameters,
then to determine the small $b$ asymptotics of the path integral 
\begin{equation}\label{justo}Z=\int_\CC D\phi\,\exp(-S)\end{equation}
is straightforward.  One has
$Z=\sum_\rho a^\rho Z_\rho$,  with $Z_\rho=\int_{\CC_\rho}D\phi \,\exp(-S)$.
On the other hand, the cycle $\CC_\rho$ was defined so that along this cycle, $h=-\mathrm{Re}\,
S $ is maximal at the critical point $\rho$.  So for small $b$,
\begin{equation}\label{nusto}Z_\rho\sim
\exp(-S_\rho).
\end{equation}
The asymptotic behavior  of $Z$ is thus given by the contributions of those critical
points that maximize $h_\rho=\mathrm{Re}(-S_\rho)$, subject to the condition that $a^\rho\not=0$. 

At this point, we can actually understand more explicitly why (\ref{nexpres}) must hold
in the physical region.   A look back to (\ref{dofo}) shows that as long as $b$ and the
$\eta_i$ are real, all critical points have the same value of $\mathrm{Re}\,S$.  
So all critical points with $a^\rho\not=0$ are equally important for small $b$ in the
physical region.  Thus, the computation of \cite{Zamolodchikov:1995aa} showing that
in the physical region the Liouville three-point function is dominated by the contribution
of the real critical point also shows that in this region, all other critical points have $a^\rho=0$.

\subsubsection{The Minisuperspace Approximation}\label{minis}

\hspace{0.25in}The Gamma function gives a practice case for some of these ideas.  (For a previous analysis along
similar lines to what we will explain, see \cite{PS}.  For previous mathematical work, see \cite{Berry,Boyd}.)
The most familiar 
integral representation of the Gamma function is
\begin{equation}\label{ggamma} \Gamma(z)=\int_0^\infty d t \,t^{z-1}\exp(-t),~~
\mathrm{Re}\,z>0.\end{equation}
A change of variables $t=e^\phi$ converts this to
\begin{equation}\label{zamma}\Gamma(z)=\int_{-\infty}^\infty d\phi\,\,\exp(-S),\end{equation}
where the ``action'' is
\begin{equation}\label{amma}S=-z\phi+e^\phi.\end{equation}
This integral is sometimes called the minisuperspace approximation \cite{Cur,Polchinski:1990mh,Seiberg:1990eb} to Liouville theory,
as it is the result of a truncation of the Liouville path integral to the constant mode of $\phi$ (and
a rescaling of $\phi$ to replace $e^{2b\phi}$ by $e^\phi$).

If $z$ is real and positive, the action $S$ has a unique real critical point at $\phi=\log\,z$, and this
is actually the absolute minimum of $S$ (on the real $\phi$ axis).
We call this critical point $\rho_0$.   Gradient flow from  $\rho_0$ keeps $\phi$
real, so the corresponding integration cycle $\CC_{\rho_0}$  is simply the real $\phi$
axis.  If $z$ is not real but $\mathrm{Re}\,z>0$, then $\CC_{\rho_0}$, defined
by gradient flow from $\rho_0$, is
not simply the real $\phi$ axis, but is equivalent to it modulo Cauchy's Residue theorem.
  The original integral (\ref{ggamma}) or (\ref{zamma}) can be approximated for $z\to\infty$
in the half-space $\mathrm{Re}\,z>0$ by an expansion near the critical point $\rho_0$, at 
which the
value of the action  is $S=-z\log z +z$.  The contribution of this critical point leads to
Stirling's formula $\Gamma(z)\sim \exp(z\log z -z+\O(\log z)),~\mathrm{Re}\,z>0$.

The Gamma function can be analytically continued beyond the half-space 
to a meromorphic function of $z$, defined in the whole complex plane with poles at
non-positive integers.  This is analogous to the fact that the exact Liouville three-point
function (\ref{toffo}) is a meromorphic function of the $\alpha_i$, even when we 
vary them to a region in which the path integral over real $\phi$ does not converge.
The analytic continuation of the Gamma function to negative $\mathrm{Re}\,z$ can be exhibited
by deforming the integration contour in (\ref{zamma}) into the complex $\phi$ plane as
$z$ varies.  To understand the behavior of the integral for $\mathrm{Re}\,z\leq 0$,
it helps to express this integral in terms of contributions of critical points.  The complex
critical points of $S$ are easily determined; they are the points $\rho_n$ with \begin{equation}\label{zonk}\phi_n=\log z +2\pi  i n,~~n\in\Z.\end{equation}
For $\mathrm{Re}\,z>0$, the integration contour defining the Gamma function
is simply $\CC_{\rho_0}$,
 but for negative $\mathrm{Re}\,z$, the integration contour has a more elaborate expansion 
 $\sum_{n\in \T}\,\CC_{\rho_n}$, where $\T$ is determined in Appendix \ref{gammastokes}.  
Once one determines $\T$, the analog of Stirling's formula for 
$\mathrm{Re}\,z\leq 0$ is immediate.

The essential point is that the integration contour in the definition of the Gamma function
can be chosen to vary smoothly as $z$ varies, but its expression as a sum of critical
point contours $\CC_{\rho_n}$ jumps in crossing the Stokes walls at $\mathrm{Re}\,z=0$.
  The present part is an attempt to understand
to what extent the machinery just sketched
actually applies to the full Liouville theory, not just the minisuperspace approximation.

\subsection{Complex Solutions Of Liouville Theory}\label{compsol}

\hspace{0.25in}The result of our investigation has not been as simple as we originally hoped.
The classical Liouville equations do not have enough complex critical points to account
for the multivaluedness (\ref{dofo}), at least not in a straightforward sense.

As soon as one allows the Liouville field $\phi$ to be complex, one meets the fact
that the classical Liouville equations are invariant under
\begin{equation}\label{doft}\phi\to \phi+ik\pi/b, ~~k\in\Z.\end{equation}
This  assertion, which extends  what we just described in the minisuperspace
approximation, actually accounts for part of the multivaluedness (\ref{dofo}).  The shift (\ref{doft})
gives $G\to G+2\pi i k(1-\sum_i\eta_i)$.

This is all we get by considering complex solutions of the Liouville equations in a simple way. For example, in the physical region, even if we allow
$\phi$ to be complex, the most general solution of Liouville's equations is the one
described in \cite{Zamolodchikov:1995aa}, modulo a shift (\ref{doft}).  We prove
this in section \ref{liouvillesolutions} by adapting standard arguments about Liouville's equations in a simple way.

Outside of the physical region, the solutions of the complex Liouville equations
are no more numerous. One can try to find some complex solutions by directly generalizing
the formulas of \cite{Zamolodchikov:1995aa} to complex parameters.
For a certain (difficult to characterize) range of the parameters $\eta_i$ and $b$,
this procedure works and gives, again, the unique solution of the complex Liouville
equations, modulo a transformation (\ref{doft}).    In other ranges of the parameters,
the formulas of \cite{Zamolodchikov:1995aa} do not generalize and one can then
argue that the complex Liouville equations have no solutions at all.

The way that the formulas of \cite{Zamolodchikov:1995aa} fail to generalize is
instructive.\footnote{This was anticipated by A. B. Zamolodchikov, whom we thank for
discussions.}  In general,
when one extends these formulas to complex values of $\eta_i$ and $b$,
branch points appear in the solution and $\phi$ is not singlevalued.  (The quantities
such as $\exp(2b\phi)$ that appear in the classical Liouville equations remain singlevalued.)
Singlevaluedness of $\phi$ places a serious constraint on the range of parameters for
which a complex critical point exists.

We will show that, after taking into account the symmetry (\ref{doft}),  ordinary, singlevalued complex solutions of Liouville's equations suffice for 
understanding the semiclassical asymptotics of the Liouville two-point function, and also
for understanding the semiclassical asymptotics of the three-point function in a somewhat larger
region than considered in \cite{Zamolodchikov:1995aa}.  In particular we will see that the old ``fixed-area'' prescription for computing correlators outside the physical region can be replaced by the machinery of complex saddlepoints, which makes the previously-subtle question of locality manifest.  But for general values of the $\eta_i$,
there are not enough singlevalued complex solutions to account for the asymptotics of the three-point function.

What then are we to make of the semiclassical limit outside of the region where solutions exist?  Rather surprisingly, we have found that allowing ourselves to use the multivalued ``solutions'' just mentioned in the semiclassical expansion enables us to account for the asymptotics of the DOZZ formula throughout the full analytic continuation in the $\eta_i$.  There is a simple prescription for how to evaluate the action of these ``solutions'', and which has them as stationary points.  This prescription agrees with the conventional Liouville action on singlevalued solutions and produces its analytic continuation when evaluated on the multivalued ``solutions''.\footnote{As explained in section \ref{gaction} the prescription is essentially to re-express the Liouville field in terms of the ``physical metric'' $g_{ij}=e^{2b\phi}\tilde{g}_{ij}$, which is always single valued.  The branch points of $\phi$ become isolated divergences of the metric that have a very specific form, and which turn out to be integrable if a ``principal value'' regularization is used.}    In particular once $\phi$ is multivalued, to evaluate the action one must pick a branch of $\phi$ at each insertion point of a heavy vertex operator
$\exp(2\eta_i\phi/b)$; allowing all possible choices, one does indeed recover the
multivaluedness expected in (\ref{dofo}).  We do not have a clear rationale for why this is allowed.  For one thing if we do not insist on expanding on cycles attached to critical points as in (\ref{expres}) then it seems clear that for any value of $\eta_i$ we can always find an integration cycle that passes only through single-valued field configurations simply by arbitrarily deforming the original cycle in the physical region in a manner that preserves the convergence as we continue in $\eta_i$.  It is only when we try to deform this cycle in such a way that the semiclassical expansion is transparent that we apparently need to consider these exotic integration cycles attached to multivalued ``solutions''.

We also attempt to probe the classical solutions that contribute to the 
three-point function of heavy primary fields by inserting a fourth light primary field.
This is not expected to significantly modify the critical points contributing
to the path integral, but should enable one to ``measure'' or observe those
critical points.  If the light primary field is ``degenerate'' in the sense of \cite{Belavin:1984vu},
then one can obtain a very concrete formula for the four-point function, and this formula supports the idea that the three-point function is dominated by the multivalued
classical solution.  When the light operator is non-degenerate the situation is more subtle, a naive use of the multivalued solution suggests an unusual singularity in the Liouville four-point function which we are able to prove does not exist.  We speculate as to the source of this discrepancy, but we have been unable to give a clear picture of how it is resolved.      

\subsection{Liouville Theory And Chern-Simons Theory}\label{lcs}

\hspace{0.25in}Since it somewhat strains the credulity to believe that the Liouville path integral
should be expanded around a multivalued classical solution, we have also looked
for another interpretation.   Virasoro conformal blocks in two dimensions have a relation
to Chern-Simons theory in three-dimensions with gauge group $SL(2,\mathbb{C})$ that was identified
long ago \cite{HVerlinde}.  An aspect of this relation is that quantization of Teichmuller
space \cite{ChF}, which is an ingredient in $SL(2,\mathbb{C})$ Chern-Simons theory,
can be used to describe Virasoro conformal blocks \cite{Teschner}.
Since Liouville theory can be constructed by gluing together
Virasoro conformal blocks for left- and right-movers,  it should also have
an expression in terms of Chern-Simons theory;
 one hopes to express Liouville theory on a Riemann surface
$\Sigma$ in terms of Chern-Simons on $\Sigma\times I$, where $I$ is a unit interval.  The boundary
conditions required at the ends of $I$ are those of \cite{HVerlinde}.  These boundary
conditions have recently been reconsidered and the relation between Liouville and Chern-Simons
theory developed in more detail \cite{GWNow}.  

Given these facts, instead of looking for complex solutions of Liouville theory on $\Sigma$,
we can look for complex solutions of $SL(2,\mathbb{C})$ Chern-Simons theory on $\Sigma\times I$
with the appropriate boundary conditions.  Here we find a simpler story than was summarized in
section \ref{compsol}.  Solutions are precisely parametrized by the integers $n$ and $m_i$ of
eqn. (\ref{dofo}) and the action depends on those parameters in precisely the expected fashion.
So a possible interpretation of the results of the present part is that if one wishes
to apply the machinery of complex saddle points and integration cycles to Liouville
theory in a conventional way, one should use the Chern-Simons description.   Possibly this reflects
the fact that the gradient flow equations of complex Chern-Simons theory are elliptic
(as analyzed in \cite{Analytic}); this is not so for complexified Liouville theory.

\subsection{Timelike Liouville Theory}\label{applic}

\hspace{0.25in}As an application of these ideas, we will consider the case of what we will call timelike
Liouville theory, or Liouville theory with large negative central
charge.  This is the case that $b$ is small and imaginary, so that $b^2<0$.  If $b$ is imaginary, then the exponential
term $\exp(2b\phi)$ in the Liouville action is of course no longer real.   One can compensate
for this by taking $\phi\to i\phi$, but then the kinetic energy of the Liouville field becomes
negative, and the Liouville field becomes timelike in the sense of string theory.
From that point of view, ordinary Liouville theory, in which the kinetic energy of $\phi$ has the usual
sign, might be called spacelike Liouville theory.  We will use that terminology occasionally.  Timelike Liouville theory has possible applications in quantum cosmology \cite{Freivogel:2006xu,Sekino:2009kv,Harlow:2010my}, and also as the worldsheet description of closed string tachyon condensation \cite{Strominger:2003fn}.

It was shown in  
\cite{Zamolodchikov:2005fy} that the DOZZ formula, when analytically continued in $b$, has a natural boundary
of holomorphy 
on the imaginary axis.   On the other hand, it was also shown that the Ward identities that
lead to the DOZZ formula have a second solution -- which we will call the timelike DOZZ
formula -- that is well-behaved on the imaginary $b$
axis, but runs into trouble if analytically continued to real $b$.  If $b$ is neither real nor
imaginary, the two formulas are both well-behaved but different.  The timelike DOZZ formula has also
been independently derived as a possible ``matter'' theory to be coupled to spacelike Liouville in \cite{KP1} and further studied in \cite{KP2,KP3}.  Its first appearance seems to be as equation (4.5) in \cite{Schomerus:2003vv}, where it appeared as an intermediate step in a proposal for the $c=1$ limit of Liouville.\footnote{In \cite{Schomerus:2003vv} it was argued that the timelike DOZZ formula should be multiplied by various nonanalytic factors in order for it to describe timelike Liouville.  This proposal seems to work only when $b=ip/q$ with $p,q\in\mathbb{Z}$, and does not allow continuation to generic $\alpha$.  These modifications are allowed because for these special values of $b$ the uniqueness argument for the timelike DOZZ formula breaks down.  Some interesting applications of timelike Liouville seem to require generic values of $b$ and $\alpha$, for example in coupling to a ``matter'' CFT with generic $c>25$, so we are interested in describing a theory that works for generic imaginary $b$.  From the point of view of this part analyticity in $b$ is also more natural to consider; since the integrand of the path integral is an analytic function of its parameters the integral should be analytic as well.  We will see below that evaluating the Liouville path integral in the timelike regime does not produce any nonanalytic factors, so they would have to be put in by hand.  Schomerus's justification of the extra factors involves wanting the three point function to reduce to the two point function, which may be an appropriate
requirement in a theory that works precisely at $b= i p/q$ (and may be related
to Virasoro minimal models).  The timelike DOZZ formula does not have this property, but we suggest an alternative interpretation in section \ref{tcft?} that does not require the new factors.}

From the perspective of the present part, with all fields and parameters potentially continued to complex
values, timelike Liouville theory and ordinary or spacelike Liouville theory are the same theory, possibly
with different integration cycles.   We will investigate this question and show that the timelike
DOZZ formula can indeed come from the same path integral that gives the ordinary or spacelike DOZZ formula,
with an extra factor that represents a change in the integration cycle.

It was shown in \cite{Strominger:2003fn,Schomerus:2003vv,Zamolodchikov:2005fy} that timelike Liouville theory does not at first seem to have all of the usual properties of a conformal
field theory; this issue was discussed further in \cite{McElgin:2007ak} but not resolved.
The simplest problem in interpreting the timelike DOZZ formula in terms of conformal field theory is that naively it appears that the two-point function is not diagonal in the conformal dimensions.  Our path integral interpretation of the timelike Liouville correlators sheds some light on this question; we will argue that the two-point function is indeed diagonal and conjecture that the problems which have been identified have to do with the existence of new degenerate fields that do not decouple in the conventional way.  This is possible because of the intrinsically nonunitary nature of timelike Liouville.  We have not been able to answer the more subtle question of which states to factorize correlators on.  For a minisuperspace analysis of this problem,
see \cite{FS}.  

\subsection{Outline}\label{outline}

\hspace{0.25in}An outline of this part is as follows.  In section \ref{slt}, we review Liouville theory.
In section \ref{liouvillesolutions}, we study complex solutions of Liouville's equations on the sphere with heavy operators.
In section \ref{anastokes}, show that the analytic continuation of the DOZZ
formula in a restricted region can be interpreted in terms of complex classical solutions.  In section \ref{4pointsection}, we study the full analytic continuation and confront the issue of the nonexistence of nonsingular solutions.  We then use
a fourth light primary field to probe the classical configurations contributing to the three-point
function of heavy primaries, confirming our explanation of the DOZZ analytic continuation in terms of singular ``solutions''.  In section \ref{csfun}, we reinterpret the question of complex
classical solutions in terms of Chern-Simons theory.  In section \ref{timelike},
we consider timelike Liouville theory.  In section \ref{conclusion} we give a brief summary of our results and suggest directions for future work.  Finally, in a series of appendices, we describe
a variety of useful technical results.

The length of the part is partly the result of an attempt to keep it self-contained.
We have written out fairly detailed accounts of a variety of results that are known but
are relatively hard to extract from the literature.  This is especially so in section \ref{slt}
and in some of the appendices.  Casual readers are welcome to skip to the conclusion, which contains the highlights in compact form.

\chapter{Review Of Liouville Theory}\label{slt}
We begin with an overview of  Liouville theory.  The goal is to present and motivate 
all the existing results that we will need in following sections; there are no new 
results here.  Some relatively modern reviews on Liouville theory are 
\cite{Teschner:2001rv,Nakayama:2004vk}; a much older one is \cite{Seiberg:1990eb}.  
Our conventions are mostly those of \cite{Zamolodchikov:1995aa}.
  
\subsection{Action, Boundary Condition, and Equation of Motion}

\hspace{0.25in}The Liouville action, obtained for example by gauge fixing a generic conformal 
field theory coupled to two-dimensional gravity \cite{Ginsparg:1993is}, is
\be
\label{liouvaction}
S_L=\frac{1}{4\pi} \int{d^2 \xi \sqrt{\widetilde{g}}\left[\partial_a \phi \partial_b \phi 
\widetilde{g}^{ab}+Q \widetilde{\mathcal{R}}\phi+4 \pi \mu e^{2b\phi}\right]}.
\ee
Here $Q=b+\frac{1}{b}$, and the exponential operator is defined in a renormalization 
scheme using $\widetilde{g}$ to measure distances.  The metric $\widetilde{g}$ is 
referred to as the ``reference'' metric ($\tilde{\mathcal R}$ is its scalar curvature), while the quantity 
$g_{ab}=e^{\frac{2}{Q}\phi}\widetilde{g}_{ab}$ is referred to as the 
``physical'' metric.  Since we are viewing Liouville theory as a complete theory 
in and of itself, the ``physical''  metric is no more physical than the reference one, 
but it is extremely useful for semiclassical intuition so we will often discuss it in what 
follows.  This theory is invariant, except for a $c$-number anomaly,
 under conformal transformations:
\begin{align}
\nonumber
z'&=w(z)\\
\phi'(z',\overline{z}')&=\phi(z,\overline{z})-\frac{Q}{2}\log \left|\frac{\partial w}{\partial z}\right|^2.
\label{transformation}
\end{align}
Here we use a complex coordinate $z=\xi^1+i \xi^2$, 
and $w(z)$ is any locally holomorphic function.  Under these transformations the 
renormalized exponential operators have conformal weights 
\be
\label{dimension}
\Delta(e^{2 \alpha \phi})=\overline{\Delta}(e^{2 \alpha \phi})=\alpha (Q-\alpha),
\ee 
as we explain in section \ref{cpo}.\footnote{In the terminology that we adopt, the scaling dimension of an operator is $\Delta+\bar{\Delta}$, which is twice the weight for a scalar operator.}
The stress tensor is
\be
\label{stresstensor}
T(z)=-(\partial \phi)^2+Q\partial^2 \phi,
\ee
and the central charge of the conformal algebra is
\be
c_L=1+6Q^2=1+6(b+b^{-1})^2. \label{centralcharge}
\ee

We will study this theory on a two-sphere.
It is convenient to take the reference metric to be the flat metric 
$ds^2=dz\,d\bar z$, with
\begin{equation}\label{logo} \phi = -2Q \log r+\O(1) \qquad \text{as }r \to \infty, ~r=|z|,\end{equation}
which ensures that the physical metric is a smooth metric on $\sf S^2$. This ensures that $\phi$ is nonsingular at infinity with respect to (\ref{transformation}).
The intuition for the condition (\ref{logo}) is that there is an operator insertion at infinity
representing the curvature of $\sf S^2$, which has been suppressed in taking the
reference metric to be flat.

Though the use of a flat reference metric is convenient, with this choice there is some
subtlety in computing the action; one must regulate the region of 
integration and introduce boundary terms.  Following \cite{Zamolodchikov:1995aa}, we let
$D$ be  a disk of radius $R$, and define the action as the large $R$ limit of
\be
\label{flataction}
S_L=\frac{1}{4\pi} \int_{D}{d^2 \xi \left[\partial_a \phi \partial_a \phi+4 \pi \mu e^{2b\phi}\right]}+\frac{Q}{\pi}\oint_{\partial D}\phi d\theta +2Q^2\log R.
\ee
The last two terms ensure finiteness of the action and also invariance under (\ref{transformation}).\footnote{One way to interpret them is to note that if we begin with the original Liouville action (\ref{liouvaction}) with round reference metric 
\be
ds^2=\frac{4}{(1+r^2)^2} \left(dr^2+r ^2 d\theta^2\right),
\ee
then the field redefinition 
\be
\phi \to \phi-Q \log\left(\frac{2}{1+r^2}\right)
\ee
produces exactly the action (\ref{flataction}) up to a finite field-independent constant.  Rather than trying to keep track of this, we will just take the action (\ref{flataction}) as our starting point.}  

The semiclassical limit $b \to 0$ is conveniently  
studied with a rescaled field $\phi_c = 2b \phi$, in terms of which the  action becomes
\be
\label{action}
b^2 S_L=\frac{1}{16\pi} \int{d^2 \xi \left[\partial_a \phi_c 
\partial_a \phi_c+16 \pi \mu b^2 e^{\phi_c}\right]+\frac{1}{2\pi}
\oint_{\partial D}\phi_c d\theta +2\log R+\mathcal{O}(b^2),}
\ee
and the boundary condition becomes 
\be
\label{infinitephi}
\phi_c(z,\overline{z}) = -2 \log(z \overline{z})+\O(1) \qquad \text{as }|z| \to \infty.
\ee
The equation of motion following from this action is
\be
\label{liouville}
\partial \overline{\partial} \phi_c = 2 \pi \mu b^2 e^{\phi_c}.
\ee
If we now define $\lambda\equiv \pi \mu b^2$ to be fixed as $b\to0$, 
then $\phi_c$ will have a fixed limit for $b\to 0$.\footnote{Intuitively this choice 
of scaling ensures that the 
radius of curvature $\lambda^{-1/2}$ of the physical metric is large in 
units of the ``microscopic'' scale $\mu^{-1}$.}  Since 
$g_{ab} = e^{\phi_c} \delta_{ab}$, the physical metric has a good limit as well.  
The equation of motion is equivalent to the condition of constant negative curvature 
of $g_{ab}$, and this is the source of the classical relationship between 
Liouville's equation (\ref{liouville}) and the uniformization of Riemann surfaces.

\subsection{Conformal Primary Operators and Semiclassical Correlators}\label{cpo}

\hspace{0.25in}Because of the unusual nature of the transformation (\ref{transformation}), we can 
guess that it will be exponentials of $\phi$ that transform with definite conformal 
weights.  Classically we see that
\be
e^{2 \alpha \phi'(z',\overline{z}')}=
\left(\frac{\partial w}{\partial z}\right)^{-\alpha Q}\left(\frac{\partial
 \overline{w}}{\partial \overline{z}}\right)^{-\alpha Q} e^{2 \alpha \phi(z,\overline{z})},
\ee
so that classically $V_\alpha \equiv e^{2\alpha \phi}$ is a \textit{primary} conformal 
operator with conformal weights $\Delta=\overline{\Delta}=\alpha Q$ \cite{Belavin:1984vu}.  
$\alpha$ is called the Liouville momentum.  Quantum mechanically, the conformal
weights of these operators are modified.
  In free field theory, normal ordering  contributes a well-known additional 
term $-\alpha^2$ to each weight.  In Liouville theory, the quantum correction
is exactly the same, since we can compute
the weight of the operator $V_\alpha$ by considering correlations in a state of our
choice.  We simply consider correlations in a state in which $\phi<<0$, thus turning off
the Liouville interactions and reducing the computation of operator scaling to the free
field case.\footnote{We will see below that this argument requires $\Re \, \alpha<Q/2$, since otherwise the backreaction of the operator will prevent $\phi<<0$ near the operator.}  So $V_\alpha$ has conformal weight $\alpha(Q-\alpha)$, as in (\ref{dimension}).

In this subsection we will discuss the properties of these operators and their correlators 
in more detail at the semiclassical level, in particular seeing how this factor emerges in the 
formula for $\Delta$.  In the following subsections we will review the exact construction of 
Liouville theory that confirms this expression for $\Delta$ beyond the semiclassical regime. 
 
We will now consider correlation functions of primary fields,
\be
\label{pathintegral}
\biggl\langle V_{\alpha_1}(z_1,\overline{z}_1)\cdots V_{\alpha_n}(z_n,\overline{z}_n)
\biggr\rangle \equiv \int \mathcal{D}\phi_c \,e^{-S_L}\prod_{i=1}^n \exp\left(\frac{\alpha_i 
\phi_c(z_i,\overline{z}_i)}{b}\right).
\ee
We would like to approximate this path integral using the method of steepest descent for small 
$b$, but to do so we must decide how the $\alpha_i$'s scale with $b$.  
The action (\ref{action}) scales like $b^{-2}$, so for an operator to have a nontrivial 
effect on the saddle points we must choose its Liouville momentum $\alpha$ to scale like $b^{-1}$.  
Thus if we want an operator to affect the saddle point, we take $\alpha=\eta/b$ and
keep $\eta$ fixed for $b\to 0$.  This gives what is conventionally called a ``heavy'' Liouville
primary field.  Asymptotically such a field has $\Delta=\eta(1-\eta)/b^2$ for $b\to 0$.  One can also 
define ``light'' operators with $\alpha =b \sigma $, where $\sigma$ is kept fixed for $b\to 0$.
Light operators have fixed  dimensions in the 
semiclassical limit.  Insertion of such an operator has no effect on the saddle point $\phi_c$, and to lowest order in $b$ can be approximated by a $b$-independent factor of $e^{\sigma_i\phi_c(z,\bar{z})}$.

Semiclassically the insertion of a  heavy operator has the effect of adding  an additional delta
function term to the action, 
leading to a new equation of motion:
\be
\label{eom}
\partial \overline{\partial} \phi_c = 2 \pi \mu b^2 e^{\phi_c}-2\pi\sum_{i} \eta_i \delta^2(\xi-\xi_i)
\ee
Let us 
assume that in the vicinity of one of the operator insertions we may ignore the 
exponential term.  This equation then becomes  Poisson's equation:\footnote{Note the the convention that $4\partial\bar{\partial} =\nabla^{2}$.}
\be\label{Poisson}\nabla^2\phi_c=-8\pi \eta_i \delta^2(\xi-\xi_i).\ee
This has the solution
\be\phi_c(z,\overline{z})=C-4 \eta_i \log|z-z_i|,\ee
so we find that in a neighborhood of a heavy operator we have
\begin{equation}
\label{nearops}
\phi_c(z,\overline{z}) = -4 \eta_i \log|z-z_i|+\O(1) \qquad \text{as }z \to z_i.
\end{equation}
We also find that that the physical metric in this region has the form:
\be
\label{metricnearop}
ds^2=\frac{1}{r^{4 \eta_i}}(dr^2+r^2d\theta^2)
\ee
We can insert this solution back into the equation of motion to check whether the 
exponential is indeed subleading. We find that this is the case if and 
only if 
\be
\label{seiberg}
\mathrm{Re}(\eta_i)<\frac{1}{2}.
\ee
If this inequality is not satisfied, then the interactions 
affect the behaviour of the field 
arbitarily close to the operator.  
In \cite{Seiberg:1990eb}, this was interpreted as the non-existence of 
local operators with $\mathrm{Re}(\eta)>\frac{1}{2}$, and the condition that 
``good'' Liouville operators have $\mathrm{Re}(\eta)<\frac{1}{2}$ is referred to as the Seiberg bound.  
The modern interpretation of this result, as we will see in the following section, 
is that both $\alpha$ and $Q-\alpha$ correspond to the \textit{same} quantum operator, 
with a nontrivial rescaling:
\be
\label{opreflection}
V_{Q-\alpha}=R(\alpha)V_\alpha.
\ee
$R(\alpha)$ is referred to as the reflection coefficient, for reasons explained in 
\cite{Polchinski:1990mh,Seiberg:1990eb}.  Either $\alpha$ or $Q-\alpha$ will always 
obey the Seiberg bound, and we can always choose that one when studying the 
semiclassical limit.  We will thus focus only on semiclassical solutions for which all 
operators have $\mathrm{Re}(\eta_i)<\frac{1}{2}$.

We will in general be interested in complex values of $\eta_i$, so the metric 
(\ref{metricnearop}) will be complex and thus not admit a simple geometric interpretation.  
For the next few paragraphs, however, we will assume that $\eta_i$ is real to enable us to 
develop some useful intuition.  We first observe that since $\eta_i<\frac{1}{2}$, we can do a 
simple change of variables to find a metric
\be
\label{condef}
ds^2=dr'^2+r'^2 d\theta'^2,
\ee
where the coordinate ranges are $r' \in [0,\infty)$ and $\theta' \in \left[0,(1-2\eta_i)2\pi\right)$.  
Thus we can interpret the effect of the operator as producing a conical singularity 
in the physical metric, with a conical deficit for $0<\eta_i<\frac{1}{2}$ and a conical 
surplus for $\eta_i<0$.  Finding real saddle points in the presence of heavy operators with 
real $\eta$'s is thus equivalent to finding metrics of constant negative curvature on the 
sphere punctured by conical singularities of various strength.  

An interesting additional constraint comes from the Gauss-Bonnet theorem.  
The integrated curvature on a sphere must be positive to produce a positive Euler character, 
so for a metric of constant negative curvature to exist on a punctured sphere the punctures 
must introduce sufficient positive curvature to cancel the negative curvature 
from the rest of the sphere.  By integrating equation (\ref{eom}) and using the boundary 
condition (\ref{infinitephi}) we find a real solution $\phi_c$ can exist only if
\be
\label{gbconstraint}
\sum_i \eta_i>1
\ee

This inequality along with the Seiberg bound leads to interesting constraints on Liouville momenta.  In particular for the case of three heavy operators on $\sf S^2$,
the 
inequalities together imply $0<\eta_i<\frac{1}{2}$.  Unless we satisfy these inequalities,
there is no real saddle point for the Liouville path integral, 
even if the $\eta_i$
are all real.  The Gauss-Bonnet constraint (\ref{gbconstraint}) also implies that there is
no real saddle point for a product of light fields on $\sf S^2$; this case amounts to setting
all $\eta_i$ to zero.  In particular, there is no real saddle point for the Liouville partition function on
$\sf S^2$.  This has traditionally been dealt with by fixing the area (calculated
in the physical metric) and then attempting to integrate over the area; the fixed area path 
integral has a real saddle point.   We will develop an alternative based on complex
saddle points.

More generally, if the  $\eta$'s are complex, then as we mentioned above a saddle point 
$\phi_c$ will in general be complex and there is no reason to impose (\ref{gbconstraint}).  

So far we have not encountered the renormalization issues mentioned at the beginning 
of the section.  But if we try to evaluate the action (\ref{action}) on a solution obeying 
(\ref{nearops}), then we find that both the kinetic term and the source term contributed by 
the heavy operator are divergent.\footnote{The exponential term is finite since we are 
assuming (\ref{seiberg}).}  To handle this, again following \cite{Zamolodchikov:1995aa}, 
we perform the action integral only over the part of the disk $D$ that excludes a disk $d_i$ 
of radius $\epsilon$ about each of the heavy operators.  We then introduce 
``semiclassically renormalized'' operators
\be\label{ponzo}
V_{\frac{\eta_i}{b}}(z_i,\overline{z}_i)\approx \epsilon^{\frac{2\eta_i^2}{b^2}} 
\exp\left(\frac{\eta_i}{2\pi}\oint_{\partial d_i}\phi_c d\theta\right).
\ee
It is easy to check that this operator multiplied by the usual integrand of the path integral
(the exponential of minus the action) has 
a finite limit as $\epsilon\to0$ when evaluated on a solution obeying (\ref{nearops}).  
The prefactor $ \epsilon^{\frac{2\eta_i^2}{b^2}}$ in (\ref{ponzo})
contributes a term $-{2\eta_i^2}/{b^2}$ to the 
scaling dimension of the operator $V_{\eta_i/b}$; 
this is a contribution of  $-{\eta_i^2}/{b^2}$ to both $\Delta$ 
and $\overline{\Delta}$, consistent with the quantum shift $-\alpha_i^2$ of the operator
weights.  We can
 thus incorporate the effects of all the heavy operators by introducing a modified action:
\begin{align}
\label{regaction}
\nonumber
b^2 \widetilde{S}_L=&\frac{1}{16\pi} \int_{D-\cup_i d_i} d^2\xi \left(\partial_a \phi_c \partial_a 
\phi_c+16 \lambda e^{\phi_c}\right)+\frac{1}{2\pi}\oint_{\partial D}\phi_c d\theta +2\log R\\
&-\sum_i \left(\frac{\eta_i}{2\pi}\oint_{\partial d_i}\phi_{c}d\theta_i +2 \eta_i^2\log \epsilon \right)
\end{align}
The equations of motion for this action automatically include both Liouville's 
equation (\ref{liouville}) and the boundary conditions (\ref{infinitephi}) and (\ref{nearops}).  
The final semiclassical expression for the expectation value of a product of heavy and light
primary fields is
\begin{align}
\label{semiclassicalcorr}
\left\langle V_{\frac{\eta_1}{b}}(z_1,\overline{z}_1)\cdots 
V_{\frac{\eta_n}{b}}(z_n,\overline{z}_n)V_{b\sigma_1 }(x_1,\overline{x}_1)\cdots V_{b \sigma_m}
(x_m,\overline{x}_m)\right\rangle 
\approx e^{-\widetilde{S}_L[\phi_\eta]}\prod_{i=1}^me^{\sigma_i \phi_\eta (x_i,\overline{x}_i)} .\end{align}
Here there are $n$ heavy operators and $m$ light operators, and $\phi_\eta$ 
is the solution of (\ref{eom}) obeying the correct boundary conditions.  In this formula effects that are $O(b^0)$ in the exponent have been kept only if they depend on the positions or conformal weights of the light operators.  We will do light operator computations in sections (\ref{dozzthreelight}, \ref{probe}, \ref{genf}, \ref{tdozzthreelight}), and we will be more careful about these corrections there.  If there is 
more than one solution, and we will find that in general there will be, then the right hand side of (\ref{semiclassicalcorr}) will include a sum (or integral) over these saddlepoints.

\subsection{DOZZ Formula}\label{dozzo}

\hspace{0.25in}In two-dimensional conformal field theory,
the expectation value of a product of three primary operators on $\sf S^2$ is 
determined up to a constant by conformal symmetry \cite{Belavin:1984vu}.  We saw in the 
previous section that the operators $V_\alpha$ are
primaries of weight $\Delta=\alpha(Q-\alpha)$, so their three-point function must be of the form
\be
\label{3point}
\langle V_{\alpha_1}(z_1,\overline{z}_1)V_{\alpha_2}(z_2,\overline{z}_2)
V_{\alpha_3}(z_3,\overline{z}_3)\rangle=
\frac{C(\alpha_1,\alpha_2,\alpha_3)}{|z_{12}|^{2(\Delta_1+\Delta_2-\Delta_3)}
|z_{13}|^{2(\Delta_1+\Delta_3-\Delta_2)}|z_{23}|^{2(\Delta_2+\Delta_3-\Delta_1)}}.
\ee
Here $z_{ij}=z_i-z_j$.  The function $C(\alpha_1,\alpha_2,\alpha_3)$ 
is the main dynamical data of any CFT.  In a CFT with only finitely many primaries,
matrix elements of $C$ are often called structure constants, but this terminology
does not seem entirely felicitous when $C$ depends on continuous variables. The 
DOZZ formula is an analytic expression for $C$  in Liouville theory 
\cite{Dorn:1994xn,Zamolodchikov:1995aa}.  This proposal satisfies all the expected conditions
in Liouville theory, and is the unique function that does so; in
particular, it is the unique solution of recursion relations that were derived in  
\cite{Teschner:1995yf,Teschner:2001rv}. Knowing  $C(\alpha_1,\alpha_2,\alpha_3)$,
along with rules for a sewing construction of higher order amplitudes can
 be viewed as an exact 
construction of the quantum Liouville theory.  

The DOZZ formula reads:
\begin{align}
\label{dozz} \nonumber
&C(\alpha_1,\alpha_2,\alpha_3)=\left[\pi \mu \gamma(b^2) 
b^{2-2b^2}\right]^{(Q-\sum{\alpha_i})/b}\\
&\times\frac{\Upsilon_0 \Upsilon_b(2\alpha_1)\Upsilon_b(2\alpha_2)
\Upsilon_b(2\alpha_3)}{\Upsilon_b(\alpha_1+\alpha_2+\alpha_3-Q)
\Upsilon_b(\alpha_1+\alpha_2-\alpha_3)
\Upsilon_b(\alpha_2+\alpha_3-\alpha_1)\Upsilon_b(\alpha_1+\alpha_3-\alpha_2)}.
\end{align}
Here $\Upsilon_b(x)$ is an entire function of $x$ defined (for real and positive $b$) by 
\begin{equation}
\label{logupsilon}
\log\Upsilon_b(x)=\int_0^\infty\frac{dt}{t}\left[(Q/2-x)^2 e^{-t}-\frac{\sinh^2((Q/2-x)\frac{t}{2})}{\sinh{\frac{tb}{2}}\sinh{\frac{t}{2b}}}\right]\qquad 0<\mathrm{Re}(x)< Q.
\end{equation}
Though this integral representation is limited to the strip $ 0<\mathrm{Re}(x)< Q$, $\Upsilon_b(x)$
has an analytic continuation to an entire function of $x$.  This follows from recursion
relations that are explained in 
 Appendix \ref{upsilonapp}, along with other properties of  $\Upsilon_b$.  $\Upsilon_0$ is defined as $\frac{d\Upsilon_b(x)}{dx}|_{x=0}$, and 
 $\gamma(x)\equiv\frac{\Gamma(x)}{\Gamma(1-x)}$.  In the following section 
 we will discuss some of the motivation for this formula, but for the moment we 
 will just make three observations:
\begin{itemize}
\item[(1)]  This expression obeys 
$C(Q-\alpha_1,\alpha_2,\alpha_3)=R(\alpha_1)C(\alpha_1,\alpha_2,\alpha_3)$ with 
\begin{align} \nonumber
R(\alpha)&=\left[\pi \mu \gamma(b^2)b^{2-2b^2}\right]^{(2\alpha-Q)/b} 
\frac{\Upsilon_b(2\alpha_1-Q)}{\Upsilon(2\alpha_1)}\\
&=\left[\pi \mu \gamma(b^2)\right]^{(2\alpha-Q)/b} \frac{b^2}
{\gamma(2\alpha/b-1-1/b^2)\gamma(2 b \alpha-b^2)};
\end{align}
this justifies the reflection formula (\ref{opreflection}).  To derive this, one uses the 
reflection symmetry
$\Upsilon_b(Q-x)=\Upsilon_b(x)$ and also the recursion relations for $\Upsilon_b$.
 
\item[(2)]  The entire expression (\ref{dozz}) is almost invariant under $b \to \frac{1}{b}$, and it becomes so if we also send $\mu \to \widetilde{\mu}$, with 
\be
\pi \widetilde{\mu} \gamma(1/b^2)=\left[\pi \mu \gamma(b^2)\right]^{\frac{1}{b^2}}
\ee
This is a weak-strong duality, in the sense that if $\mu$ scales like 
$b^{-2}$ to produce good semiclassical saddle points with finite curvature as $b\to 0$, then $\widetilde{\mu}\widetilde{b}^2=\widetilde{\mu}/b^2$ will be extremely singular in the same limit so the dual picture will not be semiclassical.

\item[(3)]  $C(\alpha_1,\alpha_2,\alpha_3)$ as defined in (\ref{dozz}) is a meromorphic 
function of the $\alpha_i$, with the only poles coming from the zeros of the $\Upsilon_b$'s 
in the denominator.  In particular it is completely well-behaved in regions
 where the inequalities (\ref{gbconstraint}) and (\ref{seiberg}) are violated.  That said, 
 the integral representation of $\Upsilon_b$ is only valid in the strip $ 0<\mathrm{Re}(x)< Q$, and 
 in the semiclassical limit,
for four of the $\Upsilon_b$'s in (\ref{dozz}),  the boundary of the strip is precisely 
where the inequality (\ref{gbconstraint}) or (\ref{seiberg}) breaks down. This can lead 
to a change in
the nature of the semiclassical limit.  In particular when all three $\alpha$'s are real 
and obey the Seiberg and Gauss-Bonnet inequalities, all seven $\Upsilon_b$'s can 
be evaluated by the integral (\ref{logupsilon}).  This is not an accident; in particular,
we will argue below that analytically continuing past the line 
$\mathrm{Re}(\eta_1+\eta_2+\eta_3)=1$ corresponds to crossing a Stokes line
 in the Liouville path integral; the number of contributing saddle points increases as we do so. 
\end{itemize}

\subsection{Four-Point Functions and Degenerate Operators}
\label{4pointreview}

\hspace{0.25in}We will for the most part be  studying the semiclassical limit of the DOZZ formula, but we 
will find it extremely helpful to also consider certain four-point functions.\footnote{The 
material discussed here is mostly not required until the final two parts of 
section \ref{4pointsection}, so the reader who is unfamiliar with the CFT techniques 
of \cite{Belavin:1984vu} may wish to stop after equation (\ref{spacelike2point}) and postpone 
the rest.} In two-dimensional CFT, the four-point function
on $\sf S^2$  is the first correlation function whose position dependence is not 
completely determined by conformal symmetry.  It is strongly constrained, but 
unfortunately there is much freedom in how to apply the constraint and there do not 
seem to be standard conventions in the literature.  We will define:
\begin{align} \nonumber
\label{4point}
&\biggl\langle V_{\alpha_1}(z_1,\overline{z}_1)V_{\alpha_2}(z_2,\overline{z}_2)
V_{\alpha_3}(z_3,\overline{z}_3)V_{\alpha_4}(z_4,\overline{z}_4)\biggr\rangle \\ 
&=|z_{13}|^{2(\Delta_4-\Delta_1-\Delta_2-\Delta_3)}|z_{14}|^{2(\Delta_2+
\Delta_3-\Delta_1-\Delta_4)}
|z_{24}|^{-4 \Delta_2}|z_{34}|^{2(\Delta_1+\Delta_2-\Delta_3-\Delta_4)}
 {G}_{1234}(y,\overline{y}),
\end{align}
with the harmonic ratio $y$ defined as:
\be
\label{harmonic}
y=\frac{z_{12}z_{34}}{z_{13}z_{24}}.
\ee
This parametrization is chosen so that the limit $z_4\to\infty$, $z_3 \to 1$, $z_2 \to y$, and $z_1 \to 0$ is simple:
\be
\lim_{z_4 \to \infty} |z_4|^{4\Delta_4}\bigl \langle V_{\alpha_1}(0,0)V_{\alpha_2}(y,\overline{y})V_{\alpha_3}(1,1)V_{\alpha_4}(z_4,\overline{z}_4)\bigr\rangle={G}_{1234}(y,\overline{y})
\ee
Using radial quantization as in \cite{Belavin:1984vu}, we can write this as 
\be
\label{bootstrap}
{G}_{1234}(y,\overline{y})=\langle \alpha_4|V_{\alpha_3}(1,1)V_{\alpha_2}(y,\overline{y})|\alpha_1\rangle.
\ee
We can also write $C$ as
\be
\label{3pointstates}
C(\alpha_1,\alpha_2,\alpha_3)=\langle \alpha_3|V_{\alpha_2}(1,1)|\alpha_1 \rangle.
\ee
In a conventional two-dimensional CFT, these two equations are the starting point 
for the conformal bootstrap program \cite{Belavin:1984vu}.  In this program,
one expresses the four-point function (\ref{bootstrap}) in terms of products of
three points functions in two different ways, either by inserting a complete set of states
between the fields  $V_{\alpha_3}(1,1)$ and $V_{\alpha_2}(y,\bar y)$ in
(\ref{bootstrap}), or by using the operator product expansion to replace the product
of those two fields with a single field.  In Liouville, the situation is more 
subtle since $\alpha$ is a continuous label with complex values and it is not immediately clear
what is meant by a complete set of states.  Similarly, in making the operator product expansion, 
one expands the product of two fields in terms of a complete set of fields, and it is again not
clear how to formulate this.
 This problem was solved by Seiberg \cite{Seiberg:1990eb}, who argued 
 using minisuperspace that the states with $\alpha=\frac{Q}{2}+iP$ are indeed 
 delta-function normalizable for real and positive $P$, and moreover that these states 
 along with their Virasoro descendants are a complete basis of normalizable states.  
 One can check the first of these assertions directly from the DOZZ formula by 
 demonstrating that\footnote{In showing this, one uses the fact that the numerator 
 of the DOZZ formula has
a simple zero for $\epsilon\to 0$, while the denominator has a double zero for $\epsilon\to 0$
and $P_1-P_2\to 0$.  One encounters the relation 
$\lim_{\epsilon\to 0}\epsilon/((P_1-P_2)^2+\epsilon^2)=\pi\delta(P_1-P_2)$.}
\be
\label{normalization}
\lim_{\epsilon \to 0}C(Q/2+iP_1,\epsilon,Q/2+iP_2)=2\pi \delta(P_1-P_2)G(Q/2+iP_1),
\ee
with the two-point normalization $G(\alpha)$ given by
\be
\label{spacelike2point}
G(\alpha)=\frac{1}{R(\alpha)}=\frac{1}{b^2}\left[\pi \mu \gamma(b^2)\right]^{(Q-2\alpha)/b} \gamma(2\alpha/b-1-1/b^2)\gamma(2 b \alpha-b^2).
\ee
Seiberg also argued semiclassically that the state $V_{\alpha_2}(y,\overline{y})|\alpha_1\rangle$ with both $\alpha$'s real and less than $Q/2$ is normalizable if and only if $\alpha_1+\alpha_2>\frac{Q}{2}$.  This follows from the Gauss-Bonnett constraint.  If we assume that $\alpha_1$ and $\alpha_2$ are in this range, then we can expand this state in terms of the normalizable states $|Q/2+iP,k,\overline{k}\rangle$.  Here $|Q/2+iP,k,\overline{k}\rangle$ is a shorthand notation for $V_{Q/2+iP}(0,0)|vac\rangle$ and its Virasoro descendants.  Similarly if $\alpha_3+\alpha_4>\frac{Q}{2}$ the state $\langle \alpha_4|V_{\alpha_3}(1,1)$ is also normalizable, and we can evaluate (\ref{4point}) by inserting a complete set of normalizable states.  Using (\ref{opreflection}), (\ref{3pointstates}), and (\ref{normalization}) this leads to 
\begin{align}
\label{factorized4point} \nonumber
{G}_{1234}(y,\overline{y})=\frac{1}{2}&\int_{-\infty}^{\infty}\frac{dP}{2\pi}
C(\alpha_1,\alpha_2,Q/2+iP)C(\alpha_3,\alpha_4,Q/2-i P)\\
&\times\mathcal{F}_{1234}(\Delta_i,\Delta_P,y)\mathcal{F}_{1234}(\Delta_i,\Delta_P,\overline{y}).
\end{align}
Here $\Delta_P=P^2+Q^2/4$, and the function $\mathcal{F}_{1234}$ is the familiar 
Virasoro conformal block \cite{Belavin:1984vu}, expressible as
\be
\label{conformalblock}
\mathcal{F}_{1234}(\Delta_i,\Delta_P,\overline{y})=y^{\Delta_P-\Delta_1-\Delta_2}\sum_{k=0}^\infty \beta^{P,k}_{12}\frac{\langle \alpha_4|V_{\alpha_3}(1,1)|Q/2+iP,k,0\rangle}{C(\alpha_3,\alpha_4,Q/2+iP)}y^k.
\ee
The sum over $k$ is heuristic; it  really represents a sum  over the full conformal family descended from $V_{Q/2+iP}$.  The power of $y$ for a given term is given by the level of the descendant being considered, so for example $L_{-1}L_{-2}|Q/2+iP\rangle$ contributes at order $y^3$.  $\beta^{P,k}_{12}$ is defined in \cite{Belavin:1984vu}, it appears here in the expansion of $V_{\alpha_2}|\alpha_1\rangle$ via
\begin{align} \nonumber
V_{\alpha_2}(y,\overline{y})|\alpha_1\rangle=\int_0^\infty \frac{dP}{2\pi}&
C(\alpha_1,\alpha_2,Q/2+iP)R(Q/2+iP)|y|^{2(\Delta_P-\Delta_1-\Delta_2)}\\
&\times\sum_{k,\overline{k}=0}^\infty \beta^{P,k}_{12}\beta^{P,\overline{k}}_{12}
y^k \overline{y}^{\overline{k}}|Q/2+iP,k,\overline{k}\rangle.
\end{align}
Both $\beta^{P,k}_{12}$ and the conformal 
block itself are universal building blocks for two-dimensional CFT's, 
and conformal invariance completely determines how they depend on
the conformal weights and central charge.

We can then define the general four-point function away from the specified region of 
$\alpha_1$, $\alpha_2$ by analytic continuation of (\ref{factorized4point}).  As 
observed in \cite{Zamolodchikov:1995aa}, this analytic continuation changes the form 
of the sum over states.  The reason is that as we continue in the $\alpha$'s, the various 
poles of the $C$'s can cross the contour of integration and begin to contribute discrete 
terms in addition to the integral in (\ref{factorized4point}).  

One final tool that will be useful for us is the computation of correlators 
that include degenerate fields.  A degenerate field in 2D CFT is a primary operator 
whose descendants form a short representation of the Virasoro algebra, and this 
implies that correlation functions involving the degenerate field obey a certain differential 
equation \cite{Belavin:1984vu}.  Such short
representations of the Virasoro algebra can arise only for certain values of the
conformal dimension. In Liouville theory the degenerate fields have
\be
\label{degenerate}
\alpha=-\frac{n}{2b}-\frac{m b}{2},
\ee
where $n$ and $m$ are nonnegative integers \cite{Teschner:1995yf}.  In particular we see that there are both light and heavy degenerate fields.  We will be especially interested in the light degenerate field $V_{-b/2}$, so we observe here that the differential equation its correlator obeys is
\begin{align} \nonumber
\Bigg(\frac{3}{2(2\Delta+1)}&\frac{\partial^2}{\partial z^2}-\sum_{i=1}^n \frac{\Delta_i}{(z-z_i)^2}-\sum_{i=1}^n \frac{1}{z-z_i}\frac{\partial}{\partial z_i} \Bigg)\\
&\times\biggl\langle V_{-b/2}(z,\bar{z})V_{\alpha_1}(z_1,\overline{z}_1)\cdots V_{\alpha_n}(z_n,\overline{z}_n)\biggr\rangle=0. \label{lightdegenerate}
\end{align}
Here $\Delta$ is the conformal weight of the field $V_{-b/2}$.  An identical equation holds for correlators involving $V_{-\frac{1}{2b}}$, with $\Delta$ now being the weight of $V_{-\frac{1}{2b}}$.  For example, by applying this equation to the three-point function $\langle V_{-b/2} V_{\alpha_1} V_{\alpha_2}\rangle$  and using also the fact that it must take the form
(\ref{3point}), one may show  that this three-point function vanishes unless $\alpha_2=\alpha_1 \pm b/2$.  (This relation is known as the degenerate fusion rule.)  We can check that the DOZZ formula indeed vanishes if we set $\alpha_3=-b/2$ and consider generic $\alpha_1, \alpha_2$, but there is in important subtlety in that if we simultaneously set $\alpha_2=\alpha_1 \pm b/2$ and $\alpha_3=-b/2$ the value of $C(\alpha_1,\alpha_2,\alpha_3)$ is indeterminate. (The numerator and denominator both vanish.) The lesson is that correlators with degenerate fields cannot always be simply obtained by specializing generic correlators to particular values. 
 
One can actually obtain a good limit for the four-point function with a degenerate operator from the integral expression (\ref{factorized4point}) \cite{Teschner:2001rv}.  The evaluation is subtle in that there are poles of $C(\alpha_1,\alpha_2,Q/2+iP)$ that cross the contour as we continue $\alpha_2 \to -b/2$, and in particular there are two separate pairs of poles that merge as $\alpha_2 \to -b/2$ into double poles at the ``allowed'' intermediate channels $\alpha(P)=\alpha_1 \pm b/2$.  If we are careful to perform the integral with $\alpha_2=-b/2+\epsilon$ and then take $\epsilon\to 0$, we find that the formula for the four-point function involving the light degenerate field $V_{-b/2}$ simplifies into a discrete
formula of the usual type \cite{Teschner:1995yf}: 
\be
\label{deg4}
{G}(y,\bar{y})=\sum_{\pm}C^{\pm}\,_{12} C_{34\pm} \mathcal{F}_{1234}(\Delta_i,\Delta_\pm,y)\mathcal{F}_{1234}(\Delta_i,\Delta_\pm,\overline{y}).
\ee
Here we have taken $\alpha_2=-b/2$, and $\pm$ corresponds to the operator $V_{\alpha_1 \pm b/2}$.  The raised index $\pm$ is defined using the two-point function (\ref{spacelike2point}), so:
\be
C^{\pm}\,_{12}=C(\alpha_1\pm b/2,\alpha_1,-b/2)R(\alpha_1\pm b/2)=C(\alpha_1,-b/2,Q-\alpha_1 \mp b/2).
\ee
As just discussed the value of the structure constant on the right cannot be determined unambiguously from the DOZZ formula, but the contour manipulation of the four-point function gives
\be
\label{degdeflimit}
C(\alpha_1,-b/2,Q-\alpha_1 \mp b/2)\equiv \lim_{\delta \to 0}\left[\lim_{\epsilon \to 0} \epsilon\, C(\alpha_1,-b/2+\delta,Q-\alpha_1\mp b/2+\epsilon-\delta)\right].
\ee
This definition agrees with a Coulomb gas computation in free field theory \cite{Zamolodchikov:1995aa}.\footnote{That computation is based on the observation that for the $\alpha_i$'s occuring in this structure constant, the power of $\mu$ appearing in the correlator is either zero or one.  This suggests computing the correlator by treating the Liouville potential as a perturbation of free field theory and then computing the appropriate perturbative contribution to produce the desired power of $\mu$.}  Explicitly, from the DOZZ formula we find
\begin{align} \nonumber
C^+\,_{12}&=-\frac{\pi \mu}{\gamma(-b^2)\gamma(2\alpha_1 b)\gamma(2+b^2-2b\alpha_1)}\\ 
C^-\,_{12}&=1
\end{align}

As shown in \cite{Belavin:1984vu}, by applying the differential equation (\ref{lightdegenerate}) to (\ref{deg4}) we can actually determine $\mathcal{F}_{1234}$ in terms of a hypergeometric function.  This involves using $SL(2,\mathbb{C})$ invariance to transform the partial differential equation (\ref{lightdegenerate}) into an ordinary differential equation, which turns out to be hypergeometric.\footnote{Hypergeometric functions will appear repeatedly in our analysis, so in Appendix \ref{hyps} we present a self-contained introduction.}  The analysis is standard and somewhat lengthy, so we will only present the result:
\be
\label{degblock}
\mathcal{F}_{1234}(\Delta_i,\Delta_\pm,y)=y^{\alpha_\mp}(1-y)^{\beta} F(A_{\mp},B_{\mp},C_{\mp},y),
\ee
with:
\begin{align*}
\Delta_{\pm}&=(\alpha_1\pm b/2)(Q-\alpha_1\mp b/2)\\
\Delta&=-\frac{1}{2}+\frac{3b^2}{4}\\
\alpha_{\mp}&=\Delta_{\pm}-\Delta-\Delta_1\\
\beta&=\Delta_{-}-\Delta-\Delta_3\\
A_{\mp}&=\mp b(\alpha_1-Q/2)+b(\alpha_3+\alpha_4-b)-1/2\\
B_{\mp}&=\mp b(\alpha_1-Q/2)+b(\alpha_3-\alpha_4)+1/2\\
C_{\mp}&=1\mp b(2\alpha_1-Q).
\end{align*}
Using this expression and formula (\ref{Fatinfinity}) from the Appendix, Teschner showed that (\ref{deg4}) will be singlevalued only if the structure constant obeys a recursion relation \cite{Teschner:1995yf}:
\begin{align} \nonumber
&\frac{C(\alpha_3, \alpha_4,\alpha_1+b/2)}{C(\alpha_3,\alpha_4,\alpha_1-b/2)}=-\frac{\gamma(-b^2)}{\pi \mu} \\ \nonumber
&\times\frac{\gamma(2\alpha_1 b)\gamma(2b \alpha_1-b^2)\gamma(b(\alpha_3+\alpha_4-\alpha_1)-b^2/2)}
{\gamma(b(\alpha_1+\alpha_4-\alpha_3)-b^2/2)\gamma(b(\alpha_1+\alpha_3-\alpha_4)-b^2/2)\gamma(b(\alpha_1+\alpha_3+\alpha_4)-1-3b^2/2)}
\end{align}
The reader can check that the DOZZ formula indeed obeys this recursion relation.  In fact, Teschner ran the logic the other way: by combining this recursion relation with a similar one derived from the four-point function with degenerate operator $V_{-\frac{1}{2b}}$, he showed that the DOZZ formula is the unique structure constant that allows both four-point functions to be singlevalued.  In this version of the logic, $C^\pm\,_{12}$ is determined by the Coulomb gas computation rather than the limit (\ref{degdeflimit}) of the DOZZ formula.  This at last justifies the DOZZ formula
 (\ref{dozz}).  

\chapter{Complex Solutions of Liouville's Equation}
\label{liouvillesolutions}
In this section we will describe the most general 
complex-valued solutions of Liouville's equation on  $\sf S^2$ with two or three heavy 
operators present.  The solutions we will present are simple extensions of the real solutions 
given for real $\eta$'s  in \cite{Zamolodchikov:1995aa}.   We will emphasize the new features 
that emerge once complex $\eta$'s are allowed and also establish the uniqueness 
of the solutions.  One interesting issue that will appear for the three-point function is 
 that for many regions of the parameters $\eta_1,\eta_2,
\eta_3$,  there are no nonsingular solutions of Liouville's equation with the desired 
properties, not
even complex-valued ones.  We will determine the analytic forms of the 
singularities that appear and comment on their genericity.

\subsection{General Form of Complex Solutions}\label{genfcs}

\hspace{0.25in}We will first determine the local form of a solution Liouville's equation with flat reference metric:
\be
\partial \overline{\partial}\phi_c=2\lambda e^{\phi_c}.
\ee
We have defined $\lambda=\pi \mu b^2$, which we hold fixed for $b\to 0$  to produce a nontrivial semiclassical limit.  It will be very convenient to parametrize $\phi_c$ in terms of
\be
\label{phifromf}
e^{\phi_c(z,\overline{z})}=\frac{1}{\lambda}\frac{1}{f(z,\overline{z})^2},
\ee
which gives equation of motion
\be
\label{fliouville}
\partial \overline{\partial}f=\frac{1}{f}(\partial f \overline{\partial}f-1).
\ee
There is a classic device  \cite{Poincare} that allows the transformation 
of this partial differential equation 
into two ordinary differential equations, using the 
fact
that the stress tensor (\ref{stresstensor}) is holomorphic.  In particular,
 the holomorphic and antiholomorphic components of the stress tensor are 
 proportional to $W=-{\partial^2 f}/{f}$ and $\tilde W=-{\overline{\partial}^2f}/{f}$ 
 respectively.  We thus have:
\begin{align}\label{monno}
&\partial^2f+W(z)f=0\\ \label{onno}
&\overline{\partial}^2f+\widetilde{W}(\overline{z})f=0
\end{align}
with $W$ and $\tilde W$ holomorphic.
In these equations, we may treat $z$ and $\overline{z}$ independently, 
so we must be able 
to write $f$ locally as a sum of the two linearly independent holomorphic 
solutions of the $W(z)$ equation with coefficients depending only on $\overline{z}$:
\be
f=u(z)\widetilde{u}(\overline{z})-v(z)\widetilde{v}(\overline{z})
\ee
Plugging this ansatz into the $\tilde W$ equation, we see that  $\tilde u$ and $\tilde v$ are
anti-holomorphic solutions of that equation.
Going back to the original Liouville equation, we find:
\be
\label{wronskians}
(u\partial v-v \partial u)(\widetilde{u}\overline{\partial}\widetilde{v}-\widetilde{v}\overline{\partial}\widetilde{u})=1.
\ee
The first factor is a constant since it is the Wronskian evaluated on two solutions of the $W(z)$ equation, and similarly the second factor is constant.  Both
Wronskian factors must be nonzero to satisfy this equation, so $u$ and $v$ are indeed
linearly independent, and similarly $\tilde u$ and $\tilde v$.  So each pair gives a basis
of the two linearly independent holomorphic or antiholomorphic solutions of the appropriate
equation.
 We thus arrive at a general form for any complex solution of Liouville's equation, valid locally as long as the reference metric is $ds^2=dz\otimes d\bar z$:
\be
\label{gensol}
e^{\phi_c}=\frac{1}{\lambda}\frac{1}{(u(z)\widetilde{u}(\overline{z})-v(z)\widetilde{v}(\overline{z}))^2},
\ee
with $u$ and $v$ obeying
\be
\label{uvode}
\partial^2 g+W(z)g=0
\ee
and $\widetilde{u}$ and $\widetilde{v}$ obeying
\be
\label{uvtode}
\overline{\partial}^2 \widetilde{g}+\widetilde{W}(\overline{z})\widetilde{g}=0.
\ee
The representation in (\ref{gensol}) is not quite unique; one can make an arbitrary invertible linear transformation of the pair
$\begin{pmatrix}u\\ v\end{pmatrix}$, with a compensating linear transformation of $\begin{pmatrix}\tilde u & \tilde v\end{pmatrix}$.

To specify a particular solution, we need to choose $W$ and $\widetilde{W}$ and also a basis for the solutions of (\ref{uvode}), (\ref{uvtode}).  These choices are constrained by the boundary conditions, in particular (\ref{infinitephi}) and (\ref{nearops}).  If this problem is undetermined then there are moduli to be integrated over, while if it is overdetermined there is no solution.  

In the following subsections, we we will show what happens explicitly in the special cases of two and three heavy operators on the sphere. But we first make some general comments valid for any number of such operators.  The presence of heavy operators requires the solution $\phi_c$ to be singular at specific points $z_i$.  In terms of $f$, we need
\be
\label{fnearops}
f(z,\overline{z}) \sim |z-z_i|^{2\eta_i} \qquad \qquad \text{as }z \to z_i.
\ee
Looking at the form (\ref{gensol}), there are two possible sources of these singularities.  The first is that at least one of $u$, $v$, $\widetilde{u}$, $\widetilde{v}$ is singular.  The second is that all four functions are nonzero but $u\tilde u - v \tilde v=0$, because of a cancellation
between the two terms.  Assuming that this cancellation happens at a place where none of $u$, $v$, $\widetilde{u}$, $\widetilde{v}$ are singular, we can expand
\be
\label{fnearsing}
f\sim A(z-z_0)+B(\overline{z}-\overline{z}_0)+\O(|z-z_0|^2) \qquad \qquad \text{as } z \to z_0.
\ee
Inserting this into (\ref{fliouville}), 
we find that $AB=1$ and thus $A$ and $B$ are both nonzero.  It thus cannot produce the desired 
behavior (\ref{fnearops}).  

We will have more to say about this type of singularity later, but for now we will focus on 
singularities that occur because some of the functions
are singular.  In order to produce the behavior  
(\ref{fnearops}) from singularities of the individual functions,
$u$ and $v$ must behave as  linear combinations of $(z-z_i)^{\eta_i}$ and $(z-z_i)^{1-\eta_i}$
for $z\to z_i$, with similar behavior for $\tilde u$, $\tilde v$.  To get this behavior,
$W$ and $\tilde W$ must have double poles at $z=z_i$, with suitably adjusted coefficients.
A double pole of $W$ or $\tilde W$ in a differential equation of the form (\ref{monno})
is called a regular singular point.  A double pole is the expected behavior of the stress
tensor at a point with insertion of a primary field.

Moreover, for the solution to be regular at the point at infinity on $\sf S^2$, we need (\ref{infinitephi}), which translates into
\be
f(z,\overline{z}) \sim |z|^2 \qquad \qquad \text{as }|z|\to \infty.
\ee
To achieve this, the two holomorphic solutions of the  differential equation  $(\partial_z^2+W)g=0$
 should behave as $1$ and $z$,
respectively,  near $z=\infty$.  Asking for this equation
to  have a solution of the form $a_1z+a_0+a_{-1}z^{-1}+\dots$ with arbitrary $a_{-1}$ and $a_0$
and no logarithms
in the expansion implies that $W$ vanishes for $z\to\infty$ at least as fast as $1/z^4$.
This is also the expected behavior of the stress tensor in the presence of finitely many
operator insertions on $\R^2$.  Given this behavior, the differential equation again has a regular singular
point at $z=\infty$.

We do not want additional singularities in $W$ or $\tilde W$ as they would lack a physical
interpretation. To be more precise, a pole in $W$ leads to a delta function or derivative of
a delta function in $\bar\partial W$.   Liouville's equation implies that
$\bar\partial W=0 $, and  a delta function correction to that equation implies the existence 
of a delta function source
term in Liouville's equation -- that is, an operator insertion of some kind.

Thus for a finite number of operator insertions, $W$ and $\widetilde{W}$ have only finitely 
many poles, all of at most second order.  In particular, $W$ and $\tilde W$
 are rational functions.  The parameters of these rational functions must be adjusted to 
 achieve the desired behavior  near operator insertions and at infinity.  We now study 
 this problem in the special cases of two or three heavy operators.

\subsection{Two-Point Solutions}\label{tps}

\hspace{0.25in}Specializing to the case of two operators, 
$W$ should have two double poles and should vanish as $1/z^4$ for $z\to \infty$;
$\tilde W$ should be similar.  Up to  constant multiples, 
these functions are determined by the positions of the poles:
\begin{align}\notag
&W(z)=\frac{w(1-w)z_{12}^2}{(z-z_1)^2(z-z_2)^2}\\
&\widetilde{W}(\overline{z})=\frac{\widetilde{w}(1-\widetilde{w})\overline{z}_{12}^2}{(\overline{z}-\overline{z}_1)^2(\overline{z}-\overline{z}_2)^2}.
\end{align}
We have picked a convenient parametrization of the constants.
In this case, the ODE's can be solved in terms of elementary functions. A particular basis of solutions is 
\begin{align}\notag
&g_1(z)=(z-z_1)^w (z-z_2)^{1-w}\\ \notag
&g_2(z)=(z-z_1)^{1-w}(z-z_2)^{w}\\ \notag
&\widetilde{g}_1(\overline{z})=(\overline{z}-\overline{z}_1)^{\widetilde{w}}(\overline{z}-\overline{z}_2)^{1-\widetilde{w}}\\
&\widetilde{g}_2(\overline{z})=(\overline{z}-\overline{z}_1)^{1-\widetilde{w}}(\overline{z}-\overline{z}_2)^{\widetilde{w}}.
\end{align}
It remains to determine $w$ and $\widetilde{w}$ in terms of $\eta_1$ and $\eta_2$ and to write the $u$'s and $v$'s in terms of this basis.  In doing this we need to make sure that (\ref{fnearops}) is satisfied, and also that the product of the Wronskians obeys (\ref{wronskians}).  Up to trivial redefinitions, the result is that we must have $\eta_1=\eta_2=w=\widetilde{w}\equiv \eta$, also having
\begin{align}
&u(z)=g_1(z)\\ \notag
&v(z)=g_2(z)\\ \notag
&\widetilde{u}(\overline{z})=\kappa\widetilde{g}_1(\overline{z})\\ \notag
&\widetilde{v}(\overline{z})=\frac{\widetilde{g}_2(\overline{z})}{\kappa (1-2\eta)^2 |z_{12}|^2}
\end{align}
This leads to the solution
\be
\label{2pointsol}
e^{\phi_c}=\frac{1}{\lambda} \frac{1}{\left(\kappa |z-z_1|^{2\eta} |z-z_2|^{2-2\eta}-\frac{1}{\kappa}\frac{1}{(1-2\eta)^2 |z_{12}|^2}|z-z_1|^{2-2\eta}|z-z_2|^{2\eta}\right)^2}.
\ee
The criterion $\eta_1=\eta_2$ is expected, since in conformal field theory, the two-point function for operators of distinct conformal weights
always vanishes.   $\kappa$ is an arbitrary complex number, but it is slightly constrained if we impose as a final condition that $f$ be nonvanishing away from the operator insertions. 
The denominator  in (\ref{2pointsol}) can vanish only if $\kappa$ lies on a certain real curve $\ell$  in the complex plane
($\ell$ is simply the real axis if $\eta$ is real).  Omitting the curve $\ell$ from the complex $\kappa$ plane, and taking
into account the fact that the sign of $\kappa$ is irrelevant, we get a moduli space of solutions that has complex dimension
one and that 
 as a complex manifold is a copy of the upper half-plane $\sf H$.

Returning now to the general form (\ref{2pointsol}), we will make two comments:
\begin{itemize}
\item[(i)] Suppose that $\eta$ is real.  To avoid singularities, we cannot have $\kappa$ be real, but we can instead choose it to be purely imaginary. $e^{\phi_c}$ will then be real but negative definite, and $\phi_c$ will be complex.  Nonetheless this situation still has a simple geometric interpretation: we can define a new metric $-e^{\phi_c}\delta_{ab}$, which is indeed a genuine metric on the sphere, and because of the sign change it has \textit{positive} curvature!  It has two conical singularities, and for positive $\eta's$ it describes the intrinsic geometry of an American football.  This observation is a special case of a general bijection between saddle points of spacelike and timelike Liouville, which will
be explored later.
\item[(ii)] Eqn. (\ref{2pointsol}) gives the most general form of $e^{\phi_c}$, but this leaves the possibility of adding to $\phi_c$
itself an integer multiple of $2\pi i$, as in eqn. (\ref{doft}).  Thus the moduli space of solutions has many components and is isomorphic
to $\sf H\times\Z$.
\end{itemize}

\subsection{Three-Point Solutions}\label{threp}

\hspace{0.25in}For the case of three heavy operators, the potentials $W$ and $\widetilde{W}$ must now be rational functions with three double poles.  Their behaviour at infinity determines them up to quadratic polynomials in the numerator, which we can further restrict by demanding the correct singularities of $u$, $v$, $\widetilde{u}$, and $\widetilde{v}$ at the operator insertions.  There will be a new challenge, however;  while we can
easily  choose a basis of solutions of (\ref{uvode}) and (\ref{uvtode})  
with the desired behavior near any one singular point, it is nontrivial to arrange to get the right behavior at all three singular points.

Insisting that the residues of the poles in $W$ and $\widetilde{W}$ have the correct forms to produce (\ref{fnearops}) leads to unique expressions for $W$ and $\widetilde{W}$: 
\begin{align} \nonumber \label{olgo}
&W(z)=\left[\frac{\eta_1(1-\eta_1)z_{12}z_{13}}{z-z_1}+\frac{\eta_2(1-\eta_2)z_{21}z_{23}}{z-z_2}+\frac{\eta_3(1-\eta_3)z_{31}z_{32}}{z-z_3}\right]\frac{1}{(z-z_1)(z-z_2)(z-z_3)}\\
&\widetilde{W}(\overline{z})=\left[\frac{\eta_1(1-\eta_1)\overline{z_{12}}\,\overline{z_{13}}}{\overline{z}-\overline{z}_1}+\frac{\eta_2(1-\eta_2)\overline{z_{21}}\,\overline{z_{23}}}{\overline{z}-\overline{z}_2}+\frac{\eta_3(1-\eta_3)\overline{z_{31}}\,\overline{z_{32}}}{\overline{z}-\overline{z}_3}\right]\frac{1}{(\overline{z}-\overline{z}_1)(\overline{z}-\overline{z}_2)(\overline{z}-\overline{z}_3)}
\end{align}
With these potentials, the differential equation of interest becomes essentially the hypergeometric equation, modulo an elementary
normalization.  So the solutions can be expressed in terms of hypergeometric functions, or equivalently, but slightly more
elegantly, in terms of Riemann's $P$ functions.\footnote{In Appendix \ref{hyps}, we give a self-contained development of the minimum facts
we need concerning hypergeometric and $P$-functions.  The reader unfamiliar with these functions is encouraged to read this appendix now.}  $P$ functions are solutions of a differential equation with three regular singularities at specified points, and with no singularity at infinity.  The equations (\ref{uvode}) and (\ref{uvtode}) are not quite of this form since they do have a regular singular point at infinity, but we can recast them into Riemann's form by defining $g(z)=(z-z_2)h(z)$ and $\widetilde{g}(\overline{z}) = (\overline{z}-\overline{z}_2) \widetilde{h}(\overline{z})$.  One can check that the equations obeyed by $h$ and $\widetilde{h}$ are special cases of Riemann's equation \ref{rde}, with the parameters\footnote{The unpleasant asymmetry of the second line follows from the definition of $h$, but a symmetric definition introduces significant complication in the formulas that follow so we will stay with this choice.}
\begin{align}\label{riempar}
\alpha&=\eta_1 &\alpha'&=1-\eta_1\\  \notag
\beta&=-\eta_2  &\beta'&=\eta_2-1\\   \notag
\gamma&=\eta_3  &\gamma'&=1-\eta_3.
\end{align}
We now observe that the boundary conditions (\ref{fnearops}) ensure that without loss of generality we can choose $u$, $v$, $\widetilde{u}$, and $\widetilde{v}$ to diagonalize the monodromy about any particular singular point, say $z_1$. Also picking a convenient normalization
of these functions, we have
\begin{align}
&u(z)=(z-z_2)P^{\eta_1}(x)\\  \notag
&v(z)=(z-z_2)P^{1-\eta_1}(x)\\ \notag
&\widetilde{u}(\overline{z})=a_1 (\overline{z}-\overline{z_2})P^{\eta_1}(\overline{x})\\ \notag
&\widetilde{v}(\overline{z})=a_2 (\overline{z}-\overline{z_2})P^{1-\eta_1}(\overline{x})
\end{align}
Here $a_1$,$a_2$ are complex numbers to be determined, $x ={z_{23}(z-z_1)}/{z_{13}(z-z_2)}$, and the $P$ functions explicitly are related to
hypergeometric functions by
\begin{align}
\label{3pointPfunctions}\notag
P^{\eta_1}(x)&=x^{\eta_1}(1-x)^{\eta_3}F(\eta_1+\eta_3-\eta_2,\eta_1+\eta_2+\eta_3-1,2\eta_1,x)\\ 
P^{1-\eta_1}(x)&=x^{1-\eta_1}(1-x)^{1-\eta_3}F(1-\eta_1+\eta_2-\eta_3,2-\eta_1-\eta_2-\eta_3,2-2\eta_1,x).
\end{align}
We can determine the product $a_1 a_2$ by imposing (\ref{wronskians}); by construction we know that $u\partial v-v \partial u$ and $\widetilde{u} \overline{\partial} \widetilde{v}-\widetilde{v} \overline{\partial} \widetilde{u}$ are both constant, so to make sure their product is 1 it is enough to demand it in the vicinity of $z=z_1$.  This is easy to do using the series expansion for the hypergeometric function near $x=0$, leading to
\be
\label{aproduct}
a_1 a_2=\frac{|z_{13}|^2}{|z_{12}|^2 |z_{23}|^2(1-2\eta_1)^2}.
\ee
It is clear from the above formulas that $f=u\tilde u-v\tilde v$ is singlevalued about $z=z_1$.  
For this to also be true near $z_2,z_3$ is a non-trivial constraint, which can be evaluated using 
 the connection formulas (\ref{connection1}).  For example,
\begin{align} \nonumber
f &|z-z_2|^{-2}=a_1P^{\eta_1}(x)P^{\eta_1}(\overline{x})-a_2 P^{1-\eta_1}(x)P^{1-\eta_1}(\overline{x})\\\nonumber
\end{align}
will be singlevalued near $z=z_3$, which corresponds to $x=1$, only if
\be
\label{aratio}
a_1a_{\eta_1,\eta_3}a_{\eta_1,1-\eta_3}=a_2 a_{1-\eta_1,\eta_3} a_{1-\eta_1,1-\eta_3}.
\ee
The connection coefficients $a_{ij}$ are given by (\ref{a13}), so combining this with (\ref{aproduct}) we find
\be
\label{a1squared}
(a_1)^2=\frac{|z_{13}|^2}{|z_{12}|^2|z_{23}|^2}\frac{\gamma(\eta_1+\eta_2-\eta_3)\gamma(\eta_1+\eta_3-\eta_2)\gamma(\eta_1+\eta_2+\eta_3-1)}{\gamma(2\eta_1)^2\gamma(\eta_2+\eta_3-\eta_1)}
\ee
Thus both $a_1$ and $a_2$ are determined (up to an irrelevant overall sign) , so the solution is completely determined.  The reader can check that with the ratio given by (\ref{aratio}) the solution is also singlevalued near $z_2$. This is a nontrivial computation using (\ref{a12}),
 but it has to work, since  the absence of monodromy around $z_1$, $z_3$, and $\infty$ implies that there must also be none around $z_2$.

The final form of the solution is thus
\be
\label{3pointsol}
e^{\phi_c}=\frac{1}{\lambda} \frac{|z-z_2|^{-4}}{\left[a_1 P^{\eta_1}(x)P^{\eta_1}(\overline{x})-a_2 P^{1-\eta_1}(x) P^{1-\eta_1}(\overline{x})\right]^2}.
\ee
In the end, this is simply the analytic continuation in $\eta_i$ of the real solution presented in \cite{Zamolodchikov:1995aa}, but our argument has established its uniqueness.

There is still a potential problem with the solution.  The coefficients $a_1$,$a_2$ were completely determined without any reference to avoiding cancellations between the terms in the denominator, and it is not at all clear that the denominator has no zeroes for generic $\eta$'s.  It is difficult to study the existence of such cancellations analytically for arbitrary $\eta$'s, but we have shown numerically that they indeed happen for generic complex $\eta$'s.  If we assume that such a singularity is present at $z=z_0$, then we saw above that its analytic form is given by (\ref{fnearsing}).

For real $\eta$'s, we can say  more. When the $\eta$'s are real, the right hand side of   (\ref{a1squared}) is real so $a_1$ is either  real or  imaginary.  If it is imaginary, then (\ref{aproduct}) shows that $a_2$ will also be  imaginary and with opposite sign for its imaginary part.  Moreover for real $\eta$'s,  $P^{\eta_1}(x)P^{\eta_1}(\overline{x})$ and $P^{1-\eta_1}(x) P^{1-\eta_1}(\overline{x})$ are strictly positive.  Thus when $a_1$ is purely imaginary, both terms in the denominator have the same phase and there can be no singularities arising from cancellation.  The metric $e^{\phi_c}\delta_{ab}$  will however be negative definite, so this will be a complex saddle point for $\phi_c$.  If we start with such $\eta$'s and allow them to have small imaginary parts then cancellations do not appear at once, but we find numerically that if we allow the imaginary parts to become large enough then cancellations in the denominator do occur.  

We can also consider the case that the $\eta$'s are real and $a_1$ is also real.  $a_2$ will then be real and with the same sign as $a_1$, so cancellations are now possible.  
We learned in section \ref{cpo} that real solutions can only occur if certain inequalities (\ref{seiberg}) and (\ref{gbconstraint}) are satisfied.
So if the $\eta$'s are real but violate the inequalities, the denominator in (\ref{3pointsol}) definitely vanishes somewhere.
On the other hand, if the $\eta$'s are real and satisfy the inequalities, then a real metric of constant negative curvature
corresponding to a real solution of Liouville's equations does exist.  It can be constructed by gluing together two hyperbolic
triangles, or in any  number of other ways.  So in this case, the denominator in (\ref{3pointsol}) is positive definite
away from the operator insertions.\footnote{We show this explicitly below in Appendix \ref{hypintegrals}.}  
This is the region studied in  \cite{Zamolodchikov:1995aa}.

We conclude with two remarks about the nature of these singularities near a zero of the denominator in the formula for $e^{\phi_c}$.  We first observe that the singularities naturally come in pairs since the denominator of (\ref{3pointsol}) is symmetric under exchanging $x$ and $\bar{x}$, so for example if we choose the $z_i$ to be real then the solution is symmetric under reflection across the real $z$-axis.  We secondly comment on the stability of these singularities:
the general local expansion (\ref{fnearsing}) near a zero involves two complex coefficients $A$ and $B$.   When these are of unequal magnitude,
the existence of a zero of $f$ is stable under small perturbations.  This is because one can associate to an isolated zero
of the complex function $f$ an integer-valued invariant, the winding number.  To define it, set $f=s e^{i\psi}$ where
$s$ is a positive function and $\psi$ is real.  Supposing that $f$ has an isolated zero at $z=z_0$, consider $e^{i\psi}$ as
a function defined on the circle $z=z_0+\epsilon e^{i\theta}$, for some small positive $\epsilon$ and real $\theta$.  The winding number is
defined as $\frac{1}{2\pi}\oint_0^{2\pi}d\theta \,d\psi/d\theta$, and is invariant under small changes in $f$.  (If $f$ is varied so that several zeroes
meet, then only the sum of the  winding numbers is invariant, in general.)  In the context of (\ref{fnearsing}), the winding number
is 1 for $|A|>|B|$, and $-1$ for $|A|<|B|$, and depends on higher terms in the expansion if $|A|=|B|$.  
In the case of a zero of the denominator in (\ref{3pointsol}), one generically has $|A|\not=|B|$ if the $\eta$'s are complex,
so isolated singularities arising by this mechanism are stable against small perturbations.  When the $\eta$'s are real,
the behavior near singularities of this type requires more examination.

\chapter{Analytic Continuation and Stokes Phenomena}\label{anastokes}
In this section, we use the complex classical solutions constructed in the previous section to interpret the analytic continuation first of the two-point function (\ref{spacelike2point}) and then of the three-point function as given by the DOZZ formula (\ref{dozz}). 
We will find that for the two-point function there is a satisfactory picture in terms of complex saddle points, which agrees with and we believe improves on the old fixed-area results in the semiclassical approximation.  For the three-point function we will find that the situation is more subtle; we will be able to ``improve'' on the fixed-area result here as well, but to understand the full analytic continuation we will need to confront the singularities at which the denominator of the solution vanishes.  For ease of presentation we postpone our discussion of those singularities until section \ref{4pointsection}, and we here focus only on the part of the analytic continuation that avoids them. We also include the case of three light operators as check at the end of the section.

\subsection{Analytic Continuation of the Two-Point Function}\label{atpf}

\hspace{0.25in}We saw in section \ref{4pointreview} that the DOZZ formula implies that the Liouville two-point function takes the form
\begin{align}
\label{exact2point} \nonumber
\langle V_{\alpha}(z_1,\overline{z}_1) &V_\alpha(z_2,\overline{z}_2)\rangle=\\
&|z_{12}|^{-4\alpha (Q-\alpha)}\frac{2\pi}{b^2}\left[\pi \mu \gamma(b^2)\right]^{(Q-2\alpha)/b} \gamma(2\alpha/b-1-1/b^2)\gamma(2 b \alpha-b^2)\delta(0)
\end{align}
The factor of $\delta(0)$ is a shorthand which reflects the continuum normalization of the operators with $\alpha=\frac{Q}{2}+iP$ and the
fact that we have taken the two fields in (\ref{exact2point}) to have the same Liouville momentum.  It may seem unphysical to study the analytic continuation of a divergent quantity, but as we will review, the divergence has a simple semiclassical origin that is independent of $\alpha$.\footnote{Indeed if we were to use the Liouville theory as part of a gravity theory where conformal symmetry is gauged, then to compute a two-point function of integrated vertex operators we would partially fix the gauge by fixing the positions of the two operators and then divide by the volume of the remaining conformal symmetries.  This would remove this divergent factor.}  

This ``exact'' result for the two-point function does not come from a real Liouville path integral, even if $\alpha$ is real.  One can easily
show that, for the two-point function on $\sf S^2$, the path integral over real Liouville
fields does not converge. Consider a smooth real field configuration  that obeys the boundary conditions (\ref{infinitephi}) and (\ref{nearops}).  The modified action (\ref{regaction}) will be finite.  Now consider adding a large negative real number $\Delta\phi_c$ to $\phi_c$.  The kinetic term will be unaffected and the exponential term will become smaller in absolute value, but the boundary terms will add an extra term $\Delta\phi_c(1-2\eta)$.  Recalling that we always choose the Seiberg bound to be satisfied, we see that by taking $\Delta \phi_c$ to be large and negative we can thus make the action as negative as we wish.  The path integral therefore cannot converge as an integral over real $\phi_c$'s \cite{Seiberg:1990eb}.\footnote{This divergence should not be confused with the factor of $\delta(0)$, which we will see has to do with an integral over a noncompact subgroup of $SL(2,\mathbb{C})$.  In particular we can make the same argument for the three-point function with three real $\alpha$'s and find the same divergence if $\sum_i \alpha_i<Q$, and since the DOZZ formula does not have any $\delta(0)$ it is clear that this is a different issue \cite{Seiberg:1990eb}.}

The original approach to resolve this divergence, proposed in \cite{Seiberg:1990eb}, was to  restrict the path integral only to field configurations obeying $\int d^2\xi e^{\phi_c}=A$.  This clearly avoids the divergence.  
However, if one tries to integrate over $A$, one would get back the original divergence, while on the other hand if one simply keeps $A$
fixed, one would not expect to get a local quantum field theory. As an alternative proposal, we claim that (\ref{exact2point}) is computed by a local path integral over a complex integration cycle. This is analogous to the suggestion \cite{GHP} of dealing with a somewhat similar
divergence in the path integral of Einstein gravity by Wick rotating the conformal factor of the metric to complex values.
To motivate our proposal,  we will show that the semiclassical limit of (\ref{exact2point}), with $\alpha$ scaling as $\eta/b$, is reproduced by a sum over the complex saddle points with two heavy operators that we constructed in section \ref{tps}.  We interpret this as suggesting that the path integral is evaluated over a cycle that is a sum of cycles attached to complex saddle points, as sketched in section \ref{ancon}.
The requisite sum is an infinite sum, somewhat like what one finds for the Gamma function for $\mathrm{Re}\,z<0$, as described in Appendix
\ref{gammastokes}. 
 We will also find that the set of contributing saddle points jumps discontinuously as $\eta$ crosses the real axis.  This again parallels
a result for the Gamma function, and we interpret it as a Stokes phenomenon.

\subsubsection{Evaluation of the Action for Two-Point Solutions}\label{estwo}
In computing the action of the two-point solution (\ref{2pointsol}), we first need to deal with taking the logarithm to get $\phi_c$.  The branch cut in the logarithm makes this a nontrivial operation.  To make the following manipulations simpler, we will here relabel $\kappa=i\widetilde{\kappa}$, so the solution becomes
\be
\label{2pointsolt}
e^{\phi_c}=-\frac{1}{\lambda\widetilde{\kappa}^2} \frac{1}{\left(|z-z_1|^{2\eta} |z-z_2|^{2-2\eta}+\frac{1}{\widetilde{\kappa}^2 (1-2\eta)^2 |z_{12}|^2}|z-z_1|^{2-2\eta} |z-z_2|^{2\eta}\right)^2}.
\ee
We choose $\tilde\kappa$ to ensure that the denominator has no zeroes.  Since we are imposing the Seiberg bound, we have $\mathrm{Re}(1-2\eta)>0$.  There is a sign choice in defining $\widetilde{\kappa}$, so we will choose it to have positive real part.  In particular note that if $\eta$ is real then we can have $\widetilde{\kappa}$ be real and positive.  Our prescription for taking the logarithm will then be
\begin{align} \nonumber
\phi_{c,N}(z,\overline{z})=&i \pi +2\pi i N-\log \lambda-2 \log {\widetilde{\kappa}}\\ \label{phic2}
&-2\log \left(|z-z_1|^{2\eta} |z-z_2|^{2-2\eta}+\frac{1}{\widetilde{\kappa}^2 (1-2\eta)^2 |z_{12}|^2}|z-z_1|^{2-2\eta} |z-z_2|^{2\eta}\right)
\end{align}    
The choice of branch for the final logarithm is inessential, in the sense that making a different choice would be
equivalent to shifting the integer $N$ in (\ref{phic2}).
  We will choose the branch  such that the final logarithm  behaves like $-4\eta \log |z-z_1|+(4\eta-4)\log|z_{12}|$ near $z_1$.  Its value away from $z_1$ is defined by continuity; there is no problem in extending this logarithm throughout the $z$-plane (punctured at $z_1$ and $z_2$).\footnote{Because of the boundary conditions (\ref{fnearops}), there cannot be monodromy of this logarithm about $z_1,z_2$ even though its argument vanishes there.}
We will have no such luck for the three-point function; in that case, zeroes of the logarithm are essential.  

  We will see momentarily that to compute the action, we need to know the leading behaviour near $z_1$ and $z_2$, so we observe that
\begin{align} 
&\phi_{c,N}(z,\overline{z}) \to -4\eta\log |z-z_i|+C_i \qquad \text{as} \,\,z\to z_i,
\end{align}
with
\begin{align} \nonumber
&C_1=2\pi i \left(N+\frac{1}{2}\right)-\log \lambda-2\log{\widetilde{\kappa}}+(4\eta-4)\log |z_{12}|\\
&C_2=2\pi i \left(N+\frac{1}{2}\right)-\log \lambda+2\log{\widetilde{\kappa}}+4\eta \log |z_{12}|+4\log (1-2\eta).
\end{align}
To verify\footnote{We thank X. Dong for a discussion of this point and for suggesting the following line of argument.}
 that the same integer $N$ appears in both $C_1$ and $C_2$, we note that this is clear for real $\eta$ and $\tilde\kappa$,
since then the final logarithm in (\ref{phic2}) has no imaginary part; in general it then follows by continuity.

Now to compute the modified action (\ref{regaction}), we use a very helpful trick from \cite{Zamolodchikov:1995aa}.  This is to compute ${d\widetilde{S}_L}/{d\eta}$ when $\widetilde{S}_L$ is evaluated on a saddle point.  {\it A priori}, there would be $\eta$ dependence both implicitly through the functional form of the saddle point and explicitly through the boundary terms in $\widetilde{S}_L$,  but the variation of (\ref{regaction}) with respect to $\phi_c$ is zero when evaluated on a solution and only the explicit $\eta$-dependence matters.  We thus have the remarkably simple equation:
\begin{align}\nonumber \label{dsdeta}
b^2\frac{d\widetilde{S}_L}{d\eta}&=-C_1-C_2\\
&=-2\pi i (2N+1)+2\log \lambda+(4-8\eta)\log |z_{12}|-4 \log (1-2\eta)
\end{align}
We can thus determine $\widetilde{S}_L[\phi_{c,N}]$ up to a constant by integrating this simple function, and we can determine the constant by comparing to an explicit evaluation of the action when $\eta=0$.  
When $\eta$ is zero, the saddle point (\ref{phic2}) becomes an $SL(2,\mathbb{C})$ transformation of a metric which is just minus the usual round sphere
\be
\label{sphere}
\phi_c=i\pi+2\pi i N-\log \lambda-2\log(1+z\overline{z}).
\ee
For this solution we can evaluate the action (\ref{regaction}) explicitly, finding $b^2 \widetilde{S}_0=2\pi i (N+\frac{1}{2})-\log\lambda-2$.  Now doing the integral of (\ref{dsdeta}) our final result for the action (\ref{regaction}) with nonzero $\eta$ is thus
\begin{align} \nonumber
b^2 \widetilde{S}_L=&2\pi i (N+1/2)(1-2\eta)+(2\eta-1)\lambda+4(\eta-\eta^2)\log |z_{12}|\\
&+2\left[(1-2\eta)\log{(1-2\eta)}-(1-2\eta)\right]. \label{2action}
\end{align}
We can observe immediately that the $z_{12}$ dependence is consistent with the two-point function of a scalar operator of weight $(\eta-\eta^2)/b^2$.  This action is independent of $\widetilde{\kappa}$, so when we integrate over it this will produce a divergent factor, which we interpret as the factor $\delta(0)$ in (\ref{exact2point}).  

Before moving on to the exact expression, we we will observe here that this action is multivalued as a function of $\eta$, with
a branch point emanating from $\eta=1/2$, where the original solution (\ref{2pointsolt}) is not well-defined.  
Under monodromy around this point, $N$ shifts by 2, so all even and likewise
all odd values of $N$ are linked by this monodromy.  Of course, to see the monodromy, we have to 
 consider paths in the $\eta$ plane that violate the Seiberg bound $\mathrm{Re}(\eta)<\frac{1}{2}$. 

\subsubsection{Comparison with Limit of Exact Two-Point Function}\label{compex}
We now compute the semiclassical asymptotics of (\ref{exact2point}).  We can easily find that
\be
\langle V_{\alpha}(z_1,\overline{z}_1) V_\alpha(z_2,\overline{z}_2)\rangle \sim \delta(0) |z_{12}|^{-4\eta(1-\eta)/b^2}\lambda^{(1-2\eta)/b^2} \left[\frac{\gamma(b^2)}{b^2}\right]^{(1-2\eta)/b^2}\gamma\left(\frac{(2\eta-1)}{b^2}\right)
\ee
The first three factors obviously match on to the result (\ref{2action}) that we found in the previous section, but the last two have more subtle semiclassical limits.  It is not hard to see that the factor involving $\gamma(b^2)$ is asymptotic for small positive $b$ to  $\exp\left\{-\frac{4(1-2\eta)\log b}{b^2}\right\}$, but to understand the final factor, we need to understand the asymptotics of the $\Gamma$ function at large complex values of its argument.  For real positive arguments, this is the well-known Stirling approximation, but for complex
arguments, the situation is more subtle:
\be\label{gamma}
\Gamma(x)= 
\begin{cases}
e^{x \log x-x+\O(\log x)} \qquad \qquad\qquad \quad \,\,\,\mathrm{Re}(x)>0\\
\frac{1}{e^{i\pi x}-e^{-i\pi x}}e^{x \log (-x)-x+\O(\log(-x))} \qquad \mathrm{Re}(x)<0
.\end{cases}
\ee
This result 
can be obtained in a variety of ways; because of the fact (see \cite{Cur,Polchinski:1990mh,Seiberg:1990eb} and section \ref{minis}) that the integral representation of the Gamma function
is a minisuperspace approximation to Liouville theory, we present in Appendix \ref{gammastokes} a derivation using the machinery
of critical points and Stokes lines. 
Using (\ref{gamma}), we find
\be
\gamma\left(\frac{(2\eta-1)}{b^2}\right)\sim \frac{1}{e^{i\pi (2\eta-1)/b^2}-e^{-i\pi (2\eta-1)/b^2}}\exp\left[\frac{(4\eta-2)}{b^2}\left(\log (1-2\eta)-2\log b-1\right)\right].
\ee
So we can write the semiclassical limit as
\begin{align}\nonumber
\langle V_{\alpha}(z_1,\overline{z}_1) &V_\alpha(z_2,\overline{z}_2)\rangle \sim \delta(0) |z_{12}|^{-4\eta(1-\eta)/b^2}\lambda^{(1-2\eta)/b^2}\\ 
&\times e^{-\frac{2}{b^2}\left[(1-2\eta)\log(1-2\eta)-(1-2\eta)\right]} \frac{1}{e^{i\pi (2\eta-1)/b^2}-e^{-i\pi (2\eta-1)/b^2}}.
\end{align}
All factors now clearly match (\ref{2action}) except for the last.  
To complete the argument, setting $y=e^{i\pi (2\eta-1)/b^2}$, we need to know that the function $1/(y-y^{-1})$ can be expanded in two
ways:
\be\label{twox}\frac{1}{y-y^{-1}}=\sum_{k=0}^\infty y^{-(2k+1)}= -\sum_{k=0}^\infty y^{2k+1}.  \ee
One expansion is valid for $|y|>1$ and one for $|y|<1$.  So either way, there is a set $T$ of integers with   
\be \label{wox}
\frac{1}{e^{i\pi (2\eta-1)/b^2}-e^{-i\pi (2\eta-1)/b^2}}=\pm \sum_{N\in T} e^{2\pi i (N \mp 1/2)(2\eta-1)/b^2}.
\ee
$T$ consists of nonnegative integers if $\mathrm{Im}\,((2\eta-1)/b^2>0$ and of nonpositive ones if $\mathrm{Im}\,((2\eta-1)/b^2<0$.    We have
to interpret the line $\mathrm{Im}\,((2\eta-1)/b^2)=0$ as a Stokes line along which the  representation of the integration cycle as a sum 
of cycles associated to critical points changes discontinuously.  If $b$ is real, the criterion
simplifies and only depends on the sign of $\mathrm{Im}\,\eta$. The sign in (\ref{wox}) has an analog for the Gamma function and can
be interpreted in terms of the orientations of critical point cycles.  

\subsubsection{Relationship to Fixed-Area Results}
We will now briefly discuss how to relate this point of view to the more traditional fixed-area technique \cite{Seiberg:1990eb}.  For this section we restrict to real $\alpha$'s.  We begin by defining the fixed-area expectation value for a generic Liouville correlator as
\be
\label{fixedarea}
\langle V_{\alpha_1}\cdots V_{\alpha_n}\rangle_A \equiv (\mu A)^{(\sum_i\alpha_i-Q)/b} \frac{1}{\Gamma\left((\sum_i\alpha_i-Q)/b\right)}\langle V_{\alpha_1}\cdots V_{\alpha_n}\rangle.
\ee
Assuming that $\mathrm{Re}(\sum_i \alpha_i-Q)>0$, an equivalent formula is
\be
\label{areaint}
\langle V_{\alpha_1}\cdots V_{\alpha_n}\rangle=\int_0^\infty \frac{dA}{A}e^{-\mu A}\langle V_{\alpha_1}\cdots V_{\alpha_n}\rangle_A.
\ee
With the $A$ dependence of $\langle V_{\alpha_1}\cdots V_{\alpha_n}\rangle_A $ being the simple power of $A$ given on the right hand side of
(\ref{fixedarea}), the $A$ integral in (\ref{areaint}) can be performed explicitly, leading back to (\ref{fixedarea}).
So far this is just a definition, but comparison of (\ref{areaint}) to the original Liouville path integral suggests an alternate proposal for how to compute the fixed-area expectation value: evaluate the Liouville path integral dropping the cosmological constant term and explicitly fixing the physical area $\int d^2 \xi e^{\phi_A}=A$.  Semiclassically we can do this using a Lagrange multiplier,\footnote{In eqn. (\ref{pox}), we set the Lagrange multiplier
to the value at which the equation has a solution. To find this value, one integrates
over the $z$-plane, evaluating  the integral of the left hand side with the help of  (\ref{logo}).} which modifies the equation of motion:
\be\label{pox}
\partial \overline{\partial}\phi_A=\frac{2\pi}{A}(\sum_i \eta_i-1)e^{\phi_A}-2\pi \sum \eta_i \delta^2(\xi-\xi_i).
\ee
The point to notice here is that when $\sum_i \eta_i<1$, if we define $\phi_{c,N}=i\pi +2\pi i N+\phi_A$ and $\lambda=(\sum_i \eta_i-1)/A$, the solutions of this equation are mapped exactly into the complex saddle points we have been discussing.  One can check explicitly for the semiclassical two-point function we just computed that the various factors on the right hand side of (\ref{fixedarea}) conspire to remove the evidence of the complex saddle points and produce the usual fixed-area result \cite{Zamolodchikov:1995aa}:
\be
\langle V_{\eta/b}(1,1) V_{\eta/b}(0,0)\rangle_A \equiv 2\pi \delta(0)G_A(\eta/b)\approx 2\pi \delta(0) e^{-\frac{1}{b^2}(1-2\eta)(\log \frac{A}{\pi}+\log (1-2\eta)-1)}.
\ee
Historically the proposal was to use (\ref{fixedarea}) in the other direction, as a way to 
define the Liouville correlator when $\sum_i\eta_i<1$, but it was unclear that this would 
be valid beyond the semiclassical approximation.  We see now how it emerges naturally 
from the analytic continuation of the Liouville path integral.

\subsection{Analytic Continuation of the Three-Point Function}
\label{3pointcontinuation}

\hspace{0.25in}We now move on to the three-point function.  We will initially focus on two particular 
regions of the parameter space of the variables $\eta_i$, $i=1,\dots,3$.  In what we will 
call Region I, we require
that  $\sum_i \mathrm{Re}(\eta_i)>1$, and that the imaginary parts  $\mathrm{Im}(\eta_i)$ 
are small enough that the solution (\ref{3pointsol}) does not have singularities 
coming from  zeroes
of the denominator.  The inequality  $\sum_i \mathrm{Re}(\eta_i)>1$ is needed to
prevent the path integral over $\phi_c$ from diverging at large negative $\phi_c$, 
as discussed above in the context of the two-point function.  When the $\eta_i$ are 
actually real and less than $1/2$, we get the physical region studied in  
\cite{Zamolodchikov:1995aa}, which is the only range 
of $\eta_i$ in which Liouville's equation has real
nonsingular solutions. 
In this sense the three-point function is a simpler case than the two-point function, since in that 
case no choice of $\eta$ allowed a real integration cycle for the path integral.  

We will also be interested in the region defined by 
\begin{align} \nonumber
&0<\mathrm{Re}(\eta_i)<\frac{1}{2}\\ \label{inequalities}
&\sum_i\mathrm{Re}(\eta_i)<1\\ \nonumber
&0<\mathrm{Re}(\eta_i+\eta_j-\eta_k) \qquad (i\neq j\neq k),
\end{align}
where again the imaginary parts are taken to be small enough that there are no  singularities from
zeroes of the denominator.  We will refer to this  as Region II.  Note that if the imaginary parts are all zero, we can see from (\ref{aproduct}) and (\ref{a1squared}) that $a_1$ and $a_2$ will be purely imaginary in this region and there will be no singularities.  The third line of (\ref{inequalities}) has not appeared before in our discussion; we call it the triangle inequality.  Its meaning is not immediately clear.  It is automatically satisfied when $\sum_i \mathrm{Re}(\eta_i)>1$ and $\mathrm{Re}(\eta_i)<\frac{1}{2}$, but when $\sum_i\mathrm{Re}(\eta_i)<1$ it becomes a nontrivial additional constraint.  

To get some intuition about this constraint, recall that in Region II with real $\eta$'s, $a_1$, $a_2$ are imaginary.  The metric $-e^{\phi_c} \delta_{ab}$ is thus well defined and has constant positive curvature.  Since we have taken the $\eta_i$ to be positive, the metric has three conical deficits.  Such metrics have been studied in both the physics and math literature \cite{Frolov:2001uf,umehara}, and they can be constructed geometrically in the following way.  Suppose
that we can construct a geodesic triangle on $\sf S^2$
 whose angles are $\theta_i=(1-2\eta_i) \pi$.  We can glue together two copies of this
 triangle  by sewing the edges together, and since the edges are geodesics they have zero extrinsic curvature and the metric will be smooth accross the junction.  The angular distance around the singular points will be $2\theta_i=(1-2\eta_i) 2\pi$, and as explained in the discussion
 of (\ref{condef}), this is the expected behavior for a classical solution with the insertion of
 primary fields of Liouville momenta $\eta_i/b$. So this gives a metric of constant positive
 curvature with the desired three singularities. For this construction to work, we need only make sure that a triangle exists with the specified angles.  First note that because of the positive curvature of $\sf S^2$, we must have $\sum_i \theta_i>\pi$, which gives $\sum_i \eta_i<1$.  We can choose one of vertices of the triangle, say the one labeled by $\eta_1$, to be  the north pole, and then the two legs connected to it must lie in great circles passing through both the north and the south pole.  If we extend these legs all the way down to the south pole, then the area between them is a ``diangle,'' as shown in Figure \ref{triangles}.  The third leg of the triangle then splits the diangle into two triangles, the original one and its complement, labelled $A$ and $B$ respectively in the figure.  The inequality $\sum_i \theta_i>\pi$ applied to the \textit{complementary} triangle then gives $\eta_2+\eta_3-\eta_1>0$, so this is the source of the triangle inequality in (\ref{inequalities}).  As we approach saturating the inequality, the complementary triangle $B$ becomes smaller and smaller and the original triangle $A$ degenerates into a diangle.  Once the inequality is violated, no metric with only the three desired singularities exists.

\begin{figure}[ht]
\begin{center}
\includegraphics[scale=.6]{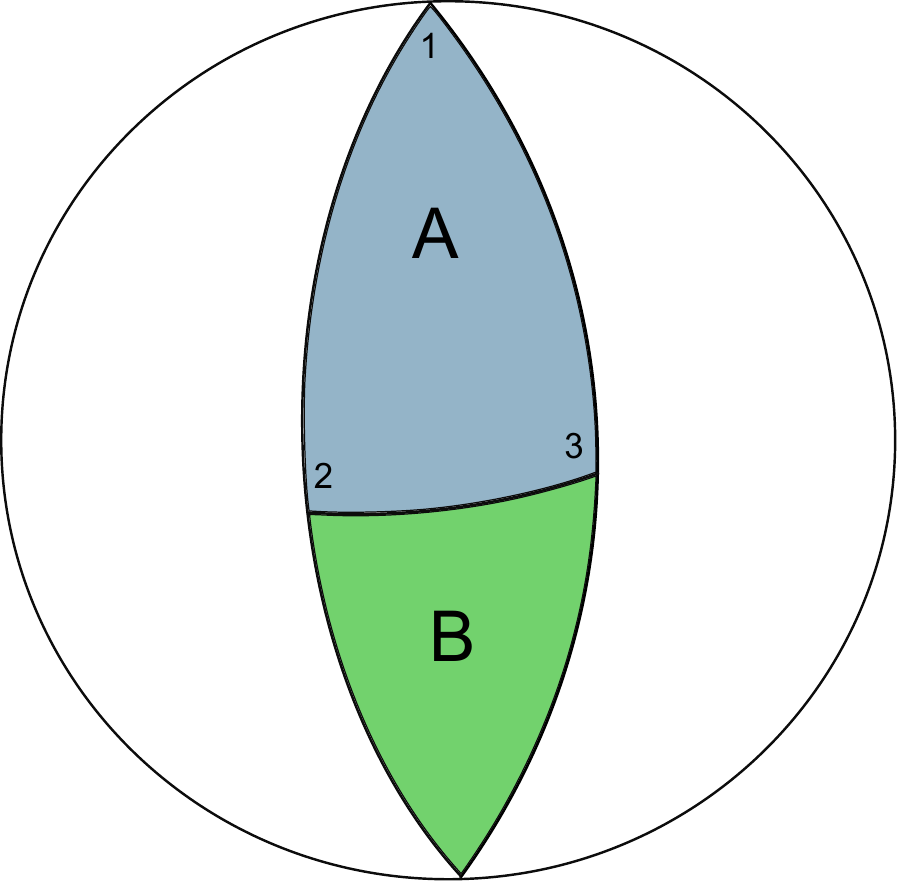}
\caption{ Spherical triangles.}
 \label{triangles}
\end{center}
\end{figure}

We will now compare the semiclassical actions of the complex saddle points (\ref{3pointsol}) in these two regions with the semiclassical limit of the DOZZ formula.  We will  see that in Region I only 
the real saddle point contributes (this is expected for reasons explained in section \ref{ancon}) while in Region II,  similarly to what we found for the two-point function, 
 infinitely many contribute.  We  interpret this change as a Stokes 
phenomenon; the condition $\mathrm{Re}(\eta_1+\eta_2+\eta_3)=1$ separating the two
regions 
evidently defines a Stokes wall.  In Region II, we will initially assume that all three operators are heavy, but in a final subsection we will treat the case that  they are light and again find evidence for a  path integral interpretation of the DOZZ formula.

\subsubsection{Evaluation of the Action for Three-Point Solutions}\label{etps}
To evaluate the action for a saddle point contributing to the three-point function,
 we can again use the trick of differentiating the action with respect to $\eta_i$.  So we need to determine the asymptotic behaviour of (\ref{3pointsol}) near $z_i$.  We denote as $\phi_{c,N}$ the
 solution corresponding to (\ref{3pointsol}), where the subscript $N$ labels the possibility of
 shifting $\phi_c$ by $2\pi i N$.    We will again have
\begin{align} 
&\phi_{c,N}(z,\overline{z}) \to -4\eta\log |z-z_i|+C_i \qquad \text{as} \,\,z\to z_i,
\end{align}
and to determine $C_i$ we again need to confront the problem of defining the logarithm of $f$.  We will first treat Region I, where we define
\begin{align} \nonumber
\phi_{c,N}(z,\overline{z})=2\pi i N&-\log \lambda -4 \log |z-z_2|\\ \label{phic31}
&-2\log\left(a_1 P^{\eta_1}(x)P^{\eta_1}(\overline{x})-a_2 P^{1-\eta_1}(x)P^{1-\eta_1}(\overline{x})\right).
\end{align}
The branch in the logarithm is chosen so that using (\ref{a1squared}) and the series expansion of $P^{\eta_1}$, we find
\begin{align} \nonumber
C_1=2\pi i N-&\log \lambda-(1-2\eta_1)\log \frac{|z_{12}|^2 |z_{13}|^2}{|z_{23}|^2}\\ \label{c1}
&-\log \frac{\gamma(\eta_1+\eta_2-\eta_3)\gamma(\eta_1+\eta_3-\eta_2)\gamma(\eta_1+\eta_2+\eta_3-1)}{\gamma(2\eta_1)^2\gamma(\eta_2+\eta_3-\eta_1)}.
\end{align}
The function $\phi_{c,N}(z,\overline{z})$ that we get by continuation away from $z_1$ will be singlevalued by the same argument as for the two-point function.  To find $C_2$ and $C_3$, we can use the connection coefficients \ref{a12}, \ref{a13}, but it is easier to just permute the indices to find
\begin{align} \nonumber
C_2=2\pi i N-&\log \lambda-(1-2\eta_2)\log \frac{|z_{12}|^2 |z_{23}|^2}{|z_{13}|^2}\\ \label{c2}
&-\log \frac{\gamma(\eta_1+\eta_2-\eta_3)\gamma(\eta_2+\eta_3-\eta_1)\gamma(\eta_1+\eta_2+\eta_3-1)}{\gamma(2\eta_2)^2\gamma(\eta_1+\eta_3-\eta_2)},\\ \nonumber
C_3=2\pi i N-&\log \lambda-(1-2\eta_3)\log \frac{|z_{23}|^2 |z_{13}|^2}{|z_{12}|^2}\\ \label{c3}
&-\log \frac{\gamma(\eta_3+\eta_2-\eta_1)\gamma(\eta_1+\eta_3-\eta_2)\gamma(\eta_1+\eta_2+\eta_3-1)}{\gamma(2\eta_3)^2\gamma(\eta_1+\eta_2-\eta_3)}.
\end{align}
As with the two-point function, we can justify the equality of $N$ in the vicinity of different points by observing that we may begin with real $\eta$'s obeying $\sum_i\eta_i>1$, for which the argument of the logarithm is real and positive.  We then continue to the desired value of $\eta$ on a path that remains in Region I.  As before, by continuity $N$ cannot change.  As a check of this claim, we observe that paths within  Region I cannot activate the branch cuts of the logarithms in these expressions for $C_i$.  Indeed, for any set of $\eta_i$'s which is in Region I, all of the arguments of $\gamma(\cdot)$ have real part between zero and one.  $\gamma(\cdot)$ has no zeros or poles in this strip, so any loop in Region I can be contracted to a point without changing the monodromy of the logarithm.  Thus there is no monodromy. 

To compute the action, we need to integrate
\be
b^2 \frac{\partial \widetilde{S}_L}{\partial \eta_i}=-C_i,
\ee
which gives
\begin{align}
\nonumber
b^2 \widetilde{S}_L=&\left(\sum_i \eta_i -1\right)\log \lambda+(\delta_1+\delta_2-\delta_3)\log |z_{12}|^2+(\delta_1+\delta_3-\delta_2)\log |z_{13}|^2\\ \nonumber
&+(\delta_2+\delta_3-\delta_1)\log |z_{23}|^2 +F(\eta_1+\eta_2-\eta_3)+F(\eta_1+\eta_3-\eta_2)+F(\eta_2+\eta_3-\eta_1)\\ \nonumber
&+F(\eta_1+\eta_2+\eta_3-1)-F(2\eta_1)-F(2\eta_2)-F(2\eta_3)-F(0)\\
&+2\pi i N(1-\sum_i \eta_i). \label{region1act}
\end{align}
Here we have
\be
\label{Fdefinition}
F(\eta)\equiv \int_{\frac{1}{2}}^\eta \log \gamma(x) dx,
\ee
with the contour staying in the strip $0<\mathrm{Re}(x)<1$, and also
$$\delta_i\equiv \eta_i(1-\eta_i).$$
The $\eta_i$-independent constant was determined in 
 \cite{Zamolodchikov:1995aa}
by explicitly evaluating the action in the case $\sum_i \eta_i=1$, with the result $b^2 \widetilde{S}_L=\sum_{i<j} 2\eta_i \eta_j \log |x_i-x_j|^2$.  The integral involved  is quite difficult
and will not be described here.\footnote{The condition that $\sum_i\eta_i=1$ means that the 
flat $SL(2,\C)$ bundle over the three-punctured sphere associated to the solution actually
has abelian monodromy.  This can perhaps be used to evaluate its action in the Chern-Simons
description that we will discuss in section \ref{csfun}.}  We can observe immediately that the $z_i$-dependence in
(\ref{region1act}) is of the correct form for a conformal three-point function.

We now evaluate the action for $\eta_i$'s in Region II.  The manipulations are similar, but we now define the branch so that
\begin{align}
C_1=&2\pi i\left(N+\frac{1}{2}\right)-\log \lambda-(1-2\eta_1)\log \frac{|z_{12}|^2 |z_{13}|^2}{|z_{23}|^2}+2\log\left(1-\sum_i\eta_i\right) \nonumber\\
&-\log \frac{\gamma(\eta_1+\eta_2-\eta_3)\gamma(\eta_1+\eta_3-\eta_2)\gamma(\eta_1+\eta_2+\eta_3)}{\gamma(2\eta_1)^2\gamma(\eta_2+\eta_3-\eta_1)}.
\end{align}
We have used $\gamma(x-1)=-\frac{1}{(x-1)^2}\gamma(x)$ to make sure that when we take the $\eta$'s to be real (and in Region II), the only imaginary parts comes from the first term.  We can again permute to find:
\begin{align}
C_2=&2\pi i\left(N+\frac{1}{2}\right)-\log \lambda-(1-2\eta_2)\log \frac{|z_{12}|^2 |z_{23}|^2}{|z_{13}|^2}+2\log\left(1-\sum_i\eta_i\right) \nonumber\\
&-\log \frac{\gamma(\eta_1+\eta_2-\eta_3)\gamma(\eta_2+\eta_3-\eta_1)\gamma(\eta_1+\eta_2+\eta_3)}{\gamma(2\eta_2)^2\gamma(\eta_1+\eta_3-\eta_2)},\\
C_3=&2\pi i\left(N+\frac{1}{2}\right)-\log \lambda-(1-2\eta_3)\log \frac{|z_{23}|^2 |z_{13}|^2}{|z_{12}|^2}+2\log\left(1-\sum_i\eta_i\right) \nonumber\\
&-\log \frac{\gamma(\eta_2+\eta_3-\eta_1)\gamma(\eta_1+\eta_3-\eta_2)\gamma(\eta_1+\eta_2+\eta_3)}{\gamma(2\eta_3)^2\gamma(\eta_1+\eta_2-\eta_3)}.
\end{align}
Finally we can again integrate this to find
\begin{align}
\nonumber
b^2 \widetilde{S}_L=&\left(\sum_i \eta_i -1\right)\log \lambda+(\delta_1+\delta_2-\delta_3)\log |z_{12}|^2+(\delta_1+\delta_3-\delta_2)\log |z_{13}|^2\\ \nonumber
&+(\delta_2+\delta_3-\delta_1)\log |z_{23}|^2 +F(\eta_1+\eta_2-\eta_3)+F(\eta_1+\eta_3-\eta_2)+F(\eta_2+\eta_3-\eta_1)\\ \nonumber
&+F(\eta_1+\eta_2+\eta_3)-F(2\eta_1)-F(2\eta_2)-F(2\eta_3)-F(0)\\
&+2\left[(1-\sum_i \eta_i)\log(1-\sum_i \eta_i)-(1-\sum_i \eta_i)\right] + 2\pi i (N+1/2)(1-\sum_i \eta_i). \label{region2act}
\end{align}
Here we determined the constant by matching to $\eta_i=0$, which as we found before gives an action $2\pi i (N+1/2)-\log \lambda-2$.

Before comparing these expressions with the asymptotics of the DOZZ formula in Regions I and II, we first comment on their multivaluedness.  In order to do this we must determine the analytic structure of the function $F(\eta)$.  It is clear from the definition (\ref{Fdefinition}) that $F(\eta)$ has branch points at each integer $\eta$.  The form of the branch points for $\eta=-n$ with $n=0,1,2,\dots$ is $-(\eta+n)\log(\eta+n)$, while for $\eta=m$ with $m=1,2,\dots$ it is $(\eta-m)\log (m-\eta)$.  We thus find that the monodromy of $F(\eta)$ around any loop in the $\eta$-plane is
\be
\label{Fmonodromy}
F(\eta) \to F(\eta)+\sum_{m=1}^\infty (\eta-m)2\pi i N_m-\sum_{n=0}^\infty (\eta+n)2\pi i N_n,
\ee
where $N_n$ and $N_m$ count the number of times the loop circles the branch points in a counterclockwise direction.  Now applying this to (\ref{region1act}), we see that continuation in the $\eta_i$ can produce far more branches than can be accounted for by nonsingular complex solutions.  In particular,  the various nonsingular solutions can only account for multivaluedness of the form $2\pi i N(1-\sum_i \eta_i)$, while continuation around a loop in the general $\eta_i$ parameter
space  can easily produce shifts of the action by terms such as  $2\pi i N(\eta_1+\eta_2-\eta_3)$.  There thus seems to be a mismatch between the branches of the action (\ref{region1act}) and the available saddle points.  One might be tempted to interpret this multivaluedness as indicating
the existence of additional solutions, but we showed in section \ref{threp} that there are no more solutions.  We will suggest a mechanism for explaining this additional multivaluedness in section \ref{4pointsection}, as part of our discussion of the singularities that appear in the case of general $\eta_i$, and another possible interpretation in section \ref{csfun}.  The situation however is simpler for continuations that stay in Region I and/or Region II.  Such a continuation will only activate the branch cuts in $F(\sum_i \eta_i-1)$, and this produces the kind of multivaluedness that can be accounted for by the known nonsingular solutions.  In particular the action (\ref{region2act}) can be gotten by analytic continuation from (\ref{region1act}) along a path that goes from Region I to Region II, with the particular saddle point we land on being determined by the number of times the path wraps around $\sum_i \eta_i=1$.

\subsubsection{Comparison with Asymptotics of the DOZZ Formula}\label{compasym}
We now compute the semiclassical limit of the DOZZ formula (\ref{dozz}) with three heavy operators in Regions I and II.\footnote{This computation  was done in 
\cite{Zamolodchikov:1995aa} in Region I with real $\eta_i$, and our computations here are simple extensions of that.}  

The semiclassical behavior of  the prefactor of the DOZZ formula is clear:
\be
\left[\lambda \gamma(b^2)b^{-2b^2}\right]^{\left(Q-\sum_i \alpha_i\right)/b} \to \exp \left[-\frac{1}{b^2}\left\{\left(\sum_i \eta_i -1\right)\log \lambda-2\left(\sum_i\eta_i-1\right)\log b\right\}\right].
\ee
To study the remaining terms, we need the $b\to 0$ behaviour of $\Upsilon_b(\eta/b)$.  In Appendix \ref{upsilonapp}, we show that
\begin{equation}
\label{etaasymp}
\Upsilon_b(\eta/b)=e^{\frac{1}{b^2}\left[F(\eta)-(\eta-1/2)^2\log b+\O(b \log b)\right]}\qquad 0<\mathrm{Re}(\eta)<1.
\end{equation}
In Region I, all of the $\Upsilon_b$'s have their arguments in the region of validity for this formula, so we find that they asymptote to:
\begin{align} \nonumber
\exp\Big[&\frac{1}{b^2}\Big\{F(2\eta_1)+F(2\eta_2)+F(2\eta_3)+F(0)\\ \nonumber
&-F(\sum_i\eta_i-1)-F(\eta_1+\eta_2-\eta_3)-F(\eta_1+\eta_3-\eta_2)-F(\eta_2+\eta_3-\eta_1)\\
&-2\left(\sum_i\eta_i-1\right)\log b\Big\}\Big].
\end{align}
Combining these two contributions, we find complete agreement with (\ref{region2act}) with $N=0$.  Thus in Region I, only one saddle point contributes and we can interpret the path integral as being evaluated on a single integration cycle passing through it.  

In Region II, the only new feature is that $\Upsilon_b\left(\sum_i \alpha_i-Q\right)$ is no longer in the region where we can apply (\ref{etaasymp}).  To deal with this, we can use the recursion relation (\ref{inverserecursion}) to move the argument back to the region where we can use (\ref{etaasymp}):
\be
\Upsilon_b\left(\left(\sum_i \eta_i-1\right)/b\right)=\gamma\left(\left(\sum_i\eta_i-1\right)/b^2\right)^{-1}b^{1-2(\sum_i \eta_i-1)/b^2} \Upsilon_b\left(\sum_i \eta_i/b\right).
\ee
Using also (\ref{gamma}) for the asymptotics of the Gamma function (and hence of $\gamma(x)=\Gamma(x)/\Gamma(1-x)$), we
finally arrive at
\begin{align} \nonumber \label{umot}
C(\eta_i/b)\sim \exp\Bigg[-&\frac{1}{b^2}\Bigg\{\left(\sum_i \eta_i-1\right)\log \lambda-F(2\eta_1)-F(2\eta_2)-F(2\eta_3)-F(0)\\ \nonumber
&+F\big(\sum_i\eta_i\big)+F(\eta_1+\eta_2-\eta_3)+F(\eta_1+\eta_3-\eta_2)+F(\eta_2+\eta_3-\eta_1)\\\nonumber
&+2\Big[\left(1-\sum_i \eta_i\right)\log\left(1-\sum_i \eta_i\right)-\left(1-\sum_i\eta_i\right)\Big]\Bigg\}\Bigg]\\
&\times \frac{1}{e^{i \pi (\sum_i \eta_i-1)/b^2}-e^{-i \pi (\sum_i \eta_i-1)/b^2}}.
\end{align}
This is in complete agreement with (\ref{region2act}), provided that as with the two-point function we interpret the final factor as coming from a sum over infinitely many complex saddle points.  Rather as before, the saddle points that contribute are $N=\{-1,-2,\dots\}$ when $\mathrm{Im}\,((\sum_i \eta-1)/b^2)<0$ and $N=\{0,1,2,\dots\}$ when $\mathrm{Im}\,((\sum_i \eta_i-1)/b^2)>0$.   The condition
$\mathrm{Im}\,((\sum_i \eta_i-1)/b^2)=0$ defines a Stokes wall.

\subsection{Three-Point Function with Light Operators}\label{dozzthreelight}

\hspace{0.25in}So far we have considered only correlators where all operators are heavy.  As a final check we will compute the semiclassical limit of the DOZZ formula (\ref{dozz}) with three three light operators of Liiouville momenta $\alpha_{i} = b\sigma_{i}$, with $\sigma_i$ fixed for $b\to 0$,
 and compare it with a semiclassical computation based on equation (\ref{semiclassicalcorr}).  This compution is essentially a repackaging of a fixed-area computation outlined in \cite{Zamolodchikov:1995aa}; we include it as an additional illustration of the machinery of complex saddle points and also because many of the details were omitted in \cite{Zamolodchikov:1995aa}.  In section \ref{timelike}, we will also use the same tools to do a new check in the context of timelike Liouville, so it is convenient to first present them in a more familiar context.

We begin by computing the asymptotics of the DOZZ formula with three light operators; in order to capture the nontrivial effects of the operators we need to compute to higher order in $b$ than before.  To order $b^0$ in the exponent the prefactor not involving $\Upsilon_b$'s becomes
\be \label{dozzprefaclight}
\left[\lambda \gamma(b^2)b^{-2b^2}\right]^{\left(Q-\sum_i \alpha_i\right)/b} = b^{-2/b^2+2\sum_i \sigma_i-4}\lambda^{1/b^2+1-\sum_i\sigma_i}e^{-2\gamma_E+O(b\log b)}
\ee
Here $\gamma_E$ is the Euler-Mascheroni constant $\gamma_E\equiv \lim_{n\to\infty}\left(\sum_{k=1}^n\frac{1}{k}-\log n\right)$.  To take the limits of the $\Upsilon_b$ functions, we need the the asymptotics of $\Upsilon_b(\sigma b)$ as $b\to 0$.  This is given by equation (\ref{sigmaasymp}):
\begin{align}
\Upsilon_{b}(b\sigma) = &\frac{Cb^{1/2 - \sigma}}{\Gamma(\sigma)}\exp{\left[-\frac{1}{4b^{2}}\log{b} + \frac{F(0)}{b^{2}} + O(b^{2}\log{b})\right]}.\label{upsilonlightlimits}
\end{align}
Here $C$ is a constant that will cancel in the final result.  This along with (\ref{upsilon0}) is sufficient to determine the asymptotics of all parts of the DOZZ formula except for the $\Upsilon_b$ involving $\sum_i \sigma_i$.  For this one we can use the recursion relation:
\be
\Upsilon_b\Big[\big(\sum_i \sigma_i -1\big)b-1/b\Big]=\gamma\Big(\sum_i\sigma_i-1-1/b^2\Big)^{-1}b^{3+2/b^2-2\sum_i \sigma_i}\Upsilon_b\Big[\big(\sum_i \sigma_i-1\big)b\Big]
\ee
To evaluate the semiclassical limit of this we need the corrections to (\ref{gamma}).  We can get these by using the machinery of Appendix C, but we can simplify the discussion using Euler's reflection formula $\Gamma(x)\Gamma(1-x)=\frac{\pi}{\sin (\pi x)}$:
\be
\gamma(x-1/b^2)=\frac{\pi}{\Gamma(1-x+1/b^2)^2 \sin\big(\pi (x-1/b^2)\big)}
\ee
The $\Gamma$ function appearing on the right hand side of this equation always has positive real part as $b\to0$, so we can simply include the first subleading terms in Stirling's formula to find
\be
\Gamma(1-x+1/b^2)=\sqrt{2\pi}b^{-2/b^2+2x-1}e^{-1/b^2}(1+\mathcal{O}(b^2)).
\ee
This then gives
\be
\gamma(\sum_i \sigma_i-1-1/b^2)=\frac{i}{e^{i\pi (\sum_i \sigma_i-1-1/b^2)}-e^{-i\pi (\sum_i \sigma_i-1-1/b^2)}}b^{4/b^2-4\sum_i \sigma_i+6}e^{2/b^2}(1+\mathcal{O}(b^2)).
\ee
Combining all these results together we can write:
\begin{align} \nonumber
C(\sigma_1 b,\sigma_2b, \sigma_3b)=&ib^{-3}\lambda^{1/b^2+1-\sum_i \sigma_i}e^{2/b^2-2\gamma_E+\mathcal{O}(b\log b)}\frac{1}{e^{i\pi (\sum_i \sigma_i-1-1/b^2)}-e^{-i\pi (\sum_i \sigma_i-1-1/b^2)}}\\
&\times \frac{\Gamma(\sigma_1+\sigma_2-\sigma_3)\Gamma(\sigma_1+\sigma_3-\sigma_2)\Gamma(\sigma_2+\sigma_3-\sigma_1)\Gamma(\sigma_1+\sigma_2+\sigma_3-1)}{\Gamma(2\sigma_1)\Gamma(2\sigma_2)\Gamma(2\sigma_3)}. \label{lightdozz}
\end{align}

We now compare this result to an appropriate refinement of (\ref{semiclassicalcorr}).  There are several subtleties to consider.  With all operators light the appropriate saddle point is the sphere (\ref{sphere}).  As with the saddle point (\ref{2pointsol}) for the two-point function with heavy operators there will be a moduli space of such solutions, in this case given by the quotient $SL(2, \mathbb{C})/SU(2)$, since the subgroup of $SL(2,\mathbb{C})$ that leaves fixed a particular round sphere metric is a copy of $SU(2)$.  The light operator insertions will depend explicitly on these moduli, so we need the general $SL(2,\mathbb{C})$ transformation of the saddlepoint (\ref{sphere}).  From (\ref{transformation}) this is given by 
\be \label{spherewithmoduli}
\phi_{c,N}(z,\bar{z})=2\pi i (N+1/2)-\log \lambda-2 \log \left(|\alpha z+\beta|^2+|\gamma z+\delta|^2\right),
\ee 
with $\alpha,\beta,\gamma, \delta\in \mathbb{C}$ and obeying $\alpha \delta-\beta \gamma=1$.  In using (\ref{semiclassicalcorr}) we will need to integrate the right hand side over all such saddlepoints.

An additional subtlety is that in (\ref{semiclassicalcorr}) all effects of the operator insertions are $\mathcal{O}(b^0)$ in the exponent.  To precisely include all effects of this order, we would need to carefully compute the renormalized fluctuation determinant about each saddle point, and also include the $\mathcal{O}(b^0)$ corrections to the action (\ref{action}).  Moreover we would need the Jacobian in transforming the integral over $\phi_c$ into an integral over the parameters $\alpha,\beta,\gamma,\delta$.  We will include the subleading terms in the action explicitly, but to simplify things we will represent the fluctuation determinant and Jacobian as a $b$-dependent prefactor $A(b)$ which is at most $\mathcal{O}(\log b)$ in the exponent.\footnote{We do NOT need to include $\mathcal{O}(b^2)$ corrections to the saddlepoint (\ref{spherewithmoduli}) even though they are present.  The reason is that the leading order saddlepoints are stationary points of the leading order action, so perturbing the solution at $\mathcal{O}(b^2)$ does not affect the action until $\mathcal{O}(b^2)$, which is beyond our interest.}  Note that neither of these things should be affected by shifting the saddlepoint by $2\pi i$ so we expect $A(b)$ to be independent of $N$.  It is also independent of $\sigma_i$ since neither effect has anything to do with the operator insertions.  With this convention, we can now write a more precise version of (\ref{semiclassicalcorr}) that is appropriate for comparison with (\ref{lightdozz}): 
\be
\label{better3corr}
\langle V_{b\sigma_{1}}(z_{1},\bar{z}_1)V_{b\sigma_{2}}(z_{2},\bar{z}_2)V_{b\sigma_{3}}(z_{3},\bar{z}_3)\rangle \approx A(b)\sum_{N\in T}e^{-S_L[\phi_{c,N}]}\int d\mu(\alpha,\beta,\gamma,\delta)\prod_{i=1}^3 e^{\sigma_i \phi_{c,N}(z_i,\bar{z}_i)}.
\ee
Here $T$ is some set of integers and 
$$d\mu(\alpha,\beta,\gamma,\delta)=4\delta^2(\alpha\delta-\beta\gamma-1)\,d^2\alpha \,d^2\beta \,d^2\gamma \,d^2\delta $$
is the invariant measure on $SL(2,\mathbb{C})$ \cite{Zamolodchikov:1995aa}.  The integrals over over the full $\alpha,\beta,\ldots$ planes.

The $\mathcal{O}(b^0)$ correction to the action (\ref{action}) is given by $\frac{1}{2\pi}\oint_{\partial D}{\phi_c d \theta}+4\log R$.  For the saddle point (\ref{spherewithmoduli}) the leading part was computed above (\ref{2action}), and now including the subleading term we find
\be
\label{betterlightS}
S_L[\phi_{c,N}]=\frac{1}{b^2}\Big[2\pi i (N+1/2)-\log \lambda-2\Big]+2\pi i (N+1/2)-\log \lambda+\mathcal{O}(b^2).
\ee
The integral over the moduli is quite difficult, we will simplify it some here and then relegate the final computation to an appendix.  Our technique is identical to that in \cite{Zamolodchikov:1995aa}.  We first note that
\begin{align} \nonumber
\int &d\mu(\alpha,\beta,\gamma,\delta)\prod_{i=1}^3 e^{\sigma_i \phi_{c,N}(z_i,\bar{z}_i)}=\lambda^{-\sum_i \sigma_i} e^{2\pi i (N+1/2)\sum_i \sigma_i} \\
&\times \int \frac{d\mu (\alpha,\beta,\gamma,\delta)}{\Big(|\alpha z_1+\beta|^2+|\gamma z_1+\delta|^2\Big)^{2\sigma_1}\Big(|\alpha z_2+\beta|^2+|\gamma z_2+\delta|^2\Big)^{2\sigma_2}\Big(|\alpha z_3+\beta|^2+|\gamma z_3+\delta|^2\Big)^{2\sigma_3}}.
\end{align}
The position dependence of this integral can be extracted by using its $SL(2,\mathbb{C})$ transformation properties; changing variables by the transformation which sends $z_1\to 0$, $z_2 \to 1$, and $z_3 \to \infty$ we find the usual three-point function behaviour
\begin{align} \label{modintegral}
\int d\mu\,\prod_{i=1}^3 e^{\sigma_i \phi_{c,N}}=&\lambda^{-\sum_i \sigma_i} e^{2\pi i (N+1/2)\sum_i \sigma_i} |z_{12}|^{2(\sigma_3-\sigma_1-\sigma_2)}|z_{23}|^{2(\sigma_1-\sigma_2-\sigma_3)}|z_{13}|^{2(\sigma_2-\sigma_1-\sigma_3)}\, I(\sigma_1,\sigma_2,\sigma_3)
\end{align}
with
\be
I(\sigma_1,\sigma_2,\sigma_3)\equiv \int \frac{d\mu (\alpha,\beta,\gamma,\delta)}{\Big(|\beta|^2+|\delta|^2\Big)^{2\sigma_1}\Big(|\alpha+\beta|^2+|\gamma+\delta|^2\Big)^{2\sigma_2}\Big(|\alpha|^2+|\gamma|^2\Big)^{2\sigma_3}}.
\ee
The result of this integral was quoted in \cite{Zamolodchikov:1995aa}, but many steps were omitted and the full evaluation is quite sophisticated.  For completeness we have included a full derivation in Appendix \ref{sl2cintegralapp}.  The result is
\be
\label{Iint}
I(\sigma_1,\sigma_2,\sigma_3)=\pi^3 \frac{\Gamma(\sigma_1+\sigma_2-\sigma_3)\Gamma(\sigma_1+\sigma_3-\sigma_2)\Gamma(\sigma_2+\sigma_3-\sigma_1)\Gamma(\sigma_1+\sigma_2+\sigma_3-1)}{\Gamma(2\sigma_1)\Gamma(2\sigma_2)\Gamma(2\sigma_3)}.
\ee
Using this along with (\ref{betterlightS}) and (\ref{modintegral}), we find that (\ref{better3corr}) gives
\begin{align} \nonumber
C(\sigma_i b)\approx &\pi^3 A(b)\lambda^{1/b^2+1-\sum_i \sigma_i}e^{2/b^2}\sum_{N\in T}e^{2\pi i (N+1/2)(\sum_i \sigma_i-1-1/b^2)}\\
&\times \frac{\Gamma(\sigma_1+\sigma_2-\sigma_3)\Gamma(\sigma_1+\sigma_3-\sigma_2)\Gamma(\sigma_2+\sigma_3-\sigma_1)\Gamma(\sigma_1+\sigma_2+\sigma_3-1)}{\Gamma(2\sigma_1)\Gamma(2\sigma_2)\Gamma(2\sigma_3)}
\end{align}
Comparing this with the DOZZ asymptotics (\ref{lightdozz}) we find complete agreement, with the saddle points included depending on the sign of $\Im(\sum_i \sigma_i-1/b^2)$.  We also see that apparently $A(b)=i\pi^{-3} b^{-3} e^{-2\gamma_E}$, which would be interesting to check by explictly treating the measure.  That it is imaginary is unsurpising given the complex integration cycle.  

\subsection{Summary}\label{summary}

\hspace{0.25in}This concludes our argument that the analytic continuation of the DOZZ formula in Regions I and II is described by the Liouville path integral evaluated on a complex integration cycle that changes as we cross Stokes lines.  The behaviour is completely analogous to that of the Gamma function as described in Appendix \ref{gammastokes}.   This has a qualitative explanation that was explained in 
section \ref{minis}.  The integral representation of the Gamma function is the zero mode part of the Liouville path integral, and the complex saddle points that we studied for Regions I and II differed only by shifting the zero mode.  What we learned in this section is that in Regions I and II there are no additional subtleties in the analytic continuation in $\eta_i$ beyond those that are already apparent in the zero mode.

\chapter[Four-Point Functions and Singular Saddles]{Four-Point Functions and the Interpretation of Singular Saddle Points}
\label{4pointsection}
We now confront the issue first raised in section \ref{threp}: for most complex values 
of the $\eta_i$, there are no nonsingular solutions of Liouville's equation with the 
desired boundary conditions.  
The candidate solution (\ref{3pointsol}) fails to be a solution because of zeroes of the 
denominator function
\be
\label{fnearsing0}
f(z,\overline{z})=A(z-z_0)+B(\overline{z}-\overline{z}_0)+\dots
.\ee
  At such a zero, $\phi_c=-2\log f-\log\lambda$ is singular, and perhaps more seriously, it is also 
generically multivalued.  Around a zero of $f$ with winding number $k$, $\phi_c$ changes
by $-4\pi i k$.   

This seems to raise a serious challenge to any attempt to interpret the full analytic 
continuation of the DOZZ formula in terms of conventional path integrals.  
In this section we will study this further.  We will make three arguments that even when
 $\phi_c$ is multivalued, the expression
(\ref{3pointsol})  still makes some sense and controls the asymptotic behaviour of 
the DOZZ formula.  We will first show that there is a minor redefinition of the action 
which agrees with the formula (\ref{regaction})  when there are no singularities but is finite 
even in the presence of zeroes of the denominator.  Moreover it correctly produces the analytic continuation of (\ref{regaction}).  We will then show that the presence of
singularities actually allows the full multivaluedness of the action (\ref{region2act}) to be 
realized by analytic continuation of the ``solutions.''  Finally we will probe the saddle points 
that dominate the three-point function
by including a fourth light operator.  For the case that we are able to implement this test -- the
case that the light operator is degenerate -- we will find
agreement with the (\ref{3pointsol}) for all values of the $\eta_i$.  We will close by commenting 
on the implications for general four-point functions.  

\subsection{Finiteness of the ``Action''}\label{gaction}

\hspace{0.25in}We begin by observing that in Region I defined at the beginning of subsection \ref{3pointcontinuation}, we included a restriction on the imaginary parts of the $\eta_i$'s to ensure that the denominator in (\ref{3pointsol}) did not vanish away from the operator insertions.  However, the formulas that followed  seemed to know nothing about this additional restriction; the multivaluedness in the expressions for $C_i$ and $\widetilde{S_L}$ cannot be activated without violating the conditions $\mathrm{Re}\left(\sum_i \eta_i\right)<1$ or $\mathrm{Re}\,\eta_i<\frac{1}{2}$, regardless of the imaginary parts of the $\eta_i$'s.  Moreover the expression (\ref{region2act}) for the action can easily be continued to values of $\eta_i$ where the denominator vanishes, and its value is perfectly finite there.  This is perhaps unexpected because 
near a zero of the denominator, one has
\be
\phi_c(z,\overline{z})\approx -2\log [A(z-z_0)+B(\overline{z}-\overline{z}_0)],
\ee
which has a logarithmic singularity as well as a branch cut discontinuity.\footnote{We consider the case that $|A|\neq |B|$, which is generically true for complex $\eta$'s.  When the $\eta_i$'s are real and $a_1$ is also real then we can have $|A|=|B|$, we will comment on this below.}
With such discontinuous behavior, the kinetic term in the Liouville action $\int d^2 \xi \partial_a \phi_c \partial_a \phi_c$ certainly diverges.  The finite analytic continuation of the action therefore cannot be computed by naive application of (\ref{regaction}).

We begin by observing that for solutions with no additional singularities we can rewrite
(\ref{regaction}) as
\be
\label{improvedact}
b^2\widetilde{S}_L=\frac{1}{\pi}\int_{D-\cup d_i} d^2 \xi \left[\partial f \overline{\partial} f/f^2+1/f^2\right]+\text{boundary terms}.
\ee 
As before, the $d_i$ are small discs  centered around $z_i$.
We propose that even in the presence of zeroes of the denominator of $e^{\phi_c}$, this is still the correct form of the action, with the integral defined 
by removing a small disc of radius $\epsilon$ centered around each zero and then taking $\epsilon\to0$.  The divergence from the discontinuity in $\phi_c$ is avoided since $f$ is continuous, but we still need to show that there is no divergence as $\epsilon \to 0$.  In particular near a zero at $z=z_0$, we have
the expansion (\ref{fnearsing0}),
so we can approximate the contribution to the integral from the vicinity of $z_0$ as
\be
\frac{1}{\pi} \int_\epsilon \frac{dr}{r} \int_0^{2\pi}d\theta\frac{AB+1}{\left(Ae^{i\theta}+B e^{-i \theta}\right)^2}.
\ee
The radial integral is logarithmically divergent, but as long as $|A| \neq |B|$ the angular integral is zero!  The higher order corrections to $f$ will produce manifestly finite corrections to the action, and in fact one can show that this definition of the action is invariant under coordinate transformations of the form $z\to z+\O(z^2)$.  This is thus analogous to the principal value prescription for computing the integral of $1/x$ across $x=0$.   We claim that the action computed this way agrees with what one gets by analytic continuation in $\eta_i$.  To justify this, we need to show that we can continue to use the trick of differentiating with respect to $\eta_i$ to calculate the action.  This requires a demonstration that a
multivalued  ``solution'' is a stationary point of the improved action.  To show this we can compute the variation of the improved action under $f\to f+\delta f$ with $\delta f$ continuous; most terms are clearly zero when evaluated on a multivalued ``solution,'' but  a potentially nontrivial boundary term  is generated by the integration by parts:
\be
\Delta \widetilde{S}_L= -\frac{\epsilon}{2\pi b^2}\int_0^{2\pi} d\theta \frac{\partial_r f}{f^2}\delta f\Bigg|_{(z-z_0)=\epsilon e^{i\theta}}.
\ee
For intuition, we observe that this boundary term is also present near each of the operator insertions.  Near the operator at $z_i$, we have $f\sim r^{2\eta_i}$, and the boundary term produces a nontrivial variation $-\frac{2\eta_i}{\epsilon^{2\eta}}\delta f$.  This variation is cancelled by the variation 
 $-\frac{\eta_i}{2\pi}\int_0^{2\pi}d\theta \phi_c$ of the regulated operator.  The point however is that for $f$ obeying (\ref{fnearsing0}), this boundary term is automatically zero by itself since the angular integral vanishes.  So in this sense, a multivalued ``solution'' is a stationary point of the action.

For orientation, perhaps we should mention that a singlevalued $\phi_c$ with singularities
away from the operator insertions can never be such a stationary point.  Indeed, generalities
about elliptic differential equations ensure that a solution of the complex Liouville equations
is smooth away from operator insertions. In the singular case, it is only because $\phi_c$ is multivalued that it may be,
in some sense, a stationary point of the action. 

 With this explanation of what the action means in the presence of singularities, we may drop the conditions on the imaginary parts of $\eta_i$ from both Regions I and II and the story of the previous section goes through unchanged.  This argument does fail  in the special cases where $|A|=|B|$, for which higher order terms near
the singularity are important and the singularity may be non-isolated.  We will view this  just as a degenerate limit of the more general situation.  In particular we can continue from Region I to anywhere else in the $\eta_i$-plane without passing through a configuration with a singularity with $|A|=|B|$, so this subtlety should not affect our picture of the analytic continuation of (\ref{dozz}).

Before we move on, we observe that there are two different kinds of multivaluedness being discussed in this section.  One is with respect to $\eta_i$, and the other is with respect to $z,\bar{z}$.  For convenience we summarize the multivaluedness of various quantities in the following table:
\begin{table}
\label{multitable}
\begin{center}
\begin{tabular}{| l | p{5cm} | p{5cm} |}
\hline
 & $z,\bar{z}$ behaviour at fixed $\eta_i$ & $\eta_i$ behaviour at fixed $z,\bar{z}$\\ \hline
$e^{\phi_c}$ & singlevalued & singlevalued\\ \hline
$b^2\tilde{S_L}$ & trivial & defined up to addition by $2\pi i (\sum_i\eta_i m_i+n)$ with $m_i$ all even or all odd\\ \hline
$C_i$ & trivial & defined up to addition by $2\pi i$ \\ \hline
$a_1$ & trivial & defined up to multiplication by a sign \\ \hline
$a_1/a_2$ & trivial & singlevalued \\ \hline
$f$ & singlevalued & defined up to multiplication by a $z,\bar{z}$-independent sign \\ \hline
$\phi_c$ & possibly singlevalued, possibly monodromy of addition by $4\pi i$ about points where $f=0$& defined up to addition by $2\pi i$ \\ \hline
\end{tabular}
\end{center}
\caption{Multivaluedness properties of quantities of interest}
\end{table}
\subsection{Multivaluedness of the Action}\label{multivac}
\hspace{0.25in}We saw in subsection \ref{3pointcontinuation} that the action (\ref{region1act}) is highly multivalued as a function of the $\eta_i$, with the multivaluedness arising from the function $F(\eta)$ defined in (\ref{Fdefinition}).  

We can now interpret this multivaluedness of the action as a consequence of 
the multivaluedness in $z,\bar{z}$ that $\phi_c$ acquires in the presence of zeroes of
$f$.  This multivaluedness does not affect the kinetic and potential terms of the action as defined in section \ref{gaction}, 
since they depend only on $f$, which is singlevalued as a function of $z,\bar{z}$. But
the terms  $-\sum_i\frac{\eta_i}{2\pi}\int_0^{2\pi}d\theta \phi_c$ that come from the regulated
operator insertions are sensitive to this multivaluedness.  Their contribution to the action is
\be
\Delta \widetilde{S}_L=-\frac{1}{b^2}\sum_i \eta_i C_i
\ee
where $C_i$ is the constant term in $\phi_c$ near the operator insertion.
Using the formulas (\ref{c1})-(\ref{c3}) for $C_i$, we see that continuing along a closed
path in the parameter space of the $\eta_i$ can shift $C_i$ by an integer multiples of  $2\pi i$,
hence shifting the action by an integer linear combination of the quantities $2\pi i\eta_i$.
We can see the same effect  in the formula (\ref{region1act}) for the action; the same
processes that cause a shift in the $C_i$ cause an equivalent shift in the 
 function $F$
in this formula, leading to the same multivaluedness.  For example, on a path on which
$\eta_1+\eta_2-\eta_3$ circles around an integer value, shifting $C_1$ and $C_2$ by $2\pi i$ and $C_3$ by $-2\pi i$, there is a corresponding shift in the action from $F(\eta_1+\eta_2-\eta_3)$.  

It is important to note that it is only because $\phi_c$ can be multivalued as a function of $z,\bar{z}$ that we can realize the full multivaluedness of the action in $\eta$.  We argued below equation (\ref{c3}) that any continuation in $\eta_i$ that passes only through continuous $\phi_c$'s cannot produce monodromy for the difference of any two $C_i$'s because of continuity.  But once we allow paths in $\eta_i$ that pass through multivalued (and thus discontinuous) $\phi_c$'s, these differences can have the nontrivial monodromy necessary to produce the full set of branches of the action.  Thus the multivaluedness of the action in $\eta_i$ has a natural interpretation once we allow solutions of the complex Liouville equations that are multivalued in $z,\bar{z}$.

\subsection{Comparison With The DOZZ Formula}\label{compdozz}

\hspace{0.25in}We  are finally ready to consider in general the semiclassical asymptotics of the  DOZZ formula (\ref{dozz}).  
The DOZZ formula is constructed from the function
$\Upsilon_b(\eta/b)$, where $\eta$ is a linear combination of the $\eta_i$.  In all, seven $\Upsilon_b$ functions appear in the numerator
or denominator of the DOZZ formula.  To evaluate the small $b$ asymptotics of this formula, one needs the small $b$ asymptotics
of the $\Upsilon_b$ functions.  This is given in (\ref{bingo}) for $\eta$ in a certain strip in the complex plane; it can be determined in 
general by using the recursion relations (\ref{recrel}) to map $\eta$ into the desired strip.   In the process, the recursion relation generates
a function that can be expanded as a sum of exponentials, as in (\ref{wox}); we interpret this  as a sum over different
complex critical points.  

For generic $\eta_i$, when evaluating the asymptotics of the DOZZ formula using the asymptotic formula (\ref{bingo}), we will need to apply the recursion relations to all of the $\Upsilon_b$'s. 
There is just one crucial difference from the derivation of eqn. (\ref{umot}).  The final factor in that formula has an expansion
in positive or negative powers of $\exp(2\pi i\sum_i\eta_i)$, where $\sum_i\eta_i$ entered because in that derivation, we had
to apply the recursion relation only to one of the $\Upsilon_b$ functions, namely $\Upsilon_b(\sum_i\eta_i/b)$.  In general, we have
to allow for the possibility that the argument of any one of the seven $\Upsilon_b$ functions in the DOZZ formula may leave the favored
strip.  So $\sum_i \eta_i$ may be replaced by the equivalent expression appearing in any one of the other $\Upsilon_b$ functions,
namely $2\eta_1$, $\eta_1+\eta_2-\eta_3$, or any permutation thereof.   

In the process, it is not quite true that the action can be shifted by $2\pi i \sum_im_i\eta_i$ for arbitrary integers $m_i$.  Rather, the
$m_i$ are either all even or all odd.  This holds because similarly the $\Upsilon_b$ functions in the DOZZ formulas are all functions of
$\sum_ic_i\eta_i/b$, where the $c_i$ are all even (the factors in the numerator of the DOZZ formula) or all odd (the factors in the denominator). 

\subsubsection{A Further Comment}\label{furthercom}

One interesting point about this is that for some values of the $\eta_i$, singlevalued complex solutions of Liouville's equations do exist. 
But even in such regions, we may need to use the recursion relations to compute the asymptotics of the DOZZ formula, and hence
we seem to need the full multivaluedness of the action, even though from the present point of view this multivaluedness seems natural
only when the classical solutions are themselves multivalued.  The reason that this happens is that in continuing in $\eta_i$ from Region I to these regions we necessarily pass through regions where $\phi_c$ is multivalued in $z$.  When we arrive at the region of interest it is then possible that although a continuous single-valued solution exist we have actually landed on a discontinuous one.\footnote{For a simple example of this phenomenon, consider the function $$h(x,\bar{x},\eta)=\log\left(\frac{1}{|x|}+\frac{\eta}{|x-1|}\right).$$  For $\eta$ real and positive we can define the branch of the logarithm so that $h$ is a continuous function with an ambiguity of an overall additive factor of $2\pi i N$.  But if we choose such a branch and then at each point $x$ continue in $\eta$ around a circle containing $\eta=0$, this will produce a shift of $2\pi i$ near $x=1$ but not near $x=0$; the resulting function will thus be discontinuous even though a continuous choice of branch exists.}  The locations and strengths of these discontinuities will depend on the path in $\eta$.  This allows the full multivaluedness of the action to be realized, since the discontinuities will not affect the kinetic term when written in terms of $f$ but they will allow independent shifts of $\phi_c$ by $2\pi i N$ near the operator insertions and infinity.  


These discontinuities are admittedly unsettling so we note here that in section \ref{csfun}, we explain a different point of view in which the full multivaluedness of the action is equally natural for any values of the $\eta_i$.  

\subsection{Degenerate Four-Point Function as a Probe}\label{probe}

\hspace{0.25in}The previous two arguments for the role of multivalued ``solutions'' in the  Liouville path integral were rather indirect. We give here
a more direct argument.  In section \ref{4pointreview}, we reviewed Teschner's formula (\ref{deg4}) for the exact four-point function of a light degenerate field $V_{-b/2}$ with three generic operators $V_{\alpha_i}$.  This expression is meromorphic in $\alpha_i$, and choosing all three $\alpha_i$'s to scale like $1/b$ we can study its semiclassical limit for any values of the $\eta_i$.  Moreover we can compare this to (\ref{semiclassicalcorr}), which says that in the semiclassical limit this correlator can be evaluated by replacing the operator $V_{-b/2}$
by the function $\exp(-b\phi)=\exp(-\phi_c/2)$, where $\phi_c$ is the saddle point determined by the three heavy operators.
If there are several relevant saddle points $\phi_{c,N}$, $N\in \T$,  with action $\tilde S_{L,N}$, then (\ref{semiclassicalcorr}) gives\footnote{As discussed below (\ref{semiclassicalcorr}), we have omitted $z_2$-independent factors that are $O(b^0)$ in the exponent.  These come from the functional determinant and corrections to the action (\ref{regaction}).  These factors will cancel between the two sides in (\ref{deg4check}) below.}
\begin{align} \nonumber
\left\langle V_{\eta_4/b}\right.&\left.(z_4,\overline{z}_4) V_{\eta_3/b}(z_3,\overline{z}_3)V_{-b/2}(z_2,\overline{z}_2)V_{\eta_1/b}(z_1,\overline{z}_1)\right\rangle  \\ \nonumber
&\approx \sum_N e^{-\phi_{c,N}(z_2,\bar{z}_2)/2} e^{-\widetilde{S}_{L,N}}.
\end{align}
Using the definitions (\ref{3point}) and (\ref{4point}) and also (\ref{phifromf}), this implies
\be
\label{scdeg4}
{G}_{1234}(x,\overline{x})\approx \sqrt{\lambda} \frac{|z_{14}||z_{34}|}{|z_{13}||z_{24}|^2}\sum_N f_N(z_2,\overline{z}_2)e^{-\widetilde{S}_{134,N}}.
\ee
$\widetilde{S}_{134,N}$ is a branch of (\ref{region1act}) without its position-dependent terms, with the branch labelled by $N$, and with the replacement $\eta_2\to\eta_4$.  Explicitly:
\begin{align}
\nonumber
b^2 \widetilde{S}_{134,N}=&\left(\eta_1+\eta_3+\eta_4 -1\right)\log \lambda+F(\eta_1+\eta_4-\eta_3)+F(\eta_1+\eta_3-\eta_4)\\ \nonumber
&+F(\eta_3+\eta_4-\eta_1)+F(\eta_1+\eta_3+\eta_4-1)-F(2\eta_1)-F(2\eta_3)\\
&-F(2\eta_4)-F(0)+2\pi i (n+m_1\eta_1 +m_3\eta_3+m_4\eta_4).
\label{s134}
\end{align}
Here $n,m_i$ are integers determined by the branch $N$.  We saw in section \ref{threp} that  $e^{\phi_{c,N}}=1/f_N^2$ is uniquely determined (independent of $N$),
so $f_N$ is uniquely determined up to sign. (The sign comes from the choice of square root in defining $a_1$.)  By comparing this with the semiclassical limit of Teschner's formula (\ref{deg4}), we can thus explicitly check the position dependence of the saddle point $f(z,\bar{z})$!

Rewriting Teschner's proposal (\ref{deg4}) with a condensed notation, we have
\be
{G}_{1234}(x,\overline{x})=C^-\,_{12}C_{34-}\left[\mathcal{F}_-(x)\mathcal{F}_-(\bar{x})+\frac{C^+\,_{12}C_{34+}}{C^-\,_{12}C_{34-}}\mathcal{F}_+(x)\mathcal{F}_+(\bar{x})\right].
\ee
Teschner's recursion relation can be rewritten as
\begin{align} \nonumber
&\frac{C_{34+} C^+\,_{12}}{C_{34-} C^-\,_{12}}=-\frac{1}{(1-b(2\alpha_1-b))^2}\\
&\times\frac{\gamma^2(2b(2\alpha_1-1))\gamma(b(\alpha_3+\alpha_4-\alpha_1-b/2))}{\gamma(b(\alpha_1+\alpha_4-\alpha_3))\gamma(b(\alpha_1+\alpha_3-\alpha_4-b/2))\gamma(b(\alpha_1+\alpha_3+\alpha_4-Q-b/2))},
\end{align}
and using (\ref{aproduct}) and (\ref{aratio}) and taking the semiclassical limit this becomes
\be 
\frac{C_{34+} C^+\,_{12}}{C_{34-} C^-\,_{12}}\to-\frac{a_2}{a_1}.
\ee
Here $a_1$ and $a_2$ are the constants in the semiclassical solution (\ref{3pointsol}), with the replacement $\eta_2\to \eta_4$.  In the same limit, we can see from (\ref{3pointPfunctions}) that
\begin{align} \nonumber
&\mathcal{F}_+(x)\to P^{1-\eta_1}(x)\\
&\mathcal{F}_-(x)\to P^{\eta_1}(x).
\end{align}
In checking this, it is useful to recall that we can send $\eta_3 \to 1-\eta_3$ in the definition of $P^{1-\eta_3}(x)$ without changing the function since this is one of Kummar's permutations from Appendix \ref{hyps}.  We thus find that in the semiclassical limit we have
\be
{G}_{1234}(x,\overline{x})= C_{34-}\left[P^{\eta_1}(x)P^{\eta_1}(\overline{x})-\frac{a_2}{a_1}P^{1-\eta_1}(x)P^{1-\eta_1}(\overline{x})+O(b)\right].
\ee
With the help of (\ref{3pointsol}), we find that this will agree with (\ref{scdeg4}) if \footnote{In deriving this formula, we neglected $O(b^0)$ terms in the exponent of $C_{34-}$.  These are the same terms that we previously neglected on the right-hand side of (\ref{scdeg4}), since the difference between $\eta_1$ and $\eta_1-b/2$ affects them only at subleading order.  So this equation really needs to be true to order $O(b^0)$ in the exponent.}
\be
\label{deg4check}
e^{-\widetilde{S}_{34-,N}}=a_{1,N} \sqrt{\lambda} \frac{|z_{14}||z_{34}|}{|z_{13}|}e^{-\widetilde{S}_{134,N}+O(b)}.
\ee
Beginning with this equation we explictly include the branch dependence of $a_1$ for the rest of the section.  Semiclassically the structure constants $C_{134}$ and $C_{34-}$ are in the same region of the $\eta_i$ plane since their $\eta_i$'s differ by something that is $\O(b^2)$, so we can assume they are both a sum over the same set of branches $N$.  This justifies our equating the sums term by term in (\ref{deg4check}).  Using (\ref{s134}), we see that:
\begin{align} \nonumber
\widetilde{S}_{134,N}-\widetilde{S}_{34-,N}=&\frac{1}{2}\log \lambda+i\pi m_1+\frac{1}{2}\Big[\log \gamma(\eta_1+\eta_4-\eta_3)+\log \gamma(\eta_1+\eta_3-\eta_4)\\
&+\log \gamma(\eta_1+\eta_3+\eta_4-1)-\log \gamma(\eta_3+\eta_4-\eta_1)-2\log \gamma(2\eta_1)\Big] \nonumber\\
&+O(b).
\end{align}
Comparing with (\ref{a1squared}), we see that (\ref{deg4check}) is clearly satisfied up to an overall branch-dependent sign.  

To see that this sign works out, we need to give a more careful argument.  First we can define
\begin{align} \nonumber
a_{1,N}=\frac{|z_{13}|}{|z_{14}||z_{34}|}\exp \Big[&\log \gamma(\eta_1+\eta_4-\eta_3)+\log \gamma(\eta_1+\eta_3-\eta_4)\\
&+\log \gamma(\eta_1+\eta_3+\eta_4-1)-\log \gamma(\eta_3+\eta_4-\eta_1)\nonumber\\
&\hspace{0.25in}-2\log \gamma(2\eta_1)+i\pi \widetilde{m}_1\Big].
\end{align}
The logarithms are defined by continuation from real $\eta$'s in Region I along a specific path, which gives an unambiguous meaning to $\widetilde{m}_1$.\footnote{It does not matter what the path is, but we need to choose one.}  The signs will match in (\ref{deg4check}) if $m_1=\widetilde{m}_1$.  To demonstrate this, recall that near $z_1$ we may write
\be
\phi_{c,N}=-4\eta_1 \log |z-z_1|+C_{1,N},
\ee
with
\begin{align}\nonumber
C_{1,N}=&-2\pi i m_1-\log \lambda -(1-2\eta_1)\log \frac{|z_{14}|^2|z_{13}|^2}{|z_{34}|^2}-\log \gamma(\eta_1+\eta_4-\eta_3)\\ \nonumber
&-\log \gamma(\eta_1+\eta_3-\eta_4)-\log \gamma(\eta_1+\eta_3+\eta_4-1)\\
&+\log \gamma(\eta_3+\eta_4-\eta_1)+2\log \gamma(2\eta_1).
\end{align}
Here the logarithms are defined by analytic continuation along the same path as in defining $a_{1,N}$.  Since $\frac{\partial \tilde{S}_{L,N}}{\partial \eta_1}=-C_{1,N}$, we are justified using $m_1$ in this formula.  Finally near $z=z_1$ we have
\be
e^{-\phi_{c,N}/2}\equiv \sqrt{\lambda}f_N= |z-z_1|^{-2\eta_1}e^{-C_{1,N}}\left[1+O(|z-z_1|)\right],
\ee
so in (\ref{scdeg4}) we should choose the branch of $f_N$, and thus of $a_{1,N}$, with $\tilde{m}_1=m_1$.   

This completes our demonstration of (\ref{deg4check}).  We consider this to be very strong evidence that at least for the case of the degenerate four-point function, the Liouville path integral is controlled by singular ``solutions'' throughout the full $\eta_i$ three-plane.  

\subsection{Four-Point Function with a General Light Operator}\label{genf}

\hspace{0.25in}The discussion of the previous section showed that a certain type of four-point function is semiclassically described by singular ``solutions'' of Liouville's equation.  More specifically, the nontrivial position dependence of the correlator (\ref{scdeg4}) was captured by the function $f_N(z_2,\bar{z}_2)$.  The effect of the singularities is rather benign, however; the correlator simply has nontrivial zeros as a function of the position of the light operator.  As argued at the end of section \ref{liouvillesolutions}, the zeros of $f_N$ are generically stable under quantum corrections and thus are actually zeros of the exact four-point function (\ref{deg4}).  There is nothing inherently wrong with such zeros, but this observation is troubling nonetheless. The reason is that these zeros are smooth only because the light operator is exactly degenerate.  If instead of the operator $e^{-\phi_c/2}$ we had considered a more general light operator $e^{\sigma \phi_c}$, then a semiclassical computation based on equation (\ref{semiclassicalcorr}) (the other three operators are still heavy) would have given
\be
\label{gen4point}
G_{1234}(x,\bar{x})\approx G_0\lambda^{-\sigma}\frac{|z_{24}|^{4\sigma}|z_{13}|^{2\sigma}}{|z_{34}|^{2\sigma}|z_{14}|^{2\sigma}}\sum_N f_N(z_2,\bar{z}_2)^{-2\sigma} e^{-\tilde{S}_{134,N}}.
\ee 

Here $G_0$ is a $\mathcal{O}(b^0)$ factor from the fluctuation determinant and the corrections to the action, both of which we expect to be independent of $z_2$, and $\tilde{S}_{134,N}$ is given by (\ref{s134}).  The problem however is that in the vicinity of a point $z_0$ where $f_N(z_2,\bar{z}_2)\approx A(z_2-z_0)+B(\bar{z}_2-\bar{z}_0)$, this correlator is generically singular and discontinuous!\footnote{One might hope that the discontinuity could cancel in the sum over the different branches $N$, but this will not work because for any given generic values of $\eta_1,\eta_2,\eta_3,\sigma$ there will be a single dominant saddlepoint that is parametrically larger as $b\to0$.}  We can quantify the nature of these singularities by using the winding number introduced at the end of section \ref{liouvillesolutions}, and we find that the semiclassical correlator has winding number $-2\sigma$ around $z_0$ if $|A|>|B|$ and winding number $2\sigma$ if $|A|<|B|$.  The winding number is not an integer because the function is discontinuous.  It cannot be changed significantly by small corrections, and since it is generically nonzero we are tempted to conclude that the exact four-point function must also be discontinuous as a function of the light operator position at finite but sufficiently small $b$!\footnote{In general it is of course possible for a smooth function to have a semiclassical approximation which is discontinuous, a simple example is $\frac{1}{\Gamma(x/\lambda)}$, which has a line of zeros turn into a branch cut as $\lambda\to0$. A more sophisticated example that we have been studying extensively in this part is $\Upsilon_b(x/b)$, which exhibits the same phenomenon.  That this does not happen for the four-point function under consideration is a special consequence of the semiclassical formula (\ref{gen4point}) for the correlator, where the nontrivial $z_2$-dependence is all in a factor that is finite as $b\to0$ and the factor that goes like $e^{-1/b^2}$ is independent of $z_2$.}  

This situation would not be entirely without precedent; in the $SL(2,\R)$ WZNW model appropriate for studying strings in $AdS_3$ \cite{Giveon:1998ns,Teschner:1997fv,Teschner:1999ug,Teschner:2001gi} it was shown in \cite{Maldacena:2000hw,Maldacena:2001km} that the exact 4-point function of certain operators has singularities when all four operators are at distinct positions.  This could be seen semiclassically from stringy instantons going ``on-shell'' and was reproduced exactly using the machinery of the Knizhnik-Zamolodchikov equation \cite{Knizhnik:1984nr}.  In that situation however the singularities were localized to isolated points and the correlator was continuous away from those points. In the remainder of this section we will give an argument that in Liouville there are in fact no singularities, isolated or otherwise, in the exact four-point function when the operator positions do not coincide.  We will then close the section with some speculation about where our semiclassical argument goes wrong.  We caution however that we will use some plausible pieces of lore that have not strictly been proven, so our argument is slightly heuristic.

We take as a starting point the exact formula (\ref{factorized4point}) for the Liouville 4-point function, which we reproduce here for convenience
\begin{align}\nonumber
G_{1234}(x,\bar{x})=&\frac{1}{2}\int_{-\infty}^{\infty}\frac{dP}{2\pi}C(\alpha_1,\alpha_2,Q/2+iP)C(\alpha_3,\alpha_4,Q/2-iP)\\\label{f4p2}
&\times\mathcal{F}_{1234}(\Delta_i,\Delta_P,x)\mathcal{F}_{1234}(\Delta_i,\Delta_P,\bar{x}).
\end{align}
This formula is strictly true only when $\Re(\alpha_1+\alpha_2)>Q/2$ and $\Re(\alpha_3+\alpha_4)>Q/2$.  Away from this region, which we will certainly be with three heavy operators obeying the Seiberg bound and one light operator, there are additional discrete terms that are residues of the finite number of poles that have crossed the contour of integration.  Looking at this expression, we see that there are only two possible sources of singularities in $x,\bar{x}$.  The first is singularities of the Virasoro conformal blocks $\mathcal{F}_{1234}$ as a function of $x$, and the second is possible divergence of the integral over $P$ for particular values of $x$.  We will address each of these issues, beginning with possible singularities of the conformal blocks.  

The conformal blocks are expected to have branch points at $x=0,1,\infty$, which correspond to the UV singularities of the correlator when the operator at $z_2$ approaches the operators at $z_1$, $z_3$, or $z_4$.  The singularity at $x=0$ is manifest from the definition (\ref{conformalblock}), and the singularity at $x=1$ arises from the nonconvergence of the series in (\ref{conformalblock}) when $|x|=1$.  When all operator weights are real and positive the fact that the radius of convergence of this series is indeed one follows from the convergence of inserting a complete set of states in unitary quantum mechanics.  The convergence for generic complex operator weights has actually never been proven in the literature, although it was conjectured to be true in \cite{Zamolodchikov:1989mz} and discussed more recently in \cite{Teschner:2001rv,Hadasz:2004cm}.  In \cite{Hadasz:2004cm} it was proven that if the radius of convergence is indeed one, then there are no other singularities with $|x|>1$ except for the singularity at infinity.  We will also not be able to prove this convergence, but we give two pieces of evidence in favor of it.  First we note that the $c\to\infty$ limit of the conformal block, which turns out to mean including only descendants of the form $(L_{-1})^n|Q/2+iP\rangle$ in the sum in (\ref{conformalblock}), can be evaluated explictly from the definition and gives \cite{Zamolodchikov:1989mz,Zamolodchikov:1985ie}
\be
\lim_{c\to\infty}\mathcal{F}_{1234}(\Delta_i,\Delta_P,x)=x^{\Delta_P-\Delta_1-\Delta_2}F(\Delta_P+\Delta_2-\Delta_1,\Delta_P+\Delta_3-\Delta_4,2\Delta_P,x).
\ee
As discussed in Appendix \ref{hyps}, this hypergeometric function is singular only at $x=0,1,\infty$.  So any additional singularities of $\mathcal{F}_{1234}$ would have to disappear in the $c\to\infty$ limit, which seems unnatural.  When $c$ is finite but one of the external legs is degenerate we can again compute the conformal block, with result (\ref{degblock}).  Again it only has singularities in the expected places.  In 27 years of studying these functions as far as we know no evidence has emerged for singularities at any other points in $x$, so from now on we assume that they do not exist.  

The other possible source of singularities in the four point function (\ref{f4p2}) is divergence of the integral over $P$.  To study this further, we need large $P$ expressions both for the structure constants and the conformal blocks.  The appropriate asymptotics for $\Upsilon_b$ are quoted (with some minor typos we correct here) as equation 14 in \cite{Teschner:2001rv}:
\be
\label{imupsilon}
\log\Upsilon_b(x)=x^2\log x-\frac{3}{2}x^2\mp\frac{i\pi}{2} x^2+\mathcal{O}(x\log x) \qquad \Im \, x\to \pm \infty.
\ee
We will not derive this, but it isn't hard to get these terms from our expression (\ref{etaasymp}) for the semiclassical limit of $\Upsilon_b$ with $x$ scaling like $1/b$.\footnote{To do this, we observe that for large $\eta$ we can use Stirling's formula to approximate $\log \gamma(x)$ inside the integral expression (\ref{Fdefinition}) for $F(\eta)$.  This is not quite the same as a full finite-$b$ derivation since in principle there could be subleading terms in $b$ that become important for sufficiently large $\eta$, but we have checked this formula numerically at finite $b$ with excellent agreement so apparently this does not happen.  The formula (\ref{etaasymp}) was valid only for $\eta$ in a certain region, but using the recursion relations to get to other regions will not affect things to the order we are working in (\ref{imupsilon}) so (\ref{imupsilon}) is valid for arbitrary $\Re(x)$.}  Using this in (\ref{dozz}), we find that at large real $P$ we have
\be
\label{ClargeP}
C(\alpha_1,\alpha_2,Q/2\pm iP)=16^{-P^2+\mathcal{O}(P \log P)}.
\ee
The structure of the conformal blocks at large $P$ was studied by Al. B. Zamolodchikov in a series of  papers \cite{Zamolodchikov1986,Zamolodchikov1987},\footnote{English translations are availiable online but hard to find.  This also especially the case for reference \cite{Zamolodchikov:1989mz}, which gives a beautiful exposition of the general formalism of \cite{Belavin:1984vu} that is more complete than anything else in the literature.  The most accessible place to find the formula quoted here seems to be in section 7 of \cite{Zamolodchikov:1995aa}, but beware of a notational difference in that our conventions are related to theirs by $1 \leftrightarrow 2$.}; he obtained the following remarkable result:
\begin{align}\label{remarkable}\nonumber
\mathcal{F}_{1234}(\Delta_i,\Delta,x)=&(16q)^{\Delta-\frac{c-1}{24}}x^{\frac{c-1}{24}-\Delta_1-\Delta_2}(1-x)^{\frac{c-1}{24}-\Delta_2-\Delta_3}\\
&\times\theta_3(q)^{\frac{c-1}{2}-4(\Delta_1+\Delta_2+\Delta_3+\Delta_4)}\left(1+\mathcal{O}(1/\Delta)\right).
\end{align}
Here $\theta_3(q)=\sum_{n=-\infty}^{\infty} q^{n^2}$ and $q=\exp\left[-\pi K(1-x)/K(x)\right]$, with
\be
K(x)=\frac{1}{2}\int_0^1 \frac{dt}{\sqrt{t(1-t)(1-x t)}}.
\ee
This $q$ can be interpreted as $\exp(i\pi \tau)$, where $\tau$ is the usual modular parameter of the elliptic curve $y^2=t(1-t)(1-xt)$.  So in particular
$\mathrm{Im}\,\tau$ is always positive and one always has $|q|<1$ when the elliptic curve is smooth (that is, for $x\not=0,1,\infty$).
For fun we note that, like most things in this part, $K(x)$ is actually a hypergeometric function: equation (\ref{hypintegral}) gives $K(x)=\frac{\pi}{2}F(1/2,1/2,1,x)$.  

The derivation of (\ref{remarkable}) uses certain reasonable assumptions about the semiclassical limits of correlation functions; we will be explicit about them in Appendix \ref{blocksapp}, where for convenience we review the origin of the leading behaviour 
\be
\mathcal{F}_{1234}(\Delta_i,\Delta_P,x)\sim\left(16q\right)^{P^2}.
\ee
This will be sufficient for our study of the integral in (\ref{f4p2}); combining it with (\ref{ClargeP}) we see that the integral will converge, for $x\not=0,1,\infty$, given
the fact that $|q|<1$.    Thus  the integral cannot generate any new singularities.  This completes our argument that the Liouville four-point function (\ref{f4p2}) cannot have any new singularities in $x$.

So what is wrong with our semiclassical argument for such singularities in the beginning of the section?  To really understand this we would have to compute the semiclassical limit of (\ref{f4p2}) and compare it to our formula (\ref{gen4point}). For the moment, this is beyond our ability.  We may guess however that the problem lies in our assumption that the factor $G_0$ is independent of $z_2$.  This was true for the degenerate computation in the previous section, but the singularity we discovered here perhaps suggests that more sophisticated renormalization of the nondegenerate light operator is required in the vicinity of any singular points of the ``solution''.  It is at first unsettling that the renormalization of the operator should depend on the positions and strengths of the other heavy operators, but we already saw in section \ref{gaction} that even the ``principal value'' prescription for evaluating the action depended on these things at distances arbitrarily close to the singular point.  Thus we expect that once an appropriate renormalization is performed, the semiclassical singularity in (\ref{gen4point}) will be smoothed out.  It would be good to be more explicit about what this renormalization is, but we will not try to do so here.  

A different perspective on this four-point function is provided by the Chern-Simons formulation of Liouville theory, which we will introduce momentarily.  In this formulation it seems clear that there are conventional nonsingular solutions that exist for any $\eta_i$ and which can be used to study the semiclassical limit of this correlator; in this version of things it seems apparent that no singularity can emerge.  
\chapter{Interpretation In Chern-Simons Theory}\label{csfun}

\subsection{Liouville Solutions And Flat Connections}\label{ytr}

\hspace{0.25in}In section \ref{genfcs}, to a solution of Liouville's equations we associated a holomorphic
differential equation
\be \label{hole} \left(\frac{\partial^2}{\partial z^2}+W(z)\right )f=0  \end{equation}
and also an antiholomorphic differential equation
\be \label{azole}\left(\frac{\partial^2}{\partial \bar z^2}+\tilde W(\bar z)\right )f=0.  \end{equation}
Locally, (\ref{hole}) has a two-dimensional space of holomorphic solutions, and (\ref{azole}) has a two-dimensional
space of antiholomorphic solutions.  We constructed a solution of Liouville's equation from a basis
$\begin{pmatrix}u\\ v\end{pmatrix}$ of holomorphic solutions of (\ref{hole}) along with a basis $\begin{pmatrix}\tilde u\\ \tilde v
\end{pmatrix}$ of antiholomorphic solutions of (\ref{azole}).  This construction applies on any Riemann surface $\Sigma$,
though we have considered only $\sf S^2$ in the present part.

Globally, in passing around a noncontractible loop in $\Sigma$, or around a point at which there is a singularity
due to insertion of a heavy operator, the pair $\begin{pmatrix}u\\ v\end{pmatrix}$ has in general non-trivial
 monodromy.  The monodromy maps this pair to another basis of the same two-dimensional space of solutions, so it takes the form
\be \label{monodromy} \begin{pmatrix}u\\ v\end{pmatrix} \to \begin{pmatrix}\hat u\\ \hat v\end{pmatrix}=M \begin{pmatrix}u\\ v\end{pmatrix},\ee
where $M$ is a constant $2\times 2$ matrix.  Actually, the determinant of $M$ is 1, so $M$ takes values in
$SL(2,\C)$.  One way to prove this is to use the fact that the Wronskian $u\partial v-v\partial u$ is independent
of $z$, so it must have the same value whether computed in the basis $u,v$ or the basis $\hat u,\hat v$.  This condition
leads to $\det M=1$.  Alternatively, we may observe that the differential equation (\ref{hole}) may be expressed in terms of
an $SL(2,\C)$ flat connection.  We introduce the complex gauge field $\A$ defined by
\be\label{gfield}\A_z=\begin{pmatrix}0&-1\\ W(z)& 0\end{pmatrix},~~\A_{\bar z}=0.\ee
Since these $2\times 2$ matrices are traceless, we can think of $\A$ as a connection with gauge group $SL(2,\C)$.\footnote{Our convention for non-abelian gauge theory is that $D_\mu=\partial_\mu+A_\mu$, so in particular $F_{\mu\nu}=[D_\mu,D_\nu]=\partial_\mu A_\nu-\partial_\nu A_\mu+[A_\mu,A_\nu]$ and the gauge transformation is $D_\mu\to g D_\mu g^{-1}$, with $g\in G$.}  On the other hand, a short calculation shows that the condition for
a pair $\begin{pmatrix}f\cr g\end{pmatrix}$ to be covariantly constant with respect
to this connection is equivalent to requiring that  $f$ is a holomorphic
solution of the equation (\ref{hole}) while $g=\partial f/\partial z$.  Thus parallel transport of this doublet around a loop, which we accomplish by multiplying by $U=P e^{-\oint A_z dz}$, is the same as analytic continuation around the same loop.  In particular if we define the matrix 
$S=\begin{pmatrix}
u & v\\
\partial u & \partial v
\end{pmatrix}
$,
then we have $US=S M^T$.  Taking the determinant of this equation, we find that $M\in SL(2,\C)$.

Similarly, the antiholomorphic differential equation (\ref{azole}) has monodromies valued in $SL(2,\C)$.  This may be proved
either by considering the Wronskian or by introducing the corresponding flat connnection $\tilde \A$, defined by
\be\label{gfieldz}\tilde \A_z=0,~~ \tilde \A_{\bar z}=\begin{pmatrix}0&\tilde W(\bar z)\\ -1& 0\end{pmatrix}.\ee
(It is sometimes convenient to take the transpose in exchanging $z$ and $\bar z$, and we have done so, though
this will not be important in the present part.)

The connections $\A$ and $\tilde \A$ have singularities near points with heavy operator insertions.  The monodromy
around these singularities can be inferred from the local behavior of the solutions of the differential equation.
For example, the solutions of (\ref{hole}) behave as $z^\eta$ and $z^{1-\eta}$ near an operator insertion at $z=0$
with Liouville momentum $\alpha=\eta/b$.  The monodromies of these functions under a circuit in the counterclockwise
direction around $z=0$ are $\exp(\pm 2\pi i\eta)$.  An invariant way of describing this, without picking a particular basis of solutions,
is to say that 
\be \label{oxo}\mathrm{Tr}\,M=2\cos (2\pi\eta). \end{equation}
Similarly, the behavior of the local solutions near $z=0$ implies that the monodromy of the antiholomorphic equation
(\ref{azole}) around $z=0$ has the same eigenvalues, and hence again obeys (\ref{oxo}). 

More generally, the  two flat connections $\A$ and $\tilde \A$ are actually gauge-equivalent\footnote{The explicit gauge transformation between them is $g=
\begin{pmatrix}
\partial u \bar{\partial}\tilde{u}+\partial v \bar{\partial}\tilde{v} & -u  \bar{\partial}\tilde{u}-v  \bar{\partial}\tilde{v}\\
 -\tilde{u}\partial u-\tilde{v}\partial v & u \tilde{u}+v \tilde{v}
\end{pmatrix}
$.} and have conjugate monodromies
around all cycles, including noncontractible cycles on $\Sigma$ (if its genus is positive) as well as cycles of the sort just considered.  This is guaranteed by the fact that
$f=u\tilde u-v\tilde v$ has no monodromy, since the Liouville field $\phi_c$ is $e^{\phi_c}=1/\lambda f^2$.  

So a solution of Liouville's equations -- real or complex -- gives us a flat $SL(2,\C)$ connection over $\Sigma$ that can be put
in the gauge (\ref{gfield}) and can also be put in the gauge (\ref{gfieldz}).
The basic idea of the present section is that, by a complex solution of Liouville's equations, we should mean in general a flat $SL(2,\C)$
connection, up to gauge transformation, which can be gauge-transformed to either of those two forms.
We do not worry about what sort of expression it has in terms of a Liouville field.

The attentive reader may notice that we have cut some corners in this explanation, because  in section \ref{liouvillesolutions} the reference metric was chosen
to be flat in deriving the holomorphic differential equations.  This is not possible globally if $\Sigma$ has genus greater than 1, and even for $\Sigma=\sf S^2$, it involves introducing an unnatural singularity at infinity. 
A more precise description is to say that $\A$ is a flat connection that  locally, after picking a local coordinate $z$, can be put in the form (\ref{gfieldz}), in such
a way that in the intersection of coordinate patches, the gauge transformation required to compare the two descriptions is lower triangular
\be\label{lowtria}g=\begin{pmatrix} * & 0 \cr * & * \end{pmatrix}.\ee
A flat connection with this property is known as an oper.  This notion is explained in section 3 of \cite{GWNow}, but
we will not need that degree of detail here. 
The global characterization of $\tilde A$ has the same form (with upper triangular matrices replacing lower triangular ones,
given the choice we made in (\ref{gfieldz})).  Our proposal then is that a classical solution of Liouville theory is
a flat connection whose holomorphic structure is that of an oper, while its antiholomorphic structure is also that of an oper.

\subsection{Some Practice}\label{practice}

\hspace{0.25in}A few elementary observations may give us some practice with these ideas.  
Let us first consider the main example of this part, namely  $\sf S^2$ with insertions of three heavy operators of Liouville momenta
$\alpha_i=\eta_i/b$, at positions $z=z_i$.  The monodromies $M_i$ around the three points will have to obey
\be\label{moncon}\Tr\,M_i=2\cos(2\pi\eta_i),~~\det\,M_i=1\end{equation}
In addition, the product of the three monodromies must equal 1:
\be\label{oncon}M_1M_2M_3 = 1.\ee
Equivalently
\be\label{zoncon}M_1M_2=M_3{}^{-1},\ee
from which it follows that
\be\label{trm}\Tr\,M_1M_2=\Tr\,M_3{}^{-1}=2\cos(2\pi\eta_3).\ee
And of course we are only interested in a flat bundle up to conjugacy
\be\label{conj}M_i\to gM_ig^{-1},~~ g\in SL(2,\C). \ee

To start with, let us just ignore the oper condition and ask
how many  choices of the $M_i$ there are, up to conjugacy, that obey the conditions
in the last paragraph.
We can partially fix the gauge invariance by setting
\be\label{zrm}M_1=\begin{pmatrix}e^{2\pi i\eta_1}& 0\cr 0 & e^{-2\pi i\eta_1}\end{pmatrix}.\end{equation}
The remaining freedom consists of diagonal gauge transformations
\be\label{dia} g=\begin{pmatrix}\lambda & 0 \cr 0 & \lambda^{-1}\end{pmatrix},\ee
where we only care about $\lambda$ up to sign, since a gauge transformation by $g=-1$ acts trivially on all gauge fields and
monodromies.  
In general, we can take
\be\label{rmz}M_2=\begin{pmatrix} p & q \cr r & s\end{pmatrix}\ee
If we look for a solution with $q=0$, we soon find that, for generic values of the $\eta_i$, once we adjust $p$ and $r$ 
to get the right values of $\Tr\,M_2$ and $\det\,M_2$, we cannot also satisfy (\ref{trm}).
So we take $q\not=0$, in which case $\lambda$ in (\ref{dia}) can be chosen uniquely, up to sign, to set $q=1$.
Then, imposing $\det \,M_2=1$, we get
\be\label{fled}M_2=\begin{pmatrix} p & 1 \cr pq-1 & q\end{pmatrix}.\ee
Now the conditions $\Tr\,M_2=2\cos(2\pi\eta_2)$, $\Tr\,M_1M_2=2\cos(2\pi\eta_3)$ give two linear equations for $p$ and $q$
which generically have a unique solution.  So $M_2$ and therefore $M_3=M_2^{-1}M_1^{-1}$ are uniquely determined.

The conclusion is that a flat bundle on the three-punctured sphere 
with  prescribed conjugacy classes of the monodromies $M_i$ is unique, up to gauge equivalence,
even if we do not require the oper conditions.\footnote{This remains true
for non-generic values of the $\eta_i$ where the derivation in the last paragraph does not quite apply, unless the $\eta_i$
equal $(0,0,0)$ or a permutation of $(0,1/2,1/2)$.  To verify this requires only one subtlety: if for some $i$, $\Tr\,M_i=\pm 2$, so that the eigenvalues of $M_i$ are equal,
then one should not assume that $M_i$ can be diagonalized; its Jordan canonical form may be $\pm\begin{pmatrix}1&1\cr 0 & 1\end{pmatrix}$.}
The unique $SL(2,\C)$ flat bundle with these monodromies can be realized by a holomorphic differential equation and also by an
antiholomorphic one.  The proof of this statement is simply that functions $W$ and $\tilde W$ with the right singularities do exist,
as in (\ref{olgo}).

Since it can be realized by both a holomorphic differential equation and an antiholomorphic one, 
 the unique $SL(2,\C)$ flat bundle on the three-punctured sphere with monodromies in the conjugacy classes determined by the $\eta_i$  is a complex solution of
Liouville's equations in the sense considered in the present section.  What one would mean by its action and why this
action is multivalued will be explained in section \ref{chern}.  

It is instructive to consider a more generic case with $s>3$ heavy operators, with parameters $\eta_i$, inserted at points $z_i\in \sf S^2$.
Now there are $s$ monodromies $M_i$, $i=1,\dots,s$.  They are $2\times 2$ matrices, constrained by
\be \label{constmat}\Tr\,M_i=2\cos(2\pi\eta_i),  ~~ \det\,M_i=1.\end{equation}
We also require
\be\label{ormo} M_1M_2\dots M_s= 1,\end{equation}
and we are only interested in the $M_i$ up to conjugacy
\be\label{tormo}M_i\to gM_ig^{-1}.\ee
A simple parameter count shows that the moduli space $\M$ of flat bundles over the $s$-punctured sphere that obey these conditions
has complex dimension $2(s-3)$.
Instead, let us ask about the subspace  $\V\subset \M$ consisting of flat bundles
that can be realized by a holomorphic differential equation (\ref{azole}).  We already know that the potential $W$ that appears in the holomorphic differential equation is unique for $s=3$.  When one
increases $s$ by 1, adding a new singularity at $z=z_i$ for some $i$,
one adds another double pole to $W$, giving a new contribution  $\Delta W=c/(z-z_i)^2+c'/(z-z_i)$.
But $c$ is determined to get the right monodromy near $z_i$, so only $c'$ is a new parameter (usually called the accessory parameter).  Hence the dimension of $\V$
is $s-3$; $\V$ is a middle-dimensional subspace of $\M$.  (For a more complete account of this standard result, see section 8 of \cite{GWNow}.)
Similarly, the subspace $\tilde \V$ of flat bundles that can be realized by an antiholomorphic differential equation is middle-dimensional.

A complex solution of Liouville theory in the sense that we consider in the present section corresponds to an intersection point of
$\V$ and $\tilde\V$.  As $\V$ and $\tilde \V$ are both middle-dimensional, it is plausible that their intersection
generically  consists of finitely many points, or possibly even that it always consists of just one point.  Unfortunately we do not know if
this is the case.  All we really know is that for any $s$, if the $\eta_i$ are real and obey the Seiberg and Gauss-Bonnet bounds,
then there is a real solution of Liouville's equations, and this corresponds to an intersection point of $\V$ and $\tilde \V$.

\subsection{Interpretation In Chern-Simons Theory}\label{chern}

\hspace{0.25in}To explain what one would mean in this language by the action of a classical solution, and why it is multivalued, the main idea
is to relate Liouville theory on a Riemann surface $\Sigma$ to Chern-Simons theory on $\Sigma\times I$.  The basic reason that there is such a relation
is that Virasoro conformal blocks (which can be understood as building blocks of Liouville theory) can be viewed as physical states
in three-dimensional Chern-Simons theory.  This was first argued in \cite{HVerlinde} and has been reconsidered much more recently
\cite{GWNow}.  

We start with an $SL(2,\C)$ connection $\A$ with Chern-Simons action
\be\label{csact}S_{\mathrm{CS}}=\frac{1}{4\pi i b^2}\int_M\Tr\,\left(\A\wedge d\A+\frac{2}{3}\A\wedge\A\wedge\A\right)\ee
on a three-manifold $M$. 
$S_{\mathrm{CS}}$ is invariant under gauge transformations that are continuously connected to the identity, but not under homotopically
non-trivial gauge transformations.  For $M=\sf S^3$, the homotopically non-trivial gauge transformations are parametrized
by $\pi_3(SL(2,\C))=\Z$.  The integer invariant of a gauge transformation is often called winding number.
 In defining $S_{\mathrm{CS}}$,  we have picked a convenient normalization, such
that under a homotopically nontrivial gauge transformation on $\sf S^3$ of winding number $n$, $S_{\mathrm{CS}}$ transforms by
\be\label{acts}S_{\mathrm{CS}}\to S_{\mathrm{CS}}+\frac{2\pi in}{b^2}.\end{equation}  With $b$ understood as the Liouville coupling parameter,
this matches the multivaluedness of Liouville theory that comes from the trivial symmetry $\phi_c\to\phi_c+2\pi i$.  
Conventionally, the Chern-Simons action is normalized to make $S_{\mathrm{CS}}$ singlevalued mod $2\pi i\Z$ and the homotopically
nontrivial gauge transformations are regarded as symmetries \cite{DJT}.  For our purposes, this would be far too
restrictive (since we do not want to assume that $b^2$ is the inverse of an integer).  Rather, in the path integral, 
we consider integration
cycles  that are not invariant under homotopically non-trivial gauge transformations,
and we do not view homotopically nontrivial gauge transformations as symmetries of the theory.   In other words, we adopt the
perspective of \cite{Analytic}.  The integration cycles are middle-dimensional in the space of $SL(2,\C)$-valued flat connections.   A basis of the possible integration cycles is given
by the cycles that arise by steepest descent from a critical point of the action.

The Yang-Mills field strength is defined, as usual, by $\F=d\A+\A\wedge \A$. 
The classical equations of motion of Chern-Simons theory are simply
\begin{equation}\label{classeom}\F=0. \ee
We take the three-manifold $M$ to be simply $M=\Sigma\times I$, where $I$ is a unit interval and $\Sigma$
is the Riemann surface on which we want to do Liouville theory.
The fundamental group of $M$ is therefore the same as that of $\Sigma$, and so a solution of  the classical
equation of motion (\ref{classeom})  is just an $SL(2,\C)$ flat connection on $\Sigma$.  This being so, one may wonder
what we have gained by introducing a third dimension.  

The answer to this question is that to do Liouville theory, we need more than an $SL(2,\C)$ flat connection on $\Sigma$.  It must
obey two conditions: {\it (i)} it can be described by a holomorphic differential equation, and {\it (ii)} it can also be described by an antiholomorphic
differential equation.  It is possible to pick boundary conditions at the two ends of $I$ so that condition {\it (i)} is imposed at one
end and  condition {\it (ii)}  at the other end. Such  boundary conditions were introduced in Chern-Simons theory in  \cite{HVerlinde} and used to relate that theory to Virasoro conformal blocks; a variant related
to Nahm's equations and other topics in mathematical physics has been described  in \cite{GWNow}.

We may call these oper or Nahm pole boundary conditions.  For the very schematic purposes of the present part, almost all that we really need
to know about them is that they completely break the $SL(2,\C)$ gauge symmetry down to the center $\pm 1$ of the gauge group.

Now let us consider the topological classification of gauge transformations on $\Sigma\times I$. The gauge transformation
is described by a map $g$ from $\Sigma\times I$ to $SL(2,\C)$.  At the left end of $\Sigma\times I$,
 $g$ must equal  1 or $-1$.   For the present part, an overall gauge transformation by the center
of $SL(2,\C)$ will be of no interest, since it acts trivially on all gauge fields, so we can assume that at the left end, $g=1$.  But then there are two choices at the right
end, namely $g=1$ and $g=-1$.  After we make this choice, the remaining freedom in describing $g$ topologically is
given by $\pi_3(SL(2,\C))=\Z$.  The homotopy classification of gauge transformations is
by $\Z\times \Z_2$, where the $\Z_2$ factor classifies the relative value of $g$ at the two ends and $\Z$ classifies
the twist by $\pi_3(SL(2,\C))$. 

This last statement is not completely trivial; we must verify that the homotopy classification is by a simple
product $\Z\times \Z_2$ and not by a nontrivial extension $0\to \Z\to \Gamma\to\Z_2\to 0$.  Concretely, the question
is whether a gauge transformation with $g=-1$ on the right end has integer or half-integer winding number.  In fact, the winding
number is always an integer.  To see this, it suffices to exhibit an example of a gauge transformation with $g=-1$ on the right
end and with integer winding number.  We can simply pick $g$ to be a function on $\Sigma\times I$ that only depends
on the second factor; such a map can be constructed from a path from $1$ to $-1$ in the group $SL(2,\C)$.  (The conclusion just
stated remains valid when monodromy defects are included, as we do momentarily.  For this, it suffices to note that by continuity the question is independent
of the values of the $\eta_i$, while if one of the $\eta_i$ vanishes, we can forget the corresponding defect.)

\subsection{Liouville Primary Fields and Monodromy Defects}\label{orty}
\begin{figure}
\centering
\includegraphics[width=10cm]{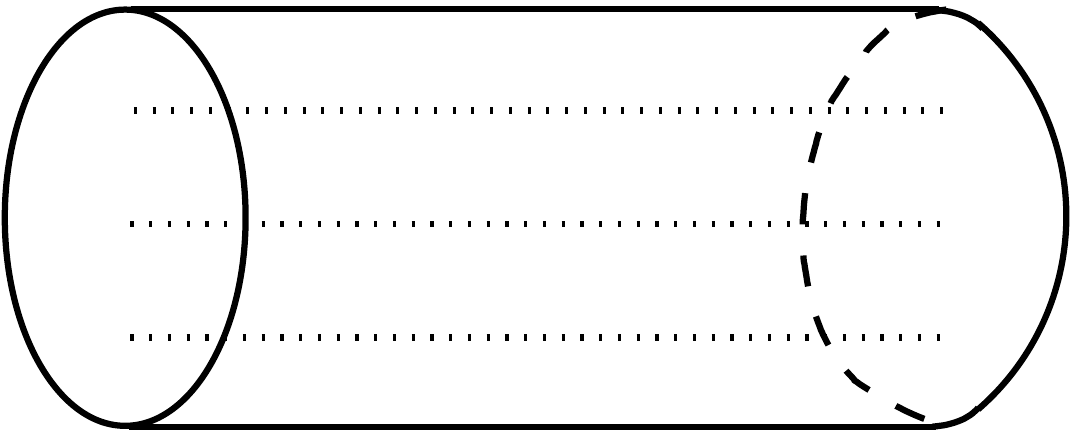}
  \caption{ \label{OneFigTwo}  Monodromy defects, depicted as horizontal dotted lines, in $\Sigma\times I$.  In the example shown, $\Sigma=\sf S^2$
and the number of monodromy defects is 3.}
\end{figure}
Some basic points about how to interpret various Liouville fields in the Chern-Simons description have been explained in \cite{GWNow}.
The main point of concern to us is how to incorporate a primary field of Liouville momentum $\alpha=\eta/b$ at a point $p\in \Sigma$.
The answer is simply that in the Chern-Simons description, the gauge field $\A$ should have an appropriate monodromy around
the codimension two locus $p\times I\subset \Sigma\times I$.  This can be achieved by requiring $\A$ to have a suitable singularity along $\Sigma$.
If $z=re^{i\theta}$  is a local coordinate that vanishes at $p$, then we require that the singular behavior of $\A$ should be
\begin{equation}\label{singb}\A=i\,d\theta\begin{pmatrix}\eta & 0 \cr 0 & -\eta \end{pmatrix}+\dots,\end{equation}
where the ellipsis  refers to less singular terms.
This singular behavior has been chosen so that the eigenvalues of the monodromy are $\exp(\pm 2\pi i\eta)$.   We call a singularity of this kind
in the gauge field a monodromy defect.  (What happens when a monodromy defect meets the Nahm pole singularity at the ends of $I$
has been analyzed in section 3.6 of \cite{wittenknots}. The details are not important here.)    If in the Liouville
description, there are several primary fields, say at points $p_i$ given by $z=z_i$, then in the Chern-Simons description we include
monodromy defects on $p_i\times I$ for each $i$ (Fig. \ref{OneFigTwo}).

Now we want to work out the topological classification of gauge transformations in the presence of monodromy defects.  
The ansatz (\ref{singb}) is only invariant under diagonal gauge transformations along $p$.    So now a gauge transformation $g$ is constrained
as follows: at the ends of $\Sigma\times I$, it equals $\pm 1$ (and we assume it to equal $+1$ on the left end), while along the monodromy
defects it is diagonal, and away from the boundary and the  monodromy defects, it takes arbitrary values in $SL(2,\C)$.

Let us look at what happens along a particular monodromy defect.  A diagonal gauge transformation can be written
\begin{equation}\label{hokey}g=\begin{pmatrix}\rho e^{i\vartheta}& 0 \cr 0 & \rho^{-1}e^{-i\vartheta}\end{pmatrix},\end{equation}
where $\rho$ is positive and $\vartheta$ is real.  Let us parametrize the interval $I$ by a parameter $y$ that equals, say, 0 at the left
end of $I$ and 1 at the right end.  We constrain $\vartheta$ to vanish at $y=0$ (where $g=1$), and to equal $\pi m$ at $y=1$,
where $m$ is even if $g=1$ on the right end of $\Sigma\times I$, and $m$ is odd if
$g=-1$ there.
We call $m$ the winding number along the defect.  It is sometimes useful to write it as
\be\label{wil}m=\frac{1}{\pi}\int_0^1 dy \,\frac{d\vartheta}{dy}.\ee
We have normalized $m$ so that wrapping once around the maximal torus of $SO(3)$
corresponds to $m=1$, while wrapping once around the maximal torus of $SU(2)$ 
corresponds to $m=2$.

So in total, with $s$ monodromy defects, the topological classification of gauge transformations is by integers
$(n,m_1,m_2,\dots,m_s)$, where $n$ is the bulk winding number and $m_i$, $i=1,\dots,s$ are winding numbers defined just
along the monodromy defects.  The $m_i$ are either all even or all odd, since their oddness or evenness is determined by the behavior of $g$.

In the presence of a monodromy defect, it is necessary to add one more term to the action.
The reason is that in the presence of a monodromy defect, 
we would like a flat bundle on the complement of the monodromy defect
that has the singularity (\ref{singb}) along the defect to be a classical solution.  Such
a flat bundle obeys
\be\label{zelve}\F=2\pi i\begin{pmatrix}\eta & 0 \cr 0 & -\eta\end{pmatrix} \delta_K.\ee
Here we write $K$ for the support of the monodromy defect, and $\delta_K$ is a two-form delta function supported on $K$.
But in the presence of the Chern-Simons action (\ref{csact}) only, the equation of motion
is simply $\F=0$ rather than (\ref{zelve}).  To get the equation we want, we must add to the
action a term
\begin{equation}\label{zovax}S_K=-\frac{1}{b^2}\int_K\,\Tr\,\A\begin{pmatrix}\eta & 0 \cr 0 & -\eta
\end{pmatrix}.\ee

Finally, we can determine how the action transforms under a gauge transformation in
the presence of a monodromy defect.  (See section 4.2.6  of \cite{Analytic} for an alternative explanation.)  We have already discussed the behavior
of the Chern-Simons term under gauge transformation, so what remains is to understand
what happens to the new interaction $S_K$.  For $K=p\times I$,
only $\A_y$, the component of $\A$ in the $y$ direction, appears in (\ref{zelve}).
Under a diagonal gauge transformation (\ref{hokey}), the diagonal matrix elements of
$\A_y$ are shifted by $\mp d\log (\rho e^{i\vartheta})/dy$.  Taking the trace and integrating
over $y$,  we find that $S_K$ is shifted by
\begin{equation}\label{ovax}\frac{2i\eta}{b^2}\int_0^1dy\,\frac{d\vartheta}{dy}=\frac{2\pi i \eta m}{b^2}.\end{equation}
(There is no contribution involving $d\log\rho$, since $\int_0^1dy (d\log \rho/dy)=0$, as $\rho=1$ at both endpoints.)

More generally, let us go back to the case of  $s$ heavy operator insertions, with Liouville
parameters $\eta_i$, $i=1,\dots,s$, inserted at points $p_i\in \Sigma$.  In Chern-Simons theory, they correspond to monodromy
defects, supported on $K_i=p_i\times I$.  In classifying gauge transformations, we introduce
a winding number $m_i$ associated to each monodromy defect. There is also a bulk winding number $n$.
The shift in the total action $S=S_{\mathrm{CS}}
+\sum_i S_{K_i}$ under a gauge transformation is
\be\label{changeaction}S\to S+\frac{2\pi i }{b^2}\left(n +\sum_{i=1}^s m_i\eta_i\right).\end{equation}

\subsection{Interpretation}\label{zonox}

\hspace{0.25in}The moral of the story is that in the Chern-Simons description, critical points are flat connections
on $\Sigma\times I$, with prescribed behavior near the ends and near monodromy defects,
modulo  {\it topologically trivial} gauge transformations.  Topologically nontrivial gauge
transformations cannot be regarded as symmetries because they do not leave the action
invariant.  Instead, they generate new critical points from old ones.

For the main example of this part -- three heavy operators on $\sf S^2$  -- all critical
points are related to each other by topologically nontrivial gauge transformations.
This means that there is a simple way to compare the path integrals over
cycles associated to different critical points.

In fact, let $\A$ be a connection that represents a critical point $\rho$.  Suppose a gauge transformation
$g$ with winding numbers $n$ and $m_1,\dots,m_s$ acts on $\A$ to produce
a new critical point $\A'$.  Let $Z_\rho$ and $Z_{\rho'}$ be the path integrals over integration cycles
associated to $\A$ and to $\A'$, respectively.  $Z_{\rho'}$ and $Z_\rho$ are not equal, since the
gauge transformation $g$ does not preserve the action.  But since $g$ transforms
the action by a simple additive $c$-number (\ref{changeaction}), there is a simple
exact formula that expresses the relation between $Z_{\rho'}$ and $Z_\rho$:
\be\label{simplex}Z_{\rho'}=Z_\rho\exp\left(-(2\pi i/b^2)(n+\sum_im_i\eta_i)\right). \ee

We will discuss the interpretation of this formula in Liouville theory in section \ref{exact} below.

\chapter{Timelike Liouville Theory}\label{timelike}
So far in this part we have analytically continued in $\alpha$ but not in $b$.  From the point of view advocated in the introduction, this is somewhat artificial; we should allow ourselves to consider the path integral with arbitrary complex values of all parameters and then study which integration cycles to use to reproduce the analytic continuation from the physical region.  For complex $b$'s with positive real part, we can simply continue the DOZZ formula and the machinery of the preceding sections is essentially unmodified.  Indeed numerical results for complex $b$ were given in \cite{Zamolodchikov:1995aa}, confirming the crossing symmetry of the four-point function based on the DOZZ formula.  As mentioned in the introduction, in various cosmological settings it is desireable to define a version of Liouville that has real central charge that is large and negative.  The most obvious way to try to do this is to continue the DOZZ formula all the way to purely imaginary $b$, since the formula (\ref{centralcharge}) will then be in the desired range \cite{Strominger:2003fn}.  This has been shown to fail rather dramatically \cite{Zamolodchikov:2005fy}, as we will discuss in the following subsection.  However, we first introduce some conventional redefinitions to simplify future formulas when $b$ is imaginary.  We begin with
\begin{align}
b&=-i \hat{b} \\
\phi&=i \hat{\phi}\\
Q&=i\left(\frac{1}{\hat{b}}-\hat{b}\right)\equiv i \hat{Q},
\end{align}
after which the action (\ref{liouvaction}) becomes
\be
\label{timelikeact}
S_L=\frac{1}{4\pi}\int d^2 \xi \sqrt{\tilde{g}}\left[-\partial_a \hat{\phi} \partial_b \hat{\phi}\tilde{g}^{ab}-\hat{Q} \tilde{R} \hat{\phi}+4\pi \mu e^{2\hat{b}\hat{\phi}}\right].
\ee
The theory with this action  is conventionally referred to as ``timelike'' Liouville theory, since the kinetic term has the wrong sign.  In this equation, superscripts are procreating at an alarming rate, so we pause to remind the reader that $\tilde{g}$ is the reference metric and does not undergo analytic continuation.  We will use ``hat'', as in $\hat{b}$, exclusively to refer to the timelike analogues of standard Liouville quantities.  The central charge is now
\be
c_L=1-6 \hat{Q}^2,
\ee 
which for small real $\hat{b}$ accomplishes our goal of large negative central charge.  The physical metric becomes $g_{ab}=e^{\frac{2}{\hat{Q}}\hat{\phi}}\tilde{g}_{ab}$, so the boundary condition on $\hat{\phi}$ at infinity is
\be
\hat{\phi}(z,\bar{z})=-2\hat{Q}\log |z|+\mathcal{O}(1).
\ee
To talk about exponential operators, it is convenient to make one final definition
\be
\alpha=i\hat{\alpha},
\ee
which gives conformal weights
\be
\label{tdimension}
\Delta\left(e^{-2\hat{\alpha}\hat{\phi}}\right)=\bar{\Delta}\left(e^{-2\hat{\alpha}\hat{\phi}}\right)=\hat{\alpha}(\hat{\alpha}-\hat{Q}).
\ee
In the presence of heavy operators $\hat{\alpha}_i=\eta_i/\hat{b}$, the generalized action (\ref{regaction}) for the rescaled field $\phi_c=2 b \phi=2\hat{b}\hat{\phi}$ with a flat reference metric is
\begin{align}\nonumber
\tilde{S}_L=-&\frac{1}{16 \pi \hat{b}^2}\int_{D-\cup_i d_i}d^2\xi \left(\partial_a \phi_c \partial_a 
\phi_c-16 \hat{\lambda} e^{\phi_c}\right)-\frac{1}{\hat{b}^2}\left(\frac{1}{2\pi}\oint_{\partial D}\phi_c d\theta +2\log R\right)\\
&+\frac{1}{\hat{b}^2}\sum_i \left(\frac{\eta_i}{2\pi}\oint_{\partial d_i}\phi_{c}d\theta_i +2 \eta_i^2\log \epsilon \right). \label{timelikeregaction}
\end{align}
Here $\hat{\lambda}=\pi\mu \hat{b}^2=-\lambda$, and in fact other than an overall sign change this is the only difference from the expression of this action in terms of the ``unhatted'' variables.  Note that $\phi_c$ and $\eta_i$ do not need to be ``hatted'' since they are the same before and after the analytic continuation.  The equation of motion is now 
\be
\label{teom}
\partial \overline{\partial} \phi_c = -2 \hat{\lambda} e^{\phi_c}-2\pi\sum_{i} \eta_i \delta^2(\xi-\xi_i),
\ee
which for positive $\mu$ is just the equation of motion for constant positive curvature with conical deficits at the heavy operators.  When $\hat{b}$ and $\eta_i$ are real and $\eta_i$ is in Region II, described by (\ref{inequalities}), this equation has a real solution.  As discussed below (\ref{inequalities}), this solution can be constructed from spherical triangles.  In the FRW/CFT application of timelike Liouville, this real saddle point is identified with the asymptotic metric in a Coleman-de Luccia bubble \cite{Freivogel:2006xu,Sekino:2009kv}.  

\subsection{The Timelike DOZZ Formula}

\hspace{0.25in}The redefinitions of the previous section make clear that at the classical level the relationship between spacelike and timelike Liouville is straightforward.  Much less clear is the question of the appropriate integration cycle for the path integral when $b$ is imaginary.  One way to attempt to specify a cycle is to try to continue the DOZZ formula from real $b$.  As just mentioned, this does not work.  We can see why by considering more carefully the analytic properties of $\Upsilon_b(x)$  in $b$ \cite{Zamolodchikov:2005fy}.  From (\ref{logupsilonapp}) we see that the defining integral for $\Upsilon_b(x)$ does not converge for any $x$ when $b$ is imaginary, which is already a sign of trouble, but this could possibly be avoided by deforming the contour.  A more sophisticated argument from \cite{Zamolodchikov:2005fy} is as follows: consider the function
\be
\label{Hdef}
H_b(x)=\Upsilon_b(x)\Upsilon_{ib}(-ix+ib),
\ee
where for the moment we take $b$ to have positive real part and negative imaginary part to ensure that both $\Upsilon$'s can be defined by the integral (\ref{logupsilonapp}).  This function is entire and has simple zeros everywhere on the lattice generated by $b$ and $1/b$, as illustrated in Figure \ref{zerolattice}.
Using the recursion relations (\ref{recrel}) we can show that $H_b$ obeys:
\begin{align} \nonumber
H_b(x+b)&=e^{\frac{i \pi}{2}(2bx-1)}H_b(x)\\
H_b(x+1/b)&=e^{\frac{i\pi}{2}(1-2x/b)}H_b(x)
\end{align}
\begin{figure}[ht]
\begin{center}
\includegraphics[scale=.9]{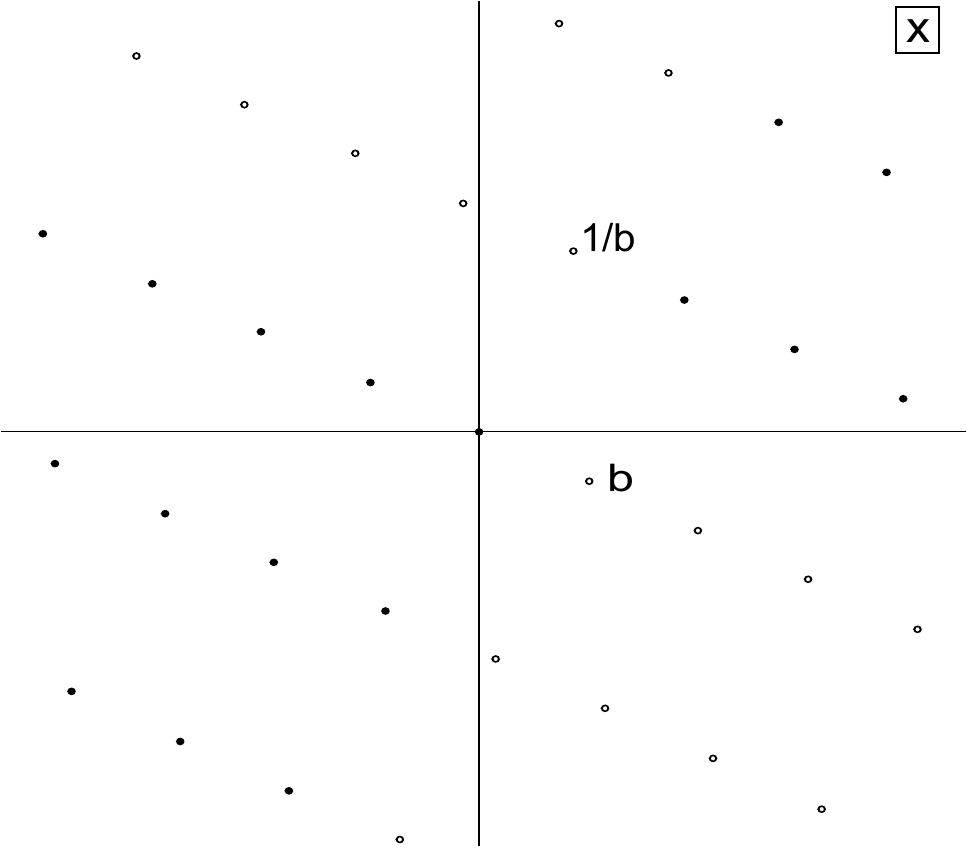}
\caption{ Zeros of $H_b(x)$.  The solid circles come from the zeros of $\Upsilon_b(x)$ while the empty circles come from zeros of $\Upsilon_{ib}(-ix+ib)$.}
 \label{zerolattice}
\end{center}
\end{figure}
It is convenient to here introduce a Jacobi $\theta$-function\footnote{Conventions for Jacobi $\theta$-functions are rather inconsistent, so we note that this definition is implemented in Mathematica as EllipticTheta$\left[1,\pi z,e^{i\pi \tau}\right]$.}
\be
\theta_1(z,\tau)=i\sum_{n=-\infty}^{\infty}(-1)^n e^{i\pi \tau (n-1/2)^2+2\pi i z(n-1/2)}\qquad \Im\,\tau>0,
\ee
which is entire in $z$ for any $\Im\, \tau>0$ and obeys
\begin{align} \nonumber
\theta_1(z+1,\tau)&=e^{-i\pi}\theta_1(z,\tau)\\ \label{thetashift}
\theta_1(z+\tau,\tau)&=e^{i\pi(1-\tau-2z)}\theta_1(z,\tau).
\end{align}
By cancelling the terms $n=1,2,...$ with $n=0,-1,...$ we see that it has a zero at $z=0$, and by applying these recursion relations we see that it has zeros for all $z=m+n\tau$ with $m,n\in \mathbb{Z}$.  In fact these zeros are simple and are the only zeros, which follows from the standard product representation of the
theta function.  This function is useful for us because we can now observe that 
\be
e^{\frac{i\pi}{2}(x^2+x/b-xb)}\theta_1(x/b,1/b^2)
\ee
obeys the same recursion relations and has the same zeros as $H_b(x)$.  Their ratio is doubly periodic and entire in $x$, and must therefore be a function only of $b$.  We can determine this function by setting $x=\frac{b}{2}+\frac{1}{2b}$ and recalling that $\Upsilon_b(Q/2)=1$.  The result is that\footnote{This formula was also derived in \cite{Schomerus:2003vv}.}
\be
\label{upsilontheta}
H_b(x)=e^{\frac{i\pi}{2}\left(x^2+\frac{x}{b}-xb+\frac{b^2}{4}-\frac{3}{4b^2}-\frac{1}{4}\right)}\frac{\theta_1(x/b,1/b^2)}{\theta_1(\frac{1}{2}+\frac{1}{2b^2},1/b^2)}.
\ee
We can now use this formula to study the behaviour of $\Upsilon_b$ near imaginary $b$; since $\Upsilon_{ib}(-ix+ib)=\frac{H_b(x)}{\Upsilon_b(x)}$, if we move $b$ up towards the postive real axis then $\Upsilon_{ib}$ will approach the region of interest.  But (\ref{upsilontheta}) reveals that doing this continuation requires $\theta_1$ to approach the real $\tau$-axis.  This is actually a natural boundary of analytic continuation for $\theta_1$, with a nonlocal and extremely violent singularity running all along the real $\tau$-axis.  The detailed form of the approach to the singularity depends strongly on $z$, so there is no possibility of cancellation between the two $\theta_1$'s in $H_b$ except for at special values of $z$.  This shows that for generic values of $x$, $\Upsilon_b$ simply cannot be continued to generic imaginary $b$ \cite{Zamolodchikov:2005fy}.  This is the origin of the failure of \cite{Strominger:2003fn} to make sense of timelike Liouville theory in this way.  

What then are we to do?  One possibility is to restrict to special values of $b$ and $\alpha$ where the continuation can still be nonsingular; this is explored in \cite{Schomerus:2003vv}.  We are interested however in generic complex values of the parameters so this will not work for us.  A very interesting proposal was made by Al. B. Zamolodchikov  \cite{Zamolodchikov:2005fy}, and also independently
by Kostov and Petkova \cite{KP1,KP2,KP3}.  A key observation is that 
although we cannot continue the DOZZ formula to imaginary $b$, we can continue Teschner's recursion relations.  For real $b$ the essentially unique solution of these recursion relations is the DOZZ formula, but for generic complex $b$ the solution is not unique since we can engineer $\hat{\alpha}$-dependent combinations of $\theta$-functions by which we can multiply any solution of the recursion relations  to produce a new solution.  But when we get to imaginary $b$, it turns out that  there is again an (almost) unique solution, which is not given by analytic continuation of the DOZZ formula.\footnote{The reason that purely real and purely imaginary $b$ have essentially unique solutions of the recursion relations is that the lattice generated by $b$ and $1/b$ becomes degenerate in these two cases and functions with two real periodicities are highly constrained.  The freedom involving multiplying by $\theta$-functions of the $\hat{\alpha}$'s goes away since in these cases some of the $\theta$ functions are always evaluated at $\tau=0$.  The original DOZZ formula (\ref{dozz}) and the formula (\ref{timelikedozz}) below are related at complex $b$ by such a factor, which is why they can not be continued into each other.  For more details see \cite{Teschner:1995yf,Zamolodchikov:2005fy}.}   This solution is not quite unique because one can multiply it by an $\hat{\alpha}$-independent arbitrary function of $b$ without affecting the recursion relations.  Fixing this normalization in a way that we explain momentarily, and which slightly differs from the choice in \cite{Zamolodchikov:2005fy}, the solution is:
\begin{align} \nonumber
\hat{C}(\hat{\alpha}_1,\hat{\alpha}_2,\hat{\alpha}_3)=&\frac{2\pi}{\hat{b}}\left[-\pi \mu \gamma(-\hat{b}^2)\hat{b}^{2+2\hat{b}^2}\right]^{(\sum_i\hat{\alpha}_i-\hat{Q})/\hat{b}}e^{-i\pi(\sum_i\hat{\alpha}_i-\hat{Q})/\hat{b}}\\
&\frac{\Upsilon_{\hat{b}}(\hat{\alpha}_1+\hat{\alpha}_2-\hat{\alpha}_3+\hat{b})\Upsilon_{\hat{b}}(\hat{\alpha}_1+\hat{\alpha}_3-\hat{\alpha}_2+\hat{b})\Upsilon_{\hat{b}}(\hat{\alpha}_2+\hat{\alpha}_3-\hat{\alpha}_1+\hat{b})}
{\Upsilon_{\hat{b}}(\hat{b})\Upsilon_{\hat{b}}(2\hat{\alpha}_1+\hat{b})\Upsilon_{\hat{b}}(2\hat{\alpha}_2+\hat{b})\Upsilon_{\hat{b}}(2\hat{\alpha}_3+\hat{b})}\nonumber\\
&\hspace{1in}\times\Upsilon_{\hat{b}}(\hat{\alpha}_1+\hat{\alpha}_2+\hat{\alpha}_3-\hat{Q}+\hat{b}).
\label{timelikedozz}
\end{align}
We will refer to this as the timelike DOZZ formula.  The power of $\pi \mu\gamma(-\hat{b}^2)$ differs slightly from C.10 in \cite{Zamolodchikov:2005fy}, but the choice we have made here is given by a scaling argument in the path integral and is required for our interpretation of timelike Liouville as being a different integration cycle of ordinary Liouville.  We have also divided C.10 from \cite{Zamolodchikov:2005fy} by a factor of $\frac{\hat{b}^3}{2\pi}\gamma(1-\hat{b}^2)\gamma(2-1/\hat{b}^2)$; as just mentioned these choices do not affect the recursion relations and can be interpreted as an ambiguity in the normalization of the operators, but we will see below in section \ref{tdozzthreelight} that the choice we have made here is supported by semiclassical computation.  Moreover in section \ref{exact} we will see exactly that it is the natural choice for our interpretation of the timelike Liouville path integral.

We can also write down an exact two-point function.  Since unlike the three-point function the two-point function (\ref{spacelike2point}) does have a good analytic continuation to imaginary $b$, it is natural to choose the timelike 2-point function to agree with this analytic continuation.  This gives
\begin{align}
\label{timelike2point}
\hat{G}(\hat{\alpha})=&-\frac{\hat{b}}{2\pi}\hat{C}(0,\hat{\alpha},\hat{\alpha})=-\frac{1}{\hat{b}^2}\left[-\pi \mu \gamma(-\hat{b}^2)\right]^{(2\hat{\alpha}-\hat{Q})/\hat{b}}e^{-i\pi(2\hat{\alpha}-\hat{Q})/\hat{b}}\gamma(2\hat{\alpha}\hat{b}+\hat{b}^2)\nonumber\\
&\times\gamma\left(\frac{1}{\hat{b}^2}-\frac{2\hat{\alpha}}{\hat{b}}-1\right).
\end{align}
Note that for real $\hat{\alpha}$, this expression is not positive-definite, as expected from the wrong-sign kinetic term.  Its relation to the three-point function is somewhat arbitrary, unlike in spacelike Liouville where there was a clear rationale for the formula (\ref{normalization}).  In particular setting one of the $\hat\alpha$'s to zero in the timelike DOZZ formula does NOT produce a $\delta$-function.  Indeed the timelike DOZZ formula has a finite and nonzero limit even when $\hat{\alpha}_1\to0,~\hat{\alpha}_2\neq \hat{\alpha}_3$.  In \cite{Zamolodchikov:2005fy}, this was observed as part of a larger issue whereby the degenerate fusion rules mentioned below equation (\ref{lightdegenerate}) are not automatically satisfied by the timelike DOZZ formula.  In \cite{McElgin:2007ak}, this was interpreted as the two-point function being genuinely non-diagonal in the operator weights.  We will not be able to explain this in a completely satisfactory way, but we will suggest a possible resolution below in section \ref{tcft?}.

\subsection{Semiclassical Tests of the Timelike DOZZ formula}

\hspace{0.25in}In this section we will show in three different cases, analogous to the three cases studied above for the spacelike DOZZ formula, that the semiclassical limits of (\ref{timelikedozz}) and (\ref{timelike2point}) are consistent with our claim that they are produced by the usual Liouville path integral on a different integration cycle.  This task is greatly simplified by observing that we can trivially reuse all of our old solutions (or ``solutions'') and expressions for the action.  Say that we have a solution $\phi_{c,N}(\eta_i,\lambda,b,z,\bar{z})$ of the original Liouville equation of motion (\ref{eom}).  It is easy to check that
\be
\label{saddlemap}
\hat{\phi}_{c,N}(\eta_i, \hat{\lambda},\hat{b},z,\bar{z})\equiv \phi_{c,N}(\eta_i,\hat{\lambda},\hat{b},z,\bar{z})-i\pi
\ee
obeys (\ref{teom}).  We can also compute the action by noting that if we define the original modified action (\ref{regaction}) as $\tilde{S}_L\left[\eta_i,\lambda,b,z_i,\bar{z}_i; \phi_{c,N}(\eta_i,\lambda)\right]$, then we have
\begin{align}\nonumber
\tilde{S}_L\left[\eta_i, -\hat{\lambda},-i\hat{b},z_i,\bar{z}_i;\hat{\phi}_{c,N}(\eta_i,\hat{\lambda})\right]\equiv \hat{\tilde{S}}_L\left[\eta_i,\hat{\lambda},\hat{b},z_i,\bar{z}_i; \hat{\phi}_{c,N}(\eta_i,\lambda)\right]\\
=-\tilde{S}_L\left[\eta_i,\hat{\lambda},\hat{b},z_i,\bar{z}_i;\phi_{c,N}(\eta_i,\hat{\lambda})\right]+\frac{i\pi}{\hat{b}^2}\left(1-\sum_i\eta_i\right).
\label{stotaction}
\end{align}
The left hand side of this is just (\ref{timelikeregaction}), so we can thus compute the action for timelike Liouville theory by simple modification of our previous results.  

\subsubsection{Two-point Function}
Using (\ref{stotaction}) and (\ref{2action}), we find that the timelike version of the saddlepoint (\ref{phic2}) has timelike action
\begin{align} \nonumber
\hat{\tilde{S}}_{L}=-\frac{1}{\hat{b}^2}\Bigg[&2\pi i N(1-2\eta)+(2\eta-1)\hat{\lambda}+2\Big((1-2\eta)\log (1-2\eta)-(1-2\eta)\Big)\\
&+2(\eta-\eta^2)\log|z_{12}|^2\Bigg]. \label{timelike2pointact}
\end{align}
The semiclassical limit of (\ref{timelike2point}) with $\alpha$ heavy is:
\begin{align}
\hat{G}(\eta)\to&\left(e^{2\pi i (1-2\eta)/\hat{b}^2}-1\right)\nonumber\\&\times\exp\left\{\frac{1}{\hat{b}^2}\left[-(1-2\eta)\log \hat{\lambda}+2\big((1-2\eta)\log (1-2\eta)-(1-2\eta)\big)\right]\right\},
\end{align}
which is matched by a sum over the two saddle points $N=0$ and $N=1$ with actions given by (\ref{timelike2pointact}).  Note that the integral over the moduli would again produce a divergence, but that unlike in the DOZZ case this divergence did not seem to be produced by the limit $\alpha_1 \to 0,\alpha_2=\alpha_3$.  Note also that there is now no Stokes line in the $\eta$ plane, there are always only two saddle points that contribute.  This is analogous to the integral representation of $1/\Gamma(z)$ as discussed in appendix C.  

\subsubsection{Three-point Function with Heavy Operators} 
Similarly for three heavy operators in Region I, the timelike version of (\ref{phic31}) has timelike action 
\begin{align}
\nonumber
\hat{\widetilde{S}}_L=-\frac{1}{\hat{b}^2}\Bigg[&-\left(1-\sum_i \eta_i\right)\log \hat{\lambda}-(\hat{\delta}_1+\hat{\delta}_2-\hat{\delta}_3)\log |z_{12}|^2-(\hat{\delta}_1+\hat{\delta}_3-\hat{\delta}_2)\log |z_{13}|^2\\ \nonumber
&-(\hat{\delta}_2+\hat{\delta}_3-\hat{\delta}_1)\log |z_{23}|^2 +F(\eta_1+\eta_2-\eta_3)+F(\eta_1+\eta_3-\eta_2)\\\nonumber
&\hspace{0.25in}+F(\eta_2+\eta_3-\eta_1)+F(\eta_1+\eta_2+\eta_3-1)-F(2\eta_1)-F(2\eta_2)\\
&\hspace{0.5in}-F(2\eta_3)-F(0)+2\pi i (N-1/2)(1-\sum_i \eta_i)\Bigg], \label{tregion1act}
\end{align}
where we have defined $\hat{\delta_i}\equiv \eta_i(\eta_i-1)$, consistent with (\ref{tdimension}).  This clearly has the right position dependence for a timelike three-point function.  The action in other regions is always an analytic continuation of this action along some path, but to be definite we also give the timelike action from Region II as well:
\begin{align}
\nonumber
\hat{\widetilde{S}}_L=-\frac{1}{\hat{b}^2}\Bigg[&-\left(1-\sum_i \eta_i\right)\log \hat{\lambda}-(\hat{\delta}_1+\hat{\delta}_2-\hat{\delta}_3)\log |z_{12}|^2-(\hat{\delta}_1+\hat{\delta}_3-\hat{\delta}_2)\log |z_{13}|^2\\ \nonumber
&-(\hat{\delta}_2+\hat{\delta}_3-\hat{\delta}_1)\log |z_{23}|^2 +F(\eta_1+\eta_2-\eta_3)+F(\eta_1+\eta_3-\eta_2)\\\nonumber
&\hspace{0.25in}+F(\eta_2+\eta_3-\eta_1)+F(\eta_1+\eta_2+\eta_3)-F(2\eta_1)-F(2\eta_2)\\\nonumber
&\hspace{0.5in}-F(2\eta_3)-2F(0)\left\{(1-\sum_i \eta_i)\log(1-\sum_i \eta_i)-(1-\sum_i \eta_i)\right\}\\&\hspace{0.75in}+2\pi i N(1-\sum_i \eta_i)\Bigg]. \label{tregion2act}
\end{align}
To compare these with the timelike DOZZ formula, we can again make use of the asymptotic formula (\ref{etaasymp}).  The terms in (\ref{timelikedozz}) that don't involve $\Upsilon_{\hat{b}}$ approach
\be
e^{\frac{1-\sum_i\eta_i}{\hat{b}^2}(i\pi+2\log \hat{b}-\log \hat{\lambda})+\mathcal{O}(1/\hat{b})},
\ee
and using (\ref{etaasymp}) we find that in Region I the $\Upsilon_{\hat{b}}$'s combine with this to give 
\begin{align} \nonumber
\hat{C}(\eta_i/\hat{b})\sim\exp \Bigg\{\frac{1}{\hat{b}^2}\Bigg[&-\left(1-\sum_i \eta_i\right)\log \hat{\lambda}+F(\eta_1+\eta_2-\eta_3)+F(\eta_1+\eta_3-\eta_2)\\ \nonumber
&+F(\eta_2+\eta_3-\eta_1)+F(\eta_1+\eta_2+\eta_3-1)-F(2\eta_1)-F(2\eta_2)\\
&-F(2\eta_3)-F(0)+i\pi(1-\sum_i \eta_i)\Bigg]\Bigg\}.
\end{align}
Comparing with (\ref{tregion1act}), we see that only the saddle point with $N=0$ contributes.  In Region II as before (\ref{umot}) we need to use (\ref{inverserecursion}) to shift one of the $\Upsilon_{\hat{b}}$'s before using the asymptotic formula (\ref{etaasymp}), giving:
\be
\Upsilon_{\hat{b}}\Big(\frac{\sum_i \eta_i-1}{\hat{b}}+2\hat{b}\Big)\sim \gamma\Big((\sum_i \eta_i-1)/\hat{b}^2\Big)^{-1}\hat{b}^{\frac{1}{\hat{b}^2}\left(2(1-\sum_i \eta_i)-(\sum_i \eta_i-1/2)^2\right)}e^{\frac{1}{\hat{b}^2}F(\sum_i\eta_i)}.
\ee
The result in Region II is
\begin{align} \nonumber
\hat{C}(\eta_i/\hat{b})\sim\exp \Bigg\{\frac{1}{\hat{b}^2}\Bigg[&-\left(1-\sum_i \eta_i\right)\log \hat{\lambda}+F(\eta_1+\eta_2-\eta_3)+F(\eta_1+\eta_3-\eta_2)\\ \nonumber
&+F(\eta_2+\eta_3-\eta_1)+F(\eta_1+\eta_2+\eta_3-1)-F(2\eta_1)-F(2\eta_2)\\ \nonumber
&-F(2\eta_3)-F(0)+2(1-\sum_i\eta_i)\log (1-\sum_i\eta_i)-2(1-\sum_i\eta_i)\\
&+i\pi(1-\sum_i \eta_i)\Bigg]\Bigg\}\Big(e^{i\pi (1-\sum_i\eta_i)/\hat{b}^2}-e^{-i\pi (1-\sum_i\eta_i)/\hat{b}^2}\Big).
\end{align}
Comparing this with (\ref{tregion2act}), we see it matches a sum over two saddle points with $N=0$ and $N=1$.  Unlike the spacelike DOZZ formula there are no Stokes walls in Region II, in complete analogy with the situation for $1/\Gamma(z)$ explained in appendix \ref{gammastokes}.

\subsubsection{Three-point Function with Light Operators}\label{tdozzthreelight}
As a final semiclassical check of the timelike DOZZ formula (\ref{timelikedozz}), we will calculate its $b\to 0$ limit when all three operators are light and compare to a semiclassical computation analogous to that from section \ref{dozzthreelight}.  We will define $\sigma_i=\frac{\alpha_i}{b}=-\frac{\hat{\alpha}_i}{\hat{b}}$, which gives $\Delta \to \sigma$ as $\hat{b}\to0$.  Manipulations similar to those leading up to (\ref{lightdozz}) now give
\begin{align} \nonumber
&\hat{C}(-\sigma_1 \hat{b},-\sigma_2 \hat{b},-\sigma_3 \hat{b})=-2\pi i\hat{b}^{-3}\hat{\lambda}^{1-\sum_i \sigma_i-1/\hat{b}^2}e^{-2/\hat{b}^2-2\gamma_E+\mathcal{O}(\hat{b} \log \hat{b})}\\&\hspace{1.75in}\times\left(e^{2\pi i (\sum_i \sigma_i-1+1/\hat{b}^2)}-1\right)\nonumber\\
&\times\frac{\Gamma(1-2\sigma_1)\Gamma(1-2\sigma_2)\Gamma(1-2\sigma_3)}{\Gamma(1+\sigma_1-\sigma_2-\sigma_3)\Gamma(1+\sigma_2-\sigma_1-\sigma_3)\Gamma(1+\sigma_3-\sigma_1-\sigma_2)\Gamma(2-\sum_i\sigma_i)}.
\end{align}
From the structure of this formula, it appears that we will be able to interpret as a sum over two complex saddle points as with Region II in the previous section.  There is a subtlety however in that to produce the $\Gamma$-functions that will emerge from the modular integral in our imminent semiclassical computation, we need to apply the Euler reflection formula $\Gamma(x)\Gamma(1-x)=\pi/\sin{\pi x}$ to each of them.  Anticipating this result, we write:
\begin{align} \nonumber
\hat{C}(&-\sigma_1 \hat{b},-\sigma_2 \hat{b},-\sigma_3 \hat{b})=\hat{b}^{-3}\hat{\lambda}^{1-\sum_i \sigma_i-1/\hat{b}^2}e^{-2/\hat{b}^2-2\gamma_E+\mathcal{O}(\hat{b} \log \hat{b})}\left(e^{2\pi i (\sum_i \sigma_i-1+1/\hat{b}^2)}-1\right)\\\nonumber
&\times\frac{\Gamma(\sigma_1+\sigma_2-\sigma_3)\Gamma(\sigma_1+\sigma_3-\sigma_2)\Gamma(\sigma_2+\sigma_3-\sigma_1)\Gamma(\sum_i\sigma_i-1)}{\Gamma(2\sigma_1)\Gamma(2\sigma_2)\Gamma(2\sigma_3)}\\\label{lighttdozz}
&\times\frac{\left(e^{2\pi i (\sigma_1+\sigma_2-\sigma_3)}-1\right)\left(e^{2\pi i (\sigma_1+\sigma_3-\sigma_2)}-1\right)\left(e^{2\pi i (\sigma_2+\sigma_3-\sigma_1)}-1\right)\left(e^{2\pi i (\sigma_1+\sigma_2+\sigma_3)}-1\right)}{\left(e^{4\pi i\sigma_1}-1\right)\left(e^{4\pi i\sigma_2}-1\right)\left(e^{4\pi i\sigma_3}-1\right)}.
\end{align}
The structure of the terms in the third line show that a much more complicated set of saddlepoints are needed to explain this result than in the spacelike case (\ref{lightdozz}).  At the end of this section we will explain why this happens. 
 
The semiclassical formula analogous to (\ref{better3corr}) for this correlation function is
\be
\label{better3corrt}
\langle e^{\sigma_1 \phi_c(z_{1},\bar{z}_1)}e^{\sigma_2 \phi_c(z_{2},\bar{z}_2)}e^{\sigma_3 \phi_c(z_{3},\bar{z}_3)}\rangle \approx A(-i\hat{b})\sum_{N\in T}e^{-S_L[\hat{\phi}_{c,N}]}\int d\mu(\alpha,\beta,\gamma,\delta)\prod_{i=1}^3 e^{\sigma_i \hat{\phi}_{c,N}(z_i,\bar{z}_i)}.
\ee
Here we have assumed that the fluctuation determinant and Jacobian parametrized by $A(-i\hat{b})$ are just the analytic continuations of their spacelike counterparts.  This should be true if our path integral interpretation is correct, and we will see momentarily that this works out.  $\hat{\phi}_{c,N}$ is the timelike ``solution'' with branch choice $N$, related to the usual spacelike ``solution'' by (\ref{saddlemap}).  Explicitly 
\be
\hat{\phi}_c(z,\bar{z})=2\pi i N(z,\bar{z})-\log \hat{\lambda}-2\log (|\alpha z+\beta|^2+|\gamma z+\delta|^2).
\ee
Based on (\ref{lighttdozz}), we have allowed $N$ to vary with position to allow the different branches of the action to be realized.  This is one of the situations discussed in section \ref{furthercom} where discontinuous ``solutions'' must be included even though single-valued solutions exist.  Computing the action (\ref{timelikeact}) and simplifying the modular integral as in section \ref{dozzthreelight} we find
\be
\hat{C}(-\sigma_1 \hat{b},-\sigma_2 \hat{b},-\sigma_3 \hat{b})\approx A(-i\hat{b}) \hat{\lambda}^{1-\sum_i \sigma_i-1/\hat{b}^2} e^{-2/\hat{b}^2}I(\sigma_1,\sigma_2,\sigma_3)\sum_{N\in T} e^{-2\pi i (\sum_i m_i\sigma_i+n/\hat{b}^2)} 
\ee
Here $-n$ is the value of $N$ at $\infty$ and $-m_i$ is its value near the various insertions.  Using (\ref{Iint}) and comparing this with (\ref{lighttdozz}), we find complete agreement provided that $A(-i \hat{b})=\hat{b}^{-3}\pi^{-3}e^{-2\gamma_E}$.  Recalling that at the end of section \ref{dozzthreelight} we found $A(b)=ib^{-3}\pi^{-3}e^{-2\gamma_E}$, this indeed works out as expected.  

The set $T$ of included branches is now rather complex; it can be read off from (\ref{lighttdozz}) but we will not try to characterize it more precisely.  We observe however that many branches that correspond to discontinuous ``solutions'' are now definitely needed.  This is different than what we found for spacelike Liouville in section \ref{dozzthreelight}, where the contributing saddle points were single-valued and continuous and just the same as in Region II for heavy operators.  The reason for this distinction is that, as explained in appendix F, the modular integral over $SL(2,\mathbb{C})$ converges only when the $\sigma$'s obey certain inequalities (\ref{appineq}).  In spacelike Liouville with the $\sigma$'s in Region II the integral is convergent, so we can evaluate it without any contour deformation.  In timelike Liouville, when the $\sigma$'s are in Region II many of the inequalities are violated and the integral must be defined by analytic continuation.  This continuation results in additional Stokes phenomena, which changes the contributing saddle-points.  

\subsection{An Exact Check}
\label{exact}
\hspace{0.25in}The checks of the previous section were semiclassical, but we will now give an exact argument that the timelike DOZZ formula (\ref{timelikedozz}) is produced by evaluating the usual Liouville path integral on a new integration cycle.  We will show that the ratio of the spacelike and timelike DOZZ formulae must have a specific form and then demonstrate that it does.

We begin by defining:
\be
\label{rhointegral}
Z_\rho(\alpha_i, z_i, \bar{z}_i)=\int_{\CC_\rho}\D \phi_cV_{\alpha_1}(z_1,\bar{z}_1)...V_{\alpha_n}(z_n,\bar{z}_n) e^{-S_L},
\ee
where here $\rho$ is a critical point of the action with heavy operators as sources, and the path integral is evaluated on the steepest descent cycle $\CC_\rho$ that passes through $\rho$.  As discussed in the introduction, this quantity is \textit{not} in general equal to the Liouville correlator; we need to sum over such cycles with integer coefficients $a^\rho$ as in (\ref{expres}).  We will now argue however that the $\rho$-dependence of $Z_\rho$ is quite simple.  First recall the exact version of the action (\ref{action})\footnote{When we discuss discontinuous ``solutions'' momentarily, the kinetic term should really be understood to be expressed in terms of $f$ in equation (\ref{improvedact}).}
\begin{align}
\label{exactimprovedact}
S_L=&\frac{1}{16\pi b^2}\int_D d^2\xi \left[(\partial \phi_c)^2+16 \lambda e^{\phi_c}\right]+\frac{1}{2\pi b^2}(1+b^2)\oint_{\partial D}\phi_c d\theta\nonumber\\&\hspace{0.25in}+\frac{2}{b^2}(1+2b^2+b^4)\log R.
\end{align}
We note that under the transformation $\phi_c \to \phi_c+ 2\pi i N$, we have $S_L\to S_L+\frac{2\pi i N}{b^2}(1+b^2)$.  Semiclassically the operator $V_\alpha$ defined by (\ref{ponzo}) transforms as $V_\alpha \to V_\alpha e^{2\pi i \alpha/b}$ under the same transformation.  Since the Seiberg bound ensures that the renormalization needed to define this operator precisely is the same as in free field theory, this is actually the exact transformation of $V_\alpha$.  Moreover the path-integral measure $\D \phi_c$ is invariant under the shift.  This means that if two $\rho$'s differ only by adding $2\pi i N$, then with a slight abuse of notation we have the simple relation 
\be
Z_{\rho+2\pi i N}=e^{2\pi i N(\sum_i \alpha_i/b-1/b^2)}Z_\rho.
\ee
This result is exact, and more generally it shows that the result of integrating over a sum of integration cycles of this type can be factored out from the correlator:
\be
Z=\sum_{N=-\infty}^{\infty}a^{\rho+2\pi i N}Z_{\rho+2\pi i N}=Z_\rho \sum_{N=-\infty}^{\infty}a^{\rho+2\pi i N}e^{2\pi i N(\sum_i \alpha_i/b-1/b^2)}.
\ee
Thus in general the ratio of $Z$'s which are computed on different cycles, both of the form $\sum_{N=-\infty}^{\infty}a^{\rho+2\pi i N}\CC_{\rho+2\pi i N}$, will be expressible as a ratio of Laurent expansions in $e^{2\pi i(\sum_i \alpha_i/b-1/b^2)}$ with integer coefficients.  This is a rather nontrivial constraint; for example it implies that the ratio is invariant under shifting any particular $\alpha_i$ by $\alpha_i\to \alpha_i+b$.  There is also a more subtle invariance of the form $b \to \frac{b}{\sqrt{1+b^2}}$ and $\alpha_i \to \frac{\alpha_i}{\sqrt{1+b^2}}$. 

Unfortunately as discussed in section \ref{4pointsection}, to understand the DOZZ formula in the full range of $\alpha_i$'s it is not sufficient to only consider integration cycles that differ by a global addition of $2\pi i N$.  We found semiclassically in (\ref{multivac}) that to fully explain the DOZZ formula it was necessary to consider discontinuous ``solutions'' that differ by different multiples of $2 \pi i$ at the different operator insertions.  To proceed further we need to assume that we can apply the machinery of the previous paragraph to these ``solutions'' and their associated ``integration cycles of steepest descent.''  The idea is that the action (\ref{exactimprovedact}), with the kinetic term expressed in terms of $f$ as in (\ref{improvedact}), changes only by an overall $c$-number if we shift the field configuration by $2\pi i N$ with a position-dependent $N\in \mathbb{Z}$.  The change in the action depends on the value of $N$ at infinity, and the contributions of operator insertions also shift in a way that depends on the value of $N$ in their vicinity.  For the reader who is uncomfortable with this, we note that in the Chern-Simons interpetation espoused in section \ref{csfun}, these additional ``solutions'' were just as valid and conventional as the usual ones.  So one could in principle rephrase what follows in Chern-Simons language, which would perhaps make it sound more plausible.  We will henceforth assume that the relationship between $Z_\rho$ and $Z_{\rho'}$ is given by a formula analogous to (\ref{simplex}) in the Chern-Simons version:
\be
Z_{\rho'}=Z_\rho e^{-\frac{2\pi i}{b^2}\big(n+\sum_i m_i \alpha_i b\big)}
\ee
Here $n$ and $m_i$ are the differences in $N$ at infinity and near the various operator insertions, and $m_i$ are either all even or all odd.  

We will now compute the ratio of the spacelike and timelike DOZZ formulas for a region of $b$ where both make sense, with the goal being to check that their ratio is consistent with this result.  Using (\ref{timelikedozz}) expressed in terms of the ``unhatted'' variables as well as (\ref{dozz}), (\ref{upsilontheta}), and (\ref{Hdef}), we find:
\begin{align} \nonumber
\frac{\hat{C}(-i \alpha_1,-i \alpha_2,-i \alpha_3)}{C(\alpha_1,\alpha_2,\alpha_3)}=&-\frac{2\pi i}{b}\lim_{\epsilon\to0}\frac{\Upsilon_b(\epsilon)}{\Upsilon_0 H_b(\epsilon)}e^{i\pi(1/b-b)(Q-\sum_i\alpha_i)}\\\nonumber
&\hspace{-.15\textwidth}\times\frac{H_b(\sum_i \alpha_i-Q)H_b(\alpha_1+\alpha_2-\alpha_3)H_b(\alpha_1+\alpha_3-\alpha_2)H_b(\alpha_2+\alpha_3-\alpha_1)}{H_b(2\alpha_1)H_b(2\alpha_2)H_b(2\alpha_3)}\\\nonumber
=&-2\pi i e^{-\frac{2\pi i}{b^2}\left(\sum_i \alpha_i b-(1+b^2)/2\right)}\\
&\hspace{-.15\textwidth}\times\frac{\theta_1(\frac{\sum_i \alpha_i-Q}{b},\frac{1}{b^2})\theta_1(\frac{\alpha_1+\alpha_2-\alpha_3}{b},\frac{1}{b^2})\theta_1(\frac{\alpha_1+\alpha_3-\alpha_2}{b},\frac{1}{b^2})\theta_1(\frac{\alpha_2+\alpha_3-\alpha_1}{b},\frac{1}{b^2})}{\theta'_1(0,\frac{1}{b^2})\theta_1(2\alpha_1/b,\frac{1}{b^2})\theta_2(2\alpha_1/b,\frac{1}{b^2})\theta_3(2\alpha_1/b,\frac{1}{b^2})}.
\end{align}
Here $\theta'_1(z,\tau)\equiv \frac{\partial \theta_1}{\partial z}(z,\tau)$.  We can simplify this a bit by using (\ref{thetashift}) to shift the argument of one of the $\theta$-functions:
\begin{align} \label{stratio}
\frac{\hat{C}(-i \alpha_1,-i \alpha_2,-i \alpha_3)}{C(\alpha_1,\alpha_2,\alpha_3)}&=\nonumber\\&2\pi i \frac{\theta_1(\frac{\sum_i \alpha_i}{b},\frac{1}{b^2})\theta_1(\frac{\alpha_1+\alpha_2-\alpha_3}{b},\frac{1}{b^2})\theta_1(\frac{\alpha_1+\alpha_3-\alpha_2}{b},\frac{1}{b^2})\theta_1(\frac{\alpha_2+\alpha_3-\alpha_1}{b},\frac{1}{b^2})}{\theta'_1(0,\frac{1}{b^2})\theta_1(2\alpha_1/b,\frac{1}{b^2})\theta_2(2\alpha_1/b,\frac{1}{b^2})\theta_3(2\alpha_1/b,\frac{1}{b^2})}.
\end{align}
We'd now like to express this as a ratio of sums of terms of the form $e^{-\frac{2\pi i}{b^2}\big(n+\sum_i m_i \alpha_i b\big)}$ with integer coefficients.  To facilitate this, we define 
\begin{align} \nonumber
\tilde{\theta}_1(z,\tau)&\equiv -i e^{-i \pi \tau/4+i\pi z}\theta_1(z,\tau)=\sum_{n=-\infty}^\infty (-1)^n e^{i\pi \tau n(n-1)+2\pi i z n}\\\label{tildethetas}
\tilde{\theta}_0(\tau)&\equiv -\frac{1}{2\pi}e^{-i \pi \tau/4}\theta_1'(0,\tau)=\sum_{n=-\infty}^\infty (-1)^n n e^{i\pi \tau n(n-1)},
\end{align}
in terms of which we have:
\be \label{stratio2}
\frac{\hat{C}(-i \alpha_1,-i \alpha_2,-i \alpha_3)}{C(\alpha_1,\alpha_2,\alpha_3)}=\frac{\tilde{\theta}_1(\frac{\sum_i \alpha_i}{b},\frac{1}{b^2})\tilde{\theta}_1(\frac{\alpha_1+\alpha_2-\alpha_3}{b},\frac{1}{b^2})\tilde{\theta}_1(\frac{\alpha_1+\alpha_3-\alpha_2}{b},\frac{1}{b^2})\tilde{\theta}_1(\frac{\alpha_2+\alpha_3-\alpha_1}{b},\frac{1}{b^2})}{\tilde{\theta}_0(\frac{1}{b^2})\tilde{\theta}_1(2\alpha_1/b,\frac{1}{b^2})\tilde{\theta}_2(2\alpha_1/b,\frac{1}{b^2})\tilde{\theta}_3(2\alpha_1/b,\frac{1}{b^2})}.
\ee
From (\ref{tildethetas}), we see that the right hand side of this equation now explicitly is a ratio of the desired form.  This completes our demonstration that the timelike DOZZ formula is given by the ordinary Liouville path integral evaluated on a different integration cycle.  Note in particular that the ratio of the two is bad both for purely real and purely imaginary $b$, which illustrates the failure to directly continue between the two.  This argument also confirms our choice of prefactor in the timelike DOZZ formula, since other choices, including the one made in C.10 from \cite{Zamolodchikov:2005fy}, would have polluted this result.  

\subsection{Is Timelike Liouville a Conformal Field Theory?}
\label{tcft?}
\hspace{0.25in}Unlike most sections which are titled by questions, in this case our answer will be an optimistic ``maybe''.  We have established that the timelike DOZZ formula is computed by evaluating the Liouville path integral on a particular choice of cycle, which means that its correlation functions will necessarily obey the usual conformal Ward identities.  So in the sense that any local path integral which computes correlators that obey the conformal Ward identities is a conformal field theory, it is clear that timelike Liouville theory fits the bill.  For example as a consequence of this our semiclassical computations confirmed the usual position dependence of the two- and three-point functions.  But the real meat of this question is understanding to what extent timelike Liouville theory fits into the standard conformal field theory framework of \cite{Belavin:1984vu}.  At least one thing that seems to work is that the derivation of the timelike DOZZ formula from the recursion relations confirms that  the four-point function with a single degenerate operator constructed in the standard way is crossing symmetric.  There has however been justifiable concern in the literature \cite{Strominger:2003fn,Zamolodchikov:2005fy,McElgin:2007ak} about the fact that the timelike DOZZ formula does not obey the degenerate fusion rules when its arguments are specialized to degenerate values.  The simplest manifestation of this is the nonvanishing of  $\hat{C}(0,\hat{\alpha}_1,\hat{\alpha}_2)$ when $\hat{\alpha}_1\neq\hat{\alpha}_2$, as discussed below (\ref{timelike2point}).  

The reason that this is troubling is that semiclassically it seems obvious that $\lim_{\alpha\to0}e^{2\alpha \phi}=1$.  If this were really true as an operator equation, it would imply that the timelike Liouville two-point function is nondiagonal in the operator dimensions.   Since the diagonal nature of this function is a consequence only of the Ward identities, and we know the Ward identities are satisfied just from the path integral, something has to give.  What the Timelike DOZZ formula seems to tell us is that sending $\alpha\to0$ in the three-point function does \textit{not} produce the identity operator, but instead produces another operator of weight zero that does not obey the degenerate fusion rule.\footnote{Recall that even in spacelike Liouville there was a subtlety with computing degenerate correlators by specializing the general correlators to degenerate values, as discussed below (\ref{lightdegenerate}).}  The existence of such an operator is usually forbidden by unitarity, but timelike Liouville is necessarily nonunitary so this does not contradict anything sacred.  We believe that this is the correct interpretation.\footnote{We thank V. Petkova for useful correspondence on this point.  He points out that this nondecoupling happens in the $c<1$ Coulomb gas formalism, which is closely related to the Timelike DOZZ formula evaluated at degenerate points.}  

As evidence for this proposal, we consider the differential equation obeyed by $u,v$ for the semiclassical three-point function with three heavy operators:
$$\partial^2u+W(z)u=0$$
with
\begin{align}
\nonumber
W(z)=&\left[\frac{\eta_1(1-\eta_1)z_{12}z_{13}}{z-z_1}+\frac{\eta_2(1-\eta_2)z_{21}z_{23}}{z-z_2}+\frac{\eta_3(1-\eta_3)z_{31}z_{32}}{z-z_3}\right]\\&\hspace{1.5in}\times\frac{1}{(z-z_1)(z-z_2)(z-z_3)}.
\end{align}
We observe that if $\eta_1\to0$, there is still a regular singular point at $z=z_1$ that only cancels if we also have $\eta_2=\eta_3$.\footnote{We are here assuming  the Seiberg bound $\Re(\eta_i)<1/2$.}  When $\eta_2\neq\eta_3$, the solution will generically have a logarithmic singularity at $z=z_1$.  In this limit the standard semiclassical solution (\ref{3pointsol}) that we reviewed previously breaks down, and a new solution needs to be constructed.  We interpret this as the three-point function of a new nontrivial operator of weight zero with two conventional Liouville operators.\footnote{The monodromy matrix $M$ of the differential equation about $z_1$ in this limit has in some basis the form $\begin{pmatrix} 1 & 0 \\ \lambda & 1 \end{pmatrix}$, with $\lambda$ some function of $\eta_2$ and $\eta_3$.  This matrix has one eigenvector with eigenvalue 1, but is not diagonalizeable.  In the Chern-Simons interpretation this means that there is still a monodromy defect in the gauge field even after we send $\eta_1\to 0$.} In the spacelike case this also could have happened, but since for real $b$ spacelike Liouville is unitary such an operator cannot exist and the $\mathcal{O}(b^0)$ corrections to the saddlepoint conspire to set the correlator to zero.  In timelike Liouville there is no reason for this conspiracy to happen, and indeed from the timelike DOZZ formula we see that it does not.  We take the fact that this extra singularity disappears only when $\eta_2=\eta_3$ as evidence that, contrary to the worries expressed in \cite{Strominger:2003fn,Zamolodchikov:2005fy,McElgin:2007ak}, the real two-point function of timelike Liouville theory is indeed diagonal. 

Perhaps a natural framework to discuss a CFT that includes an extra operator of dimension 0 that does not decouple is ``logarithmic'' CFT.  Something which remains mysterious about this interpretation however is that there does not seem to be any candidate primary operator we can express in terms of the Liouville field to play this role.  We leave this unresolved for future work, but we note that in the Chern-Simons formulation it is straightforward
to describe the nondegenerate primary of dimension 0; it corresponds to a monodromy defect
with unipotent monodromy as explained in the footnote on the previous page.

The main open question that would allow a more systematic understanding of timelike Liouville as a CFT is to identify the set of states on which we should factorize correlation functions.  In spacelike Liouville theory this question was answered by Seiberg \cite{Seiberg:1990eb}, and is formalized by the expression (\ref{factorized4point}) for the four-point function.  In that case the key insight came from study of minisuperspace and the analogy to scattering off of an exponential potential.  A similar analysis for timelike Liouville theory was initiated in \cite{FS} and studied further in \cite{McElgin:2007ak}, but the Hamiltonian is non-hermitian and subtle functional analysis seems to be called for.  We have not tried to extend the minisuperspace analysis of \cite{FS,McElgin:2007ak} to the full timelike Liouville theory, but it seems that this would be the key missing step in establishing the appropriate basis of states to factorize on.  This would then allow construction of four-point functions as in (\ref{factorized4point}) for spacelike Liouville, and one could check numerically if they are crossing symmetric.  Since in the end of the day we know that the path integral does produce crossing-symmetric four-point functions that obey the Ward identities, it seems certain that such a construction is possible; it would be good to understand it explicitly.

\chapter{Conclusion} \label{conclusion}
In this conclusion we summarize our main results and suggest a few directions for future work.  We began by trying to assign a path integral interpretation to the full analytic continuation of the DOZZ formula (\ref{dozz}) for the three-point function of Liouville primary operators.  Our technique was to compare the semiclassical limit of the DOZZ formula and various other correlators to the classical actions of complex solutions of Liouville's equation.  We found that for certain regions of the parameters the analytic continuation is well described by the machinery of Stokes walls and complex saddle points, and in particular we showed that the old transition to the ``fixed-area'' region \cite{Seiberg:1990eb} can be reinterpreted in this manifestly-local language.  The main surprise was that in order to properly account for the full analytic continuation it was necessary to include multivalued/discontinuous ``solutions'', whose actions were defined according to a simple prescription in section \ref{gaction}.  In section \ref{genf},we saw that these singularities naively suggested singularities in the four-point function, but argued that they were in fact resolved by quantum corrections.  One is tempted to declare this an example of the quantum resolution of two-dimensional gravitational singularities.  

Two situations come to mind where these ideas may be relevant.  In \cite{Freivogel:2009rf} a statistical model of bubble collisions in three-dimensional de Sitter space was constructed in which the 4-point function had singularities when the operators were not coincident, similar to the naive result in section \ref{genf}.  This theory has a yet to be well-understood relationship to dS/CFT in three dimensions, which is expected to have a Liouville sector coupled to a nonunitary CFT \cite{Harlow:2010my}.  Perhaps in a more refined version of this theory the singularity could be resolved as in section \ref{genf}.  Secondly, in three-dimensional Euclidean quantum gravity with negative cosmological constant, it was found in \cite{Maloney:2007ud} that including only real smooth solutions in the path integral produces a partition function that does not have the correct form to come from a CFT computation.  Perhaps other complex ``solutions'' need to be included?\footnote{For quantum gravity in higher dimensions it is unlikely that the path integral makes sense beyond the semiclassical expansion about any particular background, so in particular it is probably not well-defined enough for us to ask about Stokes phenomenon.}  More generally the use of singular ``solutions'' seems to be a new phenomenon in field theory and we wonder where else it could appear. 

We then discussed how the question of analytic continuation could be reformulated in the Chern-Simons description of Liouville theory, where we found that the picture of analytic continuation in terms of Stokes phenomenon is more conventional and all relevant solutions seem to be nonsingular.  It would be interesting to get a more precise picture of these solutions; since explicit formulas exist on the Liouville side it seems plausible that they may also be achievable on the Chern-Simons side.  This would allow a more concrete realization of the ideas suggested in section \ref{csfun}.

Finally we used the tools developed in the previous sections to discuss an expression (\ref{timelikedozz}), proposed in \cite{Zamolodchikov:2005fy}, for an exact three-point function in timelike Liouville theory.  We found that we could interpret this formula as being the result of performing the usual Liouville path integral on a different integration cycle, which we demonstrated both semiclassically and exactly.  We also discussed the extent to which timelike Liouville theory can be understood as a conformal field theory, arguing that it probably can but that the spectrum of states to factorize on needs to be understood before the question can be decisively settled.  Even before this question is addressed however, we already consider our results sufficient motivation to begin using the formula of \cite{Zamolodchikov:2005fy} to study the various proposed applications of timelike Liouville theory to closed string tachyon condensation \cite{Strominger:2003fn} and FRW/CFT duality \cite{Freivogel:2006xu}.

\part{}\label{Gauge Fixing Paper}

\chapter{Introduction}
\hspace{0.25in}Liouville theory has been a useful component in String Theory and 2-D quantum gravity ever since Polyakov introduced it in the context of Non-critical String theory\cite{Polyakov:1981rd}.  A complete list of the applications of Liouville theory is beyond the scope of this part. Some applications of note include include its use as a non-compact conformal field theory, a model for Higher-dimensional Euclidean gravity, and as a linear dilaton background in String Theory. It is deeply ingrained in proposed holographic duals of de Sitter space and the multi-verse including the conjectured FRW/CFT \cite{Freivogel:2006xu,Susskind:2007pv,Sekino:2009kv,Harlow:2010my}. It has also been found to have a connection to four-dimensional gauge theories with extended supersymmetry \cite{Alday:2009aq}. Work on Liouville theory has yielded exact results from , e.g. the correlation function of three primary operators which is given by the  DOZZ formula \cite{Dorn:1994xn}. Combinatorial approaches have also been made to obtain results in Liouville some examples are \cite{Ambjorn:1995dg,Ambjorn:2011rs,Ambjorn:2012kd}. It has also been used Kaluza-Klein constructions as an explicit model of how spontaneous breaking of space-time translation invariance can lead to compactification of the space-time \cite{PhysRevD.28.2583,PhysRevLett.50.1719}. Recently its path integral properties under analytic continuation have been discussed in\cite{Harlow:2011ny} including the continuation of theory to the Timelike Liouville regime\cite{springerlink:10.1007/s11232-005-0003-3,Harlow:2011ny}. 

      Timelike liouville theory possesses $\mathbb{S}^{2}$ as a real saddle point about which quantum fluctuations can occur. Computing expectation values of fields on this fluctuating geometry involves a path integral over the metric of the geometry. The gauge redundancy of this path integral must be dealt with before meaningful quantities can be computed. The issue that comes up in computing the expectation values of standard classical quantities like the distance between points in this fluctuating geometry is that even after fixing to conformal gauge by imposing  $g_{\mu\nu} = e^{2\b\phi}\tilde{g}_{\mu\nu}$, where $\phi$ is the Liouville field and $\tilde{g}_{\mu\nu}$  is a reference metric of $\mathbb{S}^{2}$, not all the gauge redundancy has been removed. The remaining gauge redundancy is due to $SL_{2}(\mathbb{C})$ which transform the reference coordinates and Liouville field transform nontrivally leaving the physical manifold invariant. This invariance means that until this redundancy is fixed, the integral over metrics is not defined. Computing quantities that depend on the physical points by characterizing them with reference coordinates is not possible because the position of two points on the physical manifold is not uniquely determined by two reference points. A $SL_{2}(\mathbb{C})$ transformation will change the position of the reference points leaving the physical points alone. Computing the distance between the physical points by integrating over the reference points is not defined until the  $SL_{2}(\mathbb{C})$ redundancy is fixed.  In this part it is shown in a perturbative analysis that after fixing to conformal gauge and expanding about the spherical saddle of the Timelike Liouville field, the remaining zero mode due to the invariance under $SL_{2}(\mathbb{C})$ coordinate transformations of the reference sphere can be dealt with but using standard Fadeev-Popov methods employing the gauge condition that the ``dipole'' of the coordinate system is a fixed vector, and then integrating over all values of this dipole. Dealing with this zero mode means that a Green's function can be obtained  and a pertubative analysis of quantities on spherical geometry under the influence of fluctuations of a semi-classical Timelike Liouville field can be carried out. 

One such quantity is the expectation value of the length of a geodesic on a spherical geometry under the influence of a semi-classical Timelike Liouville field, computed to second order in the Timelike Liouville coupling $\b$. It is shown that this quantity is well defined and doesn't suffer from any power law or logarithmic divergences as a na\"{i}ve power counting argument might suggest.

{\bf{Outline:}} In section \ref{propagator}, a Green's function is obtained by implementing the gauge constraint of fixing the coordinate dipole and integrating over the value of this dipole.
In Section \ref{dist}, the Green's function is employed to compute the expectation value of the separation between two points on sphere under the influence of a fluctuating Timelike Liouville field to second order in the coupling $\b$. Finally in Section \ref{res}, some possible further applications are looked at.

For some modern reviews on Liouville theory the reader is encouraged to look at  \cite{Ginsparg:1993is,Nakayama:2004vk,Teschner:2001rv}, some slightly older reviews include \cite{PhysRevD.26.3517, Zamolodchikov:1995aa,Seiberg:1990eb}. For information on the analytic continuation of Liouville to the Timelike regime the reader is humbly referred to \cite{springerlink:10.1007/s11232-005-0003-3,Harlow:2011ny}.
 
\chapter{The Gauge Fixed Propagator}\label{propagator}
\hspace{0.25in} When coupling a generic conformal Field to two dimensional gravity, the Liouville action 
\be\label{liouvilleaction}
S_{L} = -\frac{1}{4\pi}\int dx^2 \sqrt{\tilde{g}}\big(\tilde{g}^{ab}\partial_{a}\phi\partial_{b}\phi +Q\tilde{R}\phi + 4\pi\mu e^{2b\phi}\big)
\ee
is obtained after fixing to conformal gauge \cite{David:1988hj,Ginsparg:1993is}\footnote{This means that in the path integral over  metrics a general metric is decomposed into a conformal Liouville factor and a family of conformally inequivalent reference metrics $\tilde{g}_{\mu\nu}$\cite{David:1988hj}. In this part the only relevant reference geometry is, $\mathbb{S}^{2}$ as higher genus surfaces will not be discussed.}.  (\ref{liouvilleaction}) is invariant under conformal transformations of the coordinates 
\begin{align}\label{anomaly}
z^{\prime} &= w[z]\\
\phi^{\prime}[z^\prime,\bar{z}^{\prime}] &= \phi[z,\bar{z}] - \frac{Q}{2}\log{\Big|\frac{\partial w}{\partial z}\Big|}
\end{align}
with $Q = b + \frac{1}{b}$ and the central charge $c=1+6Q^{2}$, up to a $c$-number anomaly \cite{Zamolodchikov:1995aa}.

The Euclidean space-like Liouville partition function, with a canonically normalized Liouville field $\phi$, can be written as
   \be\label{spacelikeaction}
  \mathcal{Z} =  \int\D\phi \exp{\Big[-\frac{1}{4\pi}\int dx^2 \sqrt{\tilde{g}}\big(\tilde{g}^{ab}\partial_{a}\phi\partial_{b}\phi +Q\tilde{R}\phi + 4\pi\mu e^{2b\phi}\big)\Big]}.
   \ee
   
 This form depends on the fact that the metric can be gauge fixed in a generally covariant way to conformal gauge i.e. the \emph{Physical} metric, $g_{\mu\nu}$  can be written in terms of the product of the exponentiated Liouville factor and a \emph{Reference} metric $\tilde{g}_{\mu\nu}$ giving $g_{\mu\nu} = e^{2b\phi}\tilde{g}_{\mu\nu}$\cite{Zamolodchikov:2005fy}. To make contact with the classical Liouville equation, the $1/b^{2}$ dependence  of the central charge
which has been absorbed into the definition of the Liouville field must be taken into account in order to canonically normalize the action. In the Semi-classical limit, the action can be written in terms of classical field via the field redefinition $\phi_{c} = 2b\phi$,
\be\label{spacelikeactionclassical}
-\frac{1}{16\pi b^{2}}\int dx^2 \sqrt{\tilde{g}}\big(\tilde{g}^{ab}\partial_{a}\phi_{c}\partial_{b}\phi_{c} + 2bQ\tilde{R}\phi_{c} + 16\pi\mu b^{2} e^{\phi_{c}}\big).
\ee

Here the dominant contribution of the central charge $c\propto 1/b^{2}$ has been factored out. To make a good semi-classical limit the ``cosmological constant" $\mu$ must scale as
$1/b^{2}$. The actual cosmological constant $\lambda =\pi\mu b^{2}$, is well defined in the semi-classical $b\rightarrow 0$ limit.\footnote{The value $\lambda$ is a tunable constant in the Liouville theory. It can be changed by adding a constant linear shift to the Liouville field. The value of $\lambda$ will be set by the radius of the sphere in what follows.} Timelike Liouville results from
(\ref{spacelikeaction}) under the continuation $b \rightarrow -i\hat{b}$, $\phi \rightarrow i\hat{\phi}$ and $Q\rightarrow i\hat{Q}$ with $\hat{b} \in \mathbb{R}$. The resulting action is
    \be\label{timelikeaction}
-\frac{1}{4\pi}\int dx^2 \sqrt{\tilde{g}}\big(-\tilde{g}^{ab}\partial_{a}\hat{\phi}\partial_{b}\hat{\phi} - \hat{Q}\tilde{R}\hat{\phi} + \frac{4\lambda}{\b^{2}}
e^{2\hat{b}\hat{\phi}}\big).
   \ee
 In the semi-classical limit, $\hat{b} \rightarrow 0$, (\ref{timelikeaction}) has a large negative central charge $c = 1 -6\hat{Q}^2$ with $\hat{Q} =1/\hat{b}-\hat{b}$.
   
One cannot simply compute the partition function for  (\ref{timelikeaction}) by simply integrating  (\ref{timelikeaction}) over all fluctuations about the sphere, since the kinetic term in (\ref{timelikeaction}) is the wrong sign and the path integral is formally divergent. One must take the path integral of the partition function of (\ref{spacelikeaction}) and analytically continue it, taking Stokes Phenomenon into account employing the results of \cite{Harlow:2011ny} to define the Timelike partition function. Since all the relevant saddles of the integration cycle, not just the sphere, must be taken into account to get finite answers and reproduce exact results like the Timelike DOZZ formula\footnote{This continuation property is what allows us to use a path integral approach to compute Timelike Liouville correlation functions. As the wrong sign kinetic term of (\ref{timelikeaction}) renders the partition function integral formally divergent if Stokes Phenomenon isn't taken into account\cite{Harlow:2011ny}. In this part it will be assumed that this has already been taken into account. A complete account of Stokes Phenomenon and the Timelike partition function goes beyond the scope of this part, for a nice account one should look at
\cite{Harlow:2011ny,Witten:2010cx,Bender:1978:AMM}.}.
   
The action of Timelike Liouville has the 2 sphere, $\mathbb{S}^2$ as homogeneous real saddle point. \footnote{Viewed from the Space-like side this is a complex saddle point} The saddle point of the field is defined by the
constant Liouville field value
  \be\label{constsaddle}
  \hat{\phi} = \hat{\phi}_0 =\frac{1}{2\b}\log{\Big|\frac{\hat{Q}\b\tilde{R}}{8\lambda}\Big|}.
  \ee
  
Perturbations by ``\emph{light}" operators\footnote{The terminology ``\emph{Light}" and ``\emph{Heavy"} primary operators,
is standard in the study of Liouville theory. When computing correlators of primary operators $<e^{\alpha_1\phi_1}\ldots e^{\alpha_n\phi_n}>$, an operator is called \emph{heavy} if its
Liouville momentum $\alpha_i \sim \frac{\sigma_i}{\b}$ in the $\b\rightarrow 0$ limit, and \emph{light} if $\alpha_i \sim \b\sigma_i$ as $\b\rightarrow 0$. Heavy operators can effect the
classical saddle point, as they scale in the same way as the action while light operators give sub-leading contributions. }, which scale as $\b\sigma$ in Liouville momentum will not effect the saddle point and hence a perturbative expansion of
(\ref{timelikeaction}) about the spherical saddle point can be made without changing the saddle point, i.e. fluctuations cannot change the topology. Expanding the Liouville field as $\hat{\phi} = \hat{\phi}_0 + f$ and
expanding to quadratic order in $\b$, yields a quadratic action for $f$, which apart from an irrelevant constant $S_0$, is independent of the value of $\lambda$.
  
  \be\label{linearactionint}
-\frac{1}{4\pi}\int dx^2 \sqrt{\tilde{g}}\big(-\tilde{g}^{ab}\partial_{a}f\partial_{b}f +\tilde{R}f^2\big) + S_0.
    \footnote{From the Liouville saddle point (\ref{constsaddle}) and what later follows in Section \ref{dist} this will imply that $\lambda = 1/4$.  However to aid in the analysis $\lambda$ will be left general for now and determined later.}\ee  Expressing the action (\ref{linearactionint}) in spherical coordinates and integrating by parts yields
  \be\label{laction1}
-\frac{1}{4\pi}\int d\theta d\varphi\sin{\theta}\big\{f\Big(\frac{1}{\sin{\theta}}\partial_{\theta}\sin{\theta}\partial_{\theta} +\frac{1}{\sin^2{\theta}}\partial^2_{\varphi}
+ 2\Big)f\big\}. \footnote{Note $R= 2$ for the unit sphere.}
    \ee
\begin{figure}[ht]
\begin{center}
\includegraphics[scale=0.6]{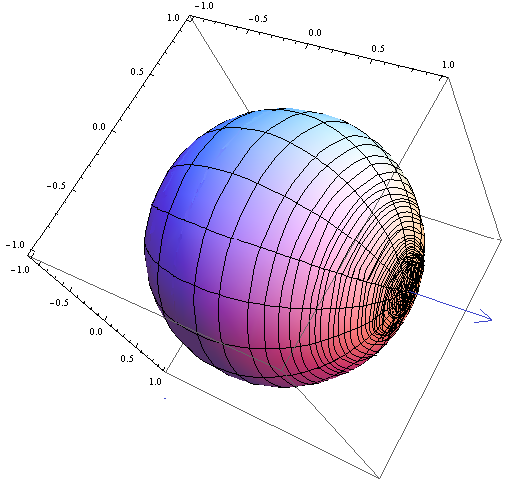}
\caption{A sphere embedded in $\mathbb{R}^{3}$  with a coordinate system possessing a dipole. There is more coordinate area on one side than the other. The blue arrow denoting the dipole vector.}
 \label{dipole}
\end{center}
\end{figure}

This action has a zero mode which must be  dealt with in order to compute quantities in perturbation theory. The zero mode corresponds to the $SL_{2}(\mathbb{C})$ conformal coordinate transformations that can be
performed on the coordinates of the reference sphere. These transformations have the effect of moving coordinate area around the sphere. The non-compact portion of this gauge redundancy can be attributed to
the overall dipole of area that the physical manifold has compared to the reference sphere; see figure \ref{dipole}. This last gauge freedom must be dealt with using a Fadeev-Popov procedure.
  
The equation of motion resulting from (\ref{laction1}) is that massive scalar field on a sphere. The calculation can be simplified by exploiting the fact that the Green's function will only depend on the geodesic separation between points on the
sphere has only one singularity and  is rotationally symmetric around that singularity\footnote{The non trivial fact that a massive scalar field on a sphere can possess a single singularity unlike the massless case which must have two, is due to mass causing field lines to be die off before they reach the other side of the sphere to form a second singularity}. This rotational symmetry implies that the Green's function will only depend on the angle between the source point and the field point, $\theta, \varphi$ and $\theta^\prime,\varphi^\prime$. This
means that the Green's function $G$ is only a function of $\cos{\beta} = \vec{x}\cdot\vec{x}^\prime = \cos{\theta}\cos{\theta^\prime} +\sin{\theta}\sin{\theta^\prime}\cos{(\varphi - \varphi^\prime)}$. Using the rotational symmetry of the differential operator and calling $\chi = \cos{\beta}$ we can rewrite the Green's function equation from (\ref{laction1}) into
  
  \be\label{laction2}
\big(\partial_{\chi}(1 - \chi^2)\partial_{\chi} + 2\big)G = \frac{1}{2\pi}\delta{[1-\chi]}.
  \ee

\begin{figure}[ht]
\begin{center}
\includegraphics[scale=0.5]{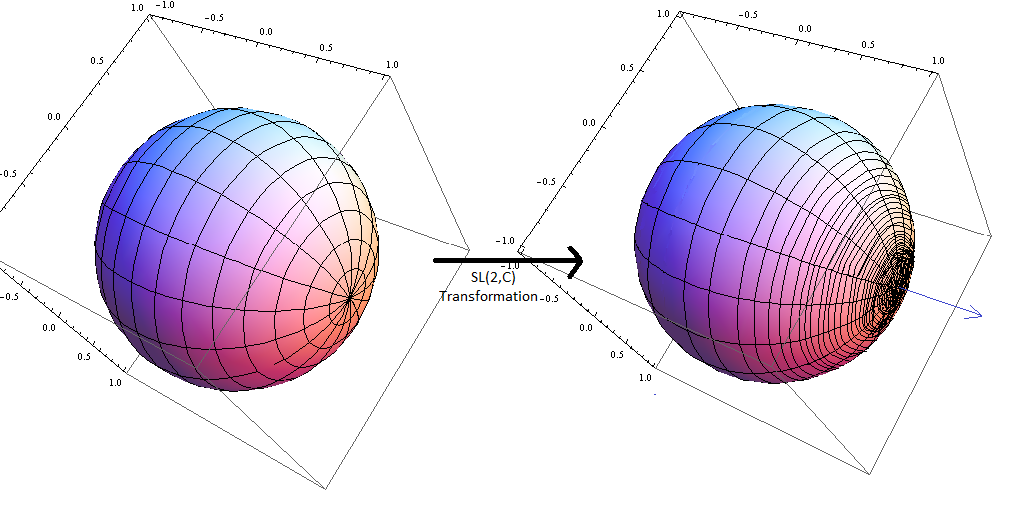}
\caption{An SL(S,C) transformation, changes the dipole of the coordinate system of the sphere}\label{dipole2}
\end{center}
\end{figure}
\hspace{0.25in}(\ref{laction2}) still possesses a zero mode. The zero mode is due to invariance of the physical geometry under $SL_{2}(\C)$ reparametrizations of the coordinates on the reference sphere which results in a compensating change of the Liouville field, leaving the physical metric invariant.  This $SL_{2}(\mathbb{C})$ transformation is a combination of a compact rotations of the coordinate system and/or a non-compact change in the dipole of coordinate system. This can be visualized as follows, if the coordinates are thought of as a charge density on a sphere embedded in $\mathbb{R}^{3}$, $SL_(2,\mathbb{C})$ transformations change the dipole of this charge density by pushing points toward one site on the sphere and repelling them from another. This increases the local charge density about one site of the sphere and decreases about another. In terms of the coordinates it increases the amount of coordinate area about one point and decreases it about another, see Figure \ref{dipole2}. If the dipole of the Liouville field is fixed as a gauge condition and the Faddeev-Popov procedure of integrating over the gauge condition is implemented, a Green's function can be obtained for use in perturbation theory. In (\ref{laction2}) the statement that there is a zero mode remaining is just the fact that a solution of (\ref{laction2}) is invariant under $G \rightarrow G+\alpha_1 P_1(\chi)= G + \alpha_1\chi $ for any value of $\alpha_1$.\footnote{There is only one $\alpha_1$ here because the coordinates have been rotated into the direction of the dipole, and $\alpha_1$ is the magnitude of the dipole vector. If this were not exploited, then $\alpha_1$ would be replaced by the vector $\vec{\alpha}$ and the zero mode statement would be $G \rightarrow G+\sum^{1}_{m=-1}\alpha_m Y^{(1,m)}[\theta,\varphi].$}
 In these rotated coordinates (\ref{laction2}) is just the equation for the Green's function for Legendre's differential equation with $l = 1$.
  
  To proceed, the Fadeev-Popov procedure is employed\cite{Faddeev196729,Faddeev:1969su} on (\ref{laction1}) with the gauge condition that the dipole over the sphere is fixed and then integrated over all values. The dipole is defined as,
  \be\label{gaugecondition}
  \vec{n} = \int d\theta\,d\varphi\sin{\theta}\Big(G\vec{x}[\theta,\varphi]\Big) .\footnote{Here $\vec{x} = |x|(\sin{\theta}\cos{\varphi}\hat{\mathbb{I}} +\sin{\theta}\sin{\varphi}\hat{\mathbb{J}}+\cos{\theta}\hat{\mathbb{K}})$ is the $\mathbb{R}^{3}$ position vector in spherical coordinates. It is imaged that the reference sphere is embedded in a larger $\mathbb{R}^{3}$.}
  \ee
  The functional delta function is inserted into the partition function for (\ref{laction1}) by inserting the identity 
  
  \be\label{fadeevpopov}
  1 = \int d^{3}\alpha\,\Delta_{FP}\,\delta^{(3)}
  \Big[ \int d\theta\,d\varphi\sin{\theta}\Big(G^{(\vec{\alpha})}\vec{x}[\theta,\varphi]\Big)- \vec{\kappa}\Big]
  \ee
into the functional integral for the partition function.
  Here $G^{(\vec{\alpha})}$ means the gauge transformed $G$ with parameters $\vec{\alpha}$, and $\Delta_{FP}$ is the Fadeev-Popov determinant.
  Following the standard methods of inserting (\ref{fadeevpopov}) into partition function with action (\ref{laction1}), transforms the linearized Timelike Liouville partition function into
  
  \begin{align}\label{funtionalintegral}
&\mathcal{N}\int d^{3}\kappa e^{ -\frac{\Lambda}{2}\vec{\kappa}\cdot\vec{\kappa}}\int d^{3}\alpha\int\D f \\&\hspace{1in}\times\exp{\Big[-\frac{1}{4\pi}\int d\theta\,d\varphi\sin{\theta}\big\{f\Big(\frac{1}{\sin{\theta}}\partial_{\theta}\sin{\theta}\partial_{\theta} +\frac{1}{\sin^2{\theta}}\partial^2_{\varphi} + 2\Big)f\big\}\Big]}\nonumber\\&\hspace{2in}\times\delta^{(3)}(\big\{\int d\theta\,d\varphi\sin{[\theta]}\vec{x} f^{(\alpha)}) \big\}- \vec{\kappa})\Delta_{FP}\nonumber.
  \end{align}
  
  Integrating over $\vec{\kappa}$ results in 
  \begin{align}\label{fgaugedint}
&\mathcal{N}\Delta_{FP}\int d^{3}\alpha\int\D f \\&\hspace{0.45in}\times\exp{\Big[-\frac{1}{4\pi}\int d\theta\,d\varphi\sin{\theta}\big\{f\Big(\frac{1}{\sin{\theta}}\partial_{\theta}\sin{\theta}\partial_{\theta} +\frac{1}{\sin^2{\theta}}\partial^2_{\varphi} + 2\Big)f\big\} -2\pi\Lambda f\vec{x}\cdot\vec{n}\big\}\Big]}.\nonumber
  \end{align}
 From the point of view of the Green's function equation obtained from (\ref{fgaugedint}), the integral (\ref{gaugecondition})  is just a c-number vector. Hence the Green function equation resulting from (\ref{fgaugedint}) can be solved by treating (\ref{gaugecondition})  as a constraint on the Green's function and then enforcing that constraint to obtain a final answer. The Green's function equation resulting from (\ref{fgaugedint}) taking into account the fact that the Green's function only depends on $\chi$ is
  \be\label{greenfunctioneqn}
  \partial_{\chi}\{(1 - \chi^2)\partial_{\chi}G\} + 2G = -\pi\Lambda n\chi + \frac{1}{2\pi}\delta(1 - \chi),
  \ee
with
\be\label{gaugecon}
n = 2\pi\int^{1}_{-1}d\chi^{\prime} \,\chi^{\prime}G[\chi^\prime].
\ee

The gauge fixing procedure has resulted in a inhomogeneous term in (\ref{greenfunctioneqn}), this term can be interpreted as a background charge that absorbs the field lines coming from
the singularity at $\chi = 1$ so that the field is smooth everywhere else on the sphere.
  The general solution to (\ref{greenfunctioneqn}) when $\chi \neq 1$  is
  \be
G = \alpha_1\chi + \Big(\frac{\pi n\Lambda}{6} + \frac{c_2}{2}\Big)\chi\log{|1 + \chi|} + \Big(\frac{\pi n\Lambda}{6} - \frac{c_2}{2}\Big)\chi\log{|1 - \chi|} - c_2.
  \ee
  
Imposing the boundary conditions of finiteness $\chi \neq 1$, smoothness of the solution at $\chi = -1$, normalizing the Green's function so (\ref{greenfunctioneqn}) is obeyed  when $\chi = 1$, and imposing the
constraint equation (\ref{gaugecon}) gives values for $\Lambda$, $\alpha_1$, $n$, $c_2$. 

The resulting Green's function is
  
  \begin{align}\label{greenfunction}
 \text{\begin{fmffile}{liouvprop}
\begin{fmfgraph*}(25,3)
\fmfleft{in}
\fmfright{out}
\fmflabel{$x$}{in}
\fmflabel{$y$}{out}
\fmf{dashes}{in,out}
\fmfdot{in,out}
\end{fmfgraph*}
\end{fmffile}}
 \hspace{6pt}&=<f[\theta_x,\varphi_x]f[\theta_y,\varphi_y]>\\&= \frac{1}{8\pi^2}\Big\{-(\log{2} + 1/2)\chi_{xy}+ \chi_{xy}\log{|1 - \chi_{xy}|} + 1\Big\}\nonumber
  \end{align}
  
with $\Lambda = -\frac{9\cdot3}{11\cdot2\pi^2} $, $n = -\frac{11}{4\cdot9\pi}$, $\alpha_1 =-\frac{1}{8\pi^2}(\log{2} + 1/2)$, and $c_2 = -\frac{1}{8\pi^2}$; and $\chi_{xy}
=\cos{\theta_x}\cos{\theta_y} +\sin{\theta_x}\sin{\theta_y}\cos{[\varphi_x - \varphi_y]}$ .

Quantities can now be computed in perturbation theory using Wicks theorem. Which relates the Green's function to the two point correlator of the field\cite{PhysRev.80.268}.  
  This is the Green's function that will be used to compute quantities in perturbation theory.
 \chapter{Perturbative Correction To The Geodesic Distance Between Two Points On The Bumpy Sphere.}\chaptermark{Geodesic Distance on the Bumpy sphere}\label{dist}
\hspace{0.25in}
Now that the zero mode has been dealt with, the perturbative correction of the expectation value of physical geodesic distance between two arbitrary points lying on a north south trajectory on the reference sphere can now be computed.\footnote{  This is more general then it seems, since one can simply rotate the coordinates of the reference sphere to move two arbitrary points onto a north-south trajectory, no generality is lost by computing north-south distances.}  The advantage of computing north/south trajectories is that the two end points will have the same value of the azimuthal angle $\varphi$ which will be  called $\varphi_0$ and this simplifies the calculation\footnote{Because $\phi$ is the standard symbol for the Liouville field, $\varphi$ will be used for the azimuthal
angle.}.

The quantity that will be studied is the expectation value of geodesic distance $L$ between two points on a north south trajectory computed up to second order in $\b$\footnote{Here the reference polar angle $\theta$ has been chosen as the parameter along the geodesic, to avoid any vielbien ambiguities.}

\begin{align}\label{geodesicdist}
L&= \Big<\int^{\theta_2}_{\theta_1}\sqrt{g_{\mu\nu}\frac{\partial x^{\mu}}{\partial\theta}\frac{\partial x^{\nu}}{\partial\theta}}\,d\theta\Big>\\&= \Big<\int^{\theta_2}_{\theta_1}\sqrt{e^{2\b\phi_{0}+2\b f}\tilde{g}_{\mu\nu}\Big\{1 + \sin^{2}{\theta}\Big(\frac{d\varphi}{\partial\theta}\Big)^{2}\Big\}}\,d\theta\Big>\nonumber\\
&=\frac{\mathcal{N}\Delta_{FP}\int d\alpha\mathcal{D} f \Big(\int^{\theta_2}_{\theta_1}\sqrt{e^{2\hat{b}\phi_0 +2\hat{b} f}\tilde{g}_{\mu\nu}\frac{\partial x^{\mu}}{\partial\theta}\frac{\partial x^{\nu}}{\partial\theta}}\,d\theta\Big)e^{-S_{\text{gauge fixed action}}}}{\mathcal{N}\Delta_{FP}\int d\alpha\mathcal{D} f e^{-S_{\text{gauge fixed action}}}}.\nonumber
\end{align}
 The unperturbed geodesic is the portion of the latitude line connecting $\theta_1$ and $\theta_2$, implying that the unperturbed geodesic $\varphi_0$ is a constant.
Variation with respect to $x^{(2)}=\varphi$ in (\ref{geodesicdist}) results in the usual geodesic equation,
\be\label{Lvar1}
\delta L = -\int^{\theta_2}_{\theta_1}d\theta\,\Big(\frac{d^{2}\varphi}{d\theta^{2}} +\Gamma^{\varphi}_{\mu\nu}\frac{dx^{\mu}}{d\theta}\frac{dx^{\nu}}{d\theta} \Big) g_{\varphi\varphi}\delta\varphi = 0.
\ee It follows that, corrections from the reference geodesic equation comes from two sources:
the explicit factor of $e^{2\b\phi}$ in (\ref{geodesicdist}), and the change in the Christoffel symbol that results from it,
\be\label{Christoffel}
\Gamma^{\lambda}_{\mu\nu} = \tilde{\Gamma}^{\lambda}_{\mu\nu} + b\tilde{g}^{\lambda\sigma}\big[(\partial_{\mu}f)\tilde{g}_{\nu\sigma} + (\partial_{\nu}f)\tilde{g}_{\sigma\mu} -
(\partial_{\sigma}f)\tilde{g}_{\mu\nu}\big].
\ee

Here $f$ is the fluctuation in the Liouville field from $\hat{\phi}_{0}$. The geodesic equations derived from (\ref{Christoffel}) are\footnote{
$\hat{\Gamma}^{\varphi}_{\varphi\theta} = \cot{\theta}$ is the only pertinent non-zero Christoffel symbol for the reference geometry of $\S^{2}$.}
\be\label{geodeqn}
\frac{d^{2}\varphi}{d\theta^{2}} + 2\tilde{\Gamma}^{\varphi}_{\varphi\theta}\Big(\frac{d\varphi}{d\theta}\Big)
=-\b\Big[2\frac{d\varphi}{d\theta}\Big(\partial_{\theta}f + \partial_{\varphi}f\Big) - 
\frac{\partial_{\varphi}f}{\sin^{2}{\theta}}\Big\{1 + \sin^{2}{\theta}\Big(\frac{d\varphi}{d\theta}\Big)^{2}\Big\}\Big].
\ee
Inserting (\ref{geodeqn}) into (\ref{Lvar1}) gives
\begin{align}\label{Lvar2}
\delta L&=-\int^{\theta_2}_{\theta_1}d\theta\,\sin^{2}{\theta}e^{2\b f}\delta\varphi\Big\{\frac{d^{2}\varphi}{d\theta^{2}} +2\cot{\theta}\frac{d\varphi}{d\theta} + \b\Big[2\frac{d\varphi}{d\theta}\Big(\partial_{\theta}f + \partial_{\varphi}f\Big) \\
&\hspace{3in}-\frac{\partial_{\varphi}f}{\sin^{2}{\theta}}\Big(1 + \sin^{2}{\theta}\Big(\frac{d\varphi}{d\theta}\Big)^{2}\Big)\Big]\Big\}\nonumber\\
&=-\int^{\theta_2}_{\theta_1}d\theta\,e^{2\b f}\delta\varphi\Big\{\frac{d}{d\theta}\Big(\sin^{2}{\theta}\frac{d\varphi}{d\theta}\Big) - \b
\partial_{\varphi}f \nonumber\\&\hspace{2in}+ \b\sin^{2}{\theta}\Big[2\frac{d\varphi}{d\theta}\partial_{\varphi}f +\Big(\frac{d\varphi}{d\theta}\Big)^{2} \partial_{\varphi}f \Big]\Big\}.\label{Lvar3}
\end{align}

Expressing the corrections in the geodesic  $\varphi[\theta]$ as,

\be\label{gcorr}
\varphi[\theta] = \varphi_0 +\b\varphi_{1}[\theta] + \b^{2}\varphi_{2}[\theta] + \ldots
\ee
and substituting (\ref{gcorr}) into (\ref{Lvar3})  leads to a set of equations of different orders in $\b$. The zeroth, first, and second order equations are respectfully,

\begin{align}
\frac{d}{d\theta}\Big(\sin^{2}{\theta}\frac{d\varphi_0}{d\theta}\Big)&=0\label{firstord}\\
\vspace{0.25in}\nonumber\\
\frac{d}{d\theta}\Big(\sin^{2}{\theta}\frac{d\varphi_1}{d\theta^{2}}\Big)=\partial_{\varphi}f-\sin^{2}{\theta}&\Big[2\frac{d\varphi_0}{d\theta}\partial_{\varphi}f + \Big(\frac{d\varphi_0}{d\theta}\Big)^{2} \partial_{\varphi}f \Big]\label{secondord}\\
\vspace{0.25in}\nonumber\\
\frac{d}{d\theta}\Big(\sin^{2}{\theta}\frac{d\varphi_2}{d\theta^{2}}\Big)=\sin^{2}{\theta}\Big[2\frac{d\varphi_1}{d\theta}&\partial_{\varphi}f + 2\frac{d\varphi_0}{d\theta} \frac{d\varphi_1}{d\theta} \partial_{\varphi}f \Big]\label{thirdord}.
\end{align}
\\
As was mentioned previously, $\varphi_0$ is a constant. This is consistent with (\ref{firstord}) and also implies that (\ref{secondord}) reduces to
\be\label{geodesicequ1}
\frac{d}{d\theta}\Big(\sin^{2}{\theta}\frac{d\varphi_1}{d\theta}\Big)=\partial_{\varphi}f.
\ee
Lastly, since $L$ is being computed to second order in $\b$, (\ref{geodesicdist}) implies that only the first order correction $\varphi_1$ is needed, and (\ref{thirdord}) is not necessary.

The classical action that generates (\ref{geodesicequ1}) up to total derivatives, is that of a forced harmonic oscillator with vanishing kinetic term\footnote{A classical action that generates (\ref{geodesicequ1}) up to total derivatives is \be\nonumber
\int^{\theta_2}_{\theta_1}\Big[\frac{1}{2}\sin^{2}{\theta}\Big(\frac{d\varphi_1}{d\theta^{2}}\Big)^{2} + f\Big]d\theta = \int^{\theta_2}_{\theta_1}\Big[\frac{1}{2}\Big\{\Big(\frac{du_1}{d\theta}\Big)^{2} - u^{2}_1\Big\} + f\Big]d\theta
\ee.
The time evolution parameter in this case be $\theta$.}.
 A good conjugate variable to describe the system is then $u = \varphi\sin{\theta}$, which rewrites (\ref{gcorr}) as
\begin{align}\label{gcorru}
u &= u_0 + \b u_1[\theta] + \b^{2}u_2[\theta] + \ldots\\ &= \varphi[\theta]\sin{\theta}  \nonumber\\&= \varphi_0 \sin{\theta} +\b\varphi_{1}[\theta]\sin{\theta} + \b^{2}\varphi_{2}[\theta] +\ldots\nonumber,
\end{align} changes (\ref{firstord}) to
\be\label{geodeqn1}
\frac{d}{d\theta}\Big(\sin{\theta}\Big[\frac{du_0}{d\theta} - \frac{u_0\cos{\theta}}{\sin{\theta}}\Big]\Big)= \sin{\theta}\Big(\frac{d^{2}u_0}{d\theta^{2}}  +u_0\Big) = 0
\ee
and changes (\ref{geodesicequ1}) into 
\be\label{geodeqn2}
\frac{d^{2}u_1}{d\theta^{2}} + u_1 =\partial_{u}f.
\ee

This rewriting makes the following computations easier.
Since the geodesics begin and end on the same value of $\varphi = \varphi_0$; our boundary conditions are that $u_{0}[\theta_2] = \varphi_0\sin{\theta_2}$, $u_{0}[\theta_1] = \varphi_0\sin{\theta_1}$, and that $u_{1}[\theta_2] = u_{1}[\theta_1] = 0$.\footnote{These boundary conditions are the correct ones as the unperturbed geodesic was just the path $\varphi[\theta]=\varphi_0$ on the reference sphere. Small fluctuations of the sphere, ``bumps", do not change the $\varphi$ position of points hence the end points do not move; therefore the correction $u_1$ should vanish at at the end points.}
These equations can formally be solved to create an expansion for $u$ up to order $\b$,
\begin{align}\label{ueqn}
u = \varphi_0\sin{\theta}\, + \, \b\Big\{\int^{\theta}_{\theta_1}d\hat{\theta}&\sin{[\theta - \hat{\theta}]}\partial_{u}f[\hat{\theta}]\nonumber\\& - \frac{\sin{[\theta_2 -\hat{\theta}]}}{\sin{[\theta_2 - \theta_1]}}\int^{\theta_2}_{\theta_1}d\hat{\theta}\sin{[\theta_2 -\hat{\theta}]}\partial_{u}f[\hat{\theta}]\Big\}.
\end{align}

Rewriting the expectation value (\ref{geodesicdist}) in terms of the variable $u$ yields

\begin{align}\label{mainquant}
L&=\Bigg< e^{b\phi_0}\int^{\theta_2}_{\theta_1}d\theta e^{\b f}\sqrt{1+\Big\{\frac{du}{d\theta} - \frac{u\cos{\theta}}{\sin{\theta}}\Big\}^{2}}\Bigg> \\ \nonumber\text{with}\hspace{1.25in}&\nonumber\\
\Big\{\frac{du}{d\theta} - \frac{u\cos{\theta}}{\sin{\theta}}\Big\}^{2}& = \frac{\b^{2}}{\sin^{2}{\theta}}\Bigg\{\frac{\sin{\theta_1}}{\sin{[\theta_2 - \theta_1]}}\int^{\theta_2}_{\theta_1}d\hat{\theta}\sin{[\theta_2 - \hat{\theta}]}\partial_{u}f[\hat{\theta}]\\&\hspace{2.5in}-\int^{\theta}_{\theta_1}d\hat{\theta}\sin{\hat{\theta}}\partial_{u}f[\hat{\theta}]\Bigg\}^{2}\nonumber.
\end{align}

The second line results from (\ref{ueqn})\footnote{Since (\ref{mainquant}) is being expanded to order $\b^{2}$, it is valid to evaluate $u$ only to first order in $\b$ as the lowest order correction under the square root sign is $\b^{2}$, (the zeroth order term drops out). Higher order corrections of $u$ will only contribute O($\b^3$) corrections.}. It is evident from (\ref{mainquant}) that $e^{b\phi_{0}} =\Big(\frac{\tilde{R}}{8\lambda}(1 - \b^{2})\Big)^{1/2}$ is the radius of the sphere which has been set to $1$ in the $\b\rightarrow0$ limit. Recalling that $\tilde{R} = 2$, this shows that the value of $\lambda = 1/4$ if $\theta_2-\theta_1$ is to be interpreted as the difference in polar angle for the unperturbed path.\footnote{ This makes sense as $\lambda^{-1/2}$ has units of radius of curvature as can be seen from  the classical equation of motion from (\ref{timelikeaction}) in the semi-classical limit. Specifically $R \propto \lambda$ \cite{Ginsparg:1993is}. Since constant shifts $\phi$ can be used to tune the value of $\lambda$,  it  sets the radius of the sphere\cite{Nakayama:2004vk}.} Expanding (\ref{mainquant}) up to  and including $\mathcal{O}[\b^{2}]$ results in

\begin{align}\label{mainquantpert}
L =\Big(\frac{1}{4\lambda}\Big)^{1/2}\Big\{&\int^{\theta_2}_{\theta_1}d\theta\Big(1  -  \frac{\b^{2}}{2}  +\frac{1}{2}\b^{2}\big<f[\theta]f[\theta]\big> \\&+\frac{\b^{2}}{2\sin^{2}{\theta}}\Big\{\frac{\sin^{2}{\theta_1}}{\sin^{2}{[\theta_2 - \theta_1]}}\int^{\theta_2}_{\theta_1}\int^{\theta_2}_{\theta_1}d\hat{\theta}\,d\bar{\theta}\sin{[\theta_2 - \hat{\theta}]}\sin{[\theta_2 - \bar{\theta}]} \nonumber\\&\hspace{0.5in} - 2\frac{\sin{\theta_1}}{\sin{[\theta_2 - \theta_1]}}\int^{\theta}_{\theta_1}\int^{\theta_2}_{\theta_1}d\hat{\theta}\,d\bar{\theta}\sin{\hat{\theta}}\sin{[\theta_2-\bar{\theta}]}\nonumber\\ &\hspace{1in}+\int^{\theta}_{\theta_1}\int^{\theta}_{\theta_1}d\hat{\theta}\,d\bar{\theta}\sin{\hat{\theta}}\sin{\bar{\theta}}\sin{\hat{\theta}}\Big\}\big<\partial_{u}f[\hat{\theta}]\partial_{u}f[\bar{\theta}]\big>\nonumber.
\footnote{The one point function $\big<f\big> =0$ by the symmetry of the linearized action(\ref{linearactionint}). Since (\ref{linearactionint}) is quadratic in $f$ having a non zero value of \big<f\big> means the field is not fluctuating about its minimum.}
\end{align}
Looking at (\ref{greenfunction})  evaluated when $\varphi_x = \varphi_y$, it is evident that the correlator made out of descendants $\big<\partial_{u}f[\theta]\partial_{u}f[\theta^\prime]\big>$ can be obtained, by taking the appropriate derivatives of (\ref{greenfunction}) and then setting $\varphi_x = \varphi_y =\varphi_0$.

\begin{align}\label{descendantcorr}
<\partial_{u}f[\hat{\theta}]\partial_{u}f[\bar{\theta}]>&= \frac{1}{8\pi^2}\Big\{-(\log{2} + 1/2)+ \log{|1 - \cos{[\hat{\theta} - \bar{\theta}]}|} \\&\hspace{1in}- \frac{\cos{[\hat{\theta} - \bar{\theta}]}}{1 - \cos{[\hat{\theta} - \bar{\theta}]}}\Big\}\nonumber.
\end{align}

There are two issues in proceeding further; first both $\big<\partial_{u}f[\theta]\partial_{u}f[\theta^\prime]\big>$ and $\big<f[\theta]f[\theta^\prime]\big>$ diverge as $\theta\rightarrow\theta^\prime$, and second, there are three non-trivial integrals that involve $\big<\partial_{u}f[\theta]\partial_{u}f[\theta^\prime]\big>$. The coincident divergence problem is treated by introducing a short distance regulator $\epsilon$, into both $\big<f[\theta]f[\theta^\prime + \epsilon]\big>$ and $\big<\partial_{u}f[\theta]\partial_{u}f[\theta^\prime+\epsilon]\big>$, evaluating (\ref{mainquantpert}) with the regulator in place, and finally taking $\epsilon\rightarrow0$\footnote{It should be noted that the $\big<f[\theta]f[\theta]\big>$ term in (\ref{mainquantpert}) is manifestly divergent. This is due to the fact that both variables are evaluated at the same point. This divergence is logarithmic, as can be seen when the regulator $\epsilon$ is added, $\big<f[\theta]f[\theta+\epsilon]\big> = \frac{1}{8\pi^2}\big(-(\log{2} + 1/2)\cos{\epsilon}+ \cos{\epsilon}\log{|1 - \cos{\epsilon}|} + 1\big)$. This $\log$, is the factor that is cancelled by the integrals involving $\big<\partial_{u}f[\theta]\partial_{u}f[\theta^\prime+\epsilon]\big>$ in (\ref{mainquantpert}). All other divergent quantities in $\big<\partial_{u}f[\theta]\partial_{u}f[\theta^\prime+\epsilon]\big>$ integral cancel internally, leaving a finite result.}. The second problem is more technical, brute force calculation of (\ref{mainquantpert}) results in a proliferation of terms to be computed. The calculation is simplified dramatically if the following trick is employed; rewriting the correlator as follows
\begin{align}\label{trick}
\big<\partial_{u}f[\hat{\theta}]\partial_{u}f[\bar{\theta} +\epsilon]\big> =\int^{\pi}_{0}\int^{\pi}_{0}d\alpha\,d\beta\,\delta[\alpha - \hat{\theta}]\delta[\beta - \bar{\theta}]\big<\partial_{u}f[\alpha]\partial_{u}f[\beta +\epsilon]\big>,
\end{align}
 and placing this into (\ref{mainquantpert}) allows all the pre-factors of  $\big<\partial_{u}f[\hat{\theta}]\partial_{u}f[\bar{\theta} +\epsilon]\big>$ to be integrated. (\ref{mainquantpert}) is reduced to

\begin{align}\label{mainquantpertint}
L =\Big(\frac{1}{4\lambda}\Big)^{1/2}\Big\{&(\theta_2 - \theta_1)\Big(1  -  \frac{\b^{2}}{2} +\frac{1}{2}\b^{2}\big<f[\theta]f[\theta+\epsilon]\big>\Big) \\&+\frac{\b^{2}}{2}\Big\{\int^{\theta_2}_{\theta_1}\int^{\theta_2}_{\theta_1}d\alpha\,d\beta\frac{\sin{[\theta_2 - \beta]}\sin{[\alpha - \theta_1]}}{\sin{[\theta_2 - \theta_1]}} \nonumber\\&\hspace{0.5in}- \int^{\theta_2}_{\theta_1}\int^{\alpha}_{\theta_1}d\alpha\,d\beta\sin{[\alpha - \beta]}\Big\}\big<\partial_{u}f[\hat{\theta}]\partial_{u}f[\bar{\theta} + \epsilon]\big>\nonumber.
\end{align}

In the $\b\rightarrow0$, $L$ should reduce to the geodesic length on the unperturbed unit sphere implying $\lambda = 1/4$. These last integrals can now be evaluated with less but still considerable effort. Once (\ref{mainquantpertint})  is evaluated at finite $\epsilon$, the Log divergence from $\big<f[\theta]f[\theta+\epsilon]\big>$ cancels the remaining Log divergence from the $\big<\partial_{u}f[\hat{\theta}]\partial_{u}f[\bar{\theta} +\epsilon]\big>$ integrals. Apart from the one Log divergence that cancels the $\big<f[\theta]f[\theta+\epsilon]\big>$ divergence, all other factors of $\log{[1-\cos{\epsilon}]}$ originating from the  $\big<\partial_{u}f[\hat{\theta}]\partial_{u}f[\bar{\theta} +\epsilon]\big>$ integrals cancel amongst themselves  in the limit $\epsilon \rightarrow0$ and the result of (\ref{mainquantpert}) is finite
\begin{align}\label{result1}
L = \,&(\theta_2 - \theta_1)(1 - \b^{2}/2) +   \frac{\b^{2}}{16\pi^{2}}\Big\{-(\log{2} + 1/2)\big\{\sin{[\theta_2 - \theta_1]}\\& + (1 - \cos{[\theta_2 - \theta_1]})\tan{\Big[\frac{\theta_2 - \theta_1}{2}\Big]}\big\} -2(\theta_2 - \theta_1) \nonumber\\&\hspace{0.5in}+ 2\tan{\Big[\frac{\theta_2 - \theta_1}{2}\Big]}\log{|1 - \cos{[\theta_2 - \theta_1]}|}\Big\}\nonumber.
\end{align}
\chapter{Results And Discussion}\label{res}

\subsection{Interpretation Of The Finiteness Of \texorpdfstring{$L$}{L} To Second Order In \texorpdfstring{$\b$}{b}.}

\hspace{0.25in}The order $\b^{2}$ correction has two contributions. The contribution proportional to $\frac{\b^{2}}{16\pi^{2}}$, is the main result of the perturbative computation.  One possible surprising result is that $L$ is finite at all for non-zero separation angle. One possible intuition due to power counting is that in higher dimensions that $L$ would have behaved much like a Wilson line and have power law divergences resulting when $\hat{\theta} = \bar{\theta}$. This divergence would result from small fluctuations of the geometry that give the geodesic infinitely small wiggles or a fractal structure, causing the distance to diverge. This does not happen here because of the restriction to two dimensions; which renders these potential divergences integrable leaving only logarithmic divergences. The remaining logarithmic divergences conspire to cancel, leaving (\ref{result1}) finite.

 An argument can be made as to why the logarithmic divergences have to cancel, leaving $L$ finite. A perturbation of the metric which changes the geometry, will result in leaving the original geodesic as a path connecting the two end points but this path will not necessarily be the shortest one. Since the new geodesic for the modified geometry will be the shortest distance between the two points, it's length must be bounded by the length of the original geodesic, which was finite. Since it must be finite, it cannot be logarithmically divergent.  It is possible, that there is some perturbative symmetry or deeper reason that causes this cancellation to happen yielding a finite result, but the author is unaware of it. It would be interesting if this cancellation of divergences would continue on in higher order terms of the quantity $L$. It is possible that a proof could be constructed for the higher order case by showing that the cancellation of higher point terms reduces to sum of repeated cancellations of the type shown here. This will have to be determined in future work.

The factor $- \frac{\b^{2}}{2}(\theta_2 - \theta_1)$  results in a $\b^{2}$ correction to the radius of the sphere. It comes from the fact that before analytic continuation to the Timelike regime, $Q = 1/b + b$\footnote{This is different in from its classical value $1/b$  because of the  requirement of the conformal weights of the primary operator $e^{2b\phi}$,  $\Delta(e^{2b\phi}) = \bar{\Delta}(e^{2b\phi}) = b(Q - b)$, and the fact that $\Delta(e^{2b\phi}) = \Bar{\Delta}(e^{2b\phi}) =1$ i.e. that $e^{2b\phi}$ transforms as $ (1,1)$ tensor so that $\int d^{2}x\sqrt{\tilde{g}}e^{2b\phi}$ is conformally invariant\cite{Ginsparg:1993is}. This conformal weight can be obtained by computing the O.P.E. of the stress tensor with the operator, $T[z]e^{2b\phi} \sim \frac{\Delta(e^{2b\phi})}{(z - w)^{2}} + \ldots$.}. This factor would be there if there was no fluctuation of the Liouville field away from the saddle point, and is independent of the gauge-fixed propagator that was derived.

\subsection{Break Down Of The Perturbation Of \texorpdfstring{$L$}{L1} For Large Separation Angle.}

\hspace{0.25in}One point of note is  that (\ref{result1}) diverges when $\theta_2 - \theta_1 \rightarrow \pi$. This is a sign that the perturbation series is breaking down, not that the distance $L$ is becoming infinite. This can be explained by noting that if the end points are taken to be the north and south poles of the sphere, the geodesic connecting them is degenerate. When the angles are not antipodal on the sphere, there is a unique unperturbed geodesic connecting them, $\varphi[\theta] = \varphi_0$, which fluctuations can be computed about. As the end points become antipodal, there are many different paths that are infinitesimally close to the true geodesic. This degeneracy means that the current expansion is not an analytic function of $\b$ as $\b\rightarrow 0$ when $\theta_2 -\theta_1 = \pi$, and hence a power series expansion around $\b\rightarrow0$ is no longer valid. This is analogous to expanding $\sqrt{x}$ around $x=0$ and noting that coefficients of the power series are infinite. The situation occurs in degenerate perturbation theory, where a perturbation breaks the degeneracy.  In the limit of the perturbation vanishing, the perturbation series begins to break down as the second order and higher terms become the same magnitude as the unperturbed states. For the present situation the result (\ref{result1}) breaks down as
\begin{equation}
(\pi - (\theta_2 - \theta_1)) \sim \frac{\hat{b}}{2\sqrt{2}\pi}\nonumber.
\end{equation}

To compute the corrections of geodesics ending on antipodal points, a resummation of series is necessary.

\subsection{The Ratio Of The Correction To The Unperturbed Distance, In The Limit Of Vanishing Angle.}

\hspace{0.25in}One other interesting fact about (\ref{result1}) is that even though the function vanishes as $\theta_2 - \theta_1 \rightarrow 0$, the ratio of the $\frac{\b^{2}}{16\pi^{2}}2\tan{\Big[\frac{\theta_2 - \theta_1}{2}\Big]}\log{|1 - \cos{[\theta_2 - \theta_1]}|}$ to the unperturbed distance $\theta_2 - \theta_1$ diverges. This is because (\ref{result1}) is not analytic at $\theta_2 - \theta_1 = 0$. This implies that (\ref{result1}) should not be trusted for very small separations of the angle. At small angles, the series breaks down as,
\be\label{planckscale}
\theta_2 - \theta_1 < \sqrt{2}e^{-\frac{8\pi^{2}}{\b^{2}}}.
\ee

Here $\sqrt{2}e^{-\frac{8\pi^{2}}{\b^{2}}}$ acts as the Planck length of the system.

 \subsection{Future Work} 

\begin{itemize}
\item \hspace{0.25in}Now that a gauge invariant propagator has been computed, many other quantities can be computed using standard techniques. Computations of the expectation value of curvature invariants or other diffeomorphism invariant quantities involving the metric, can be computed in this formalism. This can be done by expanding the Liouville factor of the metric into the saddle point contribution and the fluctuation, expanding in powers of $\b$ and using standard perturbative techniques to compute the quantity with the propagator $\big<f[\hat{\theta}]f[\bar{\theta}]\big>$ and required derivatives.

\item \hspace{0.25in}When coupling a matter CFT to Timelike Liouville theory, quantities invariant under $SL_{2}(\mathbb{C})$ transformations can be constructed. For example, The two point correlator of two fields of known scaling dimension at fixed geodesic distance.

   \hspace{0.25in} Using the results of Section \ref{dist} the correlator of two conformal matter fields of known scaling dimension at fixed geodesic distance $L$ can be computed.
\be\label{fixeddist}
<\mathcal{O}\,\mathcal{O}>  \,= \int\,d^{2}x\,d^{2}y\,\delta^{(2)}\Big[L -  \int^{y}_{x}\sqrt{g_{\mu\nu}\frac{\partial x^{\mu}}{\partial\sigma}\frac{\partial x^{\nu}}{\partial\sigma}}\,d\sigma\Big]<\mathcal{O}(x)\mathcal{O}(y)>
\ee

Here $<X(x)X(y)>$ is the correlator on the fixed reference sphere.

Na\"{i}vely there is no obstruction to extending this to $n$-point correlation functions including a delta function constraint for each pair of points, fixing there separation to a fixed physical distance.

\item \hspace{0.25in}Another quantity of note is the two point correlator of two fields of fixed scaling dimension under the influence of a probe propagator.  A scalar field under the influence of Liouville has an action of the form
\begin{align}\label{probeaction}
-\mathcal{S}_{\lambda} &= -\int d^{2}x\,\sqrt{g}\,[g^{ab}\nabla_{a}\lambda\nabla_{b}\lambda - m^{2}\lambda^{2}]\nonumber\\ &= -\int
d^{2}x\,\sqrt{\hat{g}}e^{2\b\hat{\phi}}\,[e^{-2b\hat{\phi}}\hat{g}^{ab}\nabla_{a}\lambda\nabla_{b}\lambda - m^{2}\lambda^{2}].
\end{align}

Extracting the Liouville dependence and integrating by parts, this action can be rewritten as
\begin{align}\label{probeactionuse}
-\mathcal{S}_{\lambda} &= -\int d^{2}x\,\sqrt{\hat{g}}\,[\hat{g}^{ab}\partial_{a}\lambda\partial_{b}\lambda - m^{2}e^{2\b(\hat{\phi}_{0} + f)}\lambda^{2}]\nonumber\\
&=-\int d^{2}x\,\sqrt{\hat{g}}\,[\hat{g}^{ab}\partial_{a}\lambda\partial_{b}\lambda - m^{2}e^{2\b\hat{\phi}_{0}}(1 + f + \frac{1}{2}f^{2})\lambda^{2}].
\end{align}

From (\ref{probeactionuse}) the  probe propagator and the probe Feynman rules can be obtained,

\begin{align}
<\lambda\lambda>&=\text{
\begin{fmffile}{probline}
\begin{fmfgraph*}(30,3)
\fmfleft{in}
\fmfright{out}
\fmflabel{$x$}{in}
\fmflabel{$y$}{out}
\fmf{fermion}{in,out}
\fmfdot{in,out}
\end{fmfgraph*}
\end{fmffile}}\hspace{6pt} = (\nabla^{2} + m^{2}e^{2\b\hat{\phi}_{0}})^{-1}\nonumber\\
& =\alpha_2\,\P[{\frac{1}{2}\sqrt{1+4m^{2}e^{2\b\hat{\phi}_{0}}} -1},\chi_{xy}]\nonumber\\&\hspace{0.5in}+ \alpha_3\,\Q[{\frac{1}{2}\sqrt{1+4m^{2}e^{2\b\hat{\phi}_{0}}} -1},\chi_{xy}].
\footnote{Here $\alpha_2$ and $\alpha_3$ are chosen to satisfy the boundary conditions of the field.}
\end{align}
\begin{align}&
\text{\raisebox{-0.25in}{\begin{fmffile}{propliouvert}
\begin{fmfgraph*}(20,15)
\fmfleft{i1,i2}
\fmfright{o1}
\fmf{fermion}{i1,v1}
\fmf{fermion}{i2,v1}
\fmf{dashes}{v1,o1}
\fmfdot{v1}
\fmflabel{$z$}{v1}
\end{fmfgraph*}
\end{fmffile}}} = \int\,d^{2}z\sqrt{\hat{g_z}}m^{2}e^{\hat{\phi}_0}\\
&\text{\raisebox{-0.25in}{\begin{fmffile}{propliouvert2}
\begin{fmfgraph*}(20,15)
\fmfleft{i1,i2}
\fmfright{o1,o2}
\fmf{fermion}{i1,v1}
\fmf{fermion}{i2,v1}
\fmf{dashes}{v1,o1}
\fmf{dashes}{v1,o2}
\fmfdot{v1}
\fmflabel{$z$}{v1}
\end{fmfgraph*}
\end{fmffile}}} = \frac{1}{2}\int\,d^{2}z\sqrt{\hat{g_z}}m^{2}e^{\hat{\phi}_0}.
\end{align}

One quantity that can be computed is, 
  \begin{align}\label{twopointwithprop}
  &\int d^{2}x\, d^2{}y\,\sqrt{g_x}\,\sqrt{g_y}\,<X(x)(\nabla^2 + m^2)^{-1}X(y)> \\
  &=\,\int d^{2}x\, d^{2}y\,\sqrt{\hat{g}_x}\,\sqrt{\hat{g}_y}\mathbb{Z}^{-1}\int\D\,f\, e^{f(x)}e^{f(y)}X(x)(\hat{\nabla}^2 + m^{2}e^{\phi_0}e^{f})^{-1}X(y)e^{-S_l}.
\end{align}

This is the analog of the first correction in the expansion of a Wilson line coming from a scalar mediating boson, on a fluctuating sphere. Here the coordinates, $x$, $y$, are on the reference sphere. In (\ref{twopointwithprop}) the Liouville field enters from two regimes. First from the integration measures of the coordinates
on the sphere, and second from the covariant derivative in the propagator. Taking the perturbative expansion of the Liouville field to quadratic order, (\ref{twopointwithprop}) can be
written more explicitly as
  \begin{align}\label{twopointcorr}
\int d^{2}x\, d^{2}y\,\sqrt{\hat{g}_x}\,\sqrt{\hat{g}_y}\mathbb{Z}^{-1}\int\D\,f\, [1 + &f(x) + f(y) + \frac{1}{2}f(x)^{2} + \frac{1}{2}f(y)^{2}
+f(x)f(y)]\times\nonumber\\&\times\chi(x)(\hat{\nabla}^2 + m^{2}(1 + f + \frac{1}{2}f^2))^{-1}\chi(y)e^{-S_l}.
  \end{align}
  
  where $S_l$ is the the linearized gauge fixed action Liouville action. The remaining Feynman rules obtained from (\ref{twopointcorr}) are,
\begin{align}
&\text{\raisebox{-0.25in}{\begin{fmffile}{3matterprobliouvert}
\begin{fmfgraph*}(20,15)
\fmfleft{i1,i2}
\fmfright{o1}
\fmf{fermion}{i1,v1}
\fmf{wiggly}{i2,v1}
\fmf{dashes}{v1,o1}
\fmfdot{v1}
\fmflabel{$z$}{v1}
\end{fmfgraph*}
\end{fmffile}}} = \int\,d^{z}\sqrt{\hat{g}_z}\hspace{20pt}
\text{\raisebox{-0.25in}{\begin{fmffile}{4matterprobliouvert}
\begin{fmfgraph*}(20,15)
\fmfleft{i1,i2}
\fmfright{o1,o2}
\fmf{fermion}{i1,v1}
\fmf{wiggly}{i2,v1}
\fmf{dashes}{v1,o1}
\fmf{dashes}{v1,o2}
\fmfdot{v1}
\fmflabel{$z$}{v1}
\end{fmfgraph*}
\end{fmffile}}} = \frac{1}{2}\int\,d^{z}\sqrt{\hat{g}_z}\\&
\hspace{1in}\text{\begin{fmffile}{matterprop}
\begin{fmfgraph*}(25,3)
\fmfleft{in}
\fmfright{out}
\fmflabel{$x$}{in}
\fmflabel{$y$}{out}
\fmf{wiggly}{in,out}
\fmfdot{in,out}
\end{fmfgraph*}
\end{fmffile}} \hspace{10pt}= \tilde{C}_{xy}(1 - \chi_{xy})^{-2\Delta}.
\end{align}

It follows that (\ref{twopointcorr}) corresponds to the Feynman diagrams in Figure \{\ref{one}\}.
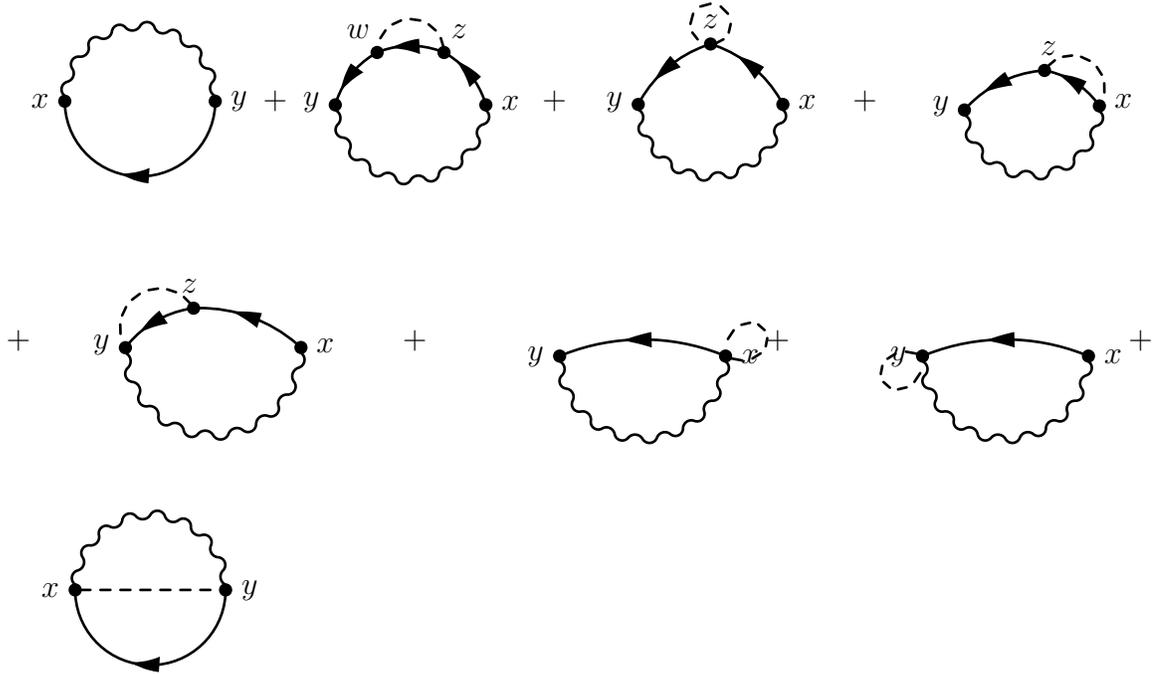
\begin{figure}[ht]\label{one}
\raisebox{-0.55in}{\begin{fmffile}{bubble1}
\begin{fmfgraph*}(30,30)
\fmfleft{x}
\fmfright{y}
\fmflabel{$x$}{v1}
\fmflabel{$y$}{v2}
\fmf{phantom,tension=4}{x,v1}
\fmf{phantom,tension=4}{v2,y}
\fmf{wiggly,left,tension=0.4}{v1,v2}
\fmf{fermion,left,tension=-0.4}{v2,v1}
\fmf{phantom}{v1,v2}
\fmfdot{v1,v2}
\end{fmfgraph*}
\end{fmffile}}
$+$ \hspace{-0.2in}
\raisebox{-0.63in}{
\begin{fmffile}{bubble2}
\begin{fmfgraph*}(30,30)
\fmfleft{x}
\fmfright{y}
\fmftop{w,z}
\fmflabel{$x$}{v1}
\fmflabel{$y$}{v2}
\fmflabel{$z$}{w1}
\fmflabel{$w$}{w2}
\fmf{phantom,tension=2}{z,w1}
\fmf{phantom,tension=2}{w2,w}
\fmf{phantom,tension=4.5}{x,v1}
\fmf{phantom,tension=4.5}{v2,y}
\fmf{wiggly,left,tension=-5.9}{v1,v2}
\fmf{fermion,right=0.2}{v1,w1}
\fmf{fermion,right=0.2}{w1,w2}
\fmf{fermion,right=0.2}{w2,v2}
\fmf{dashes,right,tension=2}{w1,w2}
\fmfdot{v1,v2,w1,w2}
\end{fmfgraph*}
\end{fmffile}}
$+$\raisebox{-0.63in}{
\begin{fmffile}{bubble3}
\begin{fmfgraph*}(30,30)
\fmfleft{x}
\fmfright{y}
\fmftop{z}
\fmftop{w}
\fmflabel{$x$}{v1}
\fmflabel{$y$}{v2}
\fmflabel{$z$}{w1}
\fmf{phantom,tension=2}{z,w1}
\fmf{phantom,tension=5}{x,v1}
\fmf{phantom,tension=5}{v2,y}
\fmf{wiggly,left,tension=-6.9}{v1,v2}
\fmf{fermion,right=0.2}{v1,w1}
\fmf{fermion,right=0.2}{w1,v2}
\fmf{dashes,right,tension=0.5}{w1,w}
\fmf{dashes,right,tension=0.5}{w,w1}
\fmfdot{v1,v2,w1}
\end{fmfgraph*}
\end{fmffile}}\hspace{0.1in}$+$
\raisebox{-0.69in}{
\begin{fmffile}{bubble4}
\begin{fmfgraph*}(30,30)
\fmfleft{x}
\fmfright{y}
\fmftop{z}
\fmfbottom{w}
\fmflabel{$x$}{v1}
\fmflabel{$y$}{v2}
\fmflabel{$z$}{w1}
\fmf{phantom,tension=2}{z,w1}
\fmf{phantom,tension=4.5}{x,v1}
\fmf{phantom,tension=5}{v2,y}
\fmf{phantom,tension=-0.6}{v2,w}
\fmf{wiggly,left,tension=-6.9}{v1,v2}
\fmf{fermion,right=0.2}{v1,w1}
\fmf{fermion,right=0.2}{w1,v2}
\fmf{dashes,left=1}{w1,v1}
\fmfdot{v1,v2,w1}
\end{fmfgraph*}
\end{fmffile}}

$+$\hspace{-0.2in}
\raisebox{-0.66in}{
\begin{fmffile}{bubble5}
\begin{fmfgraph*}(30,30)
\fmfleft{x}
\fmfright{y}
\fmftop{z}
\fmfbottom{w}
\fmflabel{$x$}{v1}
\fmflabel{$y$}{v2}
\fmflabel{$z$}{w1}
\fmf{phantom,tension=2}{z,w1}
\fmf{phantom,tension=5.5}{x,v1}
\fmf{phantom,tension=5.5}{v2,y}
\fmf{phantom,tension=-0.3}{v1,w}
\fmf{wiggly,left,tension=-6.9}{v1,v2}
\fmf{fermion,right=0.2}{v1,w1}
\fmf{fermion,right=0.2}{w1,v2}
\fmf{dashes,right=1}{w1,v2}
\fmfdot{v1,v2,w1}
\end{fmfgraph*}
\end{fmffile}}
$+$
\raisebox{-0.63in}{
\begin{fmffile}{bubble6}
\begin{fmfgraph*}(30,30)
\fmfleft{x}
\fmfright{y}
\fmflabel{$x$}{v1}
\fmflabel{$y$}{v2}
\fmf{phantom,tension=5}{x,v1}
\fmf{phantom,tension=5}{v2,y}
\fmf{wiggly,left,tension=-6.9}{v1,v2}
\fmf{fermion,right=0.2}{v1,v2}
\fmf{dashes,tension=2.5}{v1,v1}
\fmfdot{v1,v2}
\end{fmfgraph*}
\end{fmffile}}$+$
\raisebox{-0.63in}{
\begin{fmffile}{bubble7}
\begin{fmfgraph*}(30,30)
\fmfleft{x}
\fmfright{y}
\fmflabel{$x$}{v1}
\fmflabel{$y$}{v2}
\fmf{phantom,tension=5}{x,v1}
\fmf{phantom,tension=5}{v2,y}
\fmf{wiggly,left,tension=-6.9}{v1,v2}
\fmf{fermion,right=0.2}{v1,v2}
\fmf{dashes,top,tension=2.5}{v2,v2}
\fmfdot{v1,v2}
\end{fmfgraph*}
\end{fmffile}}$+$
\raisebox{-0.535in}{
\begin{fmffile}{bubble8}
\begin{fmfgraph*}(30,30)
\fmfleft{x}
\fmfright{y}
\fmflabel{$x$}{v1}
\fmflabel{$y$}{v2}
\fmf{phantom,tension=8}{x,v1}
\fmf{phantom,tension=8}{v2,y}
\fmf{dashes}{v1,v2}
\fmf{wiggly,left,tension=2}{v1,v2}
\fmf{fermion,left,tension=-2}{v2,v1}
\fmf{phantom}{v1,v2}
\fmfdot{v1,v2}
\end{fmfgraph*}
\end{fmffile}}
\caption{Feynman diagrams corresponding to (\ref{twopointcorr}). The third, sixth, and seventh diagrams correspond to renormalization of the Liouville coupling to the conformal field and
the probe mass respectively.}
\end{figure}
\item \hspace{0.25in}Lastly, there is the question of if there is a Gribov Ambiguity with constraint used to address the gauge redundancy. The current work has fixed the gauge of the $SL_{2}(\mathbb{C})$  transformations locally, but there is no guarantee that it has been fixed globally, so it is still possible that there is a non-perturbative failure of B.R.S.T. symmetry after following this procedure, i.e. the gauge fixing of the dipole may not be the unique way to fix the gauge\cite{1978NuPhB.139....1G}. A possible avenue forward could be to construct a proof showing that the fixing of the dipole is a unique gauge condition or determine whether a Gribov Ambiguity occurs.

\part{}\label{chern-simons paper}






\chapter{Introduction}

There has been increasing interest in recent years in ``vectorlike" examples \cite{Klebanov:2002ja, Sezgin:2002rt, Petkou:2003zz, Sezgin:2003pt,Girardello:2002pp, Giombi:2009wh, Giombi:2010vg,Das:2003vw,Koch:2010cy, Douglas:2010rc,Giombi:2011ya,Shenker:2011zf,Giombi:2011kc,Aharony:2011jz,Maldacena:2012sf,Maldacena:2011jn} of holography that involve dynamical fields that transform in the fundamental (rather than the adjoint) representation of a symmetry group such as $SU(N)$ at large $N$. In 3+1 bulk dimensions the bulk dynamics is described by Vasiliev higher spin gravity\cite{Vasiliev:1992av,Vasiliev:1995dn,Vasiliev:1999ba,Vasiliev:2003ev,Sezgin:2002ru,Sezgin:2003pt} .  In these systems the bulk higher spin fields correspond to singlets under the symmetry group.\footnote{See \cite{Anninos:2011ui} for a very interesting proposal for a higher spin dS/CFT duality.}   A consistent implementation of AdS/CFT requires a boundary theory that is local, and so has a stress tensor.  In a local theory we can consistently truncate a global symmetry to the singlet sector only by a local procedure, such as
promoting the symmetry to a gauge symmetry and implementing the singlet constraint by local gauge interactions.  On the other hand we do not want dynamical gauge fields that have nontrivial local gauge invariant operators that are dual to extra  ``stringy" states in the bulk beyond those conjectured by the duality.   In $2+1$ boundary dimensions there is a natural candidate discussed in the literature \cite{Giombi:2011kc} for the gauge system that does not have nontrivial local gauge dynamics, the Chern-Simons theory.   We will study this proposal in this part, focusing on the case with gauge group $SU(N)$ and matter fields that are in the fundamental $N$ dimensional vector representation.  There is an important parallel development in terms of  $1+1$ dimensional boundary systems, involving a $W_{N}$ boundary CFT that we will mention in the discussion\cite{Gaberdiel:2010pz,Gaberdiel:2011wb,Gaberdiel:2011nt,Ahn:2011pv,Chang:2011mz,Papadodimas:2011pf}.

But Chern-Simons theories have nontrivial dynamics and extra states on topologically nontrivial manifolds, and we shall see that the nontrivial Chern-Simons dynamics remain active when
coupled to matter as well. Since gauge/gravity dualities must make sense on any boundary manifold, these new states must be part of the full dual gravitational dynamics.

In this part we take a first step towards understanding this situation by analyzing the Chern-Simons theory coupled to fundamental scalars and fermions on higher genus spatial surfaces, especially the torus $T^2$.   

We begin with the warm-up example of a massive scalar field. This is not a conformal field theory (CFT) and is not dual to the bulk higher spin theory. On the other hand the mass acts as a control parameter which makes the analysis less complicated when the mass squared is of the same order of or is large compared to the inverse size of the torus. In that case we can integrate out the scalar field and the low energy theory is pure Chern-Simons theory perturbed by the operators obtained by integrating out the scalar field. We can easily state the result of our analysis. In the weak-coupling or large $k$ limit the splitting of the exact zero energy states of the pure Chern-Simons theory on torus is of order $\frac{1}{k} = \frac{\lambda}{N}$, where we have defined the 't Hooft coupling $\lambda$ as, $\lambda=\frac{N}{k}$.  

The other case we study is that of a free massless scalar field\footnote{To be precise we study the interacting fixed point theory with $\phi^6$ coupling of order $\lambda^2$ discovered in \cite{Aharony:2011jz}.}. This theory is dual to a bulk Vasiliev theory on locally $AdS_{4}$ space whose asymptotic boundary has the structure of $T^{2}\times R^{1}$.\footnote{This can be constructed by periodic identification of space-like field theory coordinates in the Poincare patch of $AdS_{4}$.} This is more complicated than the massive scalar because of the presence of the scalar zero mode. This is an approximate zero mode but it can have arbitrarily small energy and so it does not decouple from the low energy dynamics.   We study the low energy spectrum by reducing this system to an effective quantum mechanics.  We analyze this quantum mechanics and find for the $U(1)$ system the gap is $\sim \frac{1}{\sqrt{k}}$.  For the $SU(N)$ system the gap  is of order $\frac{\sqrt{\lambda}}{N}$ where the `t Hooft coupling $\lambda = N/k$. and vanishes in the large-$N$ limit.  So the bulk higher spin theory must have extremely low energy states in the classical, and small $\lambda$ limits.
These light states do not correspond to any apparent excitations of the Vasiliev fields.
They are closely analogous to the light states found in the $W\llo N$ theory
\cite{Gaberdiel:2010pz, Gaberdiel:2011zw, Gaberdiel:2011aa}.

 The critical, interacting,  $SU(N)$ scalar  theory, dual to Vasiliev gravity with a different bulk scalar boundary condition,  has parametric behavior similar to the free massive  scalar because it is gapped on the torus.  The free fermion system behaves in the same fashion.
The entropy corresponding to these light states is $S \sim N \log(k)$.

We also consider, briefly, the case where the spatial slices are Riemann surfaces of
higher genus and,  based on  results in pure Chern-Simons theory,
we find an even larger entropy, $ S \sim N^2 \log(k)$ parametrically in $N$, than
in the case of genus 1.  While we focus on the torus case for concreteness, 
it seems that the higher-genus case portends an even more radical breakdown of
the bulk description in terms of pure Vasiliev gravity.

\chapter{Perturbative Chern-Simons Matter Theory On Torus}
Pure Chern-Simons theory on a general three manifold is an exactly solvable field theory for any $k$ \cite{Witten:1988hf}.  On a torus the space of states is determined by the conformal blocks of WZW conformal field theory \cite{Witten:1988hf}. However it is often useful to understand the theory in a semiclassical weak coupling expansion at large $k$.   The classical stationary points are flat connections.  On certain manifolds the flat connections are isolated and the semiclassical expansion is in principle straightforward.  On other manifolds, including tori, the flat connections form a moduli space which must be integrated over \footnote{ References \cite{Axelrod:1991vq} and \cite{Axelrod:1993wr} study pure Chern-Simons perturbation theory. They develop the perturbation theory based on the assumptions that the flat connection is isolated and the subgroup of the gauge group which leaves the flat connection invariant is discrete. These assumptions are violated if the spatial slice is a torus or any higher genus Riemann surface. A generalization of their method should be applicable to these cases and in general to any three manifold where there is a moduli space of flat connections. } \cite{Axelrod:1991vq,Axelrod:1993wr}.   An important approach to this problem is the canonical quantization method described in the classic part \cite{Elitzur:1989nr} In In this approach the problem reduces to the quantization of the moduli space of flat connections on a spatial torus, which form a finite volume phase space.   The Hilbert space is finite dimensional and every state has exactly zero energy because the Hamiltonian of the Chern-Simons theory vanishes.   We shall follow the canonical approach to study the more complicated problem of Chern-Simons theory coupled to a scalar matter field $\phi_a$  in the fundamental representation of $SU(N)$.

 Chern-Simons gauge theory coupled to a matter field is not  a topological field theory and it has a non-vanishing Hamiltonian. The presence of the scalar field lifts the degeneracy of the flat connections.  This was first studied in  the pioneering work of Niu and Wen \cite{Wen:1990zza} which has important parallels to our work.   To see how the flat directions behave, let us write down the action of the theory. The action is given by


\begin{equation}
S = \frac{k}{4\pi} \int Tr(A\wedge dA + \frac{2}{3} A^{3}) + \int d^{3}x  \sqrt {-g} \ [ g^{{\mu}{\nu}}  (D_{\mu}{\phi})^{\dagger}(D_{\nu}{\phi}) - V(\phi^{\dagger}\phi) ]
\end{equation}
where $g_{\mu\nu}$ is the metric on the space-time, $V(\phi^{\dagger}\phi)$ is the potential, $\phi^{\dagger}\phi = \phi_a \phi_a$, and $D_{\mu}\phi = \partial_{\mu} \phi + iA_{\mu}\phi$ is the gauge covariant derivative acting on the scalar field. We take the space-time to be $T^{2}\times R^{1}$, where $R^{1}$ is the time direction. Since the Chern-Simons term is topological it does not contribute to the stress tensor of the theory. So the energy density is given by 
\begin{equation}
T^{00} \sim |D_{0}\phi|^{2} + |D_{i}\phi|^{2} + V(\phi^{\dagger}\phi)
\end{equation}
Before we proceed farther it is useful to choose a gauge. For our purpose, $A_{0} = 0$ is the convenient gauge choice. In this gauge the energy density becomes,
\begin{equation}\label{xxx}
T^{00} \sim |\dot\phi|^{2} + |D_{i}\phi|^{2} + V(\phi^{\dagger}\phi)
\end{equation}
The allowed field configurations also have to satisfy the Gauss's law constraint,
\begin{equation}
\frac{ k}{8\pi}\frac {\epsilon^{ij}}{\sqrt h} \ F^{a}_{ij} \ = \ i(\phi^{\dagger}T^{a}\dot\phi - \dot\phi^{\dagger}T^{a}\phi)
\end{equation}
where $h_{ij}$ is the metric tensor on the spatial torus and $\epsilon^{ij}$ is the completely antisymmetric symbol with $\epsilon^{12}= - \epsilon^{21} = 1$. In the following discussion we shall specialize to the case where $V(\phi^{\dagger}\phi) = M^{2}\phi^{\dagger}\phi$.  

It follows from the expression of the energy density (\ref{xxx}) that classically the lowest energy field configurations are those for which, $\dot\phi = 0$, $D_{i}\phi = 0$ and $V(\phi^{\dagger}\phi) = 0$. The simultaneous solutions of these equations also have to satisfy the Gauss's law constraint. Since $\dot\phi = 0$, it follows from the Gauss's law constraint that the spatial components of the gauge field are components of a flat connection on the torus.  Now one can solve the remaining two equations subject to the constraint that the gauge field appearing in the covariant derivative is flat. If $M\neq 0$ then the only solution is $\phi = 0$ and so the classical lowest (zero) energy field configurations are flat connections on the torus. When $M=0$, one can show that the solution is $\phi = 0$, for almost every flat connection except for those whose holonomies lie in a $SU(N-1)$ subgroup of the gauge group $SU(N)$. Although the constant mode of the massless scalar field is not an exact zero mode when coupled to gauge fields, it can have arbitrarily small energy depending on the choice of the flat connection and so it does not decouple from the low energy dynamics. This will play an important role in the following discussion. 

In the quantum theory the flat connection  degeneracy is lifted by the scalar field. To show this, we can choose a particular flat connection and expand around that. The gauge field can be decomposed as,
\begin{equation}
A = A^{f} + \frac{1}{\sqrt k} a
\end{equation}
where $A^{f}$ is a flat connection and $a$ is gauge field fluctuation. Substituting this in the action we get a term of the form,
\begin{equation}
S\supset \int d^{3}x \sqrt {-g}  [(D_{\mu}^{f} \phi)^{\dagger} D^{f\mu}\phi + M^{2}\phi^{\dagger}\phi ] + O(\frac{1}{\sqrt k})
\end{equation}
where $D_{\mu}^{f} = \partial_{\mu} + iA_{\mu}^{f}$. In the weak coupling or large $k$ limit this is the leading piece of the action containing the scalar field. This is the action of a massive scalar field moving in the background of a flat connection.  For finite $M$ we can integrate out the scalars to get an effective potential for the flat connections\footnote{Terms with higher derivatives, like Yang Mills terms, are also induced.  Their effects  are suppressed at large $k$ .  For the abelian case this can be seen in the results of \cite{Wen:1990zza,  Gukov:2004id}}. The answer is given by\footnote{Please see the appendix for a detailed derivation of the formula and explanation of various terms.},
\begin{equation}\label{vaa}
V(A_{1}^{f},A_{2}^{f}) = - \frac{1}{\sqrt 2} \frac{area(T^{2})}{\pi^{\frac{3}{2}}}\sum_{(n,m)\neq(0,0)} \frac{M^{\frac{3}{2}}K_{\frac{3}{2}}(M|m\vec a + n\vec b|)}{|m\vec a + n\vec b|^{\frac{3}{2}}} Tr(W(a)^{m}W(b)^{n})
\end{equation}
which acts as an effective potential for the flat connections. In the above formula $W(a)$ and $W(b)$ are the holonomies of the flat connection along the $a$-cycle and $b$-cycle of the torus, respectively. It is easy to see that the effective potential has minima at $(A_{1},A_{2}) = (0,0)$ and its gauge copies. So quantum mechanically only the trivial flat connection is stable and one can quantize only around the trivial flat connection.\footnote{A very similar argument was used in \cite{Luscher:1982ma} for the case of pure Yang-Mills gauge theory on the three torus $T^{3}$.} Now what is the effect of the mass of the scalar field? It is expected that if the scalar field is heavy, i.e, $M^{2}\times area(T^{2}) \gg 1$, then the low energy theory should reduce to the pure Chern-Simons theory. In particular the effective potential on the moduli space should vanish as the mass tends to infinity. It is easy to see by studying the asymptotics of the modified Bessel function for large argument that this is indeed the case with the potential. 

So we have the following picture. The scalar field creates an effective potential on the moduli space of flat connections which push the connection towards the trivial one.\footnote{We would like to mention that this is not true in a supersymmetric theory. In a supersymmetric theory the effective potential obtained by integrating out the non-zero modes will vanish due to Bose-Fermi cancellation. So we can no longer say that the low energy wave functions are localized around the trivial flat connection and its gauge copies. The dimensionally reduced quantum mechanical model does not capture the complete low energy spectrum in the supersymmetric theory.} The effect of the scalar field or the value of the effective potential depends on its mass. In the regime $M^{2}\times area(T^{2}) \gg 1$, the potential decreases exponentially like $e^{-M\sqrt {area(T^{2})}}$ and in the limit of an infinitely massive scalar field we recover the pure Chern-Simons gauge theory. For a finite mass scalar, or in general, a matter theory with gap,  we can analyze the Chern Simons theory with (\ref{vaa}) as perturbation.  We will discuss this in the next section.

  In the case of a massless scalar the scalar zero mode cannot be integrated out,  but the low energy dynamics $(E \times \sqrt{area(T^{2})}\ll 1)$ can be determined by retaining only the constant modes and studying an effective quantum mechanics.  Because the gauge fields are confined to a small neighborhood of the zero gauge field we can ignore the compactness of the flat connection moduli space in formulating the quantum mechanics.   So to compute the low energy states of the theory we can diagonalize the Hamiltonian obtained from the dimensionally reduced theory.   The study of this quantum mechanics in both the $U(1)$ and $SU(N)$  cases will occupy most of what follows.  We will turn to it after discussing the gapped theories.

Finally we briefly discuss the behavior of the system when bosons are replaced with fermions 
in the the fundamental representation.  The analysis above for the scalars applies, except
that the sign of the potential in (\ref{vaa}) is reversed.  As a result, the
minimum of the potential occurs when the gauge field holonomies
on both cycles are diagonal and equal to $-1$.  Equivalently, and more simply,
we can treat the fermions as having Scherk-Schwarz boundary conditions
along both cycles, and the gauge field holonomies as being trivial.
In either description, the free fermions expanded around their true vacuum have an energy gap
of order ${\frac{1}{{\sqrt{{area(T^2)}}}}}$ and can be integrated out, even in the absence
of a mass; the analysis of the effective theory then follows exactly that of the 
massive scalar, with a mass of order ${\frac{1}{{\sqrt{{area(T^2)}}}}}$.

\chapter{Scalar Field With Mass}
In this section we shall study the case of a scalar field with mass, $ M$ in more detail.\footnote{In this section we shall take a square torus with sides of length $R$ we will often set to 1.} A massive scalar field with mass in the region $M\sim 1/R$ or $M\gg 1/R$ is in some sense simpler because we can integrate out the scalar field, if we are interested in states  with energies $\ll 1/R$. In that case we will be left only with the pure Chern-Simons gauge theory with corrections which come from integrating out the massive scalar field.   For simplicity let us examine the $U(1)$ case first.  At large $M$ (\ref{vaa}) becomes 
\begin{equation}
V(A_1, A_2) \sim - \exp(-MR)(\cos(A_1)+\cos(A_2))
\end{equation}
where $A_1$ and $A_2$ are the eigenvalues of the holonomies around the $1$ and $2$ cycles.   The Lagrangian of the  system becomes 
\begin{equation}
L = R^{2}\Big[\frac{ k}{8\pi}\,A_{1}\frac{dA_{2}}{dt} + \exp(-MR)(\cos(A_1)+\cos(A_2))\Big]
\end{equation}
We will argue that at low energies the gauge fields are close to zero, so we can expand the cosines.  The Lagrangian becomes
\begin{equation}
L = R^{2}\Big[\frac{ k}{8\pi}\,A_{1}\frac{dA_{2}}{dt} - \hh~\exp(-MR)(A_1^2 +A_2^2)\Big]
\end{equation}
$A_1$ and $A_2$ are canonically conjugate variables.  If we define
 $P = \frac{ k}{8\pi}A_{1}$ and $Q=A_{2}$, the canonical commutation relations are given by,
\begin{equation}
[Q,P] = i , \  [\phi,\pi] = [\phi^{*},\pi^{*}] = i
\end{equation}

If we measure all energies in units of $\frac{1}{R}$ then we can set $R=1$. The Hamiltonian is given by,
\begin{equation}
H =\hh ~ \exp(-M)[(\frac{8\pi}{k})^2P^{2} + Q^{2} ]
\label{qmmass}
\end{equation}
So the low lying states are described by a harmonic oscillator with $\hbar \sim 1/k$.  The low lying states have energies\footnote{Niu and Wen \cite{Wen:1990zza} emphasized that the $\exp(-M)$ dependence showed the stability of topological order. }
\begin{equation}
E_l = \exp(-M)(\frac{8 \pi}{ k})(l+\frac{1}{2})
\end{equation}
The spread of the holonomy angle in these states $\langle A_1^2 \rangle = \langle A_2^2 \rangle \sim 1/k$ so the gauge fields are localized around the origin at large $k$.  This effect is independent of $M$, for large enough $k$.  This localization justifies expanding the cosines.

The $SU(N)$ case is similar.  We can write down the effective potential on the moduli space of flat connections created by the massive scalar field at large $M$ as\footnote{Please see the appendix for the exact expression valid for all values of the mass parameter.}, 
\begin{equation}
V \sim \exp(-M)(\sum_i\cos(\alpha_i)+\cos(\beta_i))
\end{equation}
where $\alpha_i, \beta_i$ are the eigenvalues of $(A_{1}^{f},A_{2}^{f})$.  Each pair $\alpha_i, \beta_i$ are canonically conjugate and so expanding the cosines yields a set of $N-1$ harmonic oscillators with $\hbar \sim 1/k$.  There is a residual discrete part of the Gauss's Law constraint, the Weyl group.  Here this is just the permutation group $S_{N-1}$.   A careful analysis of the measure in the pure Chern-Simons system \cite{Elitzur:1989nr,Douglas:1994ex} shows that the eigenvalues should be treated as fermions\footnote{ We would like to thank Juan Maldacena  for pointing this out to us, correcting an error in the original version of this paper.  (This does not alter our substantive conclusions.)  The fermi surface plays an important role in \cite{Aharony:2012ns} }.   So we have $N-1$ fermionic particles in a harmonic oscillator potential.

 This system is easy to study.  The low lying gaps are 
\begin{equation}
\Delta \sim \frac {\expp{-M}} {k}
 = \frac {\expp{-M} \lambda} {N}
\end{equation}
Here we have used the `t Hooft coupling appropriate for Chern Simons systems, $\lambda = N/k$.     We note that the Fermi energy of this system is given by, $E_F = e^{-M}\lambda$. 
The entropy for temperature $\expp{-M} \gg T \gg E_F$ and $k$ large  can easily be computed semiclassically.  The answer is
\begin{equation}
\label{OurEntropyA}
S  \sim \log\left [ (T~ k~ \expp{+M})^{N-1}/(N-1)! \right ]
\sim (N-1)\log \left [ T~ \expp{+M} /\lambda \right ]
\end{equation}
Note that at fixed $N$ and large $k$ the entropy goes like $N \log(k)$, parametrically
the same as the pure Chern-Simons entropy.   Here we have performed the calculation
using the harmonic oscillator representation of the system.  This approximation
is valid when the range of holonomies explored is much less than $2\pi$, which  
corresponds to temperatures much less than $\exp{-M}$.  For temperatures
\begin{equation}
\expp{-M} \ll    T\ll 1\ ,
\end{equation}
all Chern-Simons states participate in the thermal ensemble.  

We can also consider a different limit where the temperature is much less than the Fermi energy of the system. In this temperature regime the entropy of the system is given by,
\begin{equation}
S \sim  e^{M}\frac{NT}{\lambda}
\end{equation} 
This is the leading term in the Sommerfeld expansion.

In accordance with expectations from effective field theory, we have seen that there is intermediate range of temperatures in which the matter and the Kaluza-Klein modes have
decoupled, and the system is well-approximated by pure Chern-Simons theory.  

The above analysis strictly applies when $M \gg 1$.  But when $M$ decreases to of order one the only thing that changes is the detailed shape of the potential.  It still is quadratic around $A=0$.  So we still have a harmonic oscillator description of the low lying states at large $k$. This applies to a general matter sector.  As long as the matter sector on the torus has a gap, the effective potential for the flat connections will be quadratic around the origin.     Interacting critical points, like the Wilson-Fisher fixed points (often called in this context the critical SU(N) model)  have a gap on the torus, so this analysis applies to these systems.

We expect this analysis to breakdown when the gap of the matter system is $\sim 1/k$.  The massless scalar is an example.  To treat this system we will have to retain the light degrees of freedom in our effective description.  We now turn to this task.

\chapter{Field Content Of The Low Energy Theory With A Massless Scalar}
\chaptermark{Low Energy Theory With A Massless Scalar}
The energy density (\ref{xxx}) depends on the derivatives of the scalar field. So a field configuration with a spatially varying scalar field will have at least an energy of order $\frac{1}{\sqrt {area(T^{2})}}$. So at an energy scale much below this we can integrate out these modes. The scalar field also has an approximate zero mode. The constant mode of the scalar field is an exact zero mode when the gauge field configuration is such that, $D_{i}\phi = 0$. It is easy to check that this has a non-zero solution for $\phi$ only when the gauge field is vanishing or flat with the holonomy lying in a $SU(N-1)$ subgroup of the $SU(N)$.  Away form this region of the moduli space of flat connections the constant mode of the scalar field is not an exact zero mode but it can have arbitrarily small energy and so we shall keep this mode in the low energy effective quantum mechanics


The story of the gauge field goes like this. The energy density does not depend on the derivatives of the gauge field. The standard kinetic and potential terms of the gauge field involving the squares of the electric and magnetic fields is absent in this case. So it appears that a field configuration with a very large magnetic field can have energy small compared to $\frac{1}{\sqrt {area(T^{2})}}$. But this changes once we take into account the Gauss's law constraint. If we switch on a magnetic field such that the left hand side of the Gauss's law constraint is a spatially varying quantity then the scalar field on the right hand side will also have to |depend on the spatial coordinates. This requires an energy of order $\frac{1}{\sqrt {area(T^{2})}}$. So we can safely neglect (integrate out) such modes of the gauge field. 

\subsection{Abelian Gauge Theory}
The first example that we shall study is an Abelian gauge field coupled to complex scalars. We shall define the theory on the space-time manifold $T^{2}\times R^{1}$. Since the modular property of the torus will not play any role in our analysis we take a square torus with side of length $R$. The metric on the torus is given by,
\begin{equation} 
dh^{2} = dx_{1}^{2} + dx_{2}^{2}
\end{equation}
where $x_{i}\sim x_{i} + R$. To obtain the low energy effective quantum mechanics we keep only the zero momentum modes of the gauge and matter fields in the Lagrangian and the reduced Lagrangian becomes,
\begin{equation}
L = R^{2} [ \ \frac{ k}{8\pi} \ A_{1}\frac{dA_{2}}{dt} + \frac{d\phi^{*}}{dt}\frac{d\phi}{dt} - \phi^{*}\phi \ (A_{1}^{2} + A_{2}^{2}) \ ]
\end{equation}
As we have argued in the previous sections this dimensionally reduced theory can capture the very low energy ($E\ll 1/R$) states in the theory. In the $A_{0} = 0$ gauge the residual gauge transformations are the time independent $U(1)$ rotations. The reduced Lagrangian has a global $U(1)$ rotation symmetry which acts on the scalar field. Since the physical states have to be gauge invariant we shall treat this global $U(1)$ rotation symmetry of the quantum mechanics as a gauge symmetry. So the physical states are those which are invariant under the rotation. 

If we measure all energies in units of $\frac{1}{R}$ then we can set $R=1$. The Hamiltonian is given by,
\begin{equation}
H = \pi^{*}\pi + \phi^{*}\phi [ (\frac{8\pi}{ k})^{2} P^{2} + Q^{2} ]
\end{equation}
where we have defined $P = \frac{ k}{8\pi}A_{1}$ and $Q=A_{2}$. $\pi$ and $\pi^{*}$ are momenta canonically conjugate to $\phi$ and $\phi^{*}$. The canonical commutation relations are given by ,
\begin{equation}
[Q,P] = i , \  [\phi,\pi] = [\phi^{*},\pi^{*}] = i
\end{equation}
 A scaling argument shows  that the total energy of the system is proportional to $\frac{1}{\sqrt k}$.
In fact this problem is easy to solve exactly. The total wave function can be written as a product $\Psi(\phi,\phi^{*},Q) = \chi(\phi,\phi^{*})\psi(Q)$, where $\chi$ is the rotationally invariant wave function in the matter sector and $\psi$ is the wave function of the gauge sector. The gauge sector is a one dimensional harmonic oscillator and the matter sector is an isotropic two dimensional harmonic oscillator where the Gauss's Law constraint imposes rotational invariance.  The energy eigenvalues are given by
\begin{equation}
E_{j,l} \sim (2j+1)\sqrt{(\frac{16 \pi}{k})(l+\frac{1}{2})}
\end{equation}
 So at weak coupling or large $k$ this energy is well below the Kaluza-Klein scale which is of $O(1)$, which justifies our neglect of spatially varying modes.
The wave functions in the quantum mechanics are all concentrated near the origin, $A_{1} = A_{2} =0$. The width of the wave function goes like, $<A_{1}^{2}> \ = \ <A_{2}^{2}>  \ \sim \frac{1}{k}$. This is very small in the large $k$ limit which justifies our neglect of the periodicity of the holonomies $A_{1}, A_{2}$. 

\subsection{SU(N) Gauge Theory}
Our next example is $SU(N)$ Chern-Simons gauge theory coupled to massless fundamental scalar matter \footnote{The quantum mechanics arising in SU(2) gauge theory can be solved exactly. Please see the appendix for the exact solution.  }.   In the quantum mechanics problem we can treat the variables $A_{1}^{f}$ and $A_{2}^{f}$ as effectively non-compact. 

As in the U(1) case we  need to treat the dynamics of the zero mode of the scalar field by incorporating it and the gauge field dynamics in an effective dimensionally reduced quantum mechanics.  Because of the localization of the gauge field we can ignore the compactness of the flat connection moduli space.  The Lagrangian is given by
\begin{equation}
L = \frac{k}{8{\pi}} Tr(A_{1}\frac{d}{dt} A_{2})  +  \frac{d\phi^{\dagger}}{dt}\frac{d\phi}{dt}  - {\phi}^{\dagger}(A_{1}^{2}+A_{2}^{2}){\phi}
\end{equation}
Here the variables $A_{1}$ and $A_{2}$ are two arbitrary $N\times N$ Hermitian matrices not necessarily commuting, and 
$\phi$ is a complex $N$-dimensional column vector transforming in the fundamental representation of $SU(N)$. The Lagrangian has a global symmetry under which,
\begin{equation}
A_{i} \rightarrow UA_{i}U^{\dagger}, \  \phi\rightarrow U\phi
\end{equation}
where $U$ is a $SU(N)$ matrix. The $SU(N)$ global symmetry of the Lagrangian is the remnant of the $SU(N)$ gauge symmetry of the original field theory and so we should treat this $SU(N)$ symmetry as a gauge symmetry. A state in the quantum mechanics will be called physical if it is invariant under  $SU(N)$ transformations. 

To write the Lagrangian in a more manageable form we express the gauge potentials in terms of generators of the $SU(N)$ group. 
\begin{equation}
A_{i} = A_{i}^{a} T^{a} , \\  a = 1,.......,N^{2}-1
\end{equation}
where $T^{a}$ are generators in the fundamental representation which satisfy the relation, $Tr(T^{a}T^{b}) = \frac{1}{2}\delta_{ab}$. In terms of this the Lagrangian can be written as,
\begin{equation}
L = \sum_{a=1}^{N^{2}-1}P^{a}\frac{dQ^{a}}{dt} + \sum_{i=1}^{N}\frac{d\phi_{i}^{*}}{dt}\frac{d\phi_{i}}{dt} - \frac{1}{2} \sum_{a,b=1}^{N^{2}-1}M^{ab}(\phi) (P^{a}P^{b} + \hbar^{2}Q^{a}Q^{b})
\end{equation}
where, $P^{a} = A_{1}^{a}$, $Q^{a} = \frac{ k}{16\pi} A_{2}^{a}$, $\hbar = \frac{16\pi}{k}$ and $M^{ab}(\phi) = \phi^{\dagger}\lbrace T^{a},T^{b}\rbrace \phi$. It is clear from the form of the Lagrangian that $(P^{a},Q^{a})$ are canonically conjugate variables\footnote{We have made a small change definitions here relative to the U(1) case}. The Hamiltonian can be written as,
\begin{equation}
H = \sum_{i=1}^{N} \pi_{i}^{*}\pi_{i} + \frac{1}{2} \sum_{a,b=1}^{N^{2}-1}M^{ab}(\phi) (P^{a}P^{b} + \hbar^{2}Q^{a}Q^{b})
\end{equation}  
where $(\phi^{*},\pi)$ and $(\phi,\pi^{*})$ are canonically conjugate. the canonical commutation relations are given by,
\begin{equation}
[Q^{a},P^{b}] = i \delta^{ab},   [\phi_{i},\pi_{j}^{*}] = [\phi_{i}^{*},\pi_{j}] = i \delta_{ij}
\end{equation}
and rest of the commutators are zero. We can define the creation and annihilation operators in the gauge sector as,
\begin{equation}
\beta^{a} = \frac{1}{\sqrt {2\hbar}} (P^{a} - i\hbar Q^{a}), \  \beta^{a\dagger} = \frac{1}{\sqrt {2\hbar}}(P^{a} + i\hbar Q^{a})
\end{equation}
which satisfy the commutation relation,
\begin{equation}
[\beta^{a},\beta^{b\dagger}] = \delta^{ab}
\end{equation}
and rest of the commutators are zero. In terms of these operators the Hamiltonian can be rewritten as,
\begin{equation}
H = \pi_{i}^{*}\pi_{i} + \omega^{2}\phi_{i}^{*}\phi_{i} + \hbar M^{ab}(\phi) \beta^{a\dagger}{\beta}^{b}
\end{equation}
where we have defined, $\omega^{2} = \frac{N\hbar}{2}\sim \frac{N}{k}= \lambda$. We can also define the following creation and annihilation operators in the scalar sector,
\begin{equation}
\alpha_{i} = \frac{1}{\sqrt{2\omega}} (\pi_{i} - i\omega \phi_{i}), \  \alpha_{i}^{\dagger} = \frac{1}{\sqrt{2\omega}} (\pi_{i}^{*} + i\omega \phi_{i}^{*})
\end{equation}
\begin{equation}
\bar\alpha_{i} = \frac{1}{\sqrt{2\omega}} (\pi_{i}^{*} - i\omega \phi_{i}^{*}), \  \bar\alpha_{i}^{\dagger} = \frac{1}{\sqrt{2\omega}} (\pi_{i} + i\omega\phi_{i})
\end{equation}
they satisfy the commutation relation,
\begin{equation}
[\alpha_{i},\alpha_{j}^{\dagger}] = \delta_{ij}, \  [\bar\alpha_{i}, \bar\alpha_{j}^{\dagger}] = \delta_{ij} 
\end{equation}
and rest of the commutators are zero. In terms of these creation and annihilation operators the Hamiltonian can be written as,
\begin{equation}
\begin{split}
H& = \omega(\alpha_{i}^{\dagger}\alpha_{i} + \bar\alpha_{i}^{\dagger}\bar\alpha_{i}) + \frac{\hbar}{2\omega} \beta^{a\dagger}\beta^{a} + N\omega \\
  & -\frac{\hbar}{2\omega}\biggl [\bar\alpha_{i}\lbrace T^{a},T^{b}\rbrace_{ij}\alpha_{j} + \alpha_{i}^{\dagger}\lbrace T^{a},T^{b}\rbrace_{ij}\bar\alpha_{j}^{\dagger} \\  
  & - \bar\alpha_{j}^{\dagger}\lbrace T^{a},T^{b}\rbrace_{ij} \bar\alpha_{i} - \alpha_{i}^{\dagger}\lbrace T^{a},T^{b}\rbrace_{ij} \alpha_{j}\biggr] \beta^{a\dagger}\beta^{b}
\end{split}
\end{equation}
where $(i,a,b)$ are summed over.

The ground state energy of the Hamiltonian is $N\omega$. The ground state wave function is annihilated by all the annihilation operators. The unnormalized wave function can be written as,
\begin{equation}\label{groundstate}
\Psi = e^{-\omega \phi^{\dagger}\phi} e^{-\hbar TrQ^{2}} = e^{-\omega \phi^{\dagger}\phi} e^{ -\frac{TrA_{2}^{2}}{\hbar}}
\end{equation}
We can see that the width of the wave function along the gauge-field directions are of order, $\sqrt\hbar \sim \frac{1}{\sqrt k}$, and so are very small in the weak coupling or large $k$ limit.

\chapter{Singlet Sector Of The Hamiltonian}

We are interested in the states of the Hilbert space which are invariant under the global $SU(N)$ transformations of the quantum mechanical model. The singlet states in the Hilbert space can be obtained by acting on the ground-state with singlet creation operators. The "single-trace " operators are given by,
\begin{equation}
Tr\beta^{\dagger}\beta^{\dagger}, Tr\beta^{\dagger}\beta^{\dagger}\beta^{\dagger},...............,Tr(\beta^{\dagger})^{N}, \alpha^{\dagger}\bar\alpha^{\dagger}, \alpha^{\dagger}\beta^{\dagger}\bar\alpha^{\dagger},........., \alpha^{\dagger}(\beta^{\dagger})^{N}\bar\alpha^{\dagger}
\end{equation}
where we have defined the matrix $\beta^{\dagger} = \beta^{a\dagger}T^{a}$, which transforms in the adjoint representation and $\alpha^{\dagger}$ and $\bar\alpha^{\dagger}$ are column vectors transforming in the anti-fundamental and fundamental representations of the global symmetry group $SU(N)$.  Any other singlet creation operator can be written as linear combinations of products of these basis set of operators. The singlet states can be created by acting with the singlet operators on the ground state. In our case the ground state is exactly given by (\ref{groundstate}), and it is annihilated by all the annihilation operators. Now every state created by acting on the ground state with the singlet creation operators are not exact eigenfunctions of the total Hamiltonian\footnote{Our analysis will reveal that these states tend to be exact eigenstates in the large-N limit.}. States where no gauge field excitations are present are exact eigenfunctions of the total Hamiltonian, whereas the states with gauge field excitations are not in general exact eigenfunctions of the total Hamiltonian. They are exact eigenfunctions of the part of the Hamiltonian which contains no interaction term between scalar and gauge fields. This motivates us to separate the Hamiltonian into an interacting and a non-interacting part and treat the interacting part as perturbation. At this stage this separation is somewhat artificial because there is as such no small parameter in the Hamiltonian, \footnote{Although we are working in the large $k$ limit, $\frac{1}{k}$ is not a small parameter in the effective quantum mechanics problem. It is easy to show by properly scaling different variables that $\frac{1}{\sqrt k}$ is an overall multiplicative factor in the Hamiltonian.} but we shall justify this separation in the later part of this section. 

We define,
\begin{equation}
H_{0} = \omega(\alpha_{i}^{\dagger}\alpha_{i} + \bar\alpha_{i}^{\dagger}\bar\alpha_{i}) + \frac{\hbar}{2\omega} \beta^{a\dagger}\beta^{a} + N\omega
\end{equation}
and 
\begin{equation}
V = -\frac{\hbar}{2\omega}\biggl [\bar\alpha_{i}\lbrace T^{a},T^{b}\rbrace_{ij}\alpha_{j} + \alpha_{i}^{\dagger}\lbrace T^{a},T^{b}\rbrace_{ij}\bar\alpha_{j}^{\dagger}\biggr] \beta^{a\dagger}\beta^{b} = -\frac{\hbar}{2\omega} \tilde V
\end{equation}
\begin{equation}
V_{1} = \frac{\hbar}{2\omega} \biggl[\bar\alpha_{j}^{\dagger}\lbrace T^{a},T^{b}\rbrace_{ij} \bar\alpha_{i} + \alpha_{i}^{\dagger}\lbrace T^{a},T^{b}\rbrace_{ij} \alpha_{j}\biggr] \beta^{a\dagger}\beta^{b}
\end{equation}
The total Hamiltonian $H$ can be written as,
\begin{equation}
H = H_{0} + V + V_{1}
\end{equation}
It is easy to check that, $[H_{0},V_{1}] = 0$.

From the unperturbed Hamiltonian $H_{0}$ we can already see the basic dynamics of the system.  The zero point energy of the gauge fields creates a large frequency for the scalars $\omega^2 \sim N/k = \lambda$.  This is $N$ times larger than in the U(1) system.  This makes the scalar excitations heavy, with mass $\sim \omega$.  This large $\omega^2$ makes the size of $\phi^{\dagger} \phi$ smaller than in the U(1) case, $\phi^{\dagger} \phi\sim 1/\omega \sim \sqrt{k/N}$.  This is smaller by a factor of $\sqrt{N}$ than in the U(1) case.   The energy of the gauge field excitations is decreased by this factor.  The energy of such excitations is $\sim \sqrt{k/N}(1/k)  \sim \sqrt{\lambda}/N$.   This is the characteristic gap in the system.   In the large $N$ limit where we keep $\lambda$ fixed and let $N\rightarrow\infty$, the gauge field excitations are very light compared to the scalar field excitations.   We will show that the perturbative effects of the heavy scalars do not change these results.

More precisely, in the unperturbed theory the state with the smallest excitation energy can be written as,
\begin{equation}
|L> = Tr\beta^{\dagger}\beta^{\dagger}|\Omega> 
\end{equation}
where $|\Omega>$ is the ground state of  the system. The ground state is exact for the system whose wave function is given by (\ref{groundstate}). The state $|L>$ is non-degenerate and its energy is given by $\frac{\hbar}{\omega}\sim \frac{\sqrt\lambda}{N}$.   

We shall now study the effect of the terms $V_{1}$ and $V$ on the energy of this state and see that the parametric size of the gap is not changed.  

It is easy to see that the state $|L>$ is an exact eigenfunction of the Hamiltonian, $H_{0} + V_{1}$ with the same energy $\frac{\hbar}{\omega}$. 
The state is annihilated by $V_{1}$. Since $V_{1}$ commutes with the unperturbed Hamiltonian $H_{0}$ the effect of this term can be taken into account by diagonalizing $V_{1}$ restricted to degenerate eigenspaces of the unperturbed Hamiltonian $H_{0}$. Now $V_{1}$ annihilates any state which does not contain any scalar field excitation or gauge field excitations and so the energy and degeneracy of states containing only gauge field excitations or only scalar field excitations remain unchanged due to this term. We shall study the effect of this term later in this section. 

\subsection{Computation Of The Perturbation}
It is easy to check that the first order perturbation is zero. Before we write the results of the perturbation calculation we shall collect few results which are useful for our purpose. 

One can check the following results,
\begin{equation}\label{VL1}
\tilde V|L> =  \alpha^{\dagger}\beta^{\dagger}\beta^{\dagger}\bar\alpha^{\dagger} |\Omega> = |L1>
\end{equation}
\begin{equation}\label{VL2}
\tilde V|L1> = (\alpha^{\dagger}\beta^{\dagger}\bar\alpha^{\dagger})(\alpha^{\dagger}\beta^{\dagger}\bar\alpha^{\dagger}) |\Omega> + (\alpha^{\dagger}\beta^{\dagger}\beta^{\dagger}\bar\alpha^{\dagger})( \alpha^{\dagger}\bar\alpha^{\dagger}) |\Omega> + N (Tr\beta^{\dagger}\beta^{\dagger})|\Omega> 
\end{equation}
\begin{equation}\label{VL3}
\tilde V (\alpha^{\dagger}\beta^{\dagger}\bar\alpha^{\dagger})(\alpha^{\dagger}\beta^{\dagger}\bar\alpha^{\dagger}) |\Omega> = 2\alpha^{\dagger}\bar\alpha^{\dagger} (\alpha^{\dagger}\beta^{\dagger}\bar\alpha^{\dagger})(\alpha^{\dagger}\beta^{\dagger}\bar\alpha^{\dagger})|\Omega> + 2(N+1) \alpha^{\dagger}\beta^{\dagger}\beta^{\dagger}\bar\alpha^{\dagger} |\Omega>
\end{equation}
\begin{equation}\label{VL4}
\begin{split}
\tilde V (\alpha^{\dagger}\beta^{\dagger}\beta^{\dagger}\bar\alpha^{\dagger})( \alpha^{\dagger}\bar\alpha^{\dagger}) |\Omega>
& = \alpha^{\dagger}\bar\alpha^{\dagger} (\alpha^{\dagger}\beta^{\dagger}\bar\alpha^{\dagger})^{2}|\Omega> + (\alpha^{\dagger}\bar\alpha^{\dagger})^{2} (\alpha^{\dagger}\beta^{\dagger}\beta^{\dagger}\bar\alpha^{\dagger})|\Omega>  \\
 &  + (N+2) (\alpha^{\dagger}\beta^{\dagger}\beta^{\dagger}\bar\alpha^{\dagger})|\Omega>  + (N+1) \alpha^{\dagger}\bar\alpha^{\dagger} Tr\beta^{\dagger}\beta^{\dagger} |\Omega>
 \end{split}
\end{equation}
In deriving these results we have made use of the following relations,
\begin{equation}
\sum_{a=1}^{N^{2}-1}(T^{a})_{ij}(T^{a})_{kl} = \frac{1}{2} (\delta_{il}\delta_{jk} - \frac{1}{N} \delta_{ij}\delta_{kl})
\end{equation}
and 
\begin{equation}
Tr(T^{a}T^{b}) = \frac{1}{2}\delta^{ab}
\end{equation}
In (\ref{VL3}) one can neglect the $\frac{1}{N}$ piece in the large $N$ limit. The generators are taken in the fundamental representation.

The states appearing in eqns (\ref{VL1})-(\ref{VL4}) are unnormalized. The norms of these states are given by,
\begin{equation}\label{norm1}
||Tr\beta^{\dagger}\beta^{\dagger}|\Omega>||^{2}\sim N^{2}
\end{equation}
\begin{equation}\label{norm2}
||\alpha^{\dagger}\beta^{\dagger}\beta^{\dagger}\bar\alpha^{\dagger} |\Omega>||^{2} \sim N^{3}
\end{equation}
\begin{equation}
||(\alpha^{\dagger}\beta^{\dagger}\bar\alpha^{\dagger})(\alpha^{\dagger}\beta^{\dagger}\bar\alpha^{\dagger}) |\Omega>||^{2} \sim N^{4}
\end{equation}
\begin{equation}
||(\alpha^{\dagger}\beta^{\dagger}\beta^{\dagger}\bar\alpha^{\dagger})( \alpha^{\dagger}\bar\alpha^{\dagger}) |\Omega> ||^{2} \sim N^{4}
\end{equation}
\begin{equation}
||(\alpha^{\dagger}\bar\alpha^{\dagger})^{2} (\alpha^{\dagger}\beta^{\dagger}\beta^{\dagger}\bar\alpha^{\dagger})|\Omega> ||^{2} \sim N^{5}
\end{equation}
\begin{equation}
|| \alpha^{\dagger}\bar\alpha^{\dagger} (\alpha^{\dagger}\beta^{\dagger}\bar\alpha^{\dagger})^{2}|\Omega>|| ^{2} \sim N^{5}
\end{equation}
The inner product of states containing different numbers of scalar and gauge excitations are orthogonal. The inner product of the states appearing in (\ref{norm1}) and (\ref{norm2}) are given by,
\begin{equation}
\biggl((\alpha^{\dagger}\beta^{\dagger}\bar\alpha^{\dagger})(\alpha^{\dagger}\beta^{\dagger}\bar\alpha^{\dagger}) |\Omega> ,(\alpha^{\dagger}\beta^{\dagger}\beta^{\dagger}\bar\alpha^{\dagger})( \alpha^{\dagger}\bar\alpha^{\dagger}) |\Omega> \biggr) \sim N^{3}
\end{equation}
\begin{equation}
\biggl( (\alpha^{\dagger}\bar\alpha^{\dagger})^{2} (\alpha^{\dagger}\beta^{\dagger}\beta^{\dagger}\bar\alpha^{\dagger})|\Omega> ,  \alpha^{\dagger}\bar\alpha^{\dagger} (\alpha^{\dagger}\beta^{\dagger}\bar\alpha^{\dagger})^{2}|\Omega>\biggl) \sim N^{4}
\end{equation}
So these two states are orthogonal in the large-$N$ limit. In particular, the normalized states\footnote{We shall write down only powers of $N$ that appear in the normalization of the states in the large-$N$ limit. There are $O(1)$ numbers which multiply the states in the large-$N$ limit. We shall not write them because they are not important for our purpose, at least to the order we are working.} (in the large-$N$ limit) containing two gauge field excitations are of the form,
\begin{equation}\label{normal1}
\frac{1}{N}Tr(\beta^{\dagger}\beta^{\dagger})|\Omega>
\end{equation}
\begin{equation}\label{normal2}
\frac{1}{N^{\frac{3}{2}}}\alpha^{\dagger}\beta^{\dagger}\beta^{\dagger}\bar\alpha^{\dagger} |\Omega>
\end{equation}
\begin{equation}\label{normal3}
\frac{1}{N^{\frac{3}{2}}} (\alpha^{\dagger}\bar\alpha^{\dagger}) Tr\beta^{\dagger}\beta^{\dagger}|\Omega>
\end{equation}
\begin{equation}\label{normal4}
\frac{1}{N^{2}}(\alpha^{\dagger}\beta^{\dagger}\bar\alpha^{\dagger})(\alpha^{\dagger}\beta^{\dagger}\bar\alpha^{\dagger}) |\Omega>
\end{equation}
\begin{equation}\label{normal5}
\frac{1}{N^{2}}(\alpha^{\dagger}\beta^{\dagger}\beta^{\dagger}\bar\alpha^{\dagger})( \alpha^{\dagger}\bar\alpha^{\dagger}) |\Omega>
\end{equation}
\begin{equation}\label{normal6}
\frac{1}{N^{2}} (\alpha^{\dagger}\bar\alpha^{\dagger})(\alpha^{\dagger}\bar\alpha^{\dagger}) (Tr\beta^{\dagger}\beta^{\dagger})|\Omega>
\end{equation}
\begin{equation}\label{normal7}
\frac{1}{N^{\frac{5}{2}}} \alpha^{\dagger}\bar\alpha^{\dagger} (\alpha^{\dagger}\beta^{\dagger}\bar\alpha^{\dagger})^{2}|\Omega>
\end{equation}
\begin{equation}\label{normal8}
\frac{1}{N^{\frac{5}{2}}} (\alpha^{\dagger}\bar\alpha^{\dagger})^{2} (\alpha^{\dagger}\beta^{\dagger}\beta^{\dagger}\bar\alpha^{\dagger})|\Omega>
\end{equation}
There are an infinite number of such states. These states are all normalized and mutually orthogonal in the large-$N$ limit. This is true even in the interacting theory because the annihilation operators are defined with respect to the exact ground state of the interacting theory. These states are also exact eigenfunctions of the unperturbed Hamiltonian $H_{0}$. 
\subsection{1-st Order Perturbation}
It is easy to see that the the first order perturbation is zero.
\subsection{2-nd Order Perturbation}
The answer for the second order perturbation is,
\begin{equation}\label{shift}
\Delta^{(2)} = -\frac{1}{2}\frac{\hbar}{\omega}
\end{equation}
where $\Delta^{(2)}$ is the second order shift in the energy of the state $Tr\beta^{\dagger}\beta^{\dagger}|\Omega>$. We can see that the second order shift is of the same order of magnitude as the zeroth order energy of the state which is $\frac{\hbar}{\omega}$.
\subsection{3-rd Order Perturbation}
The formula for the third order energy shift is,
\begin{equation}
\Delta^{(3)} =  \sum_{k\neq n, m\neq n} \frac{V_{nk}V_{km}V_{mn}}{E_{nk}E_{nm}} - V_{nn} \sum_{k\neq n}\frac{|V_{nk}|^{2}}{E_{nk}^{2}}
\end{equation}
where, $V_{nk}= <n|V|k>$ and $E_{mn} = E_{m} - E_{n}$. The states and energies are all referred to the unperturbed Hamiltonian $H_{0}$ and all the states are normalized.

In our case, $|n> = \frac{1}{N}|L> = \frac{1}{N}Tr\beta^{\dagger}\beta^{\dagger}|\Omega>$ and so $V_{nn} = 0$. Now from eqn-(6.7) we get,
\begin{equation}
V|n> = -\frac{1}{N} \frac{\hbar}{2\omega} \alpha^{\dagger}\beta^{\dagger}\beta^{\dagger}\bar\alpha^{\dagger} |\Omega> = \frac{1}{N} |L1>
\end{equation}
Now $|L1>$ is an exact eigenfunction of $H_{0}$ and so it follows from our previous discussion that the matrix element $V_{nk}$ and $V_{mn}$ are nonzero only if $|k> = |m> \propto |L1>$. But in that case the matrix element $V_{km} =0$. So the third order perturbation vanishes.
\subsection{4-th Order Perturbation}
The formula for the fourth order perturbation is,
\begin{equation}
\Delta^{(4)} = \sum_{k_{2}\neq n, k_{3}\neq n, k_{4}\neq n}\frac{V_{nk_{2}}V_{k_{2}k_{3}}V_{k_{3}k_{4}}V_{k_{4}n}}{E_{nk_{2}}E_{nk_{3}}E_{nk_{4}}} - \sum_{k_{1}\neq n, k_{2}\neq n}\frac{V_{nk_{1}}V_{k_{1}n}}{E_{nk_{1}}^{2}} \frac{V_{nk_{2}}V_{k_{2}n}}{E_{nk_{2}}}
\end{equation}
We have not written down the terms which are multiplied by $V_{nn}$ which is zero in our case. Let us study the contribution of the first term . Using the same argument as in the case of the third order perturbation, we conclude that $|k_{2}> = |k_{4}> \propto |L1>$. if this condition is not satisfied the first term will vanish. Using this the first term can be simplified to,
\begin{equation}
\sum_{k_{3}\neq n}\frac{|V_{nL1}|^{2}|V_{L1k_{3}}|^{2}}{E_{nL1}E_{nk_{3}}E_{nL1}} 
\end{equation}
Now we shall calculate the $N$ scaling of these matrix elements in the large $N$ limit.
\begin{equation}
\begin{split}
V_{L1n} \sim <\alpha^{\dagger}\beta^{\dagger}\beta^{\dagger}\bar\alpha^{\dagger}|\frac{1}{N^{\frac{3}{2}}} V \frac{1}{N} |Tr\beta^{\dagger}\beta^{\dagger}> & =  -\frac{\hbar}{2\omega}\frac{1}{N^{\frac{5}{2}}} <\alpha^{\dagger}\beta^{\dagger}\beta^{\dagger}\bar\alpha^{\dagger}|\alpha^{\dagger}\beta^{\dagger}\beta^{\dagger}\bar\alpha^{\dagger}>  \\ & \sim -\frac{1}{\sqrt{Nk}} \frac{1}{N^{\frac{5}{2}}} (N^{3} + O(N^{2}))\\& = -\frac{1}{\sqrt k}(1 + O(\frac{1}{N}))
\end{split}
\end{equation}
Now we have to compute the matrix element $V_{k_{3}L1}$. From (\ref{VL2})we get,
\begin{align}
V|L1> = \biggl(\frac{\hbar}{2\omega}\biggr)^{2} \biggl(&(\alpha^{\dagger}\beta^{\dagger}\bar\alpha^{\dagger})(\alpha^{\dagger}\beta^{\dagger}\bar\alpha^{\dagger}) |\Omega> + (\alpha^{\dagger}\beta^{\dagger}\beta^{\dagger}\bar\alpha^{\dagger})( \alpha^{\dagger}\bar\alpha^{\dagger}) |\Omega>\nonumber\\& + N (Tr\beta^{\dagger}\beta^{\dagger})|\Omega> \biggr)
 \end{align}
Now every state appearing in the above equation is an exact eigenstate of $H_{0}$ and they are mutually orthogonal at least in the large-N limit. So the matrix element can be nonzero only if the the state $|k_{3}>$ is one of the three states appearing in the formula. Now $|k_{3}>$ cannot be the last state because it is proportional to the state $|n>$. So $|k_{3}>$ can be any one of the remaining two states. Let us first take,
\begin{equation}
|k_{3}> \propto (\alpha^{\dagger}\beta^{\dagger}\bar\alpha^{\dagger})(\alpha^{\dagger}\beta^{\dagger}\bar\alpha^{\dagger}) |\Omega>
\end{equation} 
So,
\begin{equation}
\begin{split}
V_{k_{3}L1} &\sim <(\alpha^{\dagger}\beta^{\dagger}\bar\alpha^{\dagger})(\alpha^{\dagger}\beta^{\dagger}\bar\alpha^{\dagger})|\frac{1}{N^{2}} V \frac{1}{N^{\frac{3}{2}}} |\alpha^{\dagger}\beta^{\dagger}\beta^{\dagger}\bar\alpha^{\dagger}>   \\ &  \sim - \frac{\hbar}{2\omega} \frac{1}{N^{\frac{7}{2}}} <(\alpha^{\dagger}\beta^{\dagger}\bar\alpha^{\dagger})(\alpha^{\dagger}\beta^{\dagger}\bar\alpha^{\dagger})|(\alpha^{\dagger}\beta^{\dagger}\bar\alpha^{\dagger})(\alpha^{\dagger}\beta^{\dagger}\bar\alpha^{\dagger})>  \\ & \sim - \frac{1}{\sqrt{Nk}}\frac{1}{N^{\frac{7}{2}}} (N^{4} + O(N^{3})) \sim -\frac{1}{\sqrt k} (1 + O(\frac{1}{N}))
\end{split}
\end{equation}
The same scaling holds for the other choice of $|k_{3}>$. So we can conclude that the matrix elements scale like $-\frac{1}{\sqrt k}(1+ O(\frac{1}{N}))$ and so the leading contribution in the large N limit is $-\frac{1}{\sqrt k}$.

The energy denominators are all of order $\sqrt\omega \sim \sqrt \lambda = \sqrt{\frac{N}{k}}$, because the states appearing in the formula other than $|n>$ contain scalar excitations. So the leading contribution of the matrix element in the large-$N$ limit is of order,
\begin{equation}
(-\frac{1}{\sqrt k})^{4} \frac{1}{(\sqrt\lambda)^{3}} =  \frac{1}{k^{2}} \frac{1}{\frac{N^{\frac{3}{2}}}{k^{\frac{3}{2}}}} = \frac{1}{\sqrt{Nk}}\frac{1}{N}
\end{equation}
So we can see that this contribution is $\frac{1}{N}$ suppressed compared to the second order contribution. It is easy to see that the second term in formula (\ref{normal5}) gives the same $\frac{1}{N}$ suppressed contribution.
\subsection{5-th Order Perturbation}
It is easy to convince oneself that the 5-th order perturbation also vanishes for the same reason that the third order perturbation vanished. The fifth order perturbation contains terms of two kinds. One kind of terms is multiplied by $V_{nn}$ which is identically zero in our case. The second kind of terms are all multiplied by the matrix element appearing in the third order perturbation\footnote{The formalism of time-independent perturbation theory can be used to determine the 4-th, 5-th and 6-th order perturbations. References on this formalism are presented in the classic texts\cite{LL3:1977,Sakurai:1167961}. The results for higher order perturbations are stated in \cite{Wheeler_2000}.} and so is identically zero in our case. The only term which survives is the following,
\begin{equation}
V_{nk_{1}}V_{k_{1}k_{2}}V_{k_{2}k_{3}}V_{k_{3}k_{4}}V_{k_{4}n}
\end{equation}
The energy denominator is also there and the indices except $n$ is summed over subject to the same constraint. So by following the same argument as in the previous case we conclude that $|k_{1}> = |k_{4}> \propto |L1>$. So the states $|k_{2}>$ and $|k_{3}>$ must belong to the subspace spanned by the states $(\alpha^{\dagger}\beta^{\dagger}\bar\alpha^{\dagger})(\alpha^{\dagger}\beta^{\dagger}\bar\alpha^{\dagger}) |\Omega> $ and $(\alpha^{\dagger}\beta^{\dagger}\beta^{\dagger}\bar\alpha^{\dagger})( \alpha^{\dagger}\bar\alpha^{\dagger}) |\Omega>$. These states are orthogonal in the large N limit. Now the matrix element $V_{k_{2}k_{3}}$ vanishes in this subspace. So the fifth order perturbation is identically zero.
\subsection{6-th Order Perturbation}
In this case one can show using the results stated in the previous sections that the contribution goes like,
\begin{equation}
\frac{1}{\sqrt {kN}} \frac{1}{N^{2}} , \\\\ N\rightarrow\infty
\end{equation}
\subsection{The Odd Order Perturbation Is Zero To All Orders}
Let us consider the $(2p+1)$-th order perturbation theory. The $(2p+1)$-th order perturbation contains the term,
\begin{equation}
V_{nk_{1}}V_{k_{1}k_{2}}\ldots V_{k_{p-1}k_{p}}V_{k_{p}k_{p+1}}V_{k_{p+1}k_{p+2}}...............V_{k_{2p}n}
\end{equation}
There is an energy denominator and the intermediate states $|k_{i}>$ do not take the value $|n> = Tr\beta^{\dagger}\beta^{\dagger}|\Omega>$. Let us denote by $A$ and $B$ the following matrix elements,
\begin{equation}
A = V_{nk_{1}}V_{k_{1}k_{2}}\ldots V_{k_{p-1}k_{p}}
\end{equation}
and 
\begin{equation}
B = V_{k_{p+1}k_{p+2}}\ldots V_{k_{2p}n}
\end{equation}
Since $V$ is Hermitian, the complex conjugate of $B$ can be written as,
\begin{equation}
B^{*} = V_{nk_{2p}}\ldots V_{k_{p+2}k_{p+1}}
\end{equation}
Both of these matrix elements represent the following process. The potential $V$ creates or annihilates two scalar excitations, one of type $\alpha$ and another of type $\bar\alpha$. Since we are considering only singlet states,\footnote {We are starting with the singlet state $|n>$ and and since $V$ is a singlet operator we never leave the singlet sector.} every state contains an equal number of $\alpha$ and $\bar\alpha$ excitations and so the total number of scalar excitations is always an even integer. The action of $V$ increases or decreases this integer in steps of $2$. The number of gauge excitations does not change because because $V$ contains a creation and an annihilation operator for the gauge excitations. More precisely the number operator for the gauge oscillators given by $Tr\beta^{\dagger}\beta$ commutes with $V$. Since both $A$ and $B^{*}$ represent the same physical process let us concentrate on $A$. So the systems starts at the state $|k_{p}>$ with some number of scalar excitations, say $2m$, and after a series of transitions it ends up in the state $|n>$ with zero scalar excitations. If the end state has nonzero scalar excitations then the first matrix element $V_{nk_{1}}$ vanishes and this term in the perturbation series is zero. The system can make a total of $p$ transitions and some of them are up transitions and some of them are down transitions where the number of scalar field quanta increases or decreases by 2, respectively. Let $n_{+}$ and $n_{-}$ be the number of up and down transitions. So the they have to obey the following relations,
\begin{equation}
n_{+} + n_{-} = p
\end{equation}
 and 
\begin{equation}
2(n_{+} - n_{-}) = 2m
\end{equation}
The solution is given by,
\begin{equation}
n_{+} = \frac{p+m}{2}, \\  n_{-} = \frac{p-m}{2}
\end{equation}
 Now $p$ is a fixed integer at a given order and so what can vary is the integer $m$ which determines the number the scalar excitations in the state $|k_{p}>$. Let $m=m_{0}$ be a value for which the matrix element $A$ is nonzero, i.e, the system can make $p$ transitions to reach a state with no scalar excitations. Now $n_{+}$ and $n_{-}$ are integers. So the next nearest values of $m$ for which the matrix element  $A$ is nonzero is given by $m_{0}\pm 2$. It is not $m_{0}\pm 1$ because in that case $n_{\pm}$ will be half-integers. So if $|k_{p}> = |2m_{0}>$ is one state then the nearest states are $|k_{p}'> = |2m_{0}>$ or $|k_{p}'> = |2m_{0}\pm 4>$. the same argument goes through for the amplitude $B^{*}$. Now we have the matrix element $V_{k_{p}k_{p+1}}$. This matrix element will be nonzero only if the states $|k_{p}>$ and $|k_{p+1}>$ differ by two units of scalar excitations. So $|k_{p+1}>$ has to be a state of the form $|2m_{0}\pm 2>$. But in that case we know that the matrix element $B^{*}$ will vanish, because if $|2m_{0}>$ is a valid state then the next nearest states are $|2m_{0}\pm 4>$. So in any case the total matrix element has to be zero. So the odd order perturbation contribution is zero to all orders.
 
\subsection{Gap In The System}
In the large-$N$ limit the energy of the state $Tr\beta^{\dagger}\beta^{\dagger}|\Omega>$ can be written as,
\begin{equation}
\Delta = \frac{\hbar}{\omega} \biggl( 1 + 0 - \frac{1}{2} + 0 + \frac{a_{4}}{N} + 0 + \frac{a_{6}}{N^{2}} + 0 +................\biggr)
\end{equation} 
where $a_{4}$ and $a_{6}$ are $O(1)$ numbers. This expression justifies our treatment of the potential $V$ as perturbation in the large-$N$ limit. So in the large-$N$ limit the leading term in the gap is 
\begin{equation}
\Delta = \frac{\hbar}{2\omega} \sim \frac{\sqrt\lambda}{N} 
\end{equation}

\subsection{Effect Of The Perturbation \texorpdfstring{$V_{1}$}{V1}}
The potential $V_{1}$ is a gauge singlet and it commutes with the Hamiltonian $H_{0}$. Now instead of treating $H_{0}$ as the unperturbed Hamiltonian we could have treated $H_{0} + V_{1}$ as the unperturbed Hamiltonian. It is easy to see that the exact ground state of the total Hamiltonian is also an exact eigenstate of the Hamiltonian $H_{0} + V_{1}$ with the same eigenvalue because the ground state $|\Omega>$ is annihilated by the potential $V_{1}$. In fact only states which contain both the scalar and gauge excitations are not annihilated by the potential $V_{1}$. So states containing either scalar or gauge excitations only, have the same energy when thought of as eigenstates of the Hamiltonian $H_{0}+V_{1}$. To proceed we need to consider states which contain both scalar and gauge excitations. Since $H_{0}$ and $V_{1}$ commute, to compute the change in the energy we just need to diagonalize the potential $V_{1}$ in a given eigenspace of the Hamiltonian $H_{0}$. 

	The singlet sector of $H_0$ eigenvectors is degenerate as can be seen in eqns.(\ref{normal2}) -(\ref{normal8}). We must therefore construct specific linear combinations of these $H_0$ eigenvectors to simultaneously diagonalize $V_1$. This is necessary since the naive singlet sector harmonic oscillator basis does not diagonalize $V_1$ but only reduces the operator to a block diagonal form, with the blocks corresponding to degenerate eigenspaces of $H_{0}$.

For the singlet sector in the large $N$ limit, the eigenvectors of $V_1$ that correspond to a particular block are composed of eigenvectors of $H_0$ with the same $N$ scaling of their norm. For example (\ref{normal3}) and (\ref{normal4}) both have normalizations $N^{-3/2}$ due to the fact that their inner product sans normalization goes as $N^3$ in the large $N$ limit. This implies that the block that contains these states is a $2$-dim subspace. We shall label eigenstates of this space, which are composed of linear combinations of $\alpha^{\dagger}\beta^{\dagger}\beta^{\dagger}\bar{\alpha}^{\dagger}|\Omega>$ and $\alpha^{\dagger}\bar{\alpha}^{\dagger}Tr(\beta^{\dagger}\beta^{\dagger})|\Omega>$ as $|N^3_{(i)}>$. The $N^3$-norm states are up to a normalization,

\begin{align}
|N^{3}_{(1)}> & \sim\Big [\alpha^{\dagger}\beta^{\dagger}\beta^{\dagger}\bar{\alpha}^{\dagger} \Big]|\Omega>\label{v11}\\
|N^{3}_{(2)}>&\sim\Big [\alpha^{\dagger}\beta^{\dagger}\beta^{\dagger}\bar{\alpha}^{\dagger} - N\alpha^{\dagger}\bar{\alpha}^{\dagger}Tr(\beta^{\dagger}\beta^{\dagger})\Big]|\Omega>\label{v12}
\end{align}

with eigenvalues $\frac{\hbar}{2\omega}N$ and $0$ respectively\footnote{ We are expressing eigenvalues and eigenvectors in the large $N$ limit which means the expressions are ignoring any additive terms of sub-leading order in $N$. For example the eigenvalues resulting in (\ref{v11}) and (\ref{v12}) have lower order contributions besides what is shown.}. Recalling that $\frac{\hbar}{2\omega} = \sqrt{\frac{32\pi}{ Nk}}$ we see in large $N$ the first eigenvalue goes as $\sqrt{\frac{\pi N}{2 k}}$ while the other is approximately zero.

Similarly the states with unnormalized inner products scaling as $N^4$, form a $3\times3$ block. At the $N^{4}$ level in the large $N$ limit we find a zero eigenvalue. The two nonzero eigenvalues of $V_1$ at this level being
\begin{equation}
\sqrt{\frac{8\pi}{ k}}\sqrt{4N}\hspace{1in}\frac{1}{2}\sqrt{\frac{32\pi}{\alpha k}}\frac{\sqrt{N}}{2}.
\end{equation}

These correspond to the eigenvectors
\begin{align}
|N^{4}_{(1)}> & \sim\Big [-2N^{3/2}\alpha^{\dagger}\beta^{\dagger}\bar{\alpha}^{\dagger}\alpha^{\dagger}\beta^{\dagger}\bar{\alpha}^{\dagger} +2N\alpha^{\dagger}\bar{\alpha}^{\dagger}\alpha^{\dagger}\beta^{\dagger}\beta^{\dagger}\bar{\alpha}^{\dagger}\nonumber\\ &\hspace{2.5in}+ \alpha^{\dagger}\bar{\alpha}^{\dagger}\alpha^{\dagger}\bar{\alpha}^{\dagger}Tr(\beta^{\dagger}\beta^{\dagger})\Big]|\Omega>\\
|N^{4}_{(2)}>&\sim\Big [-2N^{1/2}\alpha^{\dagger}\beta^{\dagger}\bar{\alpha}^{\dagger}\alpha^{\dagger}\beta^{\dagger}\bar{\alpha}^{\dagger} +N\alpha^{\dagger}\bar{\alpha}^{\dagger}\alpha^{\dagger}\beta^{\dagger}\beta^{\dagger}\bar{\alpha}^{\dagger}\nonumber\\&\hspace{2.5in} + \alpha^{\dagger}\bar{\alpha}^{\dagger}\alpha^{\dagger}\bar{\alpha}^{\dagger}Tr(\beta^{\dagger}\beta^{\dagger})\Big]|\Omega>
\end{align}

up to a normalization.

 To obtain the eigenvector corresponding to the zero eigenvalue, one should find the appropriate linear combination of $\alpha^{\dagger}\beta^{\dagger}\bar{\alpha}^{\dagger}\alpha^{\dagger}\beta^{\dagger}\bar{\alpha}^{\dagger}|\Omega>$, $\alpha^{\dagger}\bar{\alpha}^{\dagger}\alpha^{\dagger}\beta^{\dagger}\beta^{\dagger}\bar{\alpha}^{\dagger}|\Omega>$, and $\alpha^{\dagger}\bar{\alpha}^{\dagger}\alpha^{\dagger}\bar{\alpha}^{\dagger}Tr(\beta^{\dagger}\beta^{\dagger})|\Omega>$ that is orthogonal to $|N^{4}_{(1)}>$ and $|N^{4}_{(2)}>$.

For the $N^{5}$ level, there is also a $3\times3$ block of singlet states. This time we only have one non-zero eigenvalue in the large $N$ limit
\begin{equation}
\frac{3}{2}\sqrt{\frac{2\pi}{ k}}\sqrt{N}
\end{equation}

corresponding to the eigenvector,

\begin{equation}
|N^{5}> \sim\Big[\frac{3N^{3/2}}{2} \alpha^{\dagger}\Bar{\alpha}^{\dagger}\alpha^{\dagger}\beta^{\dagger}\bar{\alpha}^{\dagger}\alpha^{\dagger}\beta^{\dagger}\bar{\alpha}^{\dagger} + \alpha^{\dagger}\Bar{\alpha}^{\dagger}\alpha^{\dagger}\Bar{\alpha}^{\dagger}\alpha^{\dagger}\Bar{\alpha}^{\dagger}\alpha^{\dagger}Tr(\beta^{\dagger}\beta^{\dagger})\Big]|\Omega>.
\end{equation}
The zero eigenvalues correspond to vectors spanning the plane orthogonal to $|N^{5}>$ in the $\alpha^{\dagger}\Bar{\alpha}^{\dagger}\alpha^{\dagger}\beta^{\dagger}\bar{\alpha}^{\dagger}\alpha^{\dagger}\beta^{\dagger}\bar{\alpha}^{\dagger}|\Omega>$,  $\alpha^{\dagger}\Bar{\alpha}^{\dagger}\alpha^{\dagger}\Bar{\alpha}^{\dagger}\alpha^{\dagger}\Bar{\alpha}^{\dagger}\alpha^{\dagger}Tr(\beta^{\dagger}\beta^{\dagger})|\Omega>$,\\and $\alpha^{\dagger}\bar{\alpha}^{\dagger}\alpha^{\dagger}\bar{\alpha}^{\dagger}\alpha^{\dagger}\beta^{\dagger}\beta^{\dagger}\bar{\alpha}^{\dagger}|\Omega>$ basis, up to a normalization.

We can see from the above analysis that the shifts in the energies of few low lying states is always positive semi-definite. We see that the $V_1$ eigenvalues in the singlet sector do not ruin the analysis of the previous section. The eigenvalues are positive semi-definite in the large-$N$ limit and make the perturbative analysis of $V$ more robust.

\subsection{Counting Of States}
Since in the large-N limit the states containing scalar excitations are much heavier than the states containing pure gauge excitations we can compute the number of states in the low energy sector by counting only the states where there is no scalar excitation.  (The effect in equation \ref{shift}  should be representable as a shift in the pure gauge harmonic oscillator frequency.)
This is exactly the counting problem for the quantum mechanics of one hermitian matrix model with harmonic potential which has been solved in the classic paper \cite{Brezin:1977sv}.  That model in the singlet sector is exactly equivalent to $N$ free fermions in a harmonic potential. We note that the Fermi energy of the system is given by, $E_F = \sqrt\lambda$. Again this problem is easy to solve.

More precisely, we can take a limit with $k\to\infty$ and scale the temperature
so that the low-energy states of the Chern-Simons/matter-zero-mode system, and
only those states, contribute in the thermal ensemble. If we take $k$ large and
 scale the temperature $T$ according to the limit
\begin{equation}
{E_F = \sqrt\lambda} \ll T \ll m_{\rm scalar}\text{ ,}
\end{equation}
then the Boltzmann factor suppressing the contributions of the KK modes is $\exp{\{- \frac{{m_{\rm scalar}}}{T}\}}$, so the KK modes can be ignored altogether; while the number of $\beta$-oscillator  states with energies below the temperature $\delta$ grows as $\propto ( {T  \sqrt{k} })^N / N!$ and the entropy is given by,

\begin{equation}\label{ourentropyb}
S \simeq N \big [  \log (T/\sqrt{\lambda}) + O(1) \big ].
 \end{equation} 
  At this temperature, a simple semiclassical argument suffices to derive this scaling.
 The harmonic oscillator frequency goes as $\omega \equiv 1 / \sqrt{Nk}$.
  The classical partition function
for $N$ identical harmonic oscillators goes as
\begin{equation}
Z_{\rm classical}
= {\frac{1}{N!}}\int dp\llo i dq\llo i \expp{- \sum\llo i (p\llo i \sqd + \omega\sqd q\llo i\sqd) / T}\ .
\end{equation}
Upon performing the integral, one gets a $T\uu 1$ for each $p,q$ pair of 
integrals, and one gets an $\omega\uu{-1}$ from 
every $q$ integral.  The total factor is then $(\pi T / \omega)\uu N / N!$.  The log of that is
$ {\rm ln}(Z_{\rm classical}) =
N ~ {\rm ln}(T / \sqrt{\lambda})$.  This gives (\ref{ourentropyb}) exactly.
We observe that this is well-behaved in the 't Hooft limit.

We can also consider a different limit where the temperature is much less than the Fermi energy. In this regime the entropy of the system is given by,
\begin{equation}
S \sim \frac{NT}{\sqrt \lambda}
\end{equation}  
This is the leading term in the Sommerfeld expansion of the entropy.


\chapter{Discussion}

\subsection{New Light States}

We have shown that on a spatial $T^2$  the matter-Chern-Simons conformal field
theory based on a single scalar field in the
fundamental representation has a set of low-lying states with energy gaps of order ${\frac{1}{{\sqrt{
Nk}}}}$ (for the free scalar) or $\frac{1}{k}$ (for the critical scalar).  As a result, there is a divergent degeneracy of states in the limit where
the level $k$ goes to infinity, at fixed rank $N$ of the gauge group.

The Vasiliev theory successfully describes correlation functions of
higher-spin conserved currents of the
infinite$-k$ limit on $\IR^3$, as well as its partition function on $S\uu 1 \times S\uu 2$. However a consistent proposal for a gravitational dual description for 
the Chern-Simons-matter CFT analyzed in this article
should provide a bulk realization for the CFT partition function on general boundary geometry, including the light states we have found and the parametrically large (in $k$) entropy associated with them.   The entropy of our system on
$T\uu 2$ is (\ref{OurEntropyA}), which diverges for fixed $N$ and
large $k$.  This agrees with the large-$k$ entropy of the pure Chern-Simons sector
(\cite{Witten:1999ds}, \cite{Elitzur:1989nr}) 
\begin{equation}
{\rm ln}(Z) \simeq   (N-1)~ {\rm ln}(k) - {\rm ln}((N-1)!) + O(k\uu{-1})\ .
\label{CSDegenTorus}\end{equation}
which for large $N$ is
\begin{equation}
{\rm ln}(Z) \propto N ~{\rm ln}(\lambda\uu{-1})\ .
\end{equation}
The addition of matter does not affect the parametric $N~{\rm ln}(k)$ divergence of
the entropy. 

\subsection{Vasiliev As A imit Of String Theory}

It is clear that the Vasiliev theory does not by itself contain the
degrees of freedom corresponding to the large entropy of the CFT on spatial
slices of genus $g =1$.   Nor can any deformation of the theory with deforming interaction 
terms in the action or equations of motion that are proportional to positive powers of $\lambda$
(for example \cite{YinSlides, GiombiSlides}).
 There are no fundamental fields of Vasiliev theory that
could generate such a topology-dependent divergence, and any solitonic collective
excitations should have masses that scale with negative, rather than positive powers of
$\lambda$.  The proposal to derive Vasiliev gravity as a limit of string theory in which
stringy physics decouples altogether, appears to work under certain circumstances, but
not universally.  The limit $\lambda \to 0$ is not a conventional decoupling limit for
string theory like the infinite tension limit $\alpha^\prime\to 0$, where string oscillator excitations  decouple in the usual Wilsonian sense.
Rather, $\lambda\to 0$ can be thought of as a limit in which the string tension goes
to zero and each string "bit" moves as an independent particle.  However in non-simply-connected
spaces, there is a topological constraint which does not allow all the string bits
to move independently, when the string winds a noncontractible cycle.  As a result,
there is an infinite tower of independent states distinguished by their winding, but with a parametrically low cost for states with arbitrarily large winding. 

Holographic duality suggests that the large-$k$ divergence is related to an
incompleteness of the Vasiliev theory.  For a bulk with
boundary $T\uu 3$, there is a singular solution of Einstein gravity with negative cosmological
constant that is also a solution of Vasiliev gravity, in which the spatial $T\uu 2$ shrinks
to zero size.  It is natural to associate this singularity to the light states\footnote{We thank Tom Banks for conversations on this point}.   Wrapped strings and T duality resolve this singularity in standard bulk string theory situations, and we infer that a consistent ultraviolet
completion of the Vasiliev action is likely to involve string degrees of freedom to
account for the entropy.
The connection between the large-$k$ degeneracy and nonvanishing
fundamental group, for instance, may suggest an identification of our light states with the closed
string sector of the topological open-closed string theory proposed in \cite{YinSlides, TopologicalStringVasiliev}.

\subsection{Higher Genus}
For higher genus, $g\geq 2$, the quartic
interaction in the Wilson-Fisher theory
stabilizes the scalars independent of $k$ against their conformal
coupling $\frac{1}{8}~{\tt Ricci}_3~\phi\sqd$ to the Ricci scalar
interaction, and gives their energies
a gap of order $1$.  The quantum mechanical techniques discussed above should
apply here and give gaps in the gauge field sector of order $\frac{1}{k}$.
As discussed earlier, for a massive scalar field we expect that at 
temperatures $  \exp(-M)  \ll T \ll 1$ the entropy
should reduce to that of the pure Chern-Simons system.   Here  the effective $M \sim 1$ so we do not have parametric control, but the largest effect the matter field could have is if its vev where large, effectively Higgsing the system down to an SU(N-1) pure Chern-Simons theory\footnote{This is what will happen in the free massless scalar system where the  effect of the $\frac{1}{8}~{\tt Ricci}_3~\phi\sqd$ term is stabilized by the $\lambda^2 \phi^6$  discussed in \cite{Aharony:2011jz}  at a large vev  of order $\phi^2 \sim \frac{1}{\lambda}$ . \ } .  So we can use the pure Chern-Simons entropy as a good estimate of the entropy of our system at large $N$.

The entropy of the Chern-Simons theory on surfaces of genus $g$
\cite{{VerlindeProceedings}}\footnote{as cited in \cite{Moore:1989vd}}
is \begin{equation}
{\rm ln}(Z) \simeq  (g-1) (N\sqd-1)~ {\rm ln}(k) + O(k\uu 0)\ .
\label{CSDegenHigherg}\end{equation}
To understand this formula, we can use semiclassical analysis
(see, \it e.g., \rm pg. 96 of \cite{Moore:1989vd})
 to determine
the leading large-$k$ behavior of the number of states.  For a compact phase space,
the number of quantum states is given, for small Planck
constant $\hbar$, to the volume of phase space in units of $\hbar$:
\begin{equation}
n\llo{\rm states} = {\rm (const.)} \cdot \frac
{{{\tt Vol}\llo{\rm phase~space}}}{{\hbar\uu{{\frac{\tt Dim.}{2}}} }} 
~ \big [ 1 + O(\hbar) \big ] \ ,
\end{equation}

For Chern-Simons theory in canonical quantization, the phase space is the moduli space 
${\cal M}\llo{G,g}$
of flat $G$-connections on the spatial slice $\Sigma\llo g$, and the Planck constant
$\hbar$ is proportional to $\frac{1}{k}$.   The volume of the moduli space of flat connections is $k$-independent, and its dimension \cite{Witten:1988hf} is 
\begin{equation}
{\tt Dim.}({\cal M}_{G,g}) = (2g-2) ~{\rm Dim.}(G)\ .
\end{equation}
Therefore the number of quantum states, in the large-$k$ limit, is
\begin{equation}
Z = n\llo{\rm states} = {\rm (const.)} \cdot k\uu {\hh {\tt Dim.}({\cal M}_{G,g}) } 
 \big [ 1 + O(k\uu{-1}) \big ]
\end{equation}
and the entropy is
\begin{equation}
{\rm ln}(Z) = (g-1)~(N\sqd - 1) ~ {\rm ln}(k) + O(k\uu 0)\ .
\label{GenusgCSEntropy}
\end{equation}
The coefficient of the ${\rm ln}(k)$ term does not depend on the numerical, $k$-independent
factor in the volume of ${\cal M}_{G,g}$, only on its volume.  This order $N\sqd$ entropy overwhelms the entropy of the matter.  This $N\sqd~ {\rm ln}(k)$
divergence of the entropy is striking, because it is larger than any gravitational contribution
to the entropy, which would scale at most as $\frac{1}{G_N} = N$.
 
\subsection{Degrees Of Freedom}
We want to emphasize that the divergent entropy at large $k$ is not attributable
to the nonpositive scalar curvature of the boundary in the case where the boundary
is $S\uu 1 \times \Sigma\llo g, ~g\geq 1$.  It is known that CFT partition functions on
such geometries need not be convergent, and the corresponding bulk instabilities have
been studied in some cases. \cite{Maldacena:1998uz, Seiberg:1999xz, Witten:1999xp}.  However
the large-$k$ divergence of the entropy in CSM theory cannot be
an artifact of vanishing or negative
scalar curvature, as the instability is not present in some cases where
the entropy is nonetheless still logarithmically divergent with $k$.
In the case of the critical model, for instance, the unstable direction of the scalars is 
always stabilized independently of $k$, by
the quartic interaction.

In the case of the free scalar or the critical scalar, the partition function on $S\uu 3$ is
stabilized by the conformal coupling but still displays a ${\rm ln}(k)$
divergence in the free energy \cite{Kac:1988tf,Klebanov:2011gs}, 
\begin{equation}
F = - {\rm ln} (Z\llo {S\uu 3}) \simeq + \frac{N(N-1)}{2} ~ {\rm ln}(k) + O(k\uu 0)\ .
\end{equation}
This comes entirely from the Chern-Simons sector, as the conformal coupling of the scalars
allows them to contribute only terms analytic in $k$.  The value of $F = - {\rm ln}(Z\llo{S\uu 3})$
for various conformal and superconformal field theories in three dimensions has been an
object of much recent study (\cite{Jafferis:2010un, Jafferis:2011zi, 
Closset:2012vg, Myers:2010tj}), particularly the investigation of the hypothesis that $F$ is a measure
of the number of degrees of freedom of the system that decreases along renormalization
group flows, analogously to the $c$ coefficient in two dimensions \cite{Zamolodchikov:1986gt}
or the $a$ coefficient in four dimensions \cite{Cardy:1988cwa, Komargodski:2011vj}.  (A 
general proof of the equivalence between entanglement entropy in a 3-dimensional CFT and its free energy on $S\uu 3$ has been presented in \cite{Casini:2011kv}.)  With
this interpretation, we see again that there
are of order $N\sqd~{\rm ln}(k)$ degrees of freedom in the Chern-Simons-matter system \footnote{The tension between the Vasiliev bulk interpretation and $N^{2}$ degrees of freedom has also been emphasized by Klebanov (private communication)} \cite{Klebanov:2011gs}, attributable to the topological sector.

\subsection{Light States In ABJM Theory}
There have been proposals \cite{Giombi:2011kc, YinSlides} to derive Vasiliev gravity as a limit
of the ABJ theory \cite{Aharony:2008gk}.
For Chern-Simons-matter theories with ultraviolet-complete string
 duals, this same large-k divergence
 on a torus is natural when interpreted in light of string- and M- theory.  We can
for instance compactify the ABJM model on $T\uu 2$ rather than $S\uu 2$ spatial
slices, and ask what the holographic duality predicts, qualitatively, for the entropy.

\llsk Without doing a fully controlled calculation, we simply observe that the total entropy of the 
AdS should be approximately extensive in the radial direction, and that the entropy
at every point in the radial direction is divergent in the limit $k\to\infty$ with $N$ large
but fixed.  At any point in the radial direction, there are new states due to the topology that become
light at large $k$, corresponding to membranes that wrap the Hopf fiber of the $S\uu 7
 / Z\llo k$, and one direction of the longitudinal $T\uu 2$.  At large $N$ these states
 are still very heavy, but at fixed $N$, however large, the proper energy of these states,
 at any fixed point in the radius, goes to
 zero at large $k$, because the size of the Hopf fiber is $1/k$ in 11-dimensional Planck units.
The fixed-$N$, infinite-$k$ entropy contributed by any point in the radial direction diverges,
and this is visible in every duality frame.  In the type IIA duality frame, the Hopf
fiber is invisible, having been turned into the M-direction, but the
AdS radius in string units is inversely proportional to $k$, at fixed $N$.  Therefore fundamental
strings wrapping a cycle of the longitudinal torus become light, and make a divergent contribution
to the entropy.
As the longitudinal torus shrinks further towards the infrared, we T-dual to type IIB and
the T-dual radius decompactifies.  In this duality frame, there is a divergent entropy due
simply to the decompactification of the emergent T-dual dimension.

\llsk We could also ask what is the entropy of $N$ M2-branes wrapped
on $T\uu 2$ and probing a $\IC\uu 4 / \IZ\llo k$
singularity in M-theory, without taking the near-horizon limit or taking the
back-reaction into account.  This is a different approximation, but also illuminating because
we see again a naturally emerging divergent entropy at large $k$.  Reducing on
the $T\uu 2$ from M-theory to type IIB, we transform the M2-branes into $N$ particles
each carrying one unit of Kaluza-Klein momentum on the T-dual direction.  Even restricting
ourselves to normalizable states that saturate the BPS bound in this framework, we see an
entropy that diverges at large $k$.  Each of
these particles can occupy any of $k$ massless twisted sectors of the orbifold, and still
saturate the BPS bound for a Kaluza-Klein momentum unit.  Since each of $N$ interchangeable
particles can inhabit one of $k$ possible states, the total degeneracy of such
quantum states gives a contribution to the partition function of
\begin{equation}
\Delta Z \gtrsim k\uu N / N!\ ,
\end{equation}
because the symmetry factor by which one divides is no more than $N!$.
This corresponds to a contribution to the entropy of
\begin{equation}
\Delta {\rm ln}(Z)  \gtrsim N ~ {\rm ln}(k) - {\rm ln}(N!) \simeq N ~ {\rm ln}(\lambda\uu{-1}) \ , 
\end{equation}
which is remarkably similar to the Chern-Simons degeneracy (\ref{CSDegenTorus}).  

This counting is most likely an underestimate.  Though interactions between particles
may in principle lift some of these BPS vacua, a massive perturbation lifting
the flat directions allows us to reduce to Chern-Simons theory in the unhiggsed
vacuum and compute the supersymmetric index.  This classical vacuum alone
contributes to the index with the full degeneracy of the pure Chern-Simons system on
the torus for $U(N) \times U(N)$ at level $k$.

\subsection{\texorpdfstring{$N^{2}$}{N2} Entropy}

The $N\sqd$ scaling of the partition functions on $S\uu 3$ and $S\uu 1 \times
\Sigma\llo g$ with $g\geq 2$ indicates difficulties for the interpretation of the CSM theory in terms
of Vasiliev gravity.  The four-dimensional Newton constant $G_N$ as inferred from
stress tensor correlators is of order $1/N\uu 1$ in units of the AdS scale,
rather than $1/N\sqd$, so the order $N\sqd$ entropy cannot be attributed to a gravitational
effect like a horizon entropy if $L_{AdS}/N$ is indeed the true Newton constant of the
theory.  
Entropies proportional to $N\sqd$ are characteristic of matrices.  Here we see that the
vectorlike holography of Chern-Simons-matter systems rediscovers its matrixlike character.
In terms of the proposal to complete Vasiliev gravity in terms of an open-closed topological
string theory \cite{Aharony:2011jz,TopologicalStringVasiliev, YinSlides}, the $N\sqd$ scaling of
the entropy is an
indication that the graviton should reside in the closed string, rather than open string sector,
of such a theory, in accordance with the principle that it is the gravitational force that must
always carry the largest entropy \cite{Bekenstein:1980jp} and weakest interaction
\cite{ArkaniHamed:2006dz} of any sector of a quantum gravitational theory.  Reconciling
this with the identification $G\llo N \propto 1/N$ apparently dictated by stress tensor
correlation functions is a challenge for any proposal such as \cite{TopologicalStringVasiliev}.

\subsection{Higher Genus And Hyperbolic Black Holes}
To understand the bulk geometry dual
to this spatial geometry, mod out the bulk, presented in hyperbolic slicing, by the action of a
 discrete group.  This is a valid operation in any gravity theory, including Vasiliev gravity.  The correspponding bulk geometry is  a ``zero mass" hyperbolic black hole\cite{Emparan:1999gf}.
 The boundary dual of the "zero mass" black hole corresponds to the Chern-Simons Matter system on the Riemann surface at the temperature $\frac{1}{2\pi R_{\rm curv}}$, where
 $R\llo{\rm curv}$ is the curvature radius of the spatial slices \cite{Emparan:1999gf, Horowitz:2009wm, Myers:2010xs}.
 The point of unbroken gauge symmetry in the matter theory is unstable due to the ${\tt Ricci}\llo 3 \phi^2$ coupling, but interaction terms stabilize the scalar vev.
 In the critical case, for example, the $\phi\uu 4$ coupling stabilizes the
 scalars, independent of $k$.   (For the "free" scalar theory, the theory is not in fact strictly free either, due to the $\phi\uu 6$ interaction \cite{Aharony:2011jz},
which stabilizes the zero mode.)  In this case, there are no singular shrinking cycles in the bulk gravitational metric to blame for the light states but there is a finite area black hole horizon.  As mentioned above the normal geometric horizon entropy $S \sim 1/G_N \sim N$ is insufficient to account for the $N^2$ entropy found in the boundary theory.

It seems likely that tensionless winding strings are again relevant in this case.  If
we fix a point in the AdS radial direction,
the density of winding string states grows exponentially
as a function of length \cite{McGreevy:2006hk, Milnor, Margulis},
 so that there 
is a Hagedorn density with transition temperature $T_H \propto \ell\llo 0 / \alpha^\prime$,
where $\ell\llo 0$ is the proper size of the longitudinal spatial slices.
In the zero-tension limit $\alpha^\prime\to\infty$, the Hagedorn temperature goes to zero.  
At arbitrarily low temperatures, the formal entropic contribution of the winding states
exceeds the contribution of their partonic constituents, 
signaling that  the string
thermodynamics should break down in favor of an order $N\sqd$ entropy counting the
constituents, perhaps crossing over to a horizon entropy involving the closed string $G_N \sim 1/N^2$.

\subsection{RG Flow}
Understanding the renormalization group flow
of the theory to pure Chern-Simons theory may be useful for understanding the holographic dynamics of CSM theory, including the order $N\sqd$ entropy and the ${\rm ln}(k)$ divergence.
For many 3-manifolds, the
holographic dual to pure Chern-Simons theory is understood in terms of the topological
string \cite{Gopakumar:1998ki}, including cases where an order $N\sqd$ free energy is present. 
 For the
case of $S^3$ for example, there is a well-controlled dual in terms of the topological
string on the resolved conifold, where the singular behavior of the $k\to \infty$ limit
arises from the vanishing of the complexified K\"ahler parameter of the blown-up $\IC\IP\llo 1$ base of the resolved conifold, leading to unsuppressed contributions of worldsheet instantons.

\subsection{\texorpdfstring{$T^{3}$}{T3} Modular Invariance Constraints}
 The motivation to consider coupling the matter CFT to
large-$k$ Chern-Simons theory was originally to take the limit $k\to \infty$, in order to
implement a projection to the singlet sector of the operator spectrum.
\cite{Giombi:2011kc, Shenker:2011zf}.
 Given the difficulties of promoting this construction to a fully well-defined local
quantum field theory, one might wonder whether there may be some construction
the singlet-projected matter QFT without any additional states, perhaps some kind of
BF theory.  We can answer this question in
the negative.  For the partition function on $T\uu 3$, there is a simple demonstration
\cite{ChannelCovariance} that
such a construction cannot exist at all, based on modular invariance.  Treating one of the
three cycles, say $\theta^3$ as the Euclidean time direction, the singlet-projected partition
function is computed by taking the full partition function with boundary conditions such that
the matter fields are periodic up to a particular group transformation $g \in G$
around the $\theta\uu 3$ cycle, and then averaging (\it not \rm summing!) over $G$.  This
procedure is the same regardless of the shape and size of the $T\uu 3$.  However
a consistent, local quantum theory must have the same partition function when quantized
in any "channel", \it i.e. \rm with respect to the Hamiltonian and Hilbert space defined
by any foliation of the manifold.  If we switch the roles of $\theta\uu 1$ and $\theta\uu 3$,
treating the former as Euclidean time and the latter as a spatial direction, then
the average over boundary conditions on $\theta\uu 3$ generates a partition function
that is not only asymmetric with the theory in the $\theta\uu 3$ channel, but does not
have any consistent Hilbert space interpretation in the $\theta\uu 1$ channel whatsoever.

\fakeindent That is, let 
\begin{equation}
Z[L_1,  L_3 ; g_3] = {\rm partition~function~with~radii~\it L_1~\rm and~\it L_3, \rm ~~and~periodicity~\it g_3\rm ~along~\theta\uu 3}\ .
\end{equation}
Suppose a local CFT exists such that the Hilbert space on any slice always contains just exactly
the singlet sector of the full matter theory, and nothing more.  
Then the partition function for the singlet sector on a spatial slice with radius $L_1$ 
at temperature $T = \frac{1}{L_3}$ is
\begin{equation}
Z_{\rm singlet}[L_1, L_3] = \frac{1}{|G|}~\sum_{g_3\in G}~Z[L_1,  L_3 ; g_3].
\label{SingletProjection}
 \end{equation}
(If we take a continuous group $G$ with a discrete one, the formula is the same
except that the sum is replaced with an integral and the cardinality $|G|$ of
$G$ is replaced with the Haar volume.)  But there are at least two things wrong with this possibility.  First, the formula is not invariant under $L\llo1 \leftrightarrow L\llo3$.  Secondly,
if we fix $L\llo 2, L\llo 3$ and take $L\llo 1\to \infty$, the partition function does \it not \rm 
take the form of a sum of exponentials of $L\llo1$ with positive integer coefficients,
which as a consistency condition of a quantum field theory of any kind, it must.

\llsk That is to say, if any kind of Hilbert space exists at all in the $\th\uu 1$ channel, 
then it must be possible to write the partition function in the form
\begin{equation}
Z_{\rm any~consistent~theory}[L_1, L_3] = \sum_{{\rm states~in~\th\uu 1-channel}}~\expp{- L_1~ E^{(\th\uu 1-{\rm channel})}}\ ,
\label{ConsistentForm}\end{equation}
  where the energies $E^{(\th\uu 1-{\rm channel})}$ may depend on $L_3$ but
  the coefficients are 1 (or another positive integer, if there are degeneracies).  However
  the partition function (\ref{SingletProjection}) is realized in the $\th\uu 1$
  channel as an average (as opposed to a sum) of partition functions with different
  periodicities along the spatial cycle $\th\uu 3$.  Therefore it \it cannot \rm have the form
  (\ref{ConsistentForm}), unless the ground state energy of the unprojected theory in
  the $\th\uu 1$ channel would be independent of the boundary condition $g\llo 3$, which is
  not the case in general, and certainly not for free bosons or
  fermions.    Therefore the coefficient of the leading exponential of
  $L\llo1$, which in a consistent quantum theory encodes the ground state degeneracy 
  in the $\th\llo1$ channel, is fractional for this theory, signaling the nonexistence of a
  Hilbert space of any kind in this channel, let alone a Hilbert space isomorphic to the
  one in the $\th\uu 3$ channel. 
  
   This argument most directly rules out the existence of a consistent partition function
   for $T\uu 2$ spatial slices, but the inconsistency cannot be confined to
   this case alone; the existence of cobordisms -- that is, smooth geometries interpolating
   between spatial slices of different topology -- define charge-conserving maps 
   Hilbert spaces on the torus and on higher-genus Riemann surfaces.  Thus, if a local
   CFT did exist that contained only the singlet sector of the original matter CFT
   on slices of higher genus, the path integral on the interpolating manifold could
   be used to define the singlet theory on $T\uu 2$ spatial slices as well; but we know that this
   theory can have no consistent definition.
   
    This situation is similar to the case of two-dimensional CFT with global
   symmetries, where the truncation of the theory to the singlet sector is
   not consistent with modular invariance unless twisted states are added to supplement
    the Hilbert space.  The number of states that must be added increases with the cardinality
    of the group, leading to an infinite entropy when the group is continuous.  The
    light states of the Chern-Simons sector at large $k$ can be identified as analogous
    to the plethora of low-lying twisted states that appear when one tries to construct a
    modular-invariant orbifold by a group with a cardinality $|G|$ that is going to infinity. 
    
 \subsection{Light States In the \texorpdfstring{$W_N$}{WN} Models and their Gravity Duals}
 The lower dimensional duality described by a $W_N$ boundary theory \cite{Gaberdiel:2010pz,Gaberdiel:2011wb,Gaberdiel:2011nt,Ahn:2011pv,Chang:2011mz,Papadodimas:2011pf} has light states with such an origin.  These states have been described in \cite{Gaberdiel:2011aa} as twisted states in a continuous $SU(N)$ orbifold in the boundary theory, involving flat connections like the ones relevant in Chern-Simons theory.  The authors of \cite{Castro:2011iw,Gaberdiel:2012ku} 
have interpreted thses states in the bulk by \ as due to ``conical excess" solutions.  Here these states appear directly on the $S^1$ spatial geometry and are necessary for a consistent modular invariant solution and hence for the finite temperature black hole dynamics.  The light states we consider are not necessary for the bulk thermal dynamics with $S^2$ boundary.  No hint of them, or of light strings that could wrap the $T^2$ are visible there \cite{Shenker:2011zf, Giombi:2011kc}.

\subsection{dS/CFT}
One area in which Vasiliev gravity has been applied has been to the study of 
holographic cosmology, through the dS/CFT correspondence.  A
nonunitary version of the Chern-Simons-matter theory, based on replacing the
scalar bosons with scalar fermions, has been proposed as a holographic dual
for Vasiliev gravity in de Sitter space in 4 dimensions  \cite{Anninos:2011ui}.
The topology-dependent divergence of
the partition function noted in this article may have relevance for the meaning of
this correspondence, particularly for any sort of probabilistic interpretation of it \cite{Maldacena:2002vr}.

\subsection{Supersymmetric Extensions}
It would be interesting to analyze supersymmetric extensions
of the matter-Chern-Simons
theory with various amounts of supersymmetry,
from the minimal (${\cal N} = 1$ or ${\cal N}=2$) case \cite{Schwarz:2004yj}
to the almost-maximal (${\cal N}= 5,6$) case of the general ABJ \cite{Aharony:2008gk}
and ABJM \cite{Aharony:2008ug} theories,
and the maximal (${\cal N} = 8$) case
of the $k=1$ ABJM theory.
The addition of supersymmetry introduces new technical issues (for instance, exactly
flat directions on the moduli space of vacua) while promising a greater
degree of control over quantum effects.  A Vasiliev-type gravity dual has also been proposed
for the supersymmetric Chern-Simons-matter theory \cite{YinSlides}.

\subsection{Decoupling}
There is a sense in which these light states decouple at $k=\infty$.   From (\ref{qmmass}) we see that at $k = \infty$ the holonomy becomes an infinitely low-frequency degree of freedom and hence does not move.  Scattering of KK scalar modes will not change its value.  This does not mean that these light states can be removed from the theory.  To remove them would
be to fix definite values for the holonomies on the cycles of the spatial
slice.  But fixing the holonomy on spatial slices of the CFT does not
define a sensible bulk theory of any kind: With such a definition, the $T^3$ partition function would not be modular invariant;  the limit as $k \rightarrow \infty$ of physical quantities like the hyperbolic black hole entropy would not be smooth; and the higher spin correlators would not be uniquely defined independent of the order $N$ parameters
by which the holonomies are characterized.

\subsection{Condensed Matter Applications}
Finally we should note that our results may be useful in analyzing condensed matter quantum hall systems where another set of degrees of freedom become light and changes the quantum hall dynamics.  For a recent example, see \cite{barkeshli:2012ja}.

%
\end{itemize}




\chapter{Conclusion}

\hspace{0.25in}In this work we have discussed several Technical aspects and notions necessary for proposed dual models of de SitterSpace in the String theory Formalism. While what is discussed here is necessary in formulating such dual models and has more far reaching applications in String Theory at large it is by no means sufficent. Clearly a great deal more work must be done to determine if a model of de Sitter can be made in the String Theory formalism and what that model will be.

\appendix
\part{Appendix}
\chapter[Properties of the Upsilon Function]{Properties of the $\Upsilon_b$ Function}
\label{upsilonapp}
The function $\Upsilon_b$ has now become standard in the literature on Liouville theory, but for convenience we here sketch derivations of its key properties.  The function can be defined by 
\begin{equation}
\label{logupsilonapp}
\log\Upsilon_b(x)=\int_0^\infty\frac{dt}{t}\left[(Q/2-x)^2 e^{-t}-\frac{\sinh^2((Q/2-x)\frac{t}{2})}{\sinh{\frac{tb}{2}}\sinh{\frac{t}{2b}}}\right]\qquad0<\mathrm{Re}(x)<\mathrm{Re}(Q).
\end{equation}
Here $Q=b+\frac{1}{b}$.  The definition reveals that $\Upsilon_b(Q-x)=\Upsilon_b(x)$.  When $x=0$ the second term in the integral diverges logarithmically at large $t$, and at small but finite $x$ it behaves like $\log x$.  $\Upsilon_b$ therefore has a simple zero at $x=0$ as well as $x = Q$.  

To extend the function over the whole $x$-plane, we can use the identity\footnote{This identity is derived in Appendix \ref{loggammazapp}.}
\begin{equation}
\label{gammaint}
\log\Gamma(x)=\int_0^\infty \frac{dt}{t}\left[(x-1)e^{-t}-\frac{e^{-t}-e^{-x t}}{1-e^{-t}}\right] \qquad \mathrm{Re}(x)>0
\end{equation}
to show that in its range of definition $\Upsilon_b$ obeys
\begin{align}\notag \label{recrel}
\Upsilon_b(x+b)=&\gamma(bx)b^{1-2bx}\Upsilon_b(x)\\
\Upsilon_b(x+1/b)=&\gamma(x/b)b^{\frac{2x}{b}-1}\Upsilon_b(x).
\end{align}
Where:
$$\gamma(x)\equiv\frac{\Gamma(x)}{\Gamma(1-x)}$$
These recursion relations are the crucial property of $\Upsilon_b$ from the point of view of Liouville theory, among other things they are what allow a solution of Teschner's recursion relations to be expressed in terms of $\Upsilon_b$.  The recursion relations also show that the simple zeros at $x=0,Q$ induce more simple zeros at $x=-m b-n/b$ and $x=(m'+1)b+(n'+1)/b$, with $m,m'$ and $n,n'$ all non-negative integers.
It it is also useful to record the inverse recursion relations:
\begin{align}\notag
\label{inverserecursion}
\Upsilon_b(x-b)=&\gamma(bx-b^2)^{-1}\,b^{2bx-1-2b^2}\Upsilon_b(x)\\
\Upsilon_b(x-1/b)=&\gamma(x/b-1/b^2)^{-1}\,b^{1+\frac{2}{b^2}-\frac{2x}{b}}\Upsilon_b(x).
\end{align}

We will also need various semiclassical limits of $\Upsilon_b$.\footnote{We thank A. Zamolodchikov for suggesting the use of (\ref{gammaint}) in the first of these derivations.}  Rescaling $t$ by $b$ and using the identity 
$$\log x=\int_0^\infty\frac{dt}{t}\left[e^{-t}-e^{-xt}\right]\qquad\mathrm{Re}(x)>0,$$
we see that
\begin{align} \nonumber
b^2 \log &\Upsilon_b(\eta/b+b/2)=-\left(\eta-\frac{1}{2}\right)^2\log b\\
&+\int_0^\infty \frac{dt}{t}\left[\left(\eta-\frac{1}{2}\right)^2e^{-t}-\frac{2}{t}\left(1-\frac{t^2 b^4}{24}+\dots\right)\frac{\sinh^2\left[(\eta-\frac{1}{2})t/2\right]}{\sinh \frac{t}{2}}\right].
\end{align}
When $0<\mathrm{Re}(\eta)<1$, the subleading terms in the series $1+t^2 b^4+\dots$ can be integrated term by term, with only the $1$ contributing to nonvanishing order in $b$.  From the identity (\ref{gammaint}), we can find
$$F(\eta)\equiv \int_{1/2}^{\eta}\log\gamma(x)dx=\int_0^\infty\frac{dt}{t}\left[(\eta-1/2)^2e^{-t}-\frac{2}{t}\frac{\sinh^2((\eta-1/2)\frac{t}{2})}{\sinh(\frac{t}{2})}\right] \quad 0<\mathrm{Re}(\eta)<1,$$
so using this we find the asymptotic formula:
\be
\label{upsilonbetterasymp}
\Upsilon_b(\eta/b+b/2)=e^{\frac{1}{b^2}\left[-(\eta-1/2)^2 \log b+F(\eta)+\O(b^4)\right]} \qquad 0<\mathrm{Re}(\eta)<1.
\ee
In particular if we choose $\eta$ to be constant as $b\to0$ only caring about the leading terms, then we have
\begin{equation}\label{bingo}
\Upsilon_b(\eta/b)=e^{\frac{1}{b^2}\left[F(\eta)-(\eta-1/2)^2\log b+\O(b \log b)\right]}\qquad 0<\mathrm{Re}(\eta)<1,
\end{equation}
which is useful for our heavy operator calculations in section 4.

For light operator calculations we will also be interested in the situation where the argument of $\Upsilon_b$ scales like $b$.  Looking at the $b\to 0$ limit of the first recursion relation in (\ref{recrel}) we find
\be
\Upsilon_b((\sigma+1) b) \approx \frac{1}{\sigma b}\Upsilon_b(\sigma b).
\ee
One solution to this relation is
\be
\label{guess}
\Upsilon_b(\sigma b)\approx \frac{b^{-\sigma}}{\Gamma(\sigma)} h(b),
\ee
where $h(b)$ is independent of $\sigma$.  Unfortunately this solution is not unique since we can multiply it by any periodic function of $\sigma$ with period one and still obey the recursion relation.  We see however that it already has all of the correct zeros at $\sigma=0,-1,-2,...$ to match the $\Upsilon_b$ function, so we might expect that this periodic function is a constant.  This periodic function in any case is nonvanishing and has no poles, so it must be the exponential of an entire function.  If the entire function is nonconstant then it must grow as $\sigma \to \infty$, which seems to be inconsistent with the nice analytic properties of $\Upsilon_b$.  In particular (\ref{upsilonbetterasymp}) shows no sign of such singularities in $\eta$ as $\eta \to 0$.  We can derive $h(b)$ analytically, up to a $b$-independent constant which we determine numerically.  The manipulations are sketched momentarily in a footnote, the result is
\begin{equation}
\label{sigmaasymp}
\Upsilon_b(\sigma b)=\frac{C b^{1/2-\sigma}}{\Gamma(\sigma)}\exp\left[-\frac{1}{4b^2}\log b+F(0)/b^2+O(b^2 \log b)\right].
\end{equation}
The numerical agreement of this formula with the asymptotics of the integral (\ref{logupsilonapp}) is excellent;in particular we find $C=2.50663$.\footnote{To do this numerical comparison, it is very convenient to first note that for $\Re (\tilde{\sigma} b)>0$, we have $\log \Upsilon_b\left((\tilde{\sigma}+\frac{1}{2}) b\right)=-(\frac{1}{2b}-\tilde{\sigma} b)^2 \log b+\frac{1}{b^2}F(\tilde{\sigma} b^2)+I(\tilde{\sigma},b)$, with $$I(\tilde{\sigma},b)=\int_0^\infty \frac{dt}{t}\left(\frac{2}{t}-\frac{1}{\sinh (t/2)}\right)\frac{\sinh^2 \left[(\frac{1}{2b^2}-\tilde{\sigma})t/2\right]}{\sinh \frac{t}{2b^2}}.$$  This integral approaches a finite limit as $b\to 0$, which makes it easy to extract the leading terms in (\ref{sigmaasymp}) and also to do the numerical comparison with (\ref{logupsilonapp}).  Although this derivation required restrictions on $\sigma$, the final result does not since it can be continued throughout the $\sigma$ plane using the recursion relation.  Of course the asymptotic series is only useful when $\sigma$ is $\mathcal{O}(b^0)$.}
  The constant $C$ will cancel out of all of our computations since we are always computing ratios of equal numbers of $\Upsilon_b$'s.  This precise numerical agreement also confirms our somewhat vague argument for the absence of an additional periodic function in $\sigma$.  As an application of this formula we can find the asymptotics of $\Upsilon_0$ from the DOZZ formula:
\be
\label{upsilon0}
\Upsilon_0=\frac{C}{b^{1/2}}\exp\left[-\frac{1}{4b^2}\log b+F(0)/b^2+O(b^2 \log b)\right].
\ee

\chapter{Theory of Hypergeometric Functions}
\label{hyps}
This appendix will derive the results we need about hypergeometric and $P$-functions \cite{Whitwat}.  No prior exposure to either is assumed.  Our initial approach is rather pedestrian; it is aimed at producing concrete formulas (\ref{wholelotaP}-\ref{a12}) which illustrate the monodromy properties of various solutions of Riemann's hypergeometric differential equation and the ``connection coefficients'' relating them.  This ``toolbox'' approach is convenient for practical computations, but the disadvantage is that it involves complicated expressions that obscure some of the underlying symmetry.  In section \ref{hypintegrals} we give a more elegant general formulation in terms of the integral representation, which illustrates the basic logic of the previous sections in a simpler way but is less explicit about the details.  It also allows us to recast the three-point solutions of section \ref{threp} in an interesting way.
\subsection{Hypergeometric Series}
\hspace{0.25in}We begin by studying the series
\be
F(a,b,c,z)=\sum_{n=0}^{\infty}\frac{(a)_n (b)_n}{(c)_n n!}z^n.
\label{hgs}
\ee
Here $(x)_n \equiv x(x+1)\cdots(x+n-1)=\frac{\Gamma(x+n)}{\Gamma(x)}$.  It is easy to see using the ratio test that if $c$ is not a negative integer, then for any complex $a$ and $b$ the series converges absolutely for $|z|<1$, diverges for $|z|>1$, and is conditional for $|z|=1$.  It is also symmetric in $a$ and $b$, and we can observe that if either $a$ or $b$ is a nonpositive integer then the series terminates at some finite $n$.  One special case which is easy to evaluate is
$$F(1,1,2,z)=-\frac{\log(1-z)}{z},$$
which shows that the analytic continuation outside of the unit disk is not necessarily singlevalued.

We will also need the value of the series at $z=1$. We will derive this from the integral representation in section \ref{hypintegrals}:
\begin{equation}
\label{atone}
F(a,b,c,1)=\frac{\Gamma(c)\Gamma(c-a-b)}{\Gamma(c-a)\Gamma(c-b)} \qquad \mathrm{Re}(c-a-b)>0.
\end{equation}

\subsection{Hypergeometric Differential Equation}

\hspace{0.25in}By direct substitution one can check that the function $F(a,b,c,z)$ obeys the following differential equation:
\begin{equation}
z(1-z)f''+(c-(a+b+1)z)f'-abf=0.
\label{hge}
\end{equation}
This second-order equation has three regular singular points, at 0,1, and $\infty$. Since $F(a,b,c,z)$ is manifestly nonsingular at $z=0$, its analytic continuation has potential singularities only at 1 and $\infty$.  There is the possibility of a branch cut running between $1$ and $\infty$. We saw this in the special case we evaluated above, and it is standard to choose this branch cut to lie on the real axis.  We can determine the monodromy structure of a general solution of this equation by studying its asymptotic behaviour in the vicinity of the singular points. By using a power-law ansatz it is easy to see that any solution generically takes the form\footnote{When the two exponents given here become equal at one or more of the singular points, there are additional asymptotic solutions involving logs.  We will not treat this special case, although it appears for the particular choice $a=b=1/2$, $c=1$ in appendix \ref{blocksapp}.}
\be
 f(z) \sim \begin{cases}\label{hgecases}
A_0(z)+ z^{1-c}B_0(z)& \text{as $z \to 0$}\\
z^{-a}A_\infty(1/z) +z^{-b}B_\infty(1/z)& \text{as $z \to \infty$}\\
A_1(1-z)+ (1-z)^{c-a-b}B_1(1-z)& \text{as $z \to 1$}
\end{cases},
\ee
with $A_i(\cdot),B_i(\cdot)$ being holomorphic functions in a neighborhood of their argument being zero.  The solution of (\ref{hge}) defined by the series (\ref{hgs}) is a case of (\ref{hgecases}), with $A_0(z)=F(a,b,c,z)$ and $B_0=0$. We will determine the $A_i$ and $B_i$ at the other two singular points later.  These expressions confirm that a solution of (\ref{hge}) will generically have branch points at $0$, $1$ and $\infty$.

\subsection{Riemann's Differential Equation}
\hspace{0.25in}It will be very convenient for our work on Liouville to make use of Riemann's hypergeometric equation, of which (\ref{hge}) is a special case.  This more general differential equation is:
\begin{equation}
\label{rde}
\begin{split}
&f''+\left\{\frac{1-\alpha-\alpha'}{z-z_1}+\frac{1-\beta-\beta'}{z-z_2}+\frac{1-\gamma-\gamma'}{z-z_3}\right\}f'+ \\ &\left\{\frac{\alpha\alpha'z_{12}z_{13}}{z-z_1}+\frac{\beta\beta'z_{21}z_{23}}{z-z_2}+\frac{\gamma\gamma'z_{31}z_{32}}{z-z_3}\right\}\frac{f}{(z-z_1)(z-z_2)(z-z_3)}=0
\end{split}
\end{equation}
along with a constraint:
\begin{equation}
\alpha+\alpha'+\beta+\beta'+\gamma+\gamma'=1.
\end{equation}
Here $z_{ij}\equiv z_i-z_j$, the parameters $\alpha$, $\beta$, $\gamma$, $\alpha'$, $\beta'$, $\gamma'$ are complex numbers, and the constraint is imposed to make the equation nonsingular at infinity.  The points $z_i$ are regular singular points.  This is in fact the most general second-order linear differential equation with three regular singular points and no singularity at infinity.  To see that this reduces to the hypergeometric equation (\ref{hge}), one can set $z_1=0$, $z_2=\infty$, $z_3=1$, $\alpha=\gamma=0$, $\beta=a$, $\beta'=b$, and $\alpha'=1-c$. 

Solutions to Riemann's equation can always be written in terms of solutions of the hypergeometric equation; this is accomplished by first doing an $SL(2,\mathbb{C})$ transformation to send the three singular points to $0$, $1$, and $\infty$, followed by a nontrivial rescaling.  To see this explicitly, say that $g(a,b,c,z)$ is a solution of the differential equation (\ref{hge}), not necessarily the solution given by (\ref{hgs}).  Then a somewhat tedious calculation shows that
\begin{equation} \label{riemannhyp}
f=\left(\frac{z-z_1}{z-z_2}\right)^\alpha \left(\frac{z-z_3}{z-z_2}\right)^\gamma g\left(\alpha+\beta+\gamma,\alpha+\beta'+\gamma,1+\alpha-\alpha',\frac{z_{23}(z-z_1)}{z_{13}(z-z_2)}\right)
\end{equation}
is a solution of the differential equation (\ref{rde}).  

Near the singular points any solution behaves as\footnote{As before, there is a caveat that when the two exponents are equal at one or more of the singular points, there are additional solutions involving logs.}
\be
f(z) \sim \begin{cases}
A_1(z-z_1)^\alpha+B_1(z-z_1)^{\alpha'}& \text{as $z \to z_1$}\\
A_2(z-z_2)^{\beta}+B_2(z-z_2)^{\beta'}& \text{as $z \to z_2$}\\
A_3(z-z_3)^\gamma+B_3(z-z_3)^{\gamma'}& \text{as $z \to z_3$},
\end{cases}
\ee
so the monodromies are simply expressed in terms of $\alpha,\alpha', \beta, \ldots$

\subsection{Particular Solutions of Riemann's Equation}
\hspace{0.25in}We now construct explicit solutions of Riemann's equation that have simple monodromy at the three singular points in terms of the hypergeometric function (\ref{hgs}).  Given equation (\ref{riemannhyp}), the most obvious solution we can write down is
$$f^{(\alpha)}(z) \equiv \left(\frac{z-z_1}{z-z_2}\right)^\alpha \left(\frac{z-z_3}{z-z_2}\right)^\gamma F\left(\alpha+\beta+\gamma,\alpha+\beta'+\gamma,1+\alpha-\alpha',\frac{z_{23}(z-z_1)}{z_{13}(z-z_2)}\right).
$$
We denote it $f^{(\alpha)}$ because the holomorphy of the series (\ref{hgs}) at 0 ensures that any nontrivial monodromy near $z_1$ comes only from the explicit factor $(z-z_1)^\alpha$.  The differential equation is invariant under interchanging $\alpha \leftrightarrow \alpha'$, so we can easily write down another solution that is linearly independent with the first (assuming that $\alpha \neq \alpha'$):
\begin{align}
&f^{(\alpha')}(z) \equiv \left(\frac{z-z_1}{z-z_2}\right)^{\alpha'} \left(\frac{z-z_3}{z-z_2}\right)^\gamma\nonumber\\&\hspace{1.5in}\times F\left(\alpha'+\beta+\gamma,\alpha'+\beta'+\gamma,1+\alpha'-\alpha,\frac{z_{23}(z-z_1)}{z_{13}(z-z_2)}\right).\nonumber
\end{align}
This solution has the alternate monodromy around $z=0$.  

The differential equation is also invariant under $\beta \leftrightarrow \beta'$ and $\gamma \leftrightarrow \gamma'$: the former leaves the solutions $f^{(\alpha)}$, $f^{(\alpha')}$  invariant and can be ignored, but the latter apparently generates two additional solutions.  We can find even more solutions by simultaneously permuting $\{z_1,\alpha,\alpha'\} \leftrightarrow\{z_2,\beta,\beta'\} \leftrightarrow\{z_3,\gamma,\gamma'\}$, so combining these permutations we find a total of 4x6=24 solutions, known as ``Kummer's Solutions''.\footnote{This derivation of Kummer's Solutions using the symmetric equation (\ref{rde}) is quite straightforward, but if we had used the less symmetric equation (\ref{hge}) then they would seem quite mysterious.}  Since these are all solutions of the same 2nd-order linear differential equation, any three of them must be linearly dependent.  

To pin down this redundancy, it is convenient to define a particular set of six solutions, each of which has simple monodromy about one of the singular points.  The definition is somewhat arbitrary as one can change the normalization at will as well as move around the various branch cuts.  We will choose expressions that are simple when all $z$-dependence is folded into the harmonic ratio
\be
x\equiv \frac{z_{23}(z-z_1)}{z_{13}(z-z_2)}.
\ee
Our explicit definitions are the following:
\begin{align}\nonumber
P^{\alpha}(x)&=x^\alpha (1-x)^\gamma F(\alpha+\beta+\gamma,\alpha+\beta'+\gamma,1+\alpha-\alpha',x)\\\nonumber
&=x^ \alpha (1-x)^{-\alpha-\beta} F\left(\alpha+\beta+\gamma,\alpha+\beta+\gamma',1+\alpha-\alpha',\frac{x}{x-1}\right)\\\nonumber
P^{\alpha'}(x)&=x^{\alpha'} (1-x)^{\gamma'} F(\alpha'+\beta+\gamma',\alpha'+\beta'+\gamma',1+\alpha'-\alpha,x)\\\nonumber
&=x^ {\alpha'} (1-x)^{-\alpha'-\beta} F\left(\alpha'+\beta+\gamma,\alpha'+\beta+\gamma',1+\alpha'-\alpha,\frac{x}{x-1}\right)\\\nonumber
P^{\gamma}(x)&=x^\alpha (1-x)^\gamma F(\alpha+\beta+\gamma,\alpha+\beta'+\gamma,1+\gamma-\gamma',1-x)\\\nonumber
&=x^{\alpha'} (1-x)^\gamma F(\alpha'+\beta+\gamma,\alpha'+\beta'+\gamma,1+\gamma-\gamma',1-x)\\\nonumber
P^{\gamma'}(x)&=x^\alpha (1-x)^{\gamma'} F(\alpha+\beta+\gamma',\alpha+\beta'+\gamma',1+\gamma'-\gamma,1-x)\\\nonumber
&=x^{\alpha'} (1-x)^{\gamma'} F(\alpha'+\beta+\gamma',\alpha'+\beta'+\gamma',1+\gamma'-\gamma,1-x)\\\nonumber
P^{\beta}(x)&=x^\alpha (1-x)^{-\alpha-\beta}F\left(\alpha+\beta+\gamma,\alpha+\beta+\gamma',1+\beta-\beta',\frac{1}{x-1}\right)\\\nonumber
&=x^{\alpha'} (1-x)^{-\alpha'-\beta}F\left(\alpha'+\beta+\gamma,\alpha'+\beta+\gamma',1+\beta-\beta',\frac{1}{x-1}\right)\\\nonumber
P^{\beta'}(x)&=x^\alpha (1-x)^{-\alpha-\beta'}F\left(\alpha+\beta'+\gamma,\alpha+\beta'+\gamma',1+\beta'-\beta,\frac{1}{x-1}\right)\\
&= x^{\alpha'} (1-x)^{-\alpha'-\beta'}F\left(\alpha'+\beta'+\gamma,\alpha'+\beta'+\gamma',1+\beta'-\beta,\frac{1}{x-1}\right).\label{wholelotaP}
\end{align}
These formulas are somewhat intimidating, but they follow from the simple permutations just described.  For convenience in the following derivation we give two equivalent forms of each.  More symmetric integral expressions for them will be described in section \ref{hypintegrals}.  

Since only two of these can be linearly independent, there must exist coefficients $a_{ij}$ such that
\begin{align}
\nonumber
&P^{\alpha}(x)=a_{\alpha \gamma} P^{\gamma}(x)+a_{\alpha \gamma'}P^{\gamma'}(x)\\
\label{connection1}
&P^{\alpha'}(x)=a_{\alpha' \gamma} P^{\gamma}(x)+a_{\alpha' \gamma'}P^{\gamma'}(x).
\end{align}
These coefficients are called connection coefficients.  To determine them we can evaluate these two equations at $x=0$ and $x=1$, which gives
\begin{align} \nonumber
a_{\alpha\gamma}=\frac{\Gamma(1+\alpha-\alpha')\Gamma(\gamma'-\gamma)}{\Gamma(\alpha+\beta+\gamma')\Gamma(\alpha+\beta'+\gamma')}\\\nonumber
a_{\alpha\gamma'}=\frac{\Gamma(1+\alpha-\alpha')\Gamma(\gamma-\gamma')}{\Gamma(\alpha+\beta+\gamma)\Gamma(\alpha+\beta'+\gamma)}\\ \label{a13}
a_{\alpha'\gamma}=\frac{\Gamma(1+\alpha'-\alpha)\Gamma(\gamma'-\gamma)}{\Gamma(\alpha'+\beta+\gamma')\Gamma(\alpha'+\beta'+\gamma')}\\\nonumber
a_{\alpha'\gamma'}=\frac{\Gamma(1+\alpha'-\alpha)\Gamma(\gamma-\gamma')}{\Gamma(\alpha'+\beta+\gamma)\Gamma(\alpha'+\beta'+\gamma)}.
\end{align}
In solving these equations one uses (\ref{atone}).\footnote{Two identities which are useful are 
$\sin(x)\sin(y)=\sin(x+y-z)\sin(z)+\sin(z-x)\sin(z-y)$ and $\Gamma(x)\Gamma(1-x)=\frac{\pi}{\sin(\pi x)}$.}
Similarly one can find:
\begin{align} \nonumber
a_{\alpha\beta}=\frac{\Gamma(1+\alpha-\alpha')\Gamma(\beta'-\beta)}{\Gamma(\alpha+\beta'+\gamma)\Gamma(\alpha+\beta'+\gamma')}\\\nonumber
a_{\alpha\beta'}=\frac{\Gamma(1+\alpha-\alpha')\Gamma(\beta-\beta')}{\Gamma(\alpha+\beta+\gamma)\Gamma(\alpha+\beta+\gamma')}\\ \label{a12}
a_{\alpha'\beta}=\frac{\Gamma(1+\alpha'-\alpha)\Gamma(\beta'-\beta)}{\Gamma(\alpha'+\beta'+\gamma)\Gamma(\alpha'+\beta'+\gamma')}\\\nonumber
a_{\alpha'\beta'}=\frac{\Gamma(1+\alpha'-\alpha)\Gamma(\beta-\beta')}{\Gamma(\alpha'+\beta+\gamma)\Gamma(\alpha'+\beta+\gamma')}.
\end{align}

Finally we note that our expressions for the connection coefficients allow us to derive some beautiful facts about the original hypergeometric function $F(a,b,c,z)$.  First making the replacements mentioned below (\ref{rde}), we see that (\ref{connection1}) gives:
\begin{equation}
\begin{split}
F(a,b,c,z)=&\frac{\Gamma(c)\Gamma(c-a-b)}{\Gamma(c-a)\Gamma(c-b)}F(a,b,1+a+b-c,1-z)\\+&\frac{\Gamma(c)\Gamma(a+b-c)}{\Gamma(a)\Gamma(b)}(1-z)^{c-a-b}F(c-a,c-b,1+c-a-b,1-z).
\end{split}
\end{equation}

This gives explicit expressions for $A_1(1-z)$ and $B_1(1-z)$ for $F(a,b,c,z)$, as promised above.  We can also set $\alpha=0$, $\alpha'=1-c$, $\beta=0$, $\beta'=c-a-b$, $\gamma=a$, $\gamma'=b$, $z_1=0$, $z_2=1$, and $z_3=\infty$, in which cases \ref{connection1} gives:
\begin{equation}
\label{Fatinfinity}
\begin{split}
F(a,b,c,z)=&\frac{\Gamma(c)\Gamma(b-a)}{\Gamma(c-a)\Gamma(b)}(-z)^{-a}F(a,1-c+a,1-b+a,z^{-1})\\+&\frac{\Gamma(c)\Gamma(a-b)}{\Gamma(c-b)\Gamma(a)}(-z)^{-b}F(b,1-c+b,1+b-a,z^{-1}).
\end{split}
\end{equation}

This expression gives $A_\infty(1/z)$ and $B_\infty(1/z)$ for $F(a,b,c,z)$, and in fact it gives the full analytic continuation of the series (\ref{hgs}) in the region $|z|>1$, since the hypergeometric series on the right hand side converge in this region.  We can thus observe that indeed the only singular behaviour of the function $F(a,b,c,z)$ is a branch cut running from one to infinity.

\subsection{Integral Representations of Hypergeometric Functions}\label{hypintegrals}
\hspace{0.25in}We now consider the integral representations of the hypergeometric and $P$ functions.  We begin by defining
\be
I_C(a,b,c,z)=\int_C ds\,s^{a-c}(s-1)^{c-b-1}(s-z)^{-a},
\ee
where $C$ is some contour to be specified in the $s$-plane.  If we insert this integral into the hypergeometric differential equation (\ref{hge}) we get
\be
\label{totderiv}
\int_C ds \frac{d}{ds}\left[s^{a-c+1}(s-1)^{c-b}(s-z)^{-a-1}\right]=0,
\ee
so this function will be a solution of the equation as long as $C$ has the same and initial and final values for the quantity in square brackets.  As an example we can choose $C$ to run from from one to infinity, which is allowed when $\mathrm{Re} \,c> \mathrm{Re} \,b>0$.  The monodromy of the integrand as $z$ circles zero is trivial everywhere on the contour, so we expect this solution to be proportional to the original hypergeometric function (\ref{hgs}).  Indeed we have
\be
\label{hypintegral}
F(a,b,c,z)=\frac{\Gamma(c)}{\Gamma(b)\Gamma(c-b)}\int_1^\infty ds\,s^{a-c}(s-1)^{c-b-1}(s-z)^{-a}\qquad \mathrm{Re} \,c> \mathrm{Re} \,b>0.
\ee
To establish this we can use the binomial expansion 
\be
(1-z/s)^{-a}=\sum_{n=0}^\infty\frac{\Gamma(a+n)}{\Gamma(a)\Gamma(n+1)} \left(\frac{z}{s}\right)^n
\ee 
to expand the integrand.  We then change variables to $t=1/s$ and use Euler's integral for the Beta function 
\be
\beta(x,y)\equiv \frac{\Gamma(x)\Gamma(y)}{\Gamma(x+y)}=\int_0^1 dt\, t^{x-1}(1-t)^{y-1} \qquad \mathrm{Re} \,x> 0,\,\mathrm{Re} \,y>0.
\ee
This representation allows an easy determination of the value of the hypergeometric function at $z=1$.  Changing variables $s=1/t$ and using the $\beta$ function integral we find:\footnote{One might guess that this formula should also require $\mathrm{Re} \,c> \mathrm{Re} \,b>0$, but these conditions are a relic of our simple choice of contour.  The Pochhammer contour we discuss below will remove these extra conditions, but it cannot remove the condition $\mathrm{Re}\, (c-a-b)>0$ since from (\ref{hgecases}) we see that the function actually diverges at $z=1$ if this is violated.}
\be
F(a,b,c,1)=\frac{\Gamma(c)\Gamma(c-a-b)}{\Gamma(c-a)\Gamma(c-b)} \qquad \mathrm{Re}\,(c-a-b)>0.
\ee

The main point however is that by integrating on other contours it is possible to get other solutions of the hypergeometric differential equation in a straightforward way.  For example if we integrate from zero to $z$, which requires  $2>1+\mathrm{Re} \,b> \mathrm{Re} \,c>0$, then by changing variables to $w=z/s$ it is easy to see that we get 
\begin{equation}
z^{1-c}F(1+b-c,1+a-c,2-c,z)=\frac{\Gamma(2-c)}{\Gamma(1+b-c)\Gamma(1-b)}\int_0^z ds\,s^{b-c}(1-s)^{c-a-1}(z-s)^{-b},
\end{equation}
which is the other linearly independent solution of the hypergeometric differential equation with simple monodromy at $z=0$.  More generally there are four singular points of the integrand, and placing the contour between any two gives six different solutions which correspond to the six solutions that have simple monodromy at $0,1,\infty$.  

Unfortunately these simple contours require strange inequalities on $a,b,c$ to be satisfied which we certainly do not expect to hold for the general solutions we are considering in Liouville theory.  To find contours that produce solutions for arbitrary $a,b,c$ is more subtle.  The trick is to use a closed contour that winds around both points of interest twice, but in such a way that all branch cuts are crossed a net zero number of times.  This is called a Pochhammer contour, it is illustrated in Figure \ref{pochfig}.  If the inequalities we've been assuming are satisfied then we can neglect the parts of the contour that circle the endpoints and it collapses to the one that runs between the two points times a simple factor that depends on the choice of branch of the integrand. In general, if the inequalities are not satisfied, we can just use the Pochhammer contour.
\begin{figure}
\centering
{\includegraphics[scale=1]{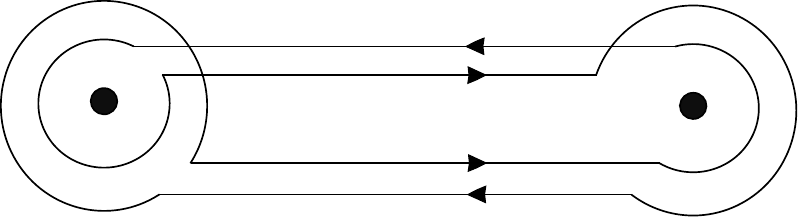}}
\caption{The Pochhammer contour.}\label{pochfig}
\end{figure}
For example we can write
\be
\label{pochhypint}
F(a,b,c,z)=\frac{\Gamma(c)}{\Gamma(b)\Gamma(c-b)}\frac{1}{(1-e^{-2\pi i b})(1-e^{2\pi i(b-c)})}\int_C ds\,s^{a-c}(s-1)^{c-b-1}(s-z)^{-a},
\ee
where $C$ is a Pochhammer contour involving $1$ and $\infty$ and the extra factor\\ $\frac{1}{(1-e^{-2\pi i b})(1-e^{2\pi i(b-c)})}$ cancels the sum over four traversals out to infinity and back.\footnote{The precise form of this factor depends on a choice of branch for the integrand. A different choice would multiply the integral by a $z$-independent constant.}  This expression gives the full analytic continuation of the hypergeometric function in all of its parameters.  

We can also construct an integral representation of an arbitrary solution of Riemann's more general differential equation.  We begin by defining
\begin{align}\nonumber
P_C=&(z-z_1)^\alpha (z-z_2)^\beta (z-z_3)^\gamma \\\label{PCint}
&\times\int _C ds (s-z_1)^{\alpha'+\beta+\gamma-1}(s-z_2)^{\alpha+\beta'+\gamma-1}(s-z_3)^{\alpha+\beta+\gamma'-1}(s-z)^{-\alpha-\beta-\gamma}.
\end{align}
By explicit substition into Riemann's equation and rewriting as a total derivative as in (\ref{totderiv}), we see that see that this integral will be a solution as long as the quantity
\be
V=(s-z_1)^{\alpha'+\beta+\gamma}(s-z_2)^{\alpha+\beta'+\gamma}(s-z_3)^{\alpha+\beta+\gamma'}(s-z)^{-\alpha-\beta-\gamma-1}
\ee
has the same value at both ends of the contour.  For a closed contour this means the contour must cross each branch cut of the integrand a net zero number of times.  We may thus choose Pochhammer type contours involving any two of the four branch points $z_1,z_2,z_3,z$.  There are six such choices, and the integral evaluated on these six different choices is proportional to our $P^{\alpha},P^{\alpha'},\ldots$ defined by (\ref{wholelotaP}).  The detailed proportionality depends on the branch choices and we will not work it out here.  

So far what we have gained in elegance over our previous formulation in terms of the series $(\ref{hgs})$ we have arguably lost in the sophistication of the contours and the branch choices.  But for $\eta$'s in the physical region this presentation allow a very nice repackaging of the Liouville solution (\ref{3pointsol}).  The quantity $f(z,\bar{z})=u\tilde{u}-v \tilde{v}$ defined by (\ref{phifromf}) has an enlightening expression in terms of the types of integrals we have been considering so far.  Our claim is that
\begin{align}\nonumber
f=&f_0|z-z_2|^2 |z-z_1|^{2\alpha} |z-z_2|^{2{\beta}} |z-z_3|^{2{\gamma}} \\
&\times\int d^2s|s-z_1|^{2(\alpha'+\beta+\gamma-1)}|s-z_2|^{2(\alpha+\beta'+\gamma-1)}|s-z_3|^{2(\alpha+\beta+\gamma'-1)}|s-z|^{-2(\alpha+\beta+\gamma)},
\end{align}
with the parameters $\alpha,\alpha',\ldots$ given by equation (\ref{riempar}) and the integral being taken over the full $s$-plane.  As long as this integral converges it is clearly monodromy invariant, and it solves the holomorphic and antiholomorphic differential equations (\ref{monno},\ref{onno}) by the same argument as just given for the integral (\ref{PCint}).  We saw in section \ref{threp} that these two properties were sufficient to uniquely determine $f$ up to an overall normalization, so this establishes our claim.  The conditions for the convergence of this integral, expressed in terms of the $\eta_i$'s, are
\begin{align}
\mathrm{Re}(\eta_1+\eta_2-\eta_3)<&1\\ \nonumber
\mathrm{Re}(\eta_1+\eta_3-\eta_2)<&1\\\nonumber
\mathrm{Re}(\eta_2+\eta_3-\eta_1)<&1\\\nonumber
\mathrm{Re}(\eta_1+\eta_2+\eta_3)>&1.
\end{align}
These are certainly obeyed in the ``physical'' region in Liouville.  Are they equivalent to it, or more precisely to Region I from section \ref{3pointcontinuation}?  Actually, they imply $0<\mathrm{Re}\,\eta_i<1$ for all $i$, while in Region I we would have had $0<\mathrm{Re}\,\eta_i<1/2$.  But for operators obeying the Seiberg bound, the integral converges only in Region I.  For more general $\eta_i$, such an expression would require a more sophisticated type of integral.  For real $\eta$'s in Region I, this expression is manifestly positive and it shows that there cannot be any zeros of $f$, something that was not clear from our old expression (\ref{3pointsol}).  We wonder if Liouville solutions for correlators with more than three heavy operators can be written in terms of generalizations of this integral.  
\def\C{{\mathcal C}}
\def\I{{\mathcal I}}

\chapter{Gamma Functions and Stokes Phenomena}\label{gammastokes}

\subsection{Generalities}
\hspace{0.25in} In this appendix, we review the  Stokes phenomena that occur for the Gamma function and its reciprocal, as they are closely related to the zero mode integrals of spacelike and timelike Liouville theory.\footnote{For a much more detailed introduction to Stokes phenomena, see section 2 of \cite{Witten:2010cx}.
For a treatment of the Gamma function along lines similar to what follows, see \cite{PS}; see \cite{Berry,Boyd} for previous mathematical work.} $\Gamma(z)$ has the following integral representation
 \cite{A.S.} for $\Re\,z>0$:
 \begin{align}\label{gamma1}
\Gamma(z) &= \int^{\infty}_{0}t^{z-1}e^{-t}dt = z^{z}\int^{\infty}_{-\infty}e^{-z(e^{\phi} - \phi)}d\phi,  
\end{align}
(With a slightly different change of coordinates by  $t = e^{\phi}$ rather then $t = ze^{\phi}$, we could have put this in the
form of the Liouville zero mode integral, as in eqn. (\ref{zamma}).)

The exponent $\I=- z (e^\phi-\phi)$ in (\ref{gamma1}) and (\ref{1overgamma}) has critical points at
\begin{equation}\label{critpts}\phi_n =2\pi i n,~~n\in \Z.\end{equation}
To each such critical point $\phi_n$, one attaches an integration contour $\C_n$.  This is a contour that passes through the critical point $\phi_n$
and along which  the exponent $\I$  has stationary phase while $\Re\,\I$ has a local maximum. More briefly, we call this a stationary phase contour. Alternatively,\footnote{In complex dimension 1, the stationary phase contour through a critical point coincides with the steepest descent contour,
but in higher dimension, the stationary phase condition is not enough to determine $\C_n$ and one must use the steepest descent condition.
  For 
much more on such matters, see section 2 of   \cite{Witten:2010cx}.  Notice that, in our case, because the function $\I$ has opposite sign in (\ref{1overgamma})
relative to (\ref{gamma1}), the cycle $\C_n$ is different in the two cases, though we do not indicate this in the notation.} the contour $\C_n$ can be defined as a contour of
steepest descent for $h=\Re\,(- z(e^\phi-\phi))$.  For a steepest descent contour $\C_n$, it is straightforward to determine the large $z$
behavior of the integral
\begin{equation}\label{dolk|}\int_{\C_n} d\phi \exp(\I). \end{equation}
The maximum of $\Re\,\I$ along the cycle $\C_n$ is, by the steepest descent condition, at $\phi_n$.  For large $z$, the integral can be
approximated by the contribution of a neighborhood of the critical point.  In our case, the value of $\I$ at a critical point is $-z(1-2\pi i n)$, so asymptotically
\begin{equation}\label{elf}\int_{\C_n} d\phi \exp(-z(e^\phi-\phi))\sim \exp(-z(1-2\pi i n)) \end{equation}
(times a subleading factor that comes from approximating the integral near the critical point).

Now let us consider the integral (\ref{gamma1}, initially assuming that $z$ is real and positive.  The Gamma function is then defined
by the integral (\ref{gamma1}), with the integration cycle $\C$ being the real $\phi$ axis.  On the real axis, there is a unique
critical point at $\phi=0$.  Moreover, for real $z$, $\I$ is real on the real axis, and the contour of steepest descent from $\phi=0$ is
simply the real axis.  Thus, if $z$ is real and positive, the integration cycle $\C$ in the definition of the Gamma function is the
same as steepest descent cycle $\C_0$, on which the asymptotics are given by (\ref{elf}).  So we get
the asymptotic behavior of the Gamma function on the real axis:
\begin{equation}\label{fork}\Gamma(z)\sim z^ze^{-z}.\end{equation}
This is essentially Stirling's formula (the factor $1/\sqrt{2\pi z}$ in Stirling's formula comes from a Gaussian approximation to 
the integral (\ref{elf}) near its critical point).  

Now let us vary $z$ away from the positive $z$ axis.  The Gamma function is still defined by the integral (\ref{gamma1}), taken along
the real $\phi$ axis, as long
as $\Re\,z>0$.  As soon as $z$ is not real, it is no longer true that the steepest descent contour $\C_0$ coincides with the real axis.
However, as long as $\mathrm{Re}\,z>0$ (we explain this condition momentarily), the steepest descent contour $\C_0$ is equivalent to the real axis, modulo a contour
deformation that is allowed by Cauchy's theorem.  Hence, Stirling's formula remains valid throughout the half-plane $\Re\,z>0$.

If we want to analytically continue the Gamma function as a function of $z$, in general we will have to vary the integration contour
$\C$ away from the real axis.  
To analytically continue beyond the region $\mathrm{Re}\,z>0$, we can let the integration contour $\C$ move away from the real $\phi$ axis,
so that the integral still converges and varies analytically with $z$.  In the case of the Gamma function, there is some restriction
on the ability to do this, since the Gamma function actually has poles at $z=0$ and along the negative $z$ axis.

Now we come to the essential subtlety that leads to Stokes phenomena.  As one varies the parameters in an integral such as (\ref{gamma1}),
the steepest descent contours $\C_n$ generically vary smoothly, but along certain ``Stokes lines'' (or Stokes walls in a problem with more
variables), they jump.  In our case, the only relevant parameter is $z$, so the Stokes lines will be defined in the $z$-plane.
For generic values of $z$, the $\C_n$ are copies of $\R$ (topologically) with both ends at
infinity in the complex $\phi$ plane.  For example, for $z$ real and positive, $\C_n$
is defined by $\Im\,z=2\pi n$ and actually is an ordinary straight line in the $\phi$ plane.  However, for special values of $z$, steepest descent  from one critical point $p$ leads (in one direction)
to another critical point $p'$.   Whether this occurs depends only on the argument of $z$, so it occurs on rays through
the origin in the $z$ plane; these rays are the Stokes lines.  As one varies $z$ across a Stokes line $\ell$, the steepest descent contour
from $p$ will jump (on one side of $\ell$, it passes by $p'$ on one side; on the other side of $\ell$, it passes by $p'$ on the other side and then
heads off in a different direction).

For the Gamma function, we can easily find the Stokes lines.  Since the steepest descent cycles have stationary phase, they
can connect one critical point $p$ to another critical point $p'$ only if the phase of $\I$ is the same at $p$ and at $p'$.
For the critical point at $\phi=2\pi i n$, the value of $\Im\,\I$ is $c_n=\Im\,(-z(1-2\pi i n))$. So $c_n=c_{n'}$  for $n\not=n'$
 if and only if $\Re\,z=0$.   We really should remove from this discussion the point $z=0$ where our integral is ill-defined for
any noncompact contour (and the Gamma function has a pole), so there are two Stokes
lines in this problem, namely the positive and negative imaginary $z$ axis.  

There is one more basic fact about this subject.  Away from Stokes lines, the steepest descent contour $\C_n$
are a basis for the possible integration cycles (on which the integral of interest converges) modulo the sort of contour deformations
that are permitted by Cauchy's theorem.  So any integration contour $\C$ -- such
as the one for the analytically continued Gamma function -- always has an expansion
\begin{equation}\label{helf}\C=\sum_n a_n \C_n \end{equation}
where the $a_n$ are integers, and the relation holds modulo contour deformations that are allowed by Cauchy's theorem. 
Since the integral over any of the $\C_n$ always has the simple asymptotics (\ref{elf}), the asymptotics of the integral over $\C$
are known if one knows the coefficients $a_n$.   As one varies $z$ in the complex plane, $\C$ will vary continuously, but the $\C_n$
jump upon crossings Stokes lines.  So the asymptotic behavior of the integral for large $z$ will jump in crossing a Stokes line.
The well-behaved problem of large $z$ asymptotics is therefore to fix an angular sector
in the complex $z$-plane between two Stokes lines and consider the behavior as $z\to\infty$ in the given angular sector.  Actually,
this would be a full picture if there were only finitely many critical points.  In the case of the Gamma function,  it will turn out that
the sum (\ref{helf}) is an infinite sum if $\Re\,z<0$ and moreover this infinite sum can diverge if $z$ is real and negative.  To get
a simple problem of large $z$ asymptotics, one must keep away from the negative $z$ axis,
where the Gamma function has its poles,
as well as from the Stokes lines.  

\subsection{Analysis Of The Gamma Function}

\hspace{0.25in}Now let us make all this concrete.
The integral (\ref{gamma1}) converges along contours that begin and end in regions where
 $\mathrm{Re}(-z(e^{\phi} -\phi))\to-\infty$.  These regions have been shaded in Fig, \ref{OldFigureOne} for the case that $z$ is real and positive.
  The $\C_n$ are the horizontal lines $\Im\,\phi=2\pi n$.  One can see by hand in this example that any integration cycle that
  begins and ends in the shaded regions is a linear combination of the $\C_n$, as in eqn. (\ref{helf}).
  The real $\phi$ axis -- which is the integration contour $\C_0$ in (\ref{gamma1}) -- coincides with $\C_0$, and is indicated in the figure
  as the Relevant Contour.

\begin{figure}
\centering
\includegraphics{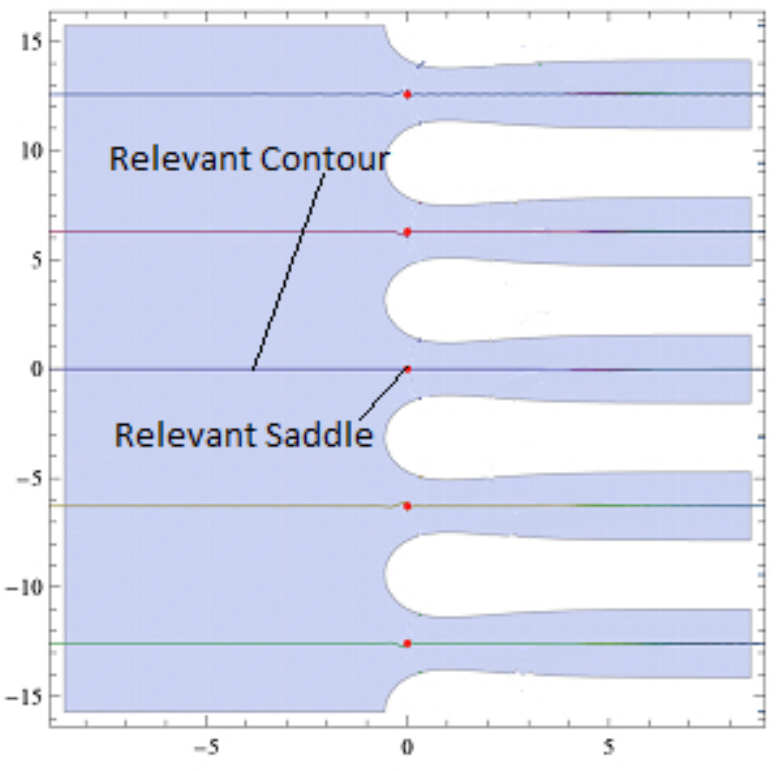}\\ \caption{ \label{OldFigureOne} For the Gamma function integral (\ref{gamma1}) to converge, the integration contour must begin and end
in regions of the $\phi$ plane with  $\mathrm{Re}(-z(e^{\phi} -\phi))\to-\infty$.  These regions are shaded here for the case
that $z$ is real and positive.  In addition, we show the critical points at $\phi_n=2\pi i n$ (represented in the figure by dots) and the steepest descent cycles $\C_n$,
which are the horizontal lines $\Im\,\phi=2\pi n$. } \end{figure} \vskip 1cm

In Fig, \ref{OldFigureTwo}, we have sketched how the steepest descent contours $\C_n$ are deformed when $z$ is no longer real but still has positive
real part.  In passing from Fig, \ref{OldFigureOne} to Fig, \ref{OldFigureTwo}, the $\C_n$ evolve continuously and it remains true that the integration contour $\C$
defining the Gamma function is just $\C_0$, modulo a deformation allowed by Cauchy's theorem.

\begin{figure}
\includegraphics{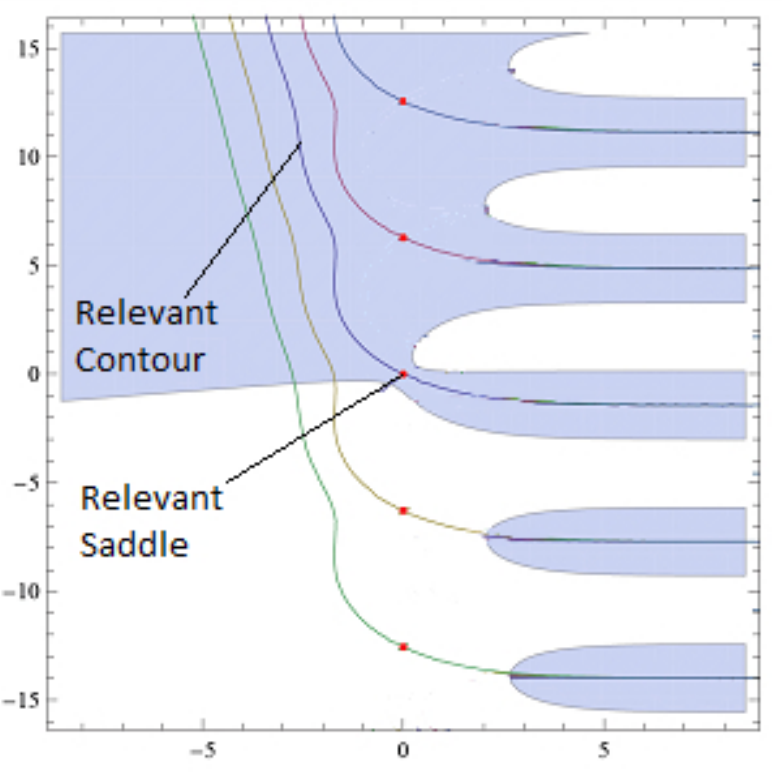}\\ \centering
\caption{ \label{OldFigureTwo} This figure is equivalent to Fig. \ref{OldFigureOne}, except that now $z$ is complex but still with $\Re\,z>0$. (In drawing
the figure, we have taken the case $\Im\,z>0$.)
The critical points are unchanged from the case that $z$ is real, but the steepest descent contours are changed.
The phrase Relevant Contour labels the contour $\C_0$ that controls the asymptotics of the Gamma function in this region.} \end{figure}
\vskip 1cm

However, the $\C_n$ jump upon crossing the Stokes lines at $\Re\,z=0$.  This is shown in Fig. \ref{OldFigureThree} for the case $\Im\,z>0$.
While in Fig, \ref{OldFigureTwo}, the steepest descent contours for the Gamma function have one end in the upper left and one end to the right, in Fig.
\ref{OldFigureThree},
they end to the right in both directions.  As a result, although nothing happens to the contour $\C$ that defines the Gamma function
in going from Fig, \ref{OldFigureTwo} to Fig. \ref{OldFigureThree}, to express $\C$ as a linear combination of the $\C_n$'s, we must in Fig, \ref{OldFigureThree} take an infinite sum
\begin{equation}\label{bloke}\C=\sum_{n\geq 0}\C_n.\end{equation}

\begin{figure}
\includegraphics{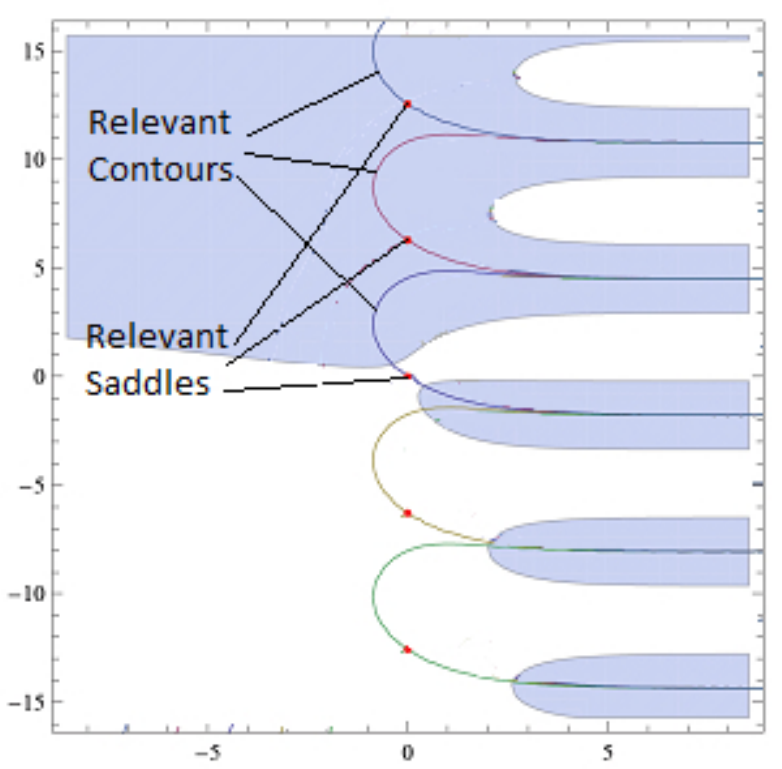}\\ \centering \label{OldFigureThree}\centering\caption{
 This is the analog of Fig, \ref{OldFigureTwo}, but now for the case that $\Re\,z<0$, $\Im\,z>0$.  (For $\Re\,z<0$, $\Im\,z<0$,
just turn the figure upside down.)  A Stokes phenomenon has occurred, relative to Fig, \ref{OldFigureTwo}.  In Fig, \ref{OldFigureTwo}, each
steepest descent curve for the Gamma function connects the shaded region in the upper left to one of the shaded regions
on the right.  This is also the behavior of the contour $\C$ that defines the Gamma function.  However, for $\Re\,z.\,\Im\,z<0$,
the steepest descent contours connect two adjacent shaded regions on the right.  To construct the contour $\C$ that controls
the Gamma function, one must take an infinite sum $\sum_{n\geq 0}\C_n$.}\end{figure}   \vskip 1cm

The dependence on $n$ of the Gamma function integral over $\C_n$ is very simple:
\begin{equation}\label{tox}\int_{\C_n}d\phi \exp(-z(e^\phi-\phi))=\exp(2\pi i n z) \int_{\C_0}d\phi\,\exp(-z(e^\phi-\phi)).\end{equation}
This is because a shift $\phi\to \phi+2\pi i n$ maps $\C_0$ to $\C_n$ and shifts $\I$ by $\I \to \I+2\pi i n z $.  This formula is
an analog for the Gamma function of eqn. (\ref{simplex}) for Liouville theory.  Using (\ref{bloke}) and (\ref{tox}), we find that
in the quadrant $\Re\,z<0$, $\Im\,z>0$, the Gamma function is
\begin{align}\label{blox}\Gamma(z)=&z^z\int_\C d\phi \exp(-z(e^\phi-\phi))=z^z\sum_{n\geq 0}\int_{\C_n} d\phi \exp(-z(e^\phi-\phi))
\\ =&z^z\sum_{n=0}^\infty\exp(2\pi i n z)\int_{\C_0} d\phi \exp(-z(e^\phi-\phi)).\end{align}
From (\ref{blox}), we can read off the asymptotic behavior of the Gamma function in the quadrant in question.  Approximating the 
integral over $\C_0$ by the value at the maximum, and performing the sum over $n$, we get
\begin{equation}\label{orox}\Gamma(z)\sim z^ze^{-z}\frac{1}{1-\exp(2\pi i z)}.\end{equation}
(Again, a prefactor analogous to the $1/\sqrt{2\pi z}$ in Stirling's formula can be found by evaluating the integral over $\C_0$
more accurately.) 
From this point of view, the poles of the Gamma function at negative integers arise not because of a problem with the integral over
$\C_0$ but because of a divergence of the geometric series. 
 The factor $1/(1-\exp(2\pi i z))$ in this formula is important only near the negative real axis.  
 
For use in the main text of the part we note that a similar analysis of the case where $\Re\,z<0$, $\Im\,z>0$ gives
\begin{equation}\Gamma(z)\sim z^ze^{-z}\frac{1}{1-\exp(-2\pi i z)},\end{equation}
and that these formula can all be combined to give 
\be
\Gamma(z)= 
\begin{cases}
e^{z \log z-z+\O(\log z)} \qquad \qquad\qquad \quad \,\,\,\mathrm{Re}(z)>0\\
\frac{1}{e^{i\pi z}-e^{-i\pi z}}e^{z \log (-z)-z+\O(\log(-z))} \qquad \mathrm{Re}(z)<0
,\end{cases}
\ee
where the logarithms are always evaluated on the principal branch.
\subsection{The Inverse Gamma Function}
\hspace{0.25in}We can play the same game for the inverse of the Gamma function, starting with the integral representation\begin{equation}\label{1overgamma}
\frac{1}{\Gamma(z)} = \frac{1}{2\pi i}\oint_{\mathcal{C}_t}t^{-z}e^{t}dt = \frac{-1}{2\pi i}z^{-z+1}\oint_{\mathcal{C}}e^{z(e^{-\phi} + \phi)}e^{-\phi}d\phi.
\end{equation}
In (\ref{1overgamma}), $\mathcal{C}_t$ starts at real $-\infty -i\epsilon$, encircles the branch cut along the negative real  $t$ axis, and ends up at $-\infty +i\epsilon$.

To arrive at the right hand side of  (\ref{1overgamma}), we have made the coordinate change $t =z e^{-\phi}$, which differs slightly from the transformation $t=ze^\phi$ used in deriving (\ref{gamma1}). 
 The critical points are still at $\phi_n=2\pi i n$.  Once again the shaded regions in Fig, \ref{OldFigureFour} are the ones in which the integral is convergent, as $\Re\,\I\to
 -\infty$, where now $\I=z(e^{-\phi}+\phi)$. The steepest descent contours are shown
 in Fig, \ref{OldFigureFour} and connect adjacent shaded regions on the left of the figure. For $z$ real and positive, the image in the $\phi$ plane of the contour $\C_t$ in
 the $t$ plane is the steepest descent contour $\C_0$ that passes through the critical point
 at $\phi=0$.  Since
 $\C$ is a steepest descent cycle, the asymptotic behavior of the integral in this region
 is just $e^\I$, with $\I$ evaluated at the critical point.
 So in this region,
 \begin{equation}\label{doft2}\frac{1}{\Gamma(z)}\sim z^{-z+1}e^z.\end{equation}
  (As always, subleading factors, including powers of $z$, can be determined by evaluating the
  integral more accurately near the critical point.)

\begin{figure}
\includegraphics{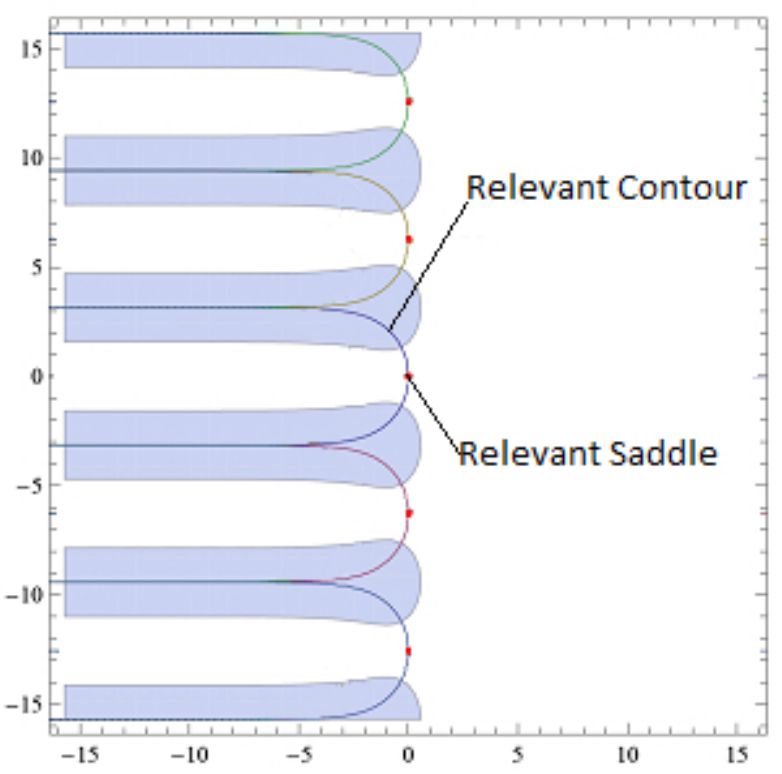}\\ \centering
\label{OldFigureFour} \caption{ For the inverse Gamma function integral, with $z$ real and positive, the critical points and steepest descent contours
are as shown here.  Regions in which the integral is convergent are again shaded. The function $1/\Gamma(z)$ is defined by an integral over a contour $\C$ that coincides with the steepest
descent contour $\C_0$ associated to the critical point at $z=0$.  This contour connects
two adjacent shaded regions on the left of the figure. } \end{figure} \vskip 1cm

\begin{figure}
\includegraphics{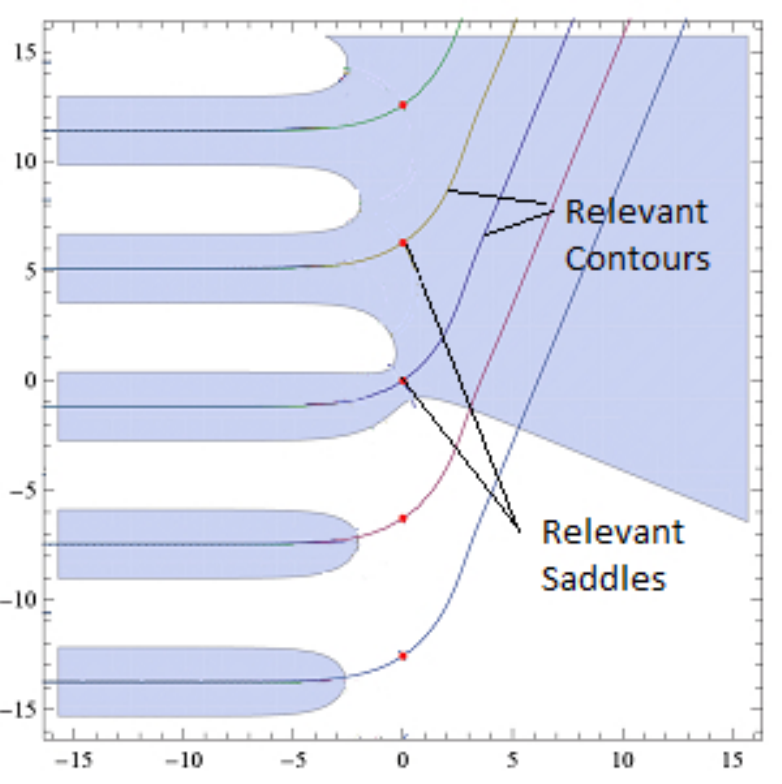}\\   \centering  \caption{
\label{OldFigureFive} For $\mathrm{Re}\,z<0$ and $\mathrm{Im}\,z>0$, the steepest descent
contours $\C_n$ connect a region on the left to the upper right, as shown here.  On the other hand,
after varying $z$, the integration contour $\C$ that defines the inverse Gamma function
still connects two adjacent regions on the left. (It is not drawn here.)  It is therefore no longer true that $\C$ equals
one of the steepest descent contours; rather, $\C$ is a difference $\C_0-\C_1$.} \end{figure}\vskip 1cm

When we vary $z$, the regions in the $\phi$ plane in which the integral converges move
up and down slightly.  The integration contour $\C$ defining the inverse Gamma function
varies smoothly and continues to connect two adjacent shaded regions.  However,
when $\Re\,z<0$, the qualitative behavior of the steepest descent contours $\C_n$ changes.
As sketched in Fig, \ref{OldFigureFive}, for such values of $z$, each $\C_n$ connects a shaded region in the left of the $\phi$
plane to the upper right.  The formula expressing $\C$ in terms of steepest descent contours
is now $\C=\C_0-\C_1$.  (Here $\C_0-\C_1$ starts in a shaded region on the left, heads to the upper right, and
then doubles back to an adjacent shaded region on the left -- thus reproducing $\C$.) Accordingly, (\ref{doft2}) is modified to
\begin{equation}\label{dofto}\frac{1}{\Gamma(z)}\sim (1-e^{2\pi i z})z^{-z+1}e^z.\end{equation}

Of course, this result for the asymptotic behavior of $1/\Gamma(z)$ is equivalent to the result
that we had earlier for the asymptotic behavior of $\Gamma(z)$.  However, seeing this
behavior directly from the Stokes phenomena that affect an integral (rather than  the inverse of
an integral) is useful background for the body of this part.

Just as in the relation between spacelike and timelike Liouville theory as studied in this
part, the Gamma function and the inverse Gamma function are essentially given by the
same integral, evaluated on different integration contours.  This fact is also related to the functional
equation obeyed by the Gamma function.
We have 
\begin{equation}\label{gammaoneminusx}
\frac{2\pi i}{\Gamma(1-z)}= \int_{\mathcal{C}_t}t^{z - 1}e^{t}dt .
\end{equation}
The integrand on the right hand side can be converted to the integrand of the Gamma function integral (\ref{gamma1}) if
we substitute $t\to -t$.  This maps $\C_t$ to another integration contour $\C'_t$ and gives
\begin{equation}\label{gammazone}
\frac{2\pi i}{\Gamma(1-z)}=\exp(i\pi(z-1))\int_{\C'_t}t^{z-1}e^{-t}d t. 
\end{equation}
So the inverse Gamma function, apart from some elementary factors and a substitution $z\to 1-z$, is given by the same
integral as the Gamma function, but with a different integration contour. 

\chapter{Semiclassical Conformal Blocks}
\label{blocksapp}
In this appendix we give a derivation, based on \cite{Zamolodchikov1986}, of an asymptotic formula for the Virasoro conformal block at large intermediate operator weight.  The original argument was somewhat terse and implicitly involved certain assumptions about the semiclassical limit of correlators, so in our view its validity has not been established completely rigorously from the definition (\ref{conformalblock}).  It has however survived stringent numerical tests \cite{Zamolodchikov:1995aa}, and in our discussion in section \ref{genf} we were comfortable assuming it to be true.  We will expand out the argument here and try to be clear about what the assumptions are.  The asymptotic formula is
\be
\mathcal{F}_{1234}(\Delta_i,\Delta,x)\sim (16q)^{\Delta},
\label{Fasymp}
\ee
with
\be
q=\exp\left[-\pi K(1-x)/K(x)\right]
\ee
and
\be
K(x)=\frac{1}{2}\int_0^1 \frac{dt}{\sqrt{t(1-t)(1-x t)}}.
\ee

The idea is to study a five-point function of a light degenerate primary operator with four primary scalar fields of generic operator weight in the semiclassical limit $c\gg 1$.  The external conformal weights $\Delta_i$ are taken to be of order $c$, and the internal weight $\Delta_p$ is initially also taken to be of this order; it will eventually be taken much larger than $c$.\footnote{This argument could be applied to any CFT with a $c$ that can be large and primary operators with the desired weights.  ''Light'' in this context just means that the operator weight of the degenerate field scales like $c^0$.  Liouville is a theory that fits the bill, but we will use general CFT language to avoid the subtleties of Liouville factorization.}  If we parameterize the central charge as $c=1+6(b+1/b)^2$, then the conformal weight of the light degenerate operator is $-\frac{1}{2}-\frac{3b^2}{4}$.  We will write the correlator as
\be
\langle\mathcal{O}_4(z_4,\bar{z}_4)\mathcal{O}_3(z_3,\bar{z}_3)\Psi(z,\bar{z})\mathcal{O}_2(z_2,\bar{z}_2)\mathcal{O}_1(z_1,\bar{z}_1)\rangle,
\ee
where $\Psi$ is the degenerate operator and we choose $|z_4|>|z_3|>|z|>|z_2|>|z_1|$.  This correlator obeys equation (\ref{lightdegenerate}), which here gives
\be
\label{deg5}
\left[\frac{1}{b^2}\partial_z^2+\sum_{i=1}^4\left(\frac{\Delta_i}{(z-z_i)^2}+\frac{1}{z-z_i}\partial_i\right)\right]\langle \mathcal{O}_4\mathcal{O}_3
\Psi\mathcal{O}_2\mathcal{O}_1\rangle=0.
\ee
We can also expand the correlator using the operator product expansion \cite{Belavin:1984vu}\footnote{We assume here that as in Liouville, the only primaries that appear in the operator product expansion are scalars.  We could drop this assumption at the cost of slightly more complicated formulas, the result (\ref{Fasymp}) would be the same.}
\be
\mathcal{O}_2(z_2,\bar{z}_2)\mathcal{O}_1(z_1,\bar{z}_1)=\sum_p C^p_{21}|z_{21}|^{2(\Delta_p-\Delta_1-\Delta_2)}\sum_{k,\tilde{k}}(z_{21})^k (\bar{z}_{21})^{\tilde{k}}\beta_{21}^{pk}\beta_{21}^{p,\tilde{k}}\mathcal{O}_p^{\{k,\tilde{k}\}}(z_1,\bar{z}_1),
\ee
which allows us to extract all dependence on $z_2$:
\begin{align}\nonumber
\langle \mathcal{O}_4\mathcal{O}_3
\Psi\mathcal{O}_2\mathcal{O}_1\rangle=&\sum_p C^p_{21}|z_{21}|^{2(\Delta_p-\Delta_1-\Delta_2)}\\
&\times\sum_{k,\tilde{k}}(z_{21})^k(\bar{z}_{21})^{\tilde{k}}\beta_{21}^{pk}\beta_{21}^{p,\tilde{k}}\langle\mathcal{O}_4\mathcal{O}_3\Psi\mathcal{O}_p^{\{k,\tilde{k}\}}\rangle.
\end{align}
In these formulae, as discussed below equation (\ref{conformalblock}) the sum over $k$ is only heuristic and more precisely includes a sum over all descendants at a given level $k$.  The operators $\mathcal{O}_p^{\{k,\tilde{k}\}}$ are the Virasoro descendants of the primary $\mathcal{O}_p$.  Now say that we define 
\be
\label{Psipdef}
\Psi_p(z,\bar{z};z_i,\bar{z}_i)\equiv\frac{\langle\mathcal{O}_4\mathcal{O}_3\Psi\mathcal{O}_p\rangle}{\langle\mathcal{O}_4\mathcal{O}_3\mathcal{O}_p\rangle.}
\ee
In the semiclassical limit we can think of the function $\Psi_p$ as the classical expectation value of the degenerate operator in the presence of the other operators, and since the light operator has weight of order $c^0$ we expect it to have a finite limit at $c\to\infty$.  In Liouville theory, this is the statement that light operators just produce $\mathcal{O}(b^0)$ factors in the correlation function as in equation (\ref{semiclassicalcorr}).  

So far we have written only exact formulas, but we now come to the first approximation: in the semiclassical limit we claim that the same formula holds also for descendants of $\mathcal{O}_p$ with the same function $\Psi_p$:
\be
\label{scfactor}
\langle\mathcal{O}_4\mathcal{O}_3\Psi\mathcal{O}_p^{\{k,\tilde{k}\}}\rangle\approx \Psi_p(z,\bar{z};z_i,\bar{z}_i)\langle\mathcal{O}_4\mathcal{O}_3\mathcal{O}_p^{\{k,\tilde{k}\}}\rangle.
\ee
The justification for this is that the correlator $\langle\mathcal{O}_4\mathcal{O}_3\Psi\mathcal{O}_p^{\{k,\tilde{k}\}}\rangle$ can be written in terms of a series of differential operators acting on $\langle\mathcal{O}_4\mathcal{O}_3\Psi\mathcal{O}_p\rangle$.  The differential operators are of the form \cite{Belavin:1984vu}
\be
\label{diffop}
\mathcal{L}_{-m}=\sum_{i=3,4,z}\left[\frac{(m-1)\Delta_i}{(z_i-z_1)^m}-\frac{1}{(z_i-z_1)^{m-1}}\partial_i\right].
\ee
Here $\Delta_z\equiv -\frac{1}{2}-\frac{3b^2}{4}$.  Similarly the correlator $\langle\mathcal{O}_4\mathcal{O}_3\mathcal{O}_p^{\{k,\tilde{k}\}}\rangle$ can be written in terms of a similar series of differential operators acting on $\langle\mathcal{O}_4\mathcal{O}_3\mathcal{O}_p\rangle$, but with the sum in (\ref{diffop}) being only over 3,4.  The point however is that in the semiclassical limit we expect 
\be
\langle\mathcal{O}_4\mathcal{O}_3\mathcal{O}_p\rangle\sim e^{-\frac{c}{6} S_{cl}}
\ee 
for some $S_{cl}$, and since we have taken $\Delta_{3,4}\sim c$ while $\Psi_p$ and $\Delta_z$ are both $\mathcal{O}(c^0)$, we see that the $i=z$ term in (\ref{diffop}) becomes unimportant, and also that in the $i=3,4$ terms we can neglect derivatives acting on $\Psi_p$.  This establishes (\ref{scfactor}).  In the semiclassical approximation we thus have
\begin{align}\nonumber
\langle \mathcal{O}_4\mathcal{O}_3
\Psi\mathcal{O}_2\mathcal{O}_1\rangle\approx&\sum_p C^p_{21}|z_{21}|^{2(\Delta_p-\Delta_1-\Delta_2)}\Psi_p(z,\bar{z};z_4,\bar{z}_4,z_3,\bar{z}_3,z_1,\bar{z}_1)\\
&\times\sum_{k,\tilde{k}}(z_{21})^k(\bar{z}_{21})^{\tilde{k}}\beta_{21}^{p,k}\beta_{21}^{p,\tilde{k}}\langle\mathcal{O}_4\mathcal{O}_3\mathcal{O}_p^{\{k,\tilde{k}\}}\rangle. 
\end{align}
We can make this formula more elegant by defining
\be
F_{1234}(\Delta_i,\Delta_p,z_i)\equiv (z_{21})^{\Delta_p-\Delta_1-\Delta_2}\sum_k(z_{21})^k \beta^{p,k}_{21}\frac{\langle\mathcal{O}_4\mathcal{O}_3\mathcal{O}_p^{\{k,0\}}\rangle}{\langle\mathcal{O}_4\mathcal{O}_3\mathcal{O}_p\rangle},
\ee
after which we get
\be
\langle \mathcal{O}_4\mathcal{O}_3
\Psi\mathcal{O}_2\mathcal{O}_1\rangle\approx\sum_p \Psi_p C^p_{21}\langle\mathcal{O}_4\mathcal{O}_3\mathcal{O}_p\rangle F_{1234}(\Delta_i,\Delta_p,z_i)F_{1234}(\Delta_i,\Delta_p,\bar{z}_i).\label{deg5exp}
\ee
We note for future convenience that $F_{1234}$ becomes the Virasoro conformal block $\mathcal{F}_{1234}$ after sending $z_4\to \infty,z_3 \to 1,z_2 \to x, z_1\to 0$.  Based on its definition, we can guess that $F_{1234}$ has a semiclassical limit \cite{Zamolodchikov1986} of the form 
\be
\label{semicF}
F_{1234}\sim e^{-\frac{c}{6}f_{cl}},
\ee
where $f_{cl}$ is called the ``semiclassical conformal block''.\footnote{In fact this exponentiation has never actually been proven directly from the definition (\ref{conformalblock}), although it has been checked to high order numerically.  We thank A. B. Zamolodchikov for a discussion of this point, and for a summary of his unpublished numerical work.}

We will now study the implications of the expressions (\ref{deg5exp}, \ref{semicF}) for the differential equation (\ref{deg5}).  It will be convenient to view $z$ and $\bar{z}$ as independent.  We observe that for generically different $\Delta_p$'s, the various terms in the sum over $p$ in (\ref{deg5exp}) have different monodromy as $z_2$ circles $z_1$.  For each $p$ the different terms in the differential equation (\ref{deg5}) have the same monodromy, so in order for the equation to be solved by (\ref{deg5exp}) it seems reasonable to expect that it must actually be solved separately for each $p$.  In the semiclassical limit the action of the derivatives with respect to $z_i$ on $\Psi_p$ is suppressed as in our discussion below (\ref{diffop}), so we find that the differential equation can be converted into an ordinary differential equation just involving $\Psi_p$:
\be
\label{psipeq}
\left[\partial_z^2+\sum_{i=1}^4\left(\frac{\delta_i}{(z-z_i)^2}-\frac{C_i}{z-z_i}\right)\right]\Psi_p=0,
\ee
with 
\be
\label{accpar}
C_i=\partial_i (S_{cl}+f_{cl}).
\ee
Here $\delta_i=b^2\Delta_i$.  In this type of differential equation the parameters $C_i$ are referred to as ``accessory parameters''; clearly if we can learn something about them then we are learning about the semiclassical conformal block.  In \cite{Zamolodchikov1986} it was shown that for $\Delta\gg c$, a combination of symmetry and the WKB approximation allows a determination of all $C_i$, and thus of the semiclassical conformal block in that limit.  

This argument begins with the observation that there are three linear relations on the accessory parameters which come from demanding that the term in round brackets in (\ref{psipeq}), which is related to the semiclassical limit of the stress tensor by the Ward identity $\langle T(z) \mathcal{O}_4\mathcal{O}_3\mathcal{O}_2\mathcal{O}_1\rangle=\sum_{i=1}^4\left(\frac{\Delta_i}{(z-z_i)^2}+\frac{1}{z-z_i}\partial_i\right)\langle  \mathcal{O}_4\mathcal{O}_3\mathcal{O}_2\mathcal{O}_1\rangle$, vanishes like $z^{-4}$ at infinity: 
\begin{align}\nonumber
&\sum_i C_i=0\\\nonumber
&\sum_i \left(C_i z_i-\delta_i\right)=0\\
&\sum_i\left(C_i z_i^2-2\delta_i z_i\right)=0.
\end{align} 
There is thus only one independent accessory parameter, which we take to be $C_2$.  If we then take the limit $z_4\to \infty,z_3 \to 1,z_2 \to x, z_1\to 0$, equation (\ref{psipeq}) becomes
\be
\label{reducedode}
\left[\partial_z^2+\frac{\delta_1}{z^2}+\frac{\delta_2}{(z-x)^2}+\frac{\delta_3}{(1-z)^2}+\frac{\delta_1+\delta_2+\delta_3-\delta_4}{z(1-z)}-\frac{C_2 x(1-x)}{z(z-x)(1-z)}\right]\Psi_p=0.
\ee
In the same limit (\ref{accpar}) simplifies to
\be
C_2=\partial_x f_{cl}.
\ee
  
One way to parametrize the effect of $C_2$ is to study the monodromy of $\Psi_p$ as $z$ circles both $x$ and $z_1$.  We will work out this relationship below, but we first note that this will give us what we want because we can also determine this monodromy from the definition of $\Psi_p$.  The reason is that as discussed in section \ref{4pointreview}, the four-point function $\langle\mathcal{O}_4\mathcal{O}_3\Psi\mathcal{O}_p\rangle$ receives contributions only from two intermediate conformal weights.  If we parametrize conformal weights as $\Delta=\alpha(b+1/b-\alpha)$ then these are $\Delta_{\pm}=\alpha_\pm(b+1/b-\alpha_\pm)$, with $\alpha_\pm=\alpha_p\pm b/2$.  These contributions behave near $z=0$ like $z^{\Delta_{\pm}-\Delta_p-\Delta_z}$, which semiclassically becomes $z^{\frac{1}{2}\left(1\pm\sqrt{1-4b^2\Delta_p}\right)}$, so their monodromy matrix in this basis is
\be
\label{Mmatrix}
M=\begin{pmatrix}
e^{i\pi\left(1+\sqrt{1-4b^2\Delta_p}\right)}  & 0\\
0 & e^{i\pi\left(1-\sqrt{1-4b^2\Delta_p}\right)}
\end{pmatrix}.
\ee 
So far everything we have said is valid for $\Delta_p\sim c$, but we now observe that if we take $\Delta_p\gg c$ then we can solve (\ref{reducedode}) in the WKB approximation, where we include only the first and last terms.  This gives approximate solution:
\be
\Psi_p\sim \exp\left[\pm\sqrt{x(1-x)C_2}\int_{z_0}^z\frac{dz'}{\sqrt{z'(1-z')(z'-x)}}\right].
\ee
Comparison with the mondromy matrix (\ref{Mmatrix}) in the same limit gives
\be
C_2\approx -\frac{\pi^2 b^2 \Delta_p}{x(1-x)K(x)^2}.
\ee
Finally we can integrate this by observing that $K(x) \partial_x K(1-x)-K(1-x)\partial_x K(x)=-\frac{\pi}{4x(1-x)}$, which follows from the Wronskian of the hypergeometric differential equation obeyed by $K(x)=\frac{\pi}{2}F(\frac{1}{2},\frac{1}{2},1,x)$, with the normalization determined by expanding near $x=0$.  This gives 
\be
f_{cl}\approx \pi b^2\Delta_p \frac{K(1-x)}{K(x)}+constant.
\ee
We can determine the constant to be $-b^2\Delta\log 16$ by matching the series expansion near $x=0$ to the normalization of the leading term in the conformal block
\be
\mathcal{F}_{1234}(\Delta_i,\Delta_p,x)=x^{\Delta_p-\Delta_1-\Delta_2}\left(1+\mathcal{O}(x)\right),
\ee
which at last gives 
\be
\mathcal{F}_{1234}\sim (16q)^\Delta.
\ee

\chapter[An Integral Expression for log(Gamma(z))]{An Integral Expression for $\log{\Gamma(z)}$}\label{loggammazapp}
\hspace{0.5in} In this appendix we derive the identity
\begin{equation}\label{lnGident}
\log{\Gamma(z)} = \int^{\infty}_{0}\frac{dt}{t}\Big[(z-1)e^{-t} -\frac{e^{-t} - e^{-zt}}{1 - e^{-t}}\Big] \hspace{0.25in}\text{$\mathrm{Re}[z] > 0.$}
\end{equation}

We by differentiating the definition (\ref{gamma1}) of the Gamma function with respect to $z$:
\begin{align}
\Gamma'(z) &= \frac{d\int^{\infty}_{0}t^{z-1}e^{-t}dt}{dz} = \int^{\infty}_{0}t^{z-1}\log{|t|}e^{-t}dt\\&= \int^{\infty}_{0}dt t^{z-1}e^{-t}\int^{\infty}_{0}\frac{ds}{s}[e^{-s} - e^{-st}] = \lim_{\epsilon \rightarrow 0} \int^{\infty}_{0}dt t^{z-1}e^{-t}\int^{\infty}_{\epsilon}\frac{ds}{s}[e^{-s} - e^{-st}]\\&= \lim_{\epsilon \rightarrow 0} \int^{\infty}_{0}dt t^{z-1}e^{-t}\int^{\infty}_{\epsilon}\frac{ds}{s}e^{-s} - \lim_{\epsilon \rightarrow 0} \int^{\infty}_{0}dt t^{z-1}\int^{\infty}_{\epsilon}\frac{ds}{s}e^{-t} e^{-st} 
\end{align}

Here we have used integral identity $\log{b} = \int^{\infty}_{0}\frac{dx}{x}(e^{-x} - e^{-bx})$. While the integral currently is convergent, we have to reexpress the lower limit in terms of $\epsilon$ in order to break the integral up into divergent pieces. We do this to allow us to enact coordinate changes on the second term. The divergences will cancel each other out. We make the coordinate change $\rho = t(1+s)$ on the second term. The change is linear in $t$. We exchange the $\rho$ and $s$ integrals and evaluate the $t$ and $\rho$ integrals giving
\begin{align}
\lim_{\epsilon \rightarrow 0} &\int^{\infty}_{0}dt t^{z-1}e^{-t}\int^{\infty}_{\epsilon}\frac{ds}{s}e^{-s} - \lim_{\epsilon \rightarrow 0}\int^{\infty}_{\epsilon}\frac{ds}{s} \int^{\infty}_{0}dt t^{z-1} e^{-t}e^{-st} \\
&=\Gamma(z)\lim_{\epsilon \rightarrow 0}\int^{\infty}_{\epsilon}\frac{ds}{s}e^{-s} - \lim_{\epsilon \rightarrow 0} \int^{\infty}_{\epsilon}\frac{ds}{s}\int^{\infty}_{0}\frac{d\rho}{1+s} \Big(\frac{\rho}{s+1}\Big)^{z-1} e^{-\rho}\\
&=\Gamma(z)\Big[\lim_{\epsilon \rightarrow 0}\int^{\infty}_{\epsilon}\frac{ds}{s}e^{-s} - \lim_{\epsilon \rightarrow 0} \int^{\infty}_{\epsilon}\frac{ds}{s(1+s)} \Big(\frac{1}{s+1}\Big)^{z-1}\Big].\label{drel}
\end{align}

We see that the Gamma function factors out of the integral representation of $\Gamma'(z)$. To continue, we need to make one more coordinate change before putting the integrals back together $1 + s = e^{\widetilde{s}}$ which yields $ds = e^{\widetilde{s}}d\widetilde{s}$. Since $\epsilon \rightarrow 0$, when $s \sim \epsilon$ then $1 +s = e^{\widetilde{s}} = 1 + \widetilde{s} + \O(\widetilde{s}^{2}) \sim 1 + \epsilon$ and $\widetilde{s} \sim \epsilon$. The lower limit remains unchanged in the second integral.
(\ref{drel}) then becomes
\begin{align}
\Gamma'(z) &=\Gamma(z)\Big[\lim_{\epsilon \rightarrow 0}\int^{\infty}_{\epsilon}\frac{ds}{s}e^{-s} - \lim_{\epsilon \rightarrow 0} \int^{\infty}_{\epsilon}\frac{ds}{s(1+s)} \Big(\frac{1}{s+1}\Big)^{z-1}\Big]\\
&= \Gamma(z)\Big[\lim_{\epsilon \rightarrow 0}\int^{\infty}_{\epsilon}\frac{ds}{s}e^{-s} - \lim_{\epsilon \rightarrow 0} \int^{\infty}_{\epsilon}\frac{d\widetilde{s}e^{\widetilde{s}}}{(e^{\widetilde{s}} - 1)e^{\widetilde{s}}} (e^{-\widetilde{s}})^{z-1}\Big]\\
&=\Gamma(z)\Big[\lim_{\epsilon \rightarrow 0}\int^{\infty}_{\epsilon}\frac{ds}{s}e^{-s} -  \lim_{\epsilon \rightarrow 0} \int^{\infty}_{\epsilon}\frac{d\widetilde{s}}{(1 - e^{-\widetilde{s}})} e^{-z\widetilde{s}}\Big]
 \end{align}

Now we can put the integrals back together and take the limit $\epsilon \rightarrow 0$, resulting in

\begin{equation}\label{digamma}
\frac{\Gamma'(z)}{\Gamma(z)} =  \int^{\infty}_{0}ds\Big[\frac{1}{s}e^{-s} - \frac{ e^{-zs}}{(1 - e^{-s})}\Big] .
\end{equation}

Integrating (\ref{digamma}) with respect to $z$ yields (\ref{lnGident})
\begin{equation}
\log{\Gamma(z)} = \int^{z}_{1}d\tilde{z}\frac{\Gamma'(\tilde{z})}{\Gamma(\tilde{z})} =\int^{\infty}_{0}\frac{ds}{s}\Big[(z - 1)e^{-s} - \frac{e^{-s} - e^{-zs}}{(1 - e^{-\widetilde{s}})}\Big]
\end{equation}

\chapter[Integral over the SL(2,C) Moduli of the Saddle Point for Three Light Operators]{Integral over the SL(2,$\mathbb{C}$) Moduli of the Saddle Point for Three Light Operators}\label{sl2cintegralapp}

In section \ref{dozzthreelight}, we encountered the following integral:
\be
\label{integral}
I(\sigma_1,\sigma_2,\sigma_3)\equiv \int \frac{d\mu (\alpha,\beta,\gamma,\delta)}{\Big(|\beta|^2+|\delta|^2\Big)^{2\sigma_1}\Big(|\alpha+\beta|^2+|\gamma+\delta|^2\Big)^{2\sigma_2}\Big(|\alpha|^2+|\gamma|^2\Big)^{2\sigma_3}}.
\ee
In the text we claimed this integral is given by
\be\label{alianswer}
I(\sigma_1,\sigma_2,\sigma_3)=\pi^3\frac{\Gamma(\sigma_1+\sigma_2-\sigma_3)\Gamma(\sigma_1+\sigma_3-\sigma_2)\Gamma(\sigma_2+\sigma_3-\sigma_1)\Gamma(\sigma_1+\sigma_2+\sigma_3-1)}{\Gamma(2\sigma_1)\Gamma(2\sigma_2)\Gamma(2\sigma_3)},
\ee
and in this appendix we will show it.  We will see along the way that the integral is divergent unless certain inequalities involving the $\sigma_i$'s are satisfied, so we will assume them as we come to them and then in the end define the integral away from those regions by analytic continuation.

Following \cite{Zamolodchikov:1995aa}, we begin by performing the coordinate change $\xi_{1} = \frac{\beta}{\delta}$, $\xi_{2} = \frac{\alpha + \beta}{\gamma + \delta}$, and $\xi_{3} = \frac{\alpha}{\gamma}$.  The measure becomes
$d\mu(\alpha,\beta,\gamma,\delta) = \frac{d^{2}\xi_{1}d^{2}\xi_{2}d^{2}\xi_{3}}{|(\xi_{1} - \xi_{2})(\xi_{2} - \xi_ {3})(\xi_{3} - \xi_{1})|^{2}}$, and the integral becomes
\begin{align} \label{lightIinitialform}\nonumber
I(\sigma_i)=&\int d^2 \xi_1 d^2 \xi_2 d^2 \xi_3 |\xi_{12}|^{-2-2\nu_3}|\xi_{23}|^{-2-2\nu_1}|\xi_{13}|^{-2-2\nu_2}\\
&\times (1+|\xi_1|^2)^{-2\sigma_1}(1+|\xi_2|^2)^{-2\sigma_2}(1+|\xi_3|^2)^{-2\sigma_3}
\end{align}
Here  $\nu_{1} = \sigma_{1} - \sigma_{2} - \sigma_{3}, \nu_{2} = \sigma_{2} - \sigma_{1} - \sigma_{3}, \nu_{3} = \sigma_{3} - \sigma_{1} - \sigma_{2}$.  This expression is invariant under the $SU(2)$ subgroup of $SL(2,\mathbb{C})$ given by $\xi_{i} \rightarrow \frac{f\xi_{i} + g}{-\bar{g}\xi_{i} + \bar{f}}$, with $|f|^{2} + |g|^{2} = 1$.  We can use this to send $\xi_{3} \to \infty$:

\begin{align}\label{lightIfinalform}
I(\sigma_1,\sigma_2,\sigma_3)=\pi\int\,d^{2}\xi_{1}d^{2}\xi_{2}|\xi_{1} - \xi_{2}|^{-2-2\nu_{3}}(1 + |\xi_{1}|^{2})^{-2\sigma_{1}}(1 + |\xi_{2}|^{2})^{-2\sigma_{2}}.
\end{align}
From here the result is quoted in \cite{Zamolodchikov:1995aa} without further explanation, we will fill in the steps.  The evaluation will involve repeated use of the defining integrals of the $\Gamma$ and $\beta$ functions, which we reproduce for convenience:
\begin{equation}\label{gammaapp}
\Gamma(x) = \int^{\infty}_{0} t^{x-1}e^{-t}dt\hspace{0.25in}\text{$\mathrm{Re}[x]>0$}
\end{equation}
\begin{equation}\label{beta}
\beta(x,y) = \frac{\Gamma(x)\Gamma(y)}{\Gamma(x+y)}=\int^{1}_{0}\lambda^{x-1}(1-\lambda)^{y-1}d\lambda=\int^{\infty}_{0}dt\frac{t^{x-1}}{(1+t)^{x+y}}
\end{equation}
\begin{center}
\text{$\mathrm{Re}[x]>0$, $\mathrm{Re}[y]>0.$}
\end{center}
We will also need a lesser-known version of the Feynman parameters which is used in closed string theory.\footnote{This identity can be easily derived by changing variables to $\tilde{t}=t|z|^2$.  
}
\begin{equation}\label{feynman}
\frac{1}{|z|^{A}} = \frac{1}{\Gamma(A/2)}\int^{\infty}_{0}dt\hspace{1pt} t^{A/2-1}e^{-t|z|^{2}}\hspace{0.1in} \text{ with $\Re\,A >0$.}
\end{equation}

Now to business.  Starting with integral (\ref{lightIfinalform}), we convert it into four real integrals defined by the coordinate change
\begin{equation}\label{real}
\xi_{1} = x +iy\hspace{1in}\xi_{2}=u+iv.
\end{equation}
Noting that $d^2\xi_{1} = \frac{-1}{2i}(d\xi_{1}\wedge d\bar{\xi_{1}}) = \frac{-1}{2i}(dx +idy)\wedge(dx-idy) = dx\wedge dy $ with a similar identity for $d^2\xi_{2}$, we find equation (\ref{lightIfinalform}) becomes
\begin{equation}\label{intreal}
I = \pi\int^{\infty}_{-\infty} dx\,dy\,du\,dv\,((x-u)^{2} + (y-v)^{2})^{-1-(\sigma_{3}-\sigma_{1}-\sigma_{2})}(1+x^{2}+y^{2})^{-2\sigma_{1}}(1+u^{2}+v^{2})^{-2\sigma_{2}}.
\end{equation}
We now use the identity (\ref{feynman}) three times with $A=4\sigma_1,4\sigma_2,2+2\nu_3$, which requires $\mathrm{Re} \,\sigma_1>0,\mathrm{Re} \,\sigma_2>0, \mathrm{Re}\, (\sigma_1+\sigma_2-\sigma_3)<1$, to get:
\begin{align}\label{realfey}
I = \pi\int^{\infty}_{-\infty} dx\,dy&\,du\,dv\,\int^{\infty}_{0}d\psi \,d\chi\, d\kappa\,\frac{\psi^{\sigma_{3}-\sigma_{2}-\sigma_{1}}\chi^{2\sigma_{1}-1}\kappa^{2\sigma_{2}-1}}{\Gamma(1+\sigma_{3}-\sigma_{1}-\sigma_{2})\Gamma(2\sigma_{1})\Gamma(2\sigma_{2})}\\\times
&\exp{[-\{\psi[(x-u)^{2}+(y-v)^{2}] +\chi(1+x^{2}+y^{2})+ \kappa(1+u^{2}+v^{2})\}]}\nonumber.
\end{align}
Collecting powers of $x,y,u,v$ we have
\begin{align}\label{realfey2}
I = \pi\int^{\infty}_{-\infty} dx&\,dy\,du\,dv\,\int^{\infty}_{0}d\psi\, d\chi\, d\kappa\,\frac{\psi^{\sigma_{3}-\sigma_{2}-\sigma_{1}}\chi^{2\sigma_{1}-1}\kappa^{2\sigma_{2}-1}}{\Gamma(1+\sigma_{3}-\sigma_{1}-\sigma_{2})\Gamma(2\sigma_{1})\Gamma(2\sigma_{2})}\\\times
&\exp{[-\{(x^{2}+y^{2})(\psi+\chi) +(u^{2}+v^{2})(\psi + \kappa) - 2\psi(ux +yv) + \chi + \kappa\}]}\nonumber.
\end{align}
Completing the square for $x$ and $y$ and factoring out $(\psi + \chi)$ gives 
\begin{align}\label{realfey3}
I = &\pi\int^{\infty}_{-\infty} dx\,dy\,du\,dv\,\int^{\infty}_{0}d\psi\, d\chi\, d\kappa\,\frac{\psi^{\sigma_{3}-\sigma_{2}-\sigma_{1}}\chi^{2\sigma_{1}-1}\kappa^{2\sigma_{2}-1}}{\Gamma(1+\sigma_{3}-\sigma_{1}-\sigma_{2})\Gamma(2\sigma_{1})\Gamma(2\sigma_{2})}\\&\hspace{0.5in}\times
\exp{[-(\psi+\chi)\Big\{\Big(x - \frac{u\psi}{\psi + \chi}\Big)^{2} + \Big(y - \frac{v\psi}{\psi + \chi}\Big)^{2}\Big\}]}\nonumber\\&\hspace{0.8in}\times\exp{-(\psi + \chi)\Big\{ (u^{2}+v^{2})\frac{\psi + \kappa}{\psi + \chi}- (u^{2}+v^{2})\Big(\frac{\psi^{2}}{(\psi + \chi)^{2}}\Big) +\frac{\chi + \kappa}{\psi + \chi}\Big\}]}\nonumber.
\end{align}
The $x,y$ integral is now a straightforward Gaussian integral.  Changing variables by the linear shift $ s = x - \frac{u\psi}{\psi + \chi}$, $t = y - \frac{v\psi}{\psi + \chi}$ and performing the integral over $s$ and $t$ we find
\begin{align}\label{realfey4}
&I = \pi^{2}\int^{\infty}_{-\infty} du\,dv\,\int^{\infty}_{0}d\psi \,d\chi\, d\kappa\,\frac{\psi^{\sigma_{3}-\sigma_{2}-\sigma_{1}}\chi^{2\sigma_{1}-1}\kappa^{2\sigma_{2}-1}}{(\psi + \chi)\Gamma(1+\sigma_{3}-\sigma_{1}-\sigma_{2})\Gamma(2\sigma_{1})\Gamma(2\sigma_{2})}\\&\hspace{1in}\times
\exp{[-\Big\{(u^{2}+v^{2})\Big[\psi + \kappa -\frac{\psi^{2}}{\psi + \chi}\Big] + \chi + \kappa\Big\}]}\nonumber.
\end{align}
We can also do the $u$ and $v$ integrals, which give\footnote{ For the $u$, $v$ integrals in (\ref{realfey4}) to converge the prefactor of the $u$ and $v$ exponetial terms in (\ref{realfey4}): $\big[\psi + \kappa - \frac{\psi^{2}}{\psi + \chi}\big] \geq 0$. It can be shown that this is so, by putting everthing over a common denominator, and noting that the term becomes a sum of positive quantities.}
\begin{align}\label{realfey5}
&I = \frac{\pi^{3}}{\Gamma(1+\sigma_{3}-\sigma_{1}-\sigma_{2})\Gamma(2\sigma_{1})\Gamma(2\sigma_{2})}\int^{\infty}_{0}d\psi\, d\chi\, d\kappa\,\frac{\psi^{\sigma_{3}-\sigma_{2}-\sigma_{1}}\chi^{2\sigma_{1}-1}\kappa^{2\sigma_{2}-1}e^{-\chi-\kappa}}{[(\psi+\kappa)(\psi+\chi)-\psi^{2}]}.
\end{align}

So far the required manipulations have been fairly obvious, but to proceed further we need to perform a rather nontrivial coordinate change on $\psi$, $\chi$, and $\kappa$.  We will motivate it by answer analysis of (\ref{alianswer}).  Summing the exponents of the $\psi$, $\chi$, and $\kappa$ terms in the numerator of (\ref{realfey5}), we get one minus the argument of the first Gamma function in (\ref{alianswer}).  This implies that in the new coordinates, $\psi$, $\chi$, and $\kappa$ must have some common factor $\rho$ in order to produce this first Gamma function. This also means $\rho$ must equal $\chi +\kappa$ due to definition (\ref{gammaapp}). Now $\chi$ and $\kappa$ are linearly independent, therefore we propose the coordinate change: $\chi=\rho \cos^{2}\theta$ and $\kappa = \rho \sin^{2}\theta$. To determine $\psi$ we see that if we take the arguments of last two Gamma functions in the numerator of (\ref{alianswer}), they add up to $2\sigma_{3}$. That means the factor $\frac{\Gamma(\sigma_{3}+\sigma_{2}-\sigma_{1})\Gamma(\sigma_{3}+\sigma_{1}-\sigma{2})}{\Gamma(2\sigma_{3})} = \beta(\sigma_{3}+\sigma_{2}-\sigma_{1},\sigma_{3}+\sigma_{1}-\sigma_{2})$ must be a factor in the integral. This Beta function will involve factors the of $\sin^2\theta$ and $\cos^{2}\theta$ from $\chi$ and $\kappa$. A quick glance at the factors of $\chi$ and $\kappa$ in (\ref{realfey5}) shows that they are  not correct.  However, if $\psi$ instead included a factor of $\cos^{2}(\theta) \sin^{2}(\theta)$, we would get the proper factors for the Beta function in terms of the integral over $\theta$. $\psi$ must then have one factor of $\rho$ and one factor of $\cos^{2}(\theta)\sin^{2}(\theta)$. Lastly $\psi$ must be linearly independent from $\chi$ and $\kappa$ since we want a one to one mapping, so $\psi$ must have a third factor which we will call $\zeta$. The coordinate change is then
\begin{equation}\label{coordinatechange}
\psi =\rho\zeta \cos^{2}(\theta) \sin^{2}(\theta)\hspace{0.5in}\chi=\rho \cos^{2}(\theta)\hspace{0.5in}\kappa = \rho \sin^{2}(\theta).
\end{equation}

The Jacobian for this is
\begin{equation}\label{jacobian}
d\psi d\chi d\kappa = 2\rho^{2}\cos^{3}(\theta) \sin^{3}(\theta) d\rho d\theta d\zeta.
\end{equation}

After the performing the transformation and collecting common terms, (\ref{realfey5}) becomes
\begin{align}\label{answer1}
&I = \frac{2\pi^{3}}{\Gamma(1+\sigma_{3}-\sigma_{1}-\sigma_{2})\Gamma(2\sigma_{1})\Gamma(2\sigma_{2})}\int^{\infty}_{0}\int^{\infty}_{0}\int^{\pi/2}_{0}d\zeta \,d\rho \,d\theta \nonumber\\&\,\hspace{0.2in}\times\frac{\rho^{\sigma_{1}+\sigma_{2}+\sigma_{3}-1-1}\zeta^{1+\sigma_{3} - \sigma_{2}-\sigma_{1} - 1}}{(1+\zeta)}(\cos\theta)^{2\sigma_{3}+2\sigma_{1}-2\sigma_{2}-1}(\sin\theta)^{2\sigma_{3} +2\sigma_{2} - 2\sigma_{1} - 1}\exp{[-\rho]}.
\end{align}
We now change the $\theta$ coordinate to $\lambda = sin^{2}\theta$ which gives $d\theta = \frac{d\lambda}{2\sqrt{\lambda(1-\lambda)}}$ and thus
\begin{align}
&I = \frac{\pi^{3}}{\Gamma(1+\sigma_{3}-\sigma_{1}-\sigma_{2})\Gamma(2\sigma_{1})\Gamma(2\sigma_{2})}\int^{\infty}_{0}\int^{\infty}_{0}\int^{1}_{0}d\zeta \,d\rho \,d\lambda \,\nonumber\\&\hspace{0.25in}\times\frac{\rho^{\sigma_{1}+\sigma_{2}+\sigma_{3}-1-1}\zeta^{1+\sigma_{3} - \sigma_{2}-\sigma_{1} - 1}}{(1+\zeta)}(1-\lambda)^{\sigma_{3}+\sigma_{1}-\sigma_{2}-1}\lambda^{\sigma_{3} +\sigma_{2} - \sigma_{1} - 1}\exp{[-\rho]}.
\end{align}
These transformations have completely factorized the integral, which we can at last evaluate using (\ref{gammaapp}) and (\ref{beta}).  The result is:
\begin{align}\label{answerfinal}
I=\pi^{3}\frac{\Gamma(\sigma_{1}+\sigma_{2}+\sigma_{3}-1)\Gamma(\sigma_{1}+\sigma_{2}-\sigma_{3})\Gamma(\sigma_{2}+\sigma_{3}-\sigma_{1})\Gamma(\sigma_{3}+\sigma_{1}-\sigma_{2})}{\Gamma(2\sigma_{1})\Gamma(2\sigma_{2})\Gamma(2\sigma_{3})}.
\end{align}

In addition to the inequalities mentioned below (\ref{intreal}), in doing these final integrals we had to assume that 
\begin{align}\label{appineq}
\mathrm{Re}(\sigma_1+\sigma_2-\sigma_3)>&0\\ \nonumber
\mathrm{Re}(\sigma_1+\sigma_3-\sigma_2)>&0\\\nonumber
\mathrm{Re}(\sigma_2+\sigma_3-\sigma_1)>&0\\\nonumber
\mathrm{Re}(\sigma_1+\sigma_2+\sigma_3)>&1. 
\end{align}
These inequalities are easy to understand from equation (\ref{lightIfinalform}); they come from convergence of the integral when $\xi_1\to \xi_2$ and when either $\xi_1$ or $\xi_2$ or both go to infinity. The inequalities $\mathrm{Re}(\sigma_1)>0$, $\mathrm{Re}(\sigma_2)>0$ that we assumed in deriving (\ref{intreal}) automatically follow from these, but the third inequality, $\mathrm{Re}(\sigma_1+\sigma_2-\sigma_3)<1$, does not.  This final inequality is somewhat mysterious becuse it breaks the symmetry between $\sigma_1,\sigma_2,\sigma_3$ and does not follow in any obvious way from the convergence of (\ref{lightIfinalform}).  In fact it is a relic of our method of evaluating the integral; in deriving (\ref{realfey}) we used (\ref{feynman}) to introduce a factor of $\frac{1}{\Gamma(1+\sigma_3-\sigma_1-\sigma_2)}$ which then cancelled out of the integral in the end.  We could have avoided this inequality by deforming the contour in (\ref{feynman}).  For another way to see that the inequality is spurious, we observe that if we had used the symmetry to send $\xi_1$ or $\xi_2$ to infinity instead of $\xi_3$ in deriving (\ref{lightIfinalform}) then a different inequality related to this one by symmetry would have appeared that we did not need to use in our evaluation.  Thus the only conditions for convergence of the integral (\ref{integral}) are given by (\ref{appineq}).

\chapter{SU(2) gauge theory}
In the case of SU(2) the potential on the moduli space has the form,
\begin{equation}
V(A_{1}^{3},A_{2}^{3}) = -\frac{2}{\pi} \sum_{(m,n)\neq (0,0)} \frac{1}{(m^{2}+n^{2})^{\frac{3}{2}}} cos(\frac{mA_{1}^{3}+nA_{2}^{3}}{2})
\end{equation}
where the SU(2) flat connection has been parametrized as $A_{i} = \frac{1}{2}A_{i}^{3}\sigma^{3}$ where $\sigma^{3}$ is the diagonal Pauli spin matrix.
So the same argument as in the Abelian case shows that we can expand around the trivial flat connection. The action for the reduced quantum mechanical problem is given by, 
\begin{equation}
{S} = \frac{ k}{8{\pi}} \int dt \ Tr(A_{1}\frac{d}{dt} A_{2})  + \int dt \  [\frac{d\phi^{\dagger}}{dt}\frac{d\phi}{dt}  - {\phi}^{\dagger}A_{i}A_{i}{\phi}]
\end{equation}
where $\phi$ is a complex two component column vector transforming in the fundamental representation of the $SU(2)$. $A_{1}$ and $A_{2}$ are $2\times 2$ traceless Hermitian matrices transforming in the adjoint representation of $SU(2)$. 
With this the Hamiltonian becomes 
\begin{equation}
H = \pi_{1}^{*}\pi_{1} + \pi_{2}^{*}\pi_{2} + {\phi}^{\dagger}A_{i}A_{i}{\phi}
\end{equation}
where $\pi_{i} (\pi_{i}^{*})$ are canonical momenta conjugate to ${\phi}_{i}({\phi}_{i}^{*})$.
The Hamiltonian again factorizes in this case. We can write 
\begin{equation}
\phi^{\dagger}A_{i}A_{i}\phi = \frac{1}{4} \phi^{\dagger}\sigma^{a}\sigma^{b}\phi A_{i}^{a}A_{i}^{b} = \frac{1}{4} \phi^{\dagger}\delta^{ab}\phi A_{i}^{a}A_{i}^{b}= \frac{1}{2}\phi^{\dagger}\phi \ Tr(A_{i}A_{i})
\end{equation}
where $A_{i} = \sum_{a=1}^{3} \frac{1}{2}A_{i}^{a}\sigma^{a}$ and $\sigma^{a}$ are generators of SU(2). So the Hamiltonian becomes 
\begin{equation}
H = \pi_{1}^{*}\pi_{1} + \pi_{2}^{*}\pi_{2} + \frac{1}{2}\phi^{\dagger}\phi \ Tr(A_{i}A_{i})
\end{equation}
We can see that the Hamiltonian has a manifestly factorized form. The Lagrangian has a global SU(2) symmetry which is the remnant of the original gauge symmetry of the field theory. So in the quantum mechanics we should set this charge to zero. This is given by the constraint, 
\begin{equation}
\frac{ k}{8\pi}i[A_{1},A_{2}]^{a} = J_{0}^{a}
\end{equation}
where $J_{0}^{a}$ is the Noether charge which generates the SU(2) rotations of the scalars. There are operator ordering ambiguities associated with the definition of the charge $J_{0}^{a}$. We shall resolve these issues in the next section where we shall discuss the case of a general $U(N)$ gauge group.

 We shall first quantize the Hamiltonian and then implement the constraint by projecting onto SU(2) invariant states in the Hilbert space.

In the $SU(2)$ case the reduced quantum mechanics model has a larger symmetry which is $SU(2)\times SU(2)$. The fields transform as 
\begin{equation}
\phi \rightarrow U_{1}\phi,\ A_{i} \rightarrow U_{2}A_{i}U_{2}^{\dagger}
\end{equation} 
where $U_{1}$ and $U_{2}$ are constant $SU(2)$ matrices. The original global gauge invariance of the quantum mechanical model is the diagonal $SU(2)$.  Now one can compute the Noether charges corresponding to these symmetries and what one finds is that left-hand side and the right-hand sides of the constraint are the generators of the individual $SU(2)$ transformations. The constraint is the statement that the physical wave functions are invariant under the diagonal symmetry transformations. Since the Hamiltonian factorizes we can start with product wave functions where each factor transforms in some definite representation of the respective $SU(2)$'s. Then the Gauss's law constraint is satisfied by picking up the singlet in the product representation.

\chapter{Effective Potential}
The method we use in this appendix is similar to that used in \cite{Aharony:2005ew}, except that we use heat-kernel method and so can be easily generalized to the case when the spatial slice is any higher genus Riemann surface.

We want to compute the determinant of the operator $-D^{2}$ where 
\begin{equation}
D_{\mu} = \partial_{\mu} + iA_{\mu}
\end{equation}
and $A_{\mu}$ is a flat gauge field of $U(N)$ in the fundamental representation.  One way of doing this is to solve the heat equation for this operator. So let us start with a quantum mechanical problem in euclidean space. The Euclidean propagator is defined as ,
\begin{equation}
G(x',s; x,0) = <x'\exp(-sH)|x>
\end{equation} 
where $s$ is a fictitious Euclidean time and $H$ is the Hamiltonian. The propagator satisfies the boundary condition :
\begin{equation}
G(x',s; x,0) \rightarrow {\delta}(x'-x) 
\end{equation} 
as $s\rightarrow 0$. The equation satisfied by the propagator is the Euclidean wave equation (Heat equation)
\begin{equation}
-\frac{\partial}{\partial s} G(x',s;x,0) = H_{x'} G(x',s;x,0)
\end{equation}
Now we expand the propagator in terms of a complete set of eigenfunctions of the Hamiltonian:
\begin{equation}
G(x',s;x,0) = \sum_{n} exp(-sE_{n}) <x'|n><n|x>
\end{equation}
It follows from this expansion that
\begin{equation}\label{mainequation}
\int_0^\infty \frac{ds}{s}\int dx\, G(x,s;x,0) = \sum_{n}(-ln E_{n}) = - ln\prod_{n} E_{n} = - ln Det (H)
\end{equation}
This is our main equation\footnote{We have done the standard thing of subtracting the infinite constant $\int^{\infty}_{0}\frac{ds}{s}e^{-s}$ in the identity $\log{b} =\int^{\infty}_{0}\frac{ds}{s}(e^{-s} - e^{-sb})$ to obtain (\ref{mainequation}). The result is obtained by a standard  renormalization.}.
So we can compute the determinant of the operator $H$ if we know the propagator of the corresponding quantum mechanical problem.
\subsection{Gauge Theory}
In our case the Hamiltonian of the quantum mechanical problem is $H= -D^{2}$. So the Hamiltonian is an $N\times N$ matrix-valued differential operator.  As a result the propagator is also an $N\times N$ matrix. Under a gauge transformation the Hamiltonian transforms as:
\begin{equation}
H_{x}\rightarrow U(x) H_{x} U(x)^{-1} 
\end{equation}
where $U(x)$ is a $U(N)$ valued gauge transformation. In the case of matrix valued differential operators and matrix valued propagators the formula for the determinant has the following form,
\begin{equation}\label{determinant}
\int_0^\infty \frac{ds}{s}\int dx\, Tr G(x,s;x,0) = - ln Det (H)
\end{equation}
where the trace is over the internal matrix indices. The form of the heat equation remains unchanged and covariance under gauge transformations requires the propagator to transform as
\begin{equation}
G(x',s;x,0) \rightarrow U(x')G(x',s;x,0)U(x)^{-1}
\end{equation} 
It is important that the trace computed in (\ref{determinant}) at the coincident points, $x=x'$, is invariant under this gauge transformation. The gauge transformation function does not depend on the fictitious euclidean time $s$.
So we have to solve the quantum mechanical problem of a particle carrying isotopic spin moving in the background of flat non-abelian gauge field. Since the background gauge field is flat at least locally the answer has the form
\begin{equation}\label{flatanswer}
G(x',s;x,0) \approx  Pexp(-i \int_x^{x'} A) \\G_{0}(x',s;x,0)
\end{equation}
where $P$ is the path ordering symbol and $G_{0}$ is the free propagator in the absence of the gauge field. We have not specified any particular path for the Wilson loop because the connection is flat and we are looking at a local patch and so the path chosen for the Wilson loop is homotopically trivial. If the space\footnote{By space we mean the three dimensional space on which the Chern-Simons matter theory lives. The fictitious Euclidean time does not play any role in our discussion. } is simply connected then this answer is exact and it shows that the propagator evaluated at coincident points, $x=x'$, is the same as the free one. So the heat kernel formula tells us that the determinant evaluated with a background gauge field is the same as the free one. This also follows from the facts that a flat connection in a simply connected space can be gauged away by a non-singular gauge transformation and the determinant is gauge invariant under such a transformation. 

In (\ref{flatanswer}) $G_{0}$ is the free propagator and so it is proportional to the identity matrix in $U(N)$ space. Since it is the free propagator it does not participate in the gauge transformation and so the $G$ defined in (\ref{flatanswer}) has the correct gauge transformation property which follows from the gauge transformation property of the Wilson line. 

\subsection{Multiply Connected Space}
If the space is not simply connected then (\ref{flatanswer}) has to corrected. Since the configuration space of the particle is multiply connected we have to work on the simply connected covering space of the configuration space.  The propagator on the original configuration space can be derived from the covering space propagator by the method of images. This works because the heat equation is a linear first order PDE. So if we can write down a solution of the heat equation which satisfies the boundary condition then that is the unique solution.

Let us denote by $M$ the three dimensional space on which the gauge theory lives. $M$ is not simply connected. Let us denote the simply connected covering space by $\hat M$. So we can write:
\begin{equation}
M = \frac{\hat M}{\pi _{1}(M)}
\end{equation}where ${\pi _{1}(M)}$ fundamental group of $M$. This equation means that there is  discrete group $\mathcal{G}$ isomorphic to the fundamental group of $M$ which acts freely on $\hat M$ and $M$ is the quotient of $\hat M$ by the action of this group.  The universal cover $\hat M$ is unique modulo diffeomorphism. We shall also assume the following things. The universal cover $\hat M$ has a metric and the group $\mathcal{G}$ is a discrete subgroup of the group of isomorphisms of the metric. As a result of this the metric induced on the quotient $M$ is the same as the metric on the cover $\hat M$\footnote{We shall relax this condition by considering conformally coupled matter fields.}. Since the operator depends on the metric if we neglect the gauge field then the heat equations are the same on the base and the covering space. Now we have to lift the gauge fields to the covering space. Since gauge fields are well defined function on the base $M$ they lift to periodic functions on the cover, i.e, the lifted gauge fields satisfy the property that, $A(x) = A({\gamma}x)$, $\forall {\gamma}\in \mathcal{G}$ and $x\in \hat M$. So the gauge fields are constant on the orbits of $\mathcal{G}$ \footnote{Tis is not a gauge invariant statement on the covering space. We shall discuss this in the following section.}. The lifted gauge fields are just the pullback of the gauge fields on base to the cover by the covering map and they are also flat on the covering space. For example in the case of $T^{2}\times R^{1}$ the covering space is $R^{2}\times R^{1}$ and the group $\mathcal{G}$ is the group of discrete translations of the plane which are isometries. The torus is the quotient of the plane by the discrete translations. The flat gauge fields on the torus lift to flat gauge fields on the plane which are periodic on the lattice. the holonomy of the gauge field along $a$ and $b$ cycles become the values of the Wilson lines  of the gauge field along the two sides of the unit cell of the lattice.  

\subsection{Torus}
We first solve the problem for $T^{2}\times R^{1}$.  The coordinates on the torus are denoted by the complex numbers $(z,\bar z)$ with periodicity $z\sim z + 2{\pi}(m + n{\tau})$ where $(m,n)\in \mathbb Z$ and ${\tau}$ is the complex structure. The coordinate along $R^{1}$ will be denoted by $x$. The metric can be written as , ${ds}^{2} = dzd\bar z + {dx}^{2}$. Since the eigenfunctions of the Hamiltonian operator are periodic on the torus and the propagator can be written in terms of the eigenfunctions of the operator it satisfies the same periodicity property:
\begin{equation}\label{periodicprop}
 G(z',\bar z',x',s;z,\bar z,x,0) = G(z'+2{\pi}(m'+n'{\tau}),\bar z' +c.c,x',s;z+2{\pi}(m+n{\tau}),\bar z+c.c,x,0)
 \end{equation}
where $c.c$ stands for complex conjugate. So the propagator on ${T^{2}\times R^{1}}$ is periodic. Now we shall work on the covering space which in this case is the complex plane. The lattice is generated by the complex numbers $(2{\pi},2{\pi}{\tau})$ and can be identified with the Abelian group $\mathbb Z \oplus \mathbb Z \sim 2{\pi}\mathbb Z \oplus 2{\pi}{\tau}\mathbb Z$. $\mathbb Z\oplus \mathbb Z$ is precisely the homotopy group of the torus and we can write, $T^{2} = \frac{R^{2}}{\mathbb Z \oplus \mathbb Z} \sim \frac{\mathbb C}{2{\pi}\mathbb Z \oplus 2{\pi}{\tau}\mathbb Z}$. So the covering space of the total geometry $T^{2}\times R^{1}$ can be written as $R^{2}\times R^{1}$ and $T^{2}\times R^{1} = \frac{R^{2}}{\Gamma}\times R^{1}$, where ${\Gamma}$ is the lattice. 

The flat gauge field  has nontrivial holonomies associated with the two noncontractible cycles in the geometry associated with the torus factor. The gauge field can be lifted to the covering space and the lifted gauge field satisfies the periodicity condition, $A(z+(m,n),\bar z+(m,n),x) = A(z,\bar z,x)$, where $(m,n)$ is a lattice translation vector. The periodicity condition is not a gauge invariant statement on the covering space. But one can think of it as a partial fixing of gauge in the covering space. The gauge transformations which survive are precisely those that have the same periodicity as the lattice. But they are also the gauge transformations which descend to the base. So the group of allowed gauge transformations on the cover are the same as the group of allowed gauge transformations on the base after this partial "gauge fixing".  

We can also think of this in the following way. The unit cell of the lattice with its sides periodically identified is identical to the torus. So any geometrical object defined on the torus can be defined on a single unit cell without any change. Once the object is defined in a single unit cell it can extended to the whole lattice by imposing the periodicity condition. In the case of gauge field this can be thought of as a gauge fixing condition. Since the covering space is simply connected there are no nontrivial flat connections on the covering space and so we can gauge it away. But the gauge transformation that we have to make does not satisfy the periodicity condition and so is not an allowed gauge transformation. 

The propagator in the covering space is :
\begin{equation}\label{covingspaceprop}
\bar G(z',\bar z',x',s;z,\bar z, x, 0) = W(z',\bar z',x';z,\bar z,x) G_{0}(z',\bar z',x',s;z,\bar z, x, 0)
\end{equation}
where $G_{0}$ is the free propagator on the cover and is proportional to the identity matrix in the $U(N)$ space. $W$ is the same Wilson line that appears in (\ref{flatanswer}). Now let us consider the case where the points $(z,z')$ belong to the same unit cell of the lattice. In that case the propagator $\bar G$ is a potential candidate for the propagator on $T^{2}\times R^{1}$. But this propagator does not satisfy the periodicity condition eqn-(12). The correct propagator can be obtained by summing over the lattice translation vectors and can be written as :
\begin{align}\label{latprop}
 G(z',\bar z',x',s;&z,\bar z, x, 0)\ =\ \sum_{{\gamma}\in{\Gamma}}\bar G(z' +{\gamma},\bar z'+\bar{\gamma},x',s;z,\bar z, x, 0) \nonumber\\ 
 &= \sum_{{\gamma}\in{\Gamma}}W(z'+{\gamma},\bar z'+\bar{\gamma},x';z,\bar z,x) G_{0}(z'+{\gamma},\bar z'+\bar{\gamma},x',s;z,\bar z, x, 0)
 \end{align}
 where $(z,z')$ belong to the same unit cell of the lattice. This is the propagator on $T^{2}\times R^{1}$. This satisfies the condition (\ref{periodicprop}). In the above formula the summation is only over the final position of the propagator. The same answer can be obtained by summing only over the initial position of the propagator. The reason for this is the following. The propagator $\bar G$ satisfies the condition that 
\begin{equation}
\bar G(z' +{\gamma},\bar z'+\bar{\gamma},x',s;z+{\gamma},\bar z + \bar{\gamma}, x, 0) = \bar G(z',\bar z',x',s;z,\bar z, x, 0),  \forall{\gamma}\in{\Gamma}
\end{equation}
This follows from the periodicity of everything under consideration along with the fact that the discrete group acts on the covering space as a group of isometries. As a result every unit cell is isometric to every other. So this result is almost trivial. So we have the following identity,
\begin{equation}
\bar G(z' +{\gamma}',\bar z'+\bar{\gamma}',x',s;z+{\gamma},\bar z + \bar{\gamma}, x, 0) =  \bar G(z' +{\gamma}' -{\gamma},\bar z'+\bar{\gamma}' - \bar{\gamma},x',s;z,\bar z, x, 0)
\end{equation}
Using this we can reduce any double sum over both the final and initial points can be reduced to a single sum over either the initial point or the final point.

Now we shall show that this also satisfies the correct boundary condition as $s\rightarrow 0$. As $s\rightarrow 0$ every term in the summation of (\ref{latprop}) is proportional to a delta function, ${\delta}^{2}(z' + {\gamma}-z){\delta}(x'-x)$, which comes from the free propagator $G_{0}$. Now according to our assumption the pair $(z,z')$ in (\ref{latprop}) refer to two points in the same unit cell. So the pair of points $z$ and $z'+{\gamma}$ can never coincide unless ${\gamma}=0$. So in the limit $s\rightarrow 0$, the only term which survives is ,$W(z',\bar z',x';z,\bar z,x) G_{0}(z',\bar z',x',s;z,\bar z, x, 0) = W(z',\bar z',x';z,\bar z,x){\delta}^{2}(z' -z){\delta}(x'-x)={\delta}^{2}(z' -z){\delta}(x'-x)$. So our solution satisfies the correct boundary condition.  

\subsection{Trace of The Propagator at Coincident Points }
To compute the determinant of the operator we have to compute the trace of the propagator at coincident points. This is the quantity $trG(z,\bar z,x,s;z,\bar z,x,0)$, where $tr$ is over the internal gauge indices. In our case the free propagator is proportional to the identity matrix where the proportionality constant is just the heat kernel of a single free complex scalar field in the covering space $R^{2}\times R^{1}$. We denote this quantity by the same symbol $G_{0}$. So we can write,
\begin{equation}
TrG(z,\bar z,x,s;z,\bar z,x,0) = \sum_{{\gamma}\in{\Gamma}}G_{0}(z+{\gamma},\bar z+\bar{\gamma},x,s;z,\bar z, x, 0)  TrW(z+{\gamma},\bar z+\bar{\gamma},x;z,\bar z,x) 
\end{equation}
The trace over the Wilson line can be expressed in terms of the trace of the products of holonomies along the $a$ and $b$ cycle in the following way. First of all the trace of the Wilson loop is independent of the choice of the point $(z,\bar z,x)$. This can be proved in the following way. 

We shall prove this in the general case where the group $\mathcal{G}$ acting on the covering space is nonabelian. This will be the case if say the spacetime manifold has the geometry ${\Sigma}_{g}\times R^{1}$ where ${\Sigma}_{g}$ is a genus $g$ Riemann surface with $g\geq 2$. In this case the the flat gauge fields are genuinely non-abelian. So let us consider the Wilson line $W({\gamma}x,x)$ where ${\gamma}\in \mathcal{G}$. $x$ is an arbitrary point on the covering space and ${\gamma}x$ is its image under the action of ${\gamma}$. In the case of torus ${\gamma}x$ is the translation of $x$ by a lattice translation vectors.

The Wilson line $W({\gamma}x,x)$ goes from $x$ to ${\gamma}x$. Let us choose another pair of points $(x',{\gamma}x')$ and consider the Wilson line $W({\gamma}x',x')$. Now let us  consider the four paths $\overrightarrow{(x,{\gamma x})}$, $\overrightarrow{({\gamma x},{\gamma}x')}$, $\overrightarrow{({\gamma}x',x')}$ and $\overrightarrow{(x',x)}$. They form a closed path and the shape of each path is arbitrary as the connection is flat. The holonomy along this closed path on the cover is zero and so we can write,
\begin{equation}
W(x,x')W(x',{\gamma}x')W({\gamma}x',{\gamma}x)W({\gamma}x,x) =1
\end{equation}
Now due to the periodicity condition,  $A(x)=A({\gamma}x)$, satisfied by the gauge field on the cover we have $W({\gamma}x',{\gamma}x)=W(x',x) = W^{-1}(x,x')$. So the zero holonomy condition reduces to 
\begin{equation}
W(x,x')W(x',{\gamma}x')W^{-1}(x,x')W({\gamma}x,x) =1
\end{equation}
This gives us 
\begin{equation}
W({\gamma}x,x)=W(x,x')W({\gamma}x',x')W^{-1}(x,x')
\end{equation}
So $TrW({\gamma}x,x)=TrW({\gamma}x',x')$. This proves our claim.

We can see form the above equation that in the Abelian case the Wilson line itself is an invariant quantity. But in the non-abelian case the Wilson lines are related by a conjugation and so the $Tr$ is an invariant quantity.

Now we again consider the case of the torus. Since the trace is independent of $(z,\bar z,x)$ we can choose a convenient value for the coordinates. Let us choose $(z,\bar z,x)$ to be the center $(0,0,0)$ and draw the lattice such that the center coincides with one vertex of a unit cell. So the the Wilson line $W(z+{\gamma},\bar z+\bar{\gamma},x;z,\bar z,x)$ becomes $W({\gamma},\bar{\gamma},0;0,0,0) $. Since the coordinate along $R^{1}$ does not play any role and the Wilson line is path-independent we can choose a path which lies on the slice $x=0$. So for ${\gamma}= m\vec a+ n\vec b$ where $(m,n)\in\mathbb Z\oplus \mathbb Z$ and $(\vec a,\vec b)$ are a set of basis vectors for the lattice, we can write$ W({\gamma},\bar{\gamma},0;0,0,0) = W(m\vec a+n\vec b)= W(a)^{m}W(b)^{n}$. $W(a)$ and $W(b)$ are the values of the Wilson line along the two sides of a unit cell. Since the sides of a unit cell represented by $(\vec a,\vec b)$ get mapped to the $a$ and $b$ cycles of the torus, $W(a)$ and $W(b)$ are precisely the holonomy of the flat connection along the two cycles of the torus. In this case they are all Abelian and so there is no ordering ambiguity. So the determinant can be written as,
\begin{equation}
-\ln Det(-D^{2}) = \sum_{(m,n)\in \mathbb Z\oplus\mathbb Z} A(m\vec a+n\vec b)Tr[W(a)^{m}W(b)^{n}]
\end{equation}
where $A(m\vec a +n\vec b)$ is given by the equation
\begin{equation}
A(m\vec a+n\vec b) = \int_{\epsilon}^\infty \frac{ds}{s}\int_{T^{2}\times R^{1}} dxdzd\bar z\,G_{0}((z,\bar z)+m\vec a+ n\vec b,x,s;(z,\bar z),x,0)
\end{equation} 
We have introduced a UV-cutoff ${\epsilon}$ in the $s$ integral. The expression for $G_{0}$ is known. The effective action obtained by integrating out the scalar field is given by $\ln Det(-D^{2})$. Now the heat-kernel of the free laplacian on $R^{d}$ is given by :
\begin{equation}\label{heatKfreeL}
G_{0}(x,s;y,0)= \frac{1}{(4{\pi}s)^{\frac{d}{2}}} exp(-\frac{(x-y)^{2}}{4s})
\end{equation}
Using this we get :
\begin{equation}
A(m\vec a+n\vec b) = \frac{vol(T^{2}\times R^{1})}{\pi} \frac{1}{|m\vec a+n \vec b|^{3}}
\end{equation}
So the final answer for the effective action is :
\begin{equation}
-S_{eff} = \frac{vol(T^{2}\times R^{1})}{\pi}\sum_{(m,n)\neq (0,0)}\frac{1}{|m\vec a+n \vec b|^{3}}Tr[W(a)^{m}W(b)^{n}]
\end{equation}
When the scalar field has a mass, $M$, the relevant operator is, $-D^{2} + M^{2}$ and we have to calculate $Det(-D^{2} + M^{2})$. The eigenvalues of the new operator is related to new by, $\lambda_{new} = \lambda_{old} + M^{2}$. So form equation $(A5)$ we conclude that,
\begin{equation}
G_{(-D^{2} + M^{2})} = e^{-M^{2}s}G_{(-D^{2})}
\end{equation}
Using this relation one can show that the effective action in the massive case is given by,
\begin{equation}
-S_{eff} = \frac{1}{\sqrt 2} \frac{vol(T^{2}\times R^{1})}{\pi^{\frac{3}{2}}}\sum_{(p,q)\neq(0,0)} \frac{M^{\frac{3}{2}}K_{\frac{3}{2}}(M|p\vec a + q\vec b|)}{|p\vec a + q\vec b|^{\frac{3}{2}}} Tr(W(a)^{p}W(b)^{q})
\end{equation}
where $K_{\frac{3}{2}}(\alpha)$ is the modified Bessel function of the second kind of order $\frac{3}{2}$. The leading asymptotic behavior of the function for $\alpha>> 1$ is given by,
\begin{equation}
K_{\frac{3}{2}}(\alpha) \sim \sqrt{\frac{\pi}{2\alpha}} e^{-\alpha}
\end{equation}  
So in the limit where $MR>>1$, the effective action can be approximated by,
\begin{equation}
-S_{eff} = \frac{vol(T^{2}\times R^{1})}{2\pi}\sum_{(p,q)\neq(0,0)} \frac{M e^{-M|p\vec a + q\vec b|}}{|p\vec a + q\vec b|^{2}} Tr(W(a)^{p}W(b)^{q})\end{equation}
$R$ is the size of the torus.
\subsection{Modular Invariance}
The effective action has been expressed in terms of the holonomy of flat connection along the two cycles of the torus, which corresponds to a particular choice of a set of basis vectors for the lattice. But the choice of cycles or the two basis vectors of the lattice is not unique. Any two choices are related by a $SL(2,\mathbb Z)$ transformation. So any automorphism of the lattice is a symmetry of the effective action. 

The $R^{1}$ part of the geometry does not play any role in the discussion. So we shall denote by $(\vec a,\vec b)$ the basis vectors of the lattice on some particular slice at some arbitrary value of $x$. We can choose a different set of basis vectors denoted by $(\vec a',\vec b')$ which are related to the old basis by,
\begin{align}\label{trans}
\vec a' = p\vec a + q\vec b \nonumber\\
 \vec b' = r\vec a + s\vec b
\end{align}
where $(p,q,r,s)$ are integers and satisfy, $ps-qr = 1$. So this is a $SL(2,\mathbb Z)$ transformation. 
Now the effective action can be written as 
\begin{align}\label{seff}
-S_{eff} &= \sum_{(m,n)\in \mathbb Z\oplus\mathbb Z} A(m\vec a + n\vec b)Tr[W(a)^{m}W(b)^{n}] \nonumber\\
&= \sum_{(m,n)\in \mathbb Z\oplus\mathbb Z} A(m\vec a' + n\vec b')Tr[W(a')^{m}W(b')^{n}] 
\end{align}
where $(\vec a',\vec b')$ are related to $(\vec a,\vec b)$ by  transformation (\ref{trans}). The new holonomies are related to the old ones by:
\begin{align}
W(a') = W(a)^{p}W(b)^{q}\nonumber\\
W(b')= W(a)^{r}W(b)^{s}
\end{align}
It is easy to check the equality in ({\ref{seff}) using these identities. So the effective action is modular invariant. This $SL(2,\mathbb Z)$ should give rise to Ward identities.

\subsection{ \texorpdfstring{$T^{2}\times S^{1}$}{T2S1}}
In this section we shall write down the answer when the field theory lives on $T^{2}\times S^{1}$. In this the cover is again $R^{2}\times R^{1}$. The group $\mathcal{G}$ is now $\mathbb Z\oplus \mathbb Z\oplus \mathbb Z$. The lattice is three dimensional with basis vectors denoted by $(\vec a,\vec b,\vec c)$. The metric is the flat metric. Since the homotopy group is $\mathbb Z\oplus \mathbb Z\oplus \mathbb Z$, the holonomies are commuting. The determinant is given by:
\begin{equation}
-lnDet(-D^{2}) = \sum_{(m,n,p)\in \mathbb Z\oplus\mathbb Z\oplus \mathbb Z} A(m\vec a+n\vec b+p\vec c)Tr[W(a)^{m}W(b)^{n}W(c)^{p}]
\end{equation}
where $A(m\vec a+n\vec b+p\vec c)$ is given by:
\begin{equation}
A(m\vec a+n\vec b+p\vec c) = \int_{\epsilon}^\infty \frac{ds}{s}\int_{T^{2}\times S^{1}} d^{3}x \,G_{0}(\vec x +m\vec a+n\vec b+p\vec c,s;\vec x,0)
\end{equation} 
where $G_{0}$ is the heat propagator that appears in the case of $T^{2}\times R^{1}$. In this case the group of automorphisms of the lattice is
 $SL(3,\mathbb Z)$ and so the effective action has this symmetry. This should also give rise to Ward identities.
 
 The final answer for the effective action in this case can be written as :
\begin{equation}
-S_{eff} = \frac{vol(T^{2}\times S^{1})}{\pi}\sum_{(m,n,p)\in\mathbb Z\oplus\mathbb Z\oplus \mathbb Z}\frac{1}{|m\vec a+n \vec b+p \vec c|^{3}}Tr[W(a)^{m}W(b)^{n}W(c)^{p}]
\end{equation}
\subsection{General Case}
In the general case the answer is:
\begin{equation}\label{logdet}
-\ln Det(-D^{2}) = \sum_{{\gamma}\in \mathcal{G}} A({\gamma})TrW({\gamma}\vec 0,\vec 0)
\end{equation}
where $A({\gamma})$ is given by,
\begin{equation}
A({\gamma}) = \int_{\epsilon}^\infty \frac{ds}{s}\int_{{\Gamma}} G_{0}({\gamma}x,s;x,0)
\end{equation}
In the general case the holonomies are non-Abelian. We have already proved that the $TrW({\gamma}x,x)$ is independent of the choice of $x$ even in the non-Abelian case. So in writing down the formula we have chosen an arbitrary value for $x$ which we have denoted by $\vec 0$. ${\gamma}\vec 0$ is the image of this point under the action of ${\gamma}\in \mathcal{G}$. In general the lattice has to be drawn on a space other than $R^{n}$. For example if we are working on the space-time geometry ${\Sigma}_{g}\times R^{1}$ where ${\Sigma}_{g}$ is a genus $g$ Riemann surface, then the lattice has to be drawn on the disc$\times R^{1}$ with hyperbolic metric on the disk. The metric on the space-time can be taken to be:
\begin{equation}
ds^{2} = \frac{4|dz|^{2}}{(1-|z|^{2})^{2}} + dx^{2}
\end{equation}
where $z$ is the coordinate on the unit disc. The free propagator has to be evaluated on this space. This answer can be found in literature. The domain of integration ${\Gamma}$ is the fundamental region for the action of group $\mathcal{G}$ on the covering space. This can be identified with a unit cell of the lattice    times whatever simply-connected non-compact direction the geometry has. 

We can choose the the point $\vec 0$ to be one of the vertices of the lattice in the same way as we did in the case of torus. The trace of the Wilson line on the covering space can again be expressed in the same way except that now the ordering has to be maintained. 

\subsection{A Short Proof Of (\ref{covingspaceprop})}
Heat equation on the covering space has the form:
\begin{equation}
-\frac{\partial}{\partial s} G(x',s;x,0) = -D_{x'}^{2} G(x',s;x,0)
\end{equation}
where $D_{{\mu}} = \partial_{\mu} + iA_{{\mu}}$ . Now one can write:
\begin{equation}
D_{{\mu}} = W(x)\partial_{{\mu}} W^{-1}(x)
\end{equation} 
where $W(x)=Pexp(-i\int_{x_{0}}^{x} A)$. $x_{0}$ is an arbitrary initial point and we have not specified any path for the integration because the connection is flat. On the simply connected covering space $W(x)$ is a well-defined function of $x$. Using (\ref{logdet}) it is easy to see that $W^{-1}(x')G(x',s;x,0)$ is the heat kernel of the free Laplacian $-{\partial}^{2}$. Therefore
\begin{equation}
G(x',s;x,0) = W(x')G_{0}(x',s;x,0)= Pexp(-i\int_{x_{0}}^{x'} A)K_{0}(x',s;x,0)
\end{equation}
Now the boundary condition as $s\rightarrow 0$ requires us to choose $x_{0} =x$. This concludes our proof.

\newpage

\bibliographystyle{JHEP}
\bibliography{bibliography}
\end{document}